\documentclass[epj,nopacs]{svjour}
\usepackage[T1]{fontenc}
\usepackage[italian,english]{babel}
\usepackage{hyperref}
\usepackage{ifpdf}
\usepackage{subfigure}
\usepackage{amssymb}
\usepackage{amsfonts}
\usepackage{epsf}
\usepackage{rotating}
\usepackage{graphicx}
\usepackage{amsmath}
\usepackage{fancyhdr}
\usepackage{lineno}
\usepackage{babel}
\usepackage{graphics}
\usepackage{pstricks}
\usepackage{color}
\usepackage{multirow}
\usepackage{nicefrac}
\usepackage{bm}
\usepackage{slashed}
\usepackage{float}
\usepackage{array}
\usepackage{boldline}
\usepackage{mathrsfs}
\usepackage{xcolor}

\usepackage{hyperref}
\hypersetup{colorlinks=true,linkcolor=blue,citecolor=red,urlcolor=lightseagreen}

\definecolor{lightseagreen}{rgb}{0.13, 0.7, 0.67}

\usepackage[numbers,sort&compress]{natbib}
\usepackage{tikz}
\usepackage{tikz-feynman}
\tikzfeynmanset{compat=1.0.0}

\newcommand{\nn}{\nonumber}
\newcommand{\lsim}{\mathrel{\mathop{\kern 0pt \rlap
			{\raise.2ex\hbox{$<$}}}
		\lower.9ex\hbox{\kern-.190em $\sim$}}}
\newcommand{\gsim}{\mathrel{\mathop{\kern 0pt \rlap
			{\raise.2ex\hbox{$>$}}}
		\lower.9ex\hbox{\kern-.190em $\sim$}}}

\newcommand{\be}{\begin{equation}}
\newcommand{\ee}{\end{equation}}
\newcommand{\bea}{\begin{eqnarray}}
\newcommand{\eea}{\end{eqnarray}}

\newcolumntype{?}{!{\vrule width 3pt}}

\newcommand*\xbar[1]{%
	\hbox{%
		\vbox{%
			\hrule height 0.65pt 
			\kern0.4ex
			\hbox{%
				\kern-0.05em
				\ensuremath{#1}%
				\kern0.0em
			}%
		}%
	}%
}

\DeclareMathSymbol{\widetildesym}{\mathord}{largesymbols}{"65}

\DeclareFontFamily{OT1}{pzc}{}
\DeclareFontShape{OT1}{pzc}{m}{it}{<-> s * [1.2] pzcmi7t}{}
\DeclareMathAlphabet{\mathpzc}{OT1}{pzc}{m}{it}

\newcommand{\y}{\mathpzc{Y}}

\newcommand{\Q}{\bm{Q}}
\newcommand{\LL}{\bm{L}}
\newcommand{\cl}{\mathpzc{l}}
\newcommand{\uq}{\mathpzc{u}}
\newcommand{\dq}{\mathpzc{d}}
\newcommand{\e}{\mathpzc{e}}
\newcommand{\T}{\mathpzc{T}}
\newcommand{\cQ}{\mathcal{Q}}
\newcommand{\X}{\mathpzc{X}}
\newcommand{\Tr}{\text{Tr}}
\newcommand{\I}{\text{I}_3}
\newcommand{\wX}{\widetilde{\X}}
\newcommand{\wl}{\widetilde{\lambda}}
\newcommand{\is}{\mbox{\Large{$i$}}}

\allowdisplaybreaks

\begin{document}
	\title{Constraining Scalar Doublet and Triplet Leptoquarks with Vacuum Stability and Perturbativity}
	\author{Priyotosh Bandyopadhyay$^{1,}$\thanks{\email{\href{bpriyo@phy.iith.ac.in}{bpriyo@phy.iith.ac.in}}} \and Shilpa Jangid$^{1,}$\thanks{\email{\href{ph19resch02006@iith.ac.in}{ph19resch02006@iith.ac.in}}} \and Anirban Karan$^{1,2,}$\thanks{\email{\href{kanirban@ific.uv.es}{kanirban@iith.ac.in}}}}                
	\institute{$^{1}$ Indian Institute of Technology Hyderabad, Kandi, Sangareddy-502284, Telangana, India \\
	$^{2}$ Instituto de F\'{i}sica Corpuscular (CSIC - Universitat de Val\`{e}ncia),
	Apt. Correus 22085, E-46071 Val\`{e}ncia, Spain}
   \date{}
	\authorrunning{{}}
	\titlerunning{{}}
	
	\abstract{We investigate the constraints on the leptoquark Yukawa couplings and the Higgs-leptoquark quartic couplings for scalar doublet leptoquark $\widetilde{R}_2$, scalar triplet leptoquark $\vec S_3$ and their combination with both three generations and one generation from perturbative unitarity and vacuum stability. Perturbative unitarity of all  the dimensionless couplings have been studied via one- and two-loop beta-functions. Introduction of new $SU(2)_L$ multiplets in terms of these leptoquarks fabricate Landau  poles at two-loop level in the gauge coupling $g_2$ at $10^{19.7}$ GeV and $10^{14.4}$ GeV, respectively for $\vec S_3$ and $\widetilde{R}_2+\vec S_3$ models with three generations. However, such Landau pole cease to exits for $\widetilde{R}_2$ and any of these extensions with both one and two generations till Planck scale. The Higgs-leptoquark quartic couplings acquire sever constraints to protect Planck scale perturbativity, whereas leptoquark Yukawa couplings gets some upper bound in order to respect Planck scale stability of Higgs Vacuum. The Higgs quartic coupling at two-loop constraints the leptoquark Yukawa couplings for $\widetilde{R}_2,\vec S_3, \,\widetilde{R}_2+\vec S_3$ with values  $\lsim  1.30, 3.90, 1.00$ with three generations. In the effective potential approach, the presence of any of these leptoquarks with any number of generations pushes the metastable vacuum of the Standard Model to the stable region.}
	\maketitle

\flushbottom
\setcounter{tocdepth}{1}
\setcounter{tocdepth}{2}
\setcounter{tocdepth}{3}
\tableofcontents

\section{Introduction}
\label{sec:intro}

During last few decades the Standard Model (SM) has been extremely successful in establishing itself as a well-accepted model providing beautiful theoretical description of elementary particles. After discovery of the Higgs boson~\cite{Aad:2012tfa, Chatrchyan:2012xdj}, the last undetected particle of the SM, followed by precise measurement of its properties at LHC the particle spectrum of the SM became complete. However, due to incapability of explaining various experimental facts like matter-antimatter asymmetry, dark matter relic density, masses of neutrinos, Higgs mass hierarchy, several flavour anomalies, etc., SM is considered as an incomplete theory. This motivates one to extend the SM with some beyond Standard Model (BSM) particles or new gauge groups or additional discrete symmetries. Various New Physics (NP) models augmented with heavy fermions and bosons have been very well-studied in the literature. Leptoquarks \cite{Dorsner:2016wpm} lie under the category of bosonic extension of the SM, but with lepton and baryon number. 

Though the notion of leptoquark \cite{Pati:1973uk,Pati:1974yy} is there in literature for nearly fifty years, it has got much attention in recent times due to its prospect of addressing various flavour anomalies \cite{Marzocca:2018wcf,Gherardi:2020qhc,Crivellin:2017zlb,Crivellin:2019dwb,Aydemir:2019ynb,Becirevic:2018afm,Bigaran:2019bqv,Mandal:2018kau,Iguro:2020keo,Lee:2021jdr,Bordone:2020lnb,Borschensky:2021hbo,Browder:2021hbl,Sheng:2021iss,Cornella:2021sby,Crivellin:2021lix,Angelescu:2018tyl,Angelescu:2021lln,Arnan:2019olv,ColuccioLeskow:2016dox,Crivellin:2020mjs,Saad:2020ihm,Saad:2020ucl,Babu:2020hun,Chang:2021axw,Zhang:2021dgl}, unexplained with SM. Simply speaking, leptoquarks are some hypothetical particles having both lepton number and baryon number. They are electromagnetically charged and colour triplet (fundamental or anti-fundamental) under $SU(3)_C$ gauge group. Under $SU(2)_L$ gauge group, they could be singlet, doublet and triplet as well. According to Lorentz representation, they might be scalar as well as vector. These leptoquarks emerge naturally in several higher gauge theories unifying matters \cite{Pati:1973uk,Pati:1974yy,Georgi:1974my,Georgi:1974sy,Dimopoulos:1979es,Farhi:1980xs,Schrempp:1984nj,Wudka:1985ef,Nilles:1983ge,Haber:1984rc,Assad:2017iib,Perez:2021ddi,Murgui:2021bdy}. In literature, numerous efforts have been devoted to studying the phenomenology of these leptoquarks at colliders \cite{Bandyopadhyay:2018syt,Bhaskar:2020gkk,Bhaskar:2021pml,Bhaskar:2021gsy,DaRold:2021pgn,Hiller:2021pul,Haisch:2020xjd,Chandak:2019iwj,Bhaskar:2020kdr,Alves:2018krf,Dorsner:2019vgp,Mandal:2018qpg,Padhan:2019dcp,Baker:2019sli,Nadeau:1993zv,Atag:1994hk,Atag:1994np,Buchmuller:1986zs,Hewett:1987yg,Hewett:1987bh,Cuypers:1995ax,Blumlein:1996qp,Belyaev,Kramer:1997hh,Plehn:1997az,Eboli:1997fb,Kramer:2004df,Hammett:2015sea,Mandal:2015vfa,Asadi:2021gah,Bandyopadhyay:2021pld}, especially at the LHC. Mainly focusing on the angular distributions, distinguishing features of scalar and vector leptoquarks carrying different SM gauge quantum numbers have also been explored at electron-proton \cite{Bandyopadhyay:2020jez}, electron-photon \cite{Bandyopadhyay:2020klr} and proton-proton \cite{Bandyopadhyay:2020wfv,Dutta:2021wid} colliders. On the other hand, lots of experimental searches for these leptoquarks have been performed at electron-positron \cite{Behrend:1986jz,Bartel:1987de,Kim:1989qz,Abreu:1998fw}, electron-proton \cite{Collaboration:2011qaa,Abramowicz:2019uti}, proton-antiproton \cite{Alitti:1991dn,Abazov:2011qj,Aaltonen:2007rb} and proton-proton \cite{Aad:2020iuy,CMS:2020gru,Aad:2021rrh,CMS:2018qqq,CMS:2020wzx} colliders, but no sign of them has yet been confirmed. Kaon and lepton Physics have implemented strong constraints on the coupling of leptoquarks to first generation of quarks and leptons \cite{Mandal:2019gff,Davidson:1993qk,Dorsner:2016wpm}. ATLAS and CMS have performed generation-wise thorough analyses on the allowed mass range of scalar and vector leptoquarks. These studies \cite{Aad:2020iuy,CMS:2020wzx,CMS:2018qqq} suggest that if there exist any leptoquark it must have mass above 1.5 TeV with the coupling to quarks and leptons below the electromagnetic coupling constant 
\footnote{Though bounds on third generation scalar leptoquark are a bit relaxed \cite{CMS:2020gru,Aad:2021rrh} and manipulating the branching fractions of the leptoquark to different generations of quarks and leptons, one can lower the bound of 1.5 TeV mass}.

Now, the 125.5 GeV mass of the observed Higgs boson indicates that its vacuum cannot be completely stable all the way up to Planck scale or even GUT scale  \cite{Isidori:2001bm}. In order for the Higgs potential to be bounded from below, the self-quartic-coupling ($\lambda_h$) of the Higgs boson must be positive. However, it is found that the negative quantum correction from top quark pushes $\lambda_h$ to negative values after the energy scale of $10^{10}$ GeV and thus the stability of SM gets hampered. Technically speaking, it is generally considered that the SM is in a metastable state. In these circumstances, the presence of some BSM scalar extensions i.e simplest extension via singlet  \cite{singletex, Gonderinger:2009jp, Costa:2014qga,Haba:2015rha,Guo:2010hq,Barger:2008jx,Khan:2014kba,Baek:2012uj, HiggsDM1,BLscalar}, $SU(2)_L$ doublet \cite{2HDMs,Chakrabarty:2015yia,Swiezewska:2015paa,Gopalakrishna:2015wwa, Honorez:2010re,HiggsDM, khan1, 2HDMpheno} or triplet representation of $SU(2)_L$ \cite{Tripletex,Khan:2016sxm} are required to restore the stability of vacuum by neutralizing the destabilizing effect of top quark. On the other hand, the inclusion of additional fermionic particles worsen the case by further lowering the energy scale until which $\lambda_h$ remains positive. To avoid the stability issue, these models are also often extended with  additional scalar particles \cite{exwfermion,Coriano:2015sea,DelleRose:2015bms,Jangid:2020dqh,Garg:2017iva}. However, it is important to note that fermions with $SU(2)_L$ gauge charge, pushes  for non-perturbativity, thus gives  constraints on  the number of generation for the Planck scale perturbativity.  \cite{Bandyopadhyay:2020djh}. This motivates us to investigate the stability of vacuum in presence of scalar leptoquarks which is not very well explored so far.

Furthermore, it is expected that every dimensionless parameter of a fundamental model should be bounded above in order to assure the perturbative expansion of the correlation functions. Now, the presence of leptoquark will tamper the perturbativity of the theory by imposing extra contributions on the renormalization group (RG) evolution of different SM coupling. Therefore, it is of paramount importance to scrutinize the perturbativity of a model while studying the stability of its vacuum.

Along with perturbativity, the effects of scalar singlet leptoquark $S_1$ in addressing the issue of vacuum stability has already been discussed in Ref. \cite{Bandyopadhyay:2016oif}.  In this paper, we study the stability and perturbativity of the models with scalar triplet leptoquark $\vec S_3$ and scalar doublet leptoquark $\widetilde R_2$. Since leptoquarks possess colour charge as well as the hypercharge, their presence affects the RG evolution of all the couplings in quite different way than usual scalars. Moreover, doublet and triplet leptoquarks originate more positive effects, required for stability, than the singlet one as they contain two and three different components respectively. On the similar ground such models are often more constrained by perturbativity. In addition, we study the BSM scenario having both the leptoquarks  $\widetilde R_2$ and $\vec S_3$ simultaneously. This model gained a lot more interest due to its prospect of generating Majorana mass term for neutrinos at one- and two-loop along with some other beautiful features \cite{Dorsner:2016wpm,AristizabalSierra:2007nf,Dorsner:2017wwn,Babu:2019mfe,Pas:2015hca,Chua:1999si,Mahanta:1999xd,Babu:2010vp}.

The paper is organized in the following way. In the very next section (Sec. \ref{sec:model}), a brief illustration of all the leptoquark models, considered for this paper, is presented. Section \ref{sec:pert} deals with perturbativity of these models in terms of different gauge couplings, top and leptoquark Yukawa couplings and Higgs-leptoquark quartic couplings. In the subsequent section (Sec. \ref{sec:stab}), we scrutinize the stability of Higgs vacuum for all of these leptoquark models by studying the evolution of $\lambda_{h}$ with the energy scale. Furthermore, we investigated the stability issue following the Coleman-Weinberg effective potential approach. In Sec. \ref{sec:ph} we describe the phenomenology of leptoquarks in light of direct and indirect bounds on their parameter space. Finally, we conclude in section \ref{sec:concl}.

\section{Leptoquark models}
\label{sec:model}

This section illustrates the theoretical description of the leptoquarks $\widetilde R_2$ and $\vec S_3$. At first, we consider the model with scalar doublet leptoquark $\widetilde R_2\,(\bm{3,2,}1/6)$, where the numbers in bracket denote the $SU(3)_c\bigotimes SU(2)_L\bigotimes U(1)_Y$ nature of it. Since, this leptoquark is a doublet under $SU(2)_L$, it has two components with the electromagnetic charges ${2}/{3}$ and ${-1}/{3}$, and we designate them as $\widetilde R_2^{2/3}$ and $\widetilde R_2^{-1/3}$. The corresponding Lagrangian is given by:
\begin{equation}
\label{eq:lag2}
\begin{split}
\mathscr L_2\supset(D^\mu\widetilde R_2)^\dagger (D_\mu\widetilde R_2)-(m_2^2+\lambda_2 H^\dagger H)(\widetilde R_2^\dagger \widetilde R_2)\\-\widetilde\lambda_2 H^\dagger\widetilde R_2\,\widetilde R_2^\dagger H-[Y_2 \,\xbar \dq_R \,(\widetilde R_2^T i\sigma_2)\,\LL_L+h.c.]~, 
\end{split}
\end{equation}
where, $D^\mu$ signifies the covariant derivative related to the kinetic term of fields, $m_2$ is the mass of the leptoquark $\widetilde R_2$ before electroweak symmetry breaking (EWSB), $\lambda_2$ and $\widetilde{\lambda}_2$ are the couplings for quartic interaction terms of $\widetilde R_2$ with scalar doublet field $H$, the $3\times 3$ matrix $Y_2$ indicates the coupling of $\widetilde R_2$ with quarks and leptons. After EWSB, the scalar field $H$ gives rise to Higgs boson $h$ and the two components of $\widetilde R_2$ get additional contributions in their masses from the quartic coupling terms. It is important to mention that the generation indices have been suppressed here. However, to get the full mathematical description of this model, one has to add the SM Lagrangian as well. In our notation, we denote the SM Yukawa couplings for the charged leptons, up-type quarks and down-type quarks as $Y_\cl$, $Y_\uq$ and $Y_\dq$ respectively. The SM Higgs potential is given by:
\begin{equation}
\label{eq:higgs_pot}
V_0=-\,\mu_h^{} |H|^2+\lambda_h |H|^4 \quad \text{with} \quad H=\frac{1}{\sqrt 2}\begin{pmatrix}
0\\
v+h
\end{pmatrix}~,
\end{equation}
under unitary gauge, where the tree-level mass of Higgs boson becomes: $M_h^{}=\sqrt{2\,\mu_h^{}}$ and the vacuum expectation value (VEV) of the scalar field can be expressed as: $v=\sqrt{\nicefrac{\mu_h^{}}{\lambda_h}}$~. After EWSB the squared masses for the leptoquarks $\widetilde R_2^{2/3}$ and $\widetilde R_2^{1/3}$ respectively become:
\begin{align}
\label{eq:mass2}
m^2_{2,2/3} \ &=\ m_2^2+\frac{1}{2}\lambda_2 v^2,\qquad\qquad \nn \\ 
m^2_{2,1/3}\ &=\ m_2^2+\frac{1}{2}(\lambda_2+ \widetilde\lambda_2) v^2,
\end{align}
and thus the two components of the doublet no longer remain degenerate and acquire a mass gap of $\frac{1}{2}\wl_2 v^2$.

In principle, there could be some other gauge invariant dimension four terms for $\widetilde R_2$, like $\epsilon_{\alpha\beta\gamma} (H^T i\sigma_2 \widetilde{R}_{2,\alpha})(\widetilde{R}_{2,\beta}^T i\sigma_2\\ \widetilde{R}_{2,\gamma})$ or $(\widetilde{R}_2^\dagger \widetilde{R}_2)(\widetilde{R}_2^\dagger \widetilde{R}_2)$.  The first term does not conserve baryon and lepton number separately; additionally it initiates proton decay via the mode $p\to\pi^+\pi^+e^-\nu\nu$ \cite{Arnold:2012sd,Kovalenko:2002eh,Dorsner:2016wpm} which in turn forces the leptoquark mass to be very high to reach the experimental value of proton lifetime. So, one should either neglect the term or assume that it is forbidden by some other symmetry. For example, if we impose a $Z_2^{l}$ discrete symmetry under which all the SM leptons as well as the leptoquarks are odd, but other particles like quarks and the scalar doublet H are even, this particular term will be prohibited. The same effect can be achieved by imposing $Z_2^q$ discrete symmetry too for which the quarks and leptoquarks are odd and all the other particles are even. On the other hand, the second term does not affect any other SM couplings up to two-loop level. So, for simplicity, we ignore it too.

In the second scenario, we study the extension of SM with scalar triplet leptoquark $\vec{S}_3\,(\bm{\xbar{3},3},1/3)$. The three excitations of this multiplet posses the electromagnetic charges ${4}/{3}$, ${1}/{3}$ and ${-2}/{3}$, and therefore we name them as $S_3^{4/3}$, $S_3^{1/3}$ and $S_3^{-2/3}$ respectively. The Lagrangian for this leptoquark is given by:
\begin{align}
\label{eq:lag3}
\mathscr{L}_3\ & \supset \ \Tr[(D^\mu S_3^{ad})^\dagger(D_\mu S_3^{ad})]-\widetilde\lambda_3\, H^\dagger S_3^{ad}\,(S_3^{ad})^\dagger H \nn \\ 
&-\,(m_3^2 + \lambda_3\, H^\dagger H)\, \Tr[(S_3^{ad})^\dagger\, S_3^{ad}]\nn\\ 
& +[ Y_3\; \xbar \Q_L^{\,c}\,(i\sigma_2\,S_3^{ad})\,\LL_L^{} +\,h.c.]~,
\end{align}
where $S_3^{ad}$ signifies $\vec S_3$ in adjoint representation, $m_3$ is the mass of $\vec S_3$ before EWSB, $\lambda_3$ and $\widetilde{\lambda}_3$ are the couplings for quartic interaction terms of this leptoquark with Higgs boson and $Y_3$ indicates its coupling with different quarks and leptons. It is interesting to notice that the term $H^\dagger (S_3^{ad})^\dagger S_3^{ad} H$ is absent in the Lagrangian, given by Eq. \eqref{eq:lag3}, since it is not an independent term. It can be easily checked that: $H^\dagger \,[S_3^{ad}\,(S_3^{ad})^\dagger+(S_3^{ad})^\dagger S_3^{ad}]\,H=(H^\dagger H)\, \Tr[(S_3^{ad})^\dagger S_3^{ad}]$ under unitary gauge. After EWSB the squared masses for leptoquarks $S_3^{4/3}$, $S_3^{1/3}$ and $S_3^{2/3}$ become:	
\begin{align}
\label{eq:mass3}
m^2_{3,4/3}\ & = \ m_3^2+\frac{1}{2}\lambda_3 v^2,\qquad \nn \\ 
m^2_{3,2/3}\ &= \ m_3^2+\frac{1}{2}(\lambda_3+ \widetilde\lambda_3) v^2,\qquad \nn \\ 
m^2_{3,1/3} \ & =\ m_3^2+\frac{1}{2}(\lambda_3+\frac{1}{2} \widetilde\lambda_3) v^2,
\end{align}
which lift the degeneracy among these three states like the previous scenario. In this case, apart from the leptoquark self-quartic interactions, i.e. $\Tr\big[(S_3^{ad})^\dagger S_3^{ad}(S_3^{ad})^\dagger S_3^{ad}\big]$ and $\Tr\big[(S_3^{ad})^\dagger S_3^{ad}\big]^2$, which we neglect for simplicity like in doublet leptoquark scenario, there could exist diquark term like $\xbar \Q_L^{\,c}\,(i\sigma_2)\,(S_3^{ad})^\dagger\,\Q_L^{}$ allowed by gauge symmetry. However, this term neither respects baryon and lepton number separately nor protects proton from decaying through $p\to e^+\pi^0$ or  $p\to\pi^+\bar\nu_e$ \cite{Barr:1989fi,Nath:2006ut,Dorsner:2005fq,Dorsner:2016wpm}. In the same fashion like $\widetilde{R}_2$ case, here also one can impose $Z_2^l$ or $Z_2^q$ symmetry to forbid this term. For our analysis, we neglect it too.

%

Lastly, we consider the scenario having both the leptoquarks $\widetilde R_2$ and $\vec S_3$. The relevant part of the Lagrangian for this model is given by: 
\begin{equation}
\mathscr L_{23}= \mathscr L_{2}+\mathscr L_{3} - [\kappa_h^{} H^\dagger S_3^{ad} \widetilde R_2+h.c.]~.
\end{equation}
The interesting feature of this model is that besides the individual interaction terms for doublet and triplet leptoquarks it contains one additional dimension three term which couples the doublet leptoquark to the triplet one through Higgs boson. As earlier cases, we have not considered the leptoquark self-quartic couplings.

In this scenario, it is important to notice that though $S_3^{4/3}$ remains as mass eigenstate, the other components of $\widetilde{R}_2$ and $\vec S_3$ do not. For instance, the squared mass matrix for $\widetilde{R}_2^{1/3}$ and $S_3^{1/3}$ becomes:
\begin{equation}
\label{eq:mass13}
M^2_{1/3}=\begin{pmatrix}
m^2_{2,1/3} & \frac{1}{2}\kappa_h^{} v\\
\frac{1}{2}\kappa_h^{*} v & m^2_{3,1/3}
\end{pmatrix},
\end{equation}
where, $\kappa_h^{*}$ indicates the complex conjugate of $\kappa_h^{}$. Therefore, these two flavour states mix together to produce the energy eigenstates as:
\begin{eqnarray}
\begin{pmatrix}
\Omega_1\\
\Omega_2
\end{pmatrix}=\begin{pmatrix}
\cos\theta_1^{}&\sin\theta_1^{}\, e^{i\phi_1}\\
-\sin\theta_1^{} \, e^{-i\phi_1}& \cos\theta_1^{}
\end{pmatrix} \begin{pmatrix}
\widetilde{R}_2^{1/3}\\
S_3^{1/3}
\end{pmatrix}, 
\end{eqnarray}
where, the mixing angle $\theta_1^{}$ and the CP violating phase $\phi_1$ are  given by:
\begin{equation}
\tan2\theta_1^{}=-\,\bigg(\frac{v\;|\kappa_h|}{m_{3,1/3}^2-m_{2,1/3}^2}\bigg) \text { and } e^{i\phi_1}=\frac{\kappa_h^*}{|\kappa_h|}~.
\end{equation}
The squared masses for the energy eigenstates $\Omega_{1,2}$ are given by:
\begin{align}
m^2(\Omega_{1,2})\ & = \frac{1}{2}\Big[\big(m_{3,1/3}^2+m_{2,1/3}^2\big) \nn \\ 
& \mp\sqrt{\big(m_{3,1/3}^2-m_{2,1/3}^2\big)^2+v^2\,|\kappa_h|^2}\;\Big]~.
\end{align}

Similarly, the squared mass matrix for $\widetilde{R}_2^{2/3}$ and $S_3^{2/3}$ becomes:
\begin{equation}
\label{eq:mass23}
M^2_{2/3}=\begin{pmatrix}
m^2_{2,1/3} & -\frac{1}{\sqrt 2}\kappa_h^{} v\\
-\frac{1}{\sqrt 2}\kappa_h^{*} v & m^2_{3,1/3}
\end{pmatrix},
\end{equation}
and these two flavour states also mix together to produce the energy eigenstates as:
\begin{eqnarray}
\begin{pmatrix}
\chi_1\\
\chi_2
\end{pmatrix}=\begin{pmatrix}
\cos\theta_2^{}&\sin\theta_2^{}\, e^{i\phi_2}\\
-\sin\theta_2^{}\, e^{-i\phi_2}& \cos\theta_2^{}
\end{pmatrix} \begin{pmatrix}
\widetilde{R}_2^{2/3}\\
S_3^{2/3}
\end{pmatrix}, 
\end{eqnarray}
where, the mixing angle $\theta_2^{}$ and the CP violating phase $\phi_2$ are  given by:
\begin{equation}
\tan2\theta_2^{}=\bigg(\frac{\sqrt 2 \,v\;|\kappa_h|}{m_{3,2/3}^2-m_{2,2/3}^2}\bigg) \text { and } e^{i\phi_2}=\frac{\kappa_h^*}{|\kappa_h|}~.
\end{equation}
The squared masses for the energy eigenstates $\Omega_{1,2}$ are given by:
\begin{align}
m^2(\chi_{1,2})\ & =\ \frac{1}{2}\Big[\big(m_{3,2/3}^2+m_{2,2/3}^2\big)  \nn\\
& \mp\sqrt{\big(m_{3,2/3}^2-m_{2,2/3}^2\big)^2+2\,v^2\,|\kappa_h|^2}\;\Big]~.
\end{align}
As a special case if $\kappa_h$ becomes zero, i.e. no mixing among doublet and triplet, then the mass and flavour states remain the same, i.e. the mixing angle becomes zero. On the other hand, if masses of doublet and triplet flavour eigenstates become same, the mixing angles turn to $\pi/4$ and mass deferences of $v|\kappa_h|$ and $\sqrt 2 v|\kappa_h|$ are generated among the mass-eigenstates with charge $1/3$ and $2/3$ respectively.

Now, regarding the generation of leptoquarks, we follow two different conventions: a) there is one generation of leptoquark that couples to one generation  of quark and lepton only, b) there exist three generations of leptoquarks, each one of which couples to one generation of quark and lepton only. Both the conventions have different pros and cons while considering several low energy and collider bounds on leptoquarks. However, for our analysis we study both of them. For the first scenario, we consider only diagonal coupling of the leptoquarks given by: $Y_\gamma^{rs}=Y_\phi\,\text{diag}(1,0,0)$ with $r,s$ being the generation indices for quarks and leptons and $\gamma\in \{2,3\}$. Obviously, one can choose $\text{diag}(0,1,0)$ or $\text{diag}(0,0,1)$ as well. In the second case, we assume $Y_{\gamma,i}^{rs}=Y_\phi\,\delta^{ir}\delta^{is}$ with $i$ being the generation of leptoquark. In this scenario, the terms $\lambda_\gamma$ and $\widetilde{\lambda}_\gamma$ also become $3\times3$ matrices, but we consider them diagonal too restricting the generation mixing of the leptoquarks.

\section{Perturbativity:}
\label{sec:pert}

In this section we study the perturbativity of the theory with respect to different dimensionless couplings. It is well known that  expansion of amplitude or cross-section in perturbative series is plausible only when the expansion parameter is less than unity. Therefore, the constraints that must be satisfied by different couplings in order to respect the perturbativity of the theory are the following \cite{Bandyopadhyay:2016oif,Jangid:2020dqh,Bandyopadhyay:2020djh}:
\begin{equation}
\label{eq:pbnd}
|\lambda_\alpha|\leq 4\pi, \quad |\widetilde\lambda_\gamma|\leq 4\pi, \quad |g_k|\leq 4\pi \quad |Y_l^{rs}|\leq \sqrt{4\pi},
\end{equation}   
where, $\lambda_\alpha$ and $\widetilde{\lambda}_\gamma$ with $\alpha\in \{h,2,3\}$ and $\gamma\in \{2,3\}$ indicate the quartic couplings of the Higgs boson with leptoquarks as well as the self-quartic coupling of the Higgs boson, $g_k$ with $k\in\{1,2,3\}$ signify the gauge couplings corresponding to $U(1)_Y$, $SU(2)_L$ and $SU(3)_C$ gauge symmetry respectively and $Y_l^{rs}$ with $l\in\{2,3,\cl,\uq,\dq\}$ represent the $(r,s)$ element of the Yukawa (or Yukawa like) coupling matrices for quarks and leptons. We generate two-loop beta functions for different couplings through \texttt{SARAH}~\cite{Staub:2013tta,Staub:2015kfa} in $\overline{\text{MS}}$ scheme and analyse them. We use the usual definition of beta function as:
\begin{equation}
\beta(x)=\frac{\partial x}{\partial (\log \mu)},
\end{equation}
while considering the running of any coupling parameter $x$ with the energy scale $\mu$. The running different parameters in generalised filed theory with dimensional regularization \cite{tHooft:1972tcz} in $\overline{\text{MS}}$ scheme have already been addressed in Refs. \cite{Machacek:1983fi,Machacek:1983tz,Machacek:1984zw,Martin:1993zk}. The RG evolution of various parameters under SM have been discussed in \cite{Chetyrkin:2012rz,JuarezW:2017tmo,Degrassi:2012ry, Buttazzo:2013uya}

\subsection{Gauge couplings}
First we discuss the renormalization group (RG) evolution of the gauge couplings. Since the doublet and triplet leptoquarks posses all the three gauge charges, namely weak hypercharge, isospin and colour, the running of all the gauge couplings will differ from SM. However, in some scenarios the weak coupling constant $g_2$ gradually increases to hit the Landau pole at some energy scale which eventually leads to sudden divergences in the other two gauge couplings also. Therefore, we present the running of $g_2$ at the beginning.

\subsubsection{Beta function of $g_2$: a brief review}
It is well established that for any non-Abelian gauge group $G=SU(N)$ the one-loop beta function of the gauge coupling $g$ is given by:
\begin{equation}
\label{eq:beta_non-ab}
\beta(g)^{1-loop}=\frac{g^3}{16\pi^2}\,\bigg[\frac{4}{3}n_f\,T(R_f)+\frac{1}{3}n_s\,T(R_s)-\frac{11}{3}C_2(G)\bigg]
\end{equation} 
where, $n_f$ is number of Dirac fermionic multiplets in representation $R_f$, $n_s$ is number of complex scalar multiplets in representation $R_s$, $C_2(G)$ is the quadratic Casimir of the gauge group $G$ and equals to $N$ since the gauge fields lie in the adjoint representation of $G$ and finally $T(R_{f/s})$ are other Casimir invariants defined by: $\Tr\,(\T^a_{R_{f/s}} \T^b_{R_{f/s}})=T(R_{f/s})\delta^{ab}$ with $\T^{a,b}_{R_{f/s}}$ being the generator of the Lie algebra in the representation $R_{f/s}$. At this point, it is worth mentioning that one should replace the factor $4/3$ by $2/3$ in Eq. \eqref{eq:beta_non-ab} while dealing with Weyl or Majorana fermions and, similarly, the factor $1/3$ must be replaced by $1/6$ for real scalar multiplets. 

If we consider the one-loop beta function of weak coupling constant $g_2$ in SM, the corresponding gauge group will be $SU(2)_L$. Hence, the fermionic contribution would come from twelve Weyl fermionic doublets: a) three generations of leptonic doublets and b) nine quark doublets (three generations and three colours). However, since all of them are Weyl fermions due to left chiral nature of the weak interaction, one must take $2/3$ factor instead of $4/3$ as the coefficient of the term $n_f\,T(R_f)$ in Eq. \eqref{eq:beta_non-ab}. On the other hand, there is only one charged scalar doublet interacting weakly in SM. Moreover, $T(R_{f/s})=1/2$ for all the fermions and scalar under $SU(2)_L$ gauge group as all of them are in fundamental representation. Thus one-loop beta function of $g_2$ in SM becomes:
\begin{align}
\label{eq:g2sm1l}
&\beta(g_2)^{1-loop}_{SM} =  \frac{g^3_2}{16\pi^2}\,\bigg[\Big(\frac{2}{3}\times 12\times \frac{1}{2}\Big)+\Big(\frac{1}{3}\times 1\times \frac{1}{2}\Big) \nn\\  & \qquad \qquad \qquad - \Big(\frac{11}{3}\times 2\Big)\bigg]=-\,\frac{19}{6}\,\Big(\frac{g^3_2}{16\pi^2}\Big)~.
\end{align}

Now, if we add one generation of scalar doublet leptoquark $\widetilde R_2$ to the SM, we can express the one-loop beta function of $g_2$ as:
\begin{equation}
\label{eq:del_b_r2}
\beta(g_2)^{1-loop}_{\widetilde R_2,1-gen} = \beta(g_2)^{1-loop}_{SM} + \Delta\beta(g_2)^{1-loop}_{\widetilde R_2}~,
\end{equation}
where, the term $\Delta\beta(g_2)^{1-loop}_{\widetilde R_2}$ signifies the sole contribution from single generation of leptoquark $\widetilde R_2$. Since $\widetilde R_2$ is a complex scalar in fundamental representation of $SU(2)_L$ having three colour choices, we find:
\begin{align}
&\Delta\beta(g_2)^{1-loop}_{\widetilde R_2}= \frac{g^3_2}{16\pi^2}\,\Big(\frac{1}{3}\times 3\times \frac{1}{2}\Big)=\frac{1}{2}\,\Big(\frac{g^3_2}{16\pi^2}\Big),\nn\\ 
\label{eq:g2r211l}
&\implies \beta(g_2)^{1-loop}_{\widetilde R_2,1-gen} = -\,\frac{8}{3}\,\Big(\frac{g^3_2}{16\pi^2}\Big).
\end{align}
Consequently, for the extension of SM with three generations of $\widetilde R_2$, the one-loop beta function of $g_2$ becomes:
\begin{align}
\beta(g_2)^{1-loop}_{\widetilde R_2,3-gen} \ & = \ \beta(g_2)^{1-loop}_{SM} + 3\,\Delta\beta(g_2)^{1-loop}_{\widetilde R_2} \nn\\ 
& =-\,\frac{5}{3}\,\Big(\frac{g^3_2}{16\pi^2}\Big).
\end{align}\label{eq:g2r231l}

Similarly, the one-loop beta function for $g_2$ in SM plus one generation of scalar triplet leptoquark $\vec{S}_3$ can also be expressed as:
\begin{equation}
\label{eq:del_b_s3}
\beta(g_2)^{1-loop}_{\vec S_3,1-gen} = \beta(g_2)^{1-loop}_{SM} + \Delta\beta(g_2)^{1-loop}_{\vec S_3}~,
\end{equation}
where, $\Delta\beta(g_2)^{1-loop}_{\vec S_3}$ contains the solo contribution of $\vec S_3$ with one generation. However, since $\vec{S}_3$ is a complex scalar triplet under $SU(2)_L$, it will be in adjoint representation; hence, $T(R_{\vec S_3})=2$
\footnote{If the generators $\T^{a,b}_R$ of $SU(N)$ Lie algebra are in adjoint representation, then $T(R)=N$.}.

 Furthermore, there will be three copies of $\vec{S}_3$ depending on the colour charges. Thus, one finds the contribution of $\vec S_3$ in the  one-loop beta function of $g_2$ as:
\begin{align}
& \Delta\beta(g_2)^{1-loop}_{\vec S_3}\  = \frac{g^3_2}{16\pi^2}\,\Big(\frac{1}{3}\times 3\times 2\Big)=2\,\Big(\frac{g^3_2}{16\pi^2}\Big), \nn \\ 
& \implies \beta(g_2)^{1-loop}_{\vec S_3,1-gen} = -\,\frac{7}{6}\,\Big(\frac{g^3_2}{16\pi^2}\Big),
\end{align}\label{eq:g2s311l}
and the one-loop beta function of $g_2$ with SM plus three generations of $\vec S_3$ necessarily becomes:
\begin{equation}
\label{eq:g2s331l}
\beta(g_2)^{1-loop}_{\vec S_3,3-gen} = \beta(g_2)^{1-loop}_{SM} + 3\, \Delta\beta(g_2)^{1-loop}_{\vec S_3}=\frac{17}{6}\,\Big(\frac{g^3_2}{16\pi^2}\Big).
\end{equation}

If the SM is extended with both $\widetilde{R}_2$ and $\vec S_3$, the one-loop beta function of $g_2$ can be calculated as:
\begin{align}
\label{eq:g2r2s311l}
&\beta(g_2)^{1-loop}_{\widetilde R_2+\vec S_3,1-gen} = \beta(g_2)^{1-loop}_{SM} + \Delta\beta(g_2)^{1-loop}_{\widetilde R_2} \nn \\
& \qquad \qquad \qquad + \Delta\beta(g_2)^{1-loop}_{\vec S_3}=-\,\frac{2}{3}\,\Big(\frac{g^3_2}{16\pi^2}\Big),&\\
\label{eq:g2r2s331l}
&\beta(g_2)^{1-loop}_{\widetilde R_2+\vec S_3,3-gen} = \beta(g_2)^{1-loop}_{SM} + 3\,\Delta\beta(g_2)^{1-loop}_{\widetilde R_2}\nn \\
& \qquad \qquad \qquad+ 3\,\Delta\beta(g_2)^{1-loop}_{\vec S_3}=\frac{13}{3}\,\Big(\frac{g^3_2}{16\pi^2}\Big).&
\end{align}

Now, we use \texttt{SARAH} to generate the two-loop contributions. For convenience, we define:
\begin{equation}
\X_a=Y_a Y_a^\dagger \quad \text{and} \quad \widetilde\X_a= Y_a^\dagger \,Y_a~,
\end{equation} 
where, $a\in\{2,3,\cl,\uq,\dq\}$. Thus the beta function of $g_2$ up to two-loops order for different the models we are working with becomes as follows:
\begin{align}
\label{eq:g2sm}
&\beta(g_2)^{2-loop}_{SM}=-\,\frac{19}{6}\Big(\frac{g_{2}^{3}}{16\pi^2}\Big)+\frac{g_{2}^{3}}{(16\pi^2)^2}\bigg[\frac{9}{10}\,g_1^2+\frac{35}{6}\,g_2^2 \nn \\
&\qquad \qquad  \qquad+12\,g_3^2-\frac{3}{2}\Tr\,\Big(\,\frac{1}{3}\X_\cl+\X_\uq+\X_\dq\Big)\bigg],\\
\label{eq:g2r21}
&\beta(g_2)^{2-loop}_{\widetilde{R}_2,1-gen}=-\,\frac{8}{3}\Big(\frac{g_{2}^{3}}{16\pi^2}\Big)+\frac{g_{2}^{3}}{(16\pi^2)^2}\bigg[g_1^2+\frac{37}{3}\,g_2^2 \nn \\ 
&\qquad \qquad \qquad  +20\,g_3^2-\frac{3}{2}\Tr\,\Big(\,\frac{1}{3}\X_\cl+\X_\uq+\X_\dq+\X_2\Big)\bigg],\\
\label{eq:g2r23}
&\beta(g_2)^{2-loop}_{\widetilde{R}_2,3-gen}=-\,\frac{5}{3}\Big(\frac{g_{2}^{3}}{16\pi^2}\Big)+\frac{g_{2}^{3}}{(16\pi^2)^2}\bigg[\frac{6}{5}\,g_1^2+\frac{76}{3}\,g_2^2 \nn \\ 
& \qquad \qquad  +36\,g_3^2-\frac{3}{2}\Tr\,\Big(\,\frac{1}{3}\X_\cl+\X_\uq+\X_\dq+\sum_{i=1}^3\X_{2,i}\Big)\bigg],\\
\label{eq:g2s31}
&\beta(g_2)^{2-loop}_{\vec S_3,1-gen}=-\,\frac{7}{6}\Big(\frac{g_{2}^{3}}{16\pi^2}\Big)+\frac{g_{2}^{3}}{(16\pi^2)^2}\bigg[\frac{5}{2}\,g_1^2+\frac{371}{6}\,g_2^2 \nn \\ 
& \qquad \qquad +44\,g_3^2-\frac{3}{2}\Tr\,\Big(\,\frac{1}{3}\X_\cl+\X_\uq+\X_\dq+3\,\X_{3}\Big)\bigg],\\
\label{eq:g2s33}
&\beta(g_2)^{2-loop}_{\vec S_3,3-gen}=\frac{17}{6}\Big(\frac{g_{2}^{3}}{16\pi^2}\Big)+\frac{g_{2}^{3}}{(16\pi^2)^2}\bigg[\frac{57}{10}\,g_1^2+\frac{1043}{6}\,g_2^2 \nonumber\\
& \qquad +108\,g_3^2-\frac{3}{2}\,\Tr\,\Big(\,\frac{1}{3}\X_\cl+\X_\uq+\X_\dq+3\sum_{i=1}^{3}\X_{3,i}\Big)\bigg],\\
\label{eq:g2r2s31}
&\beta(g_2)^{2-loop}_{\widetilde{R}_2+\vec S_3,1-gen}=-\,\frac{2}{3}\Big(\frac{g_{2}^{3}}{16\pi^2}\Big)+\frac{g_{2}^{3}}{(16\pi^2)^2}\bigg[\frac{13}{5}\,g_1^2\nonumber\\
&+\frac{205}{3}\,g_2^2+52\,g_3^2-\frac{3}{2}\,\Tr\,\Big(\,\frac{1}{3}\X_\cl+\X_\uq+\X_\dq+\X_{2}+3\,\X_3\Big)\bigg],\\
\label{eq:g2r2s33}
&\beta(g_2)^{2-loop}_{\widetilde{R}_2+\vec S_3,3-gen}=\frac{13}{3}\Big(\frac{g_{2}^{3}}{16\pi^2}\Big)+\frac{g_{2}^{3}}{(16\pi^2)^2}\bigg[6\,g_1^2+\frac{580}{3}\,g_2^2\nonumber\\
&+132\,g_3^2-\frac{3}{2}\,\Tr\,\Big(\,\frac{1}{3}\X_\cl+\X_\uq+\X_\dq+\sum_{i=1}^{3}\X_{2,i}+3\sum_{i=1}^{3}\X_{3,i}\Big)\bigg].
\end{align}

\begin{table}[h!]
	\centering
	\begin{tabular}{|c|c|c|c|c|}
		\hline
		$g_1$&$g_2$& $g_3$&$Y_\uq^{33}$&$\lambda_h$\\
		\hline
		 0.46256\footnotemark&0.64779&1.1666&0.93690&0.12604\\
		\hline
	\end{tabular}
	\caption{Initial values for different SM parameters required for RG evolution at EW scale.}
	\label{tab:SMint}
\end{table}
\footnotetext{In this paper, we have used $SU(5)$ normalization for $g_1$ since \texttt{SARAH} inherently use this convention. However, to achieve results involving usual $g_1$ coupling, one has to replace $g_1$ by $\sqrt{\frac{5}{3}}\,g_1$ throughout the paper. In that case the initial value for $g_1$ would become 0.358297.}

In the Eqs. \eqref{eq:g2r23}, \eqref{eq:g2s33} and $\eqref{eq:g2r2s33}$ the index $i$ represents the generation of leptoquark. Now, as we have defined $\Delta \beta(g_2)$ for one-loop in Eqs. \eqref{eq:del_b_r2} and \eqref{eq:del_b_s3}, one can define it for two-loops also in similar fashion. Then one can easily verify that the above described two-loop beta functions obey the following relations:
\begin{align}
&\beta(g_2)^{2-loop}_{\widetilde{R}_2,3-gen}=\beta(g_2)^{2-loop}_{SM}+\sum_{i=1}^{3}\Big[\Delta\beta(g_2)^{2-loop}_{\widetilde{R}_2}\Big]_{\is}~,\\
&\beta(g_2)^{2-loop}_{\vec S_3,3-gen}=\beta(g_2)^{2-loop}_{SM}+\sum_{i=1}^{3}\Big[\Delta\beta(g_2)^{2-loop}_{\vec S_3}\Big]_{\is}~,\\
&\beta(g_2)^{2-loop}_{\widetilde{R}_2+\vec S_3,1-gen}=\beta(g_2)^{2-loop}_{SM}+\Delta\beta(g_2)^{2-loop}_{\widetilde{R}_2} \nn \\
&\qquad \qquad \qquad \qquad +\Delta\beta(g_2)^{2-loop}_{\vec S_3}~,\\
&\beta(g_2)^{2-loop}_{\widetilde{R}_2+\vec S_3,3-gen}=\beta(g_2)^{2-loop}_{SM}+\sum_{i=1}^{3}\Big[\Delta\beta(g_2)^{2-loop}_{\widetilde{R}_2}\Big]_{\is}\nn\\
&\qquad \qquad \qquad \qquad  +\sum_{i=1}^{3}\Big[\Delta\beta(g_2)^{2-loop}_{\vec S_3}\Big]_{\is}~.
\end{align}
where $\big[\beta\big]_{\is}$ for any parameter indicates beta function of that parameter with the replacement of $f(Y_\gamma,\,\X_\gamma,\,\wX_\gamma,\,\lambda_\gamma,\\\widetilde{\lambda}_\gamma)$ to $f(Y_{\gamma,i}\,,\,\X_{\gamma,i}\,,\,\wX_{\gamma,i}\,,\,\lambda_\gamma^{ii}\,,\,\widetilde{\lambda}_\gamma^{ii})$ with $\gamma\in\{2,3\}$ and $i$ representing the generation. Similarly notation is applicable for $\Delta \beta$ also.

\subsubsection{Scale variation of $g_2$}

\begin{figure*}[h!]
	\centering
	\subfigure[1-loop]{\includegraphics[width=0.45\textwidth]{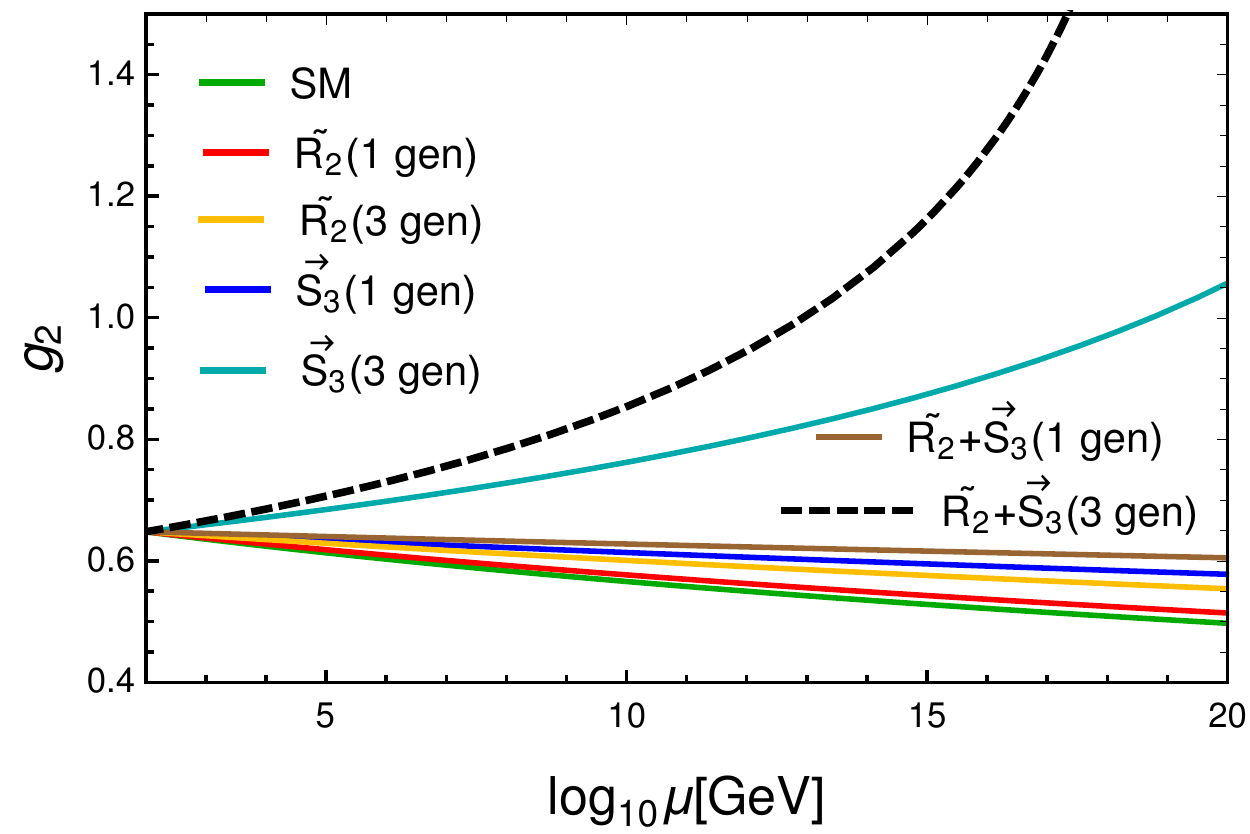}\label{g21}}\hfil
	\subfigure[2-loop]{\includegraphics[width=0.45\textwidth]{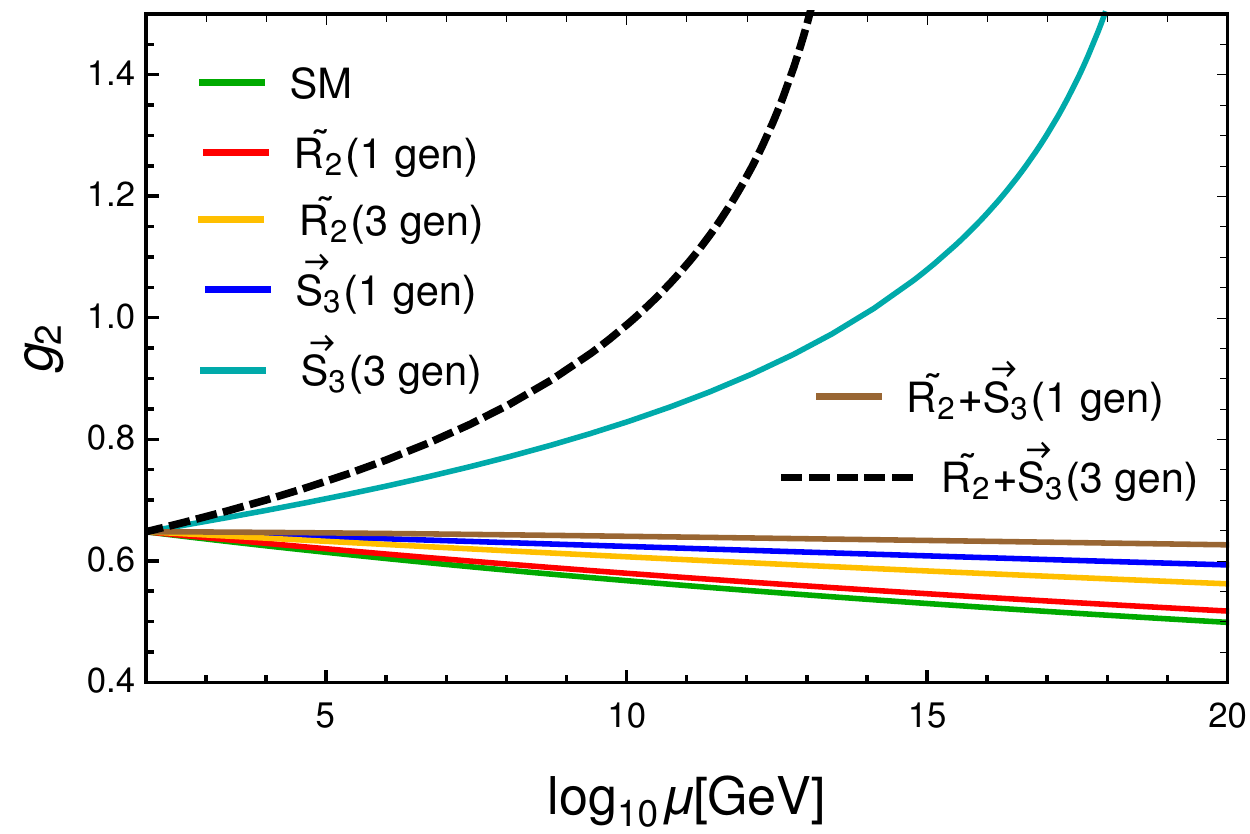}\label{g22}}
	\subfigure[]{\includegraphics[width=0.45\textwidth]{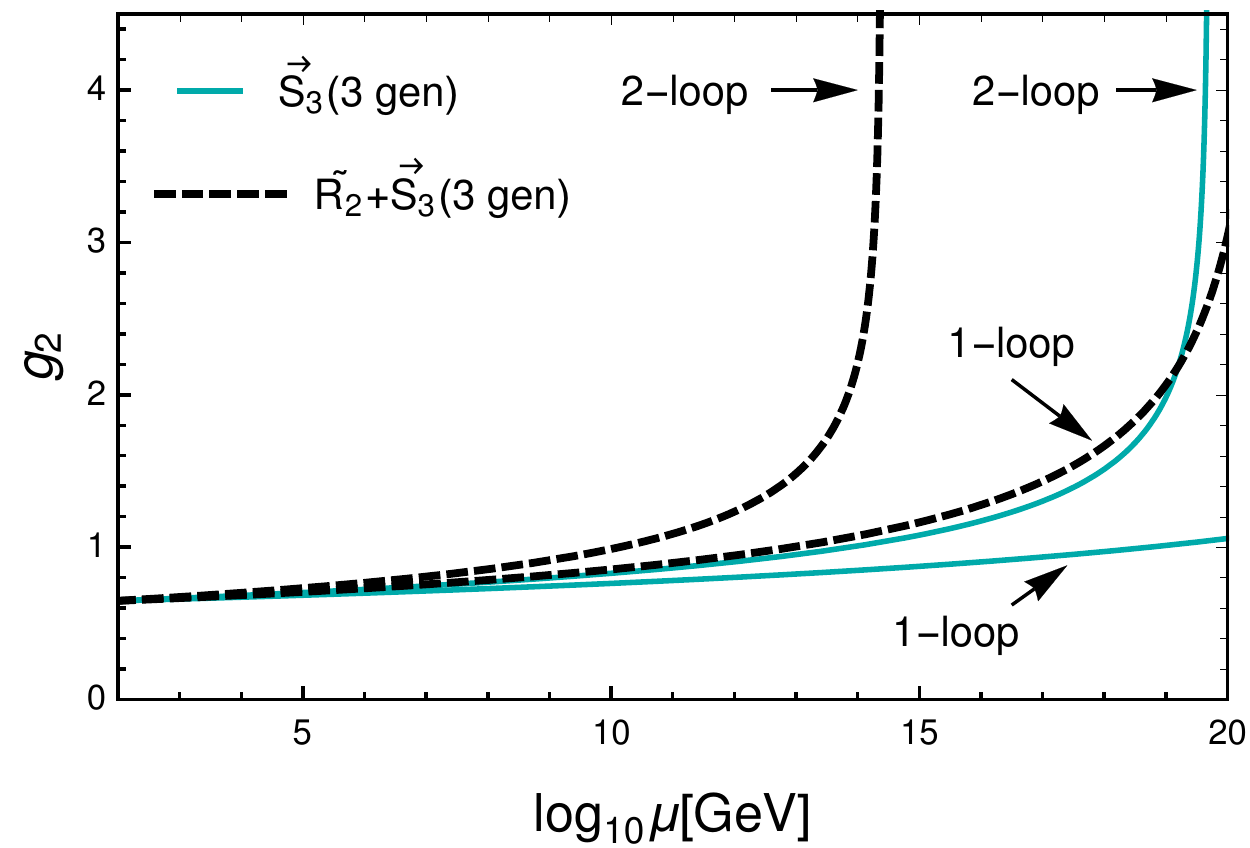}\label{g23}}
	\caption{Running of gauge coupling $g_2$ with the energy scale $\mu$ for SM and different leptoquark scenarios with one-loop and two-loop. The green curve represents SM; the red and yellow ones signify extension of SM with one and three generations of $\widetilde{R}_2$ respectively; addition of one and three generations of $\vec S_3$ to SM are indicated by the blue and cyan lines respectively; the brown and dashed black curves illustrate SM extension with one and three generations of both $\widetilde{R}_2$ and $\vec S_3$ respectively. The initial values for SM parameters are listed in Tab. \ref{tab:SMint} along with $Y_\phi=0.1$.}
	\label{fig:g2}
\end{figure*}

Using the above Eqs.
\footnote{Actually, one needs to consider running of all the couplings in a model simultaneously, since the above expressions for two-loop beta functions are coupled equations.} \eqref{eq:g2sm}-\eqref{eq:g2r2s33}, we plot the dependence of coupling $g_2$ for different models on the energy scale $\mu$ in Fig. \ref{fig:g2}. While Fig. \ref{g21} depicts the behaviour of $g_2$ at one-loop, Fig. \ref{g22} illustrates the same for two-loop. The SM is represented by the green curve; the red and yellow lines depict extension of SM with one and three generations of $\widetilde{R}_2$ respectively; the blue and cyan lines indicate addition of one and three generations of $\vec S_3$ respectively to SM; finally the brown and dashed black curves illustrate SM extension with one and three generations of both doublet and triplet leptoquarks respectively. The initial two-loop values at the electroweak (EW) scale for gauge couplings $g_1, g_2, g_3$, Higgs quartic coupling $\lambda_h$ and top-quark Yukawa coupling $Y_\uq^{33}$ are given in Tab. \ref{tab:SMint} with the contributions from other Yukawa couplings are neglected~\cite{Degrassi:2012ry, Buttazzo:2013uya}. Though the plots are made assuming $Y_\phi$ to be 0.1, they do not change significantly with the alteration of $Y_\phi$ since the dominant contribution in the two-loop beta function of $g_2$ comes from different gauge couplings, as can realized from Eqs. \eqref{eq:g2sm} - \eqref{eq:g2r2s33}. This statement also holds for other gauge couplings as well.

In  Fig. \ref{fig:g2}(a) and  Fig. \ref{fig:g2}(b) we present the variation of $g_2$ with respect to the scale $\mu$ at  one- and two-loop level for the mentioned leptoquark scenarios. 
The ordering of different curves in the Fig. \ref{fig:g2} is mainly controlled by the one-loop beta functions for different models which are presented in Eqs. \eqref{eq:g2sm1l}-\eqref{eq:g2r2s331l}. The one loop beta function of $g_2$ under SM is $-\frac{19}{6}(\frac{g_2^3}{16\pi^2})$ which gets enhanced to $-\frac{8}{3}(\frac{g_2^3}{16\pi^2})$ and $-\frac{5}{3}(\frac{g_2^3}{16\pi^2})$ for one and three generations of $\widetilde{R}_2$ respectively. However, due to more components of leptoquarks in $\vec S_3$ the positive contributions will be more. The one loop beta functions of $g_2$ for $\vec S_3$ with one and three generations become $-\frac{7}{6}(\frac{g_2^3}{16\pi^2})$ and $\frac{17}{6}(\frac{g_2^3}{16\pi^2})$. For the combined scenario of these two leptoquarks the positive effects are even stronger. For one and three generations of the combined case, the beta function becomes $-\frac{2}{3}(\frac{g_2^3}{16\pi^2})$ and $\frac{13}{3}(\frac{g_2^3}{16\pi^2})$ respectively. It is interesting to note that except three generations of $\vec S_3$ (cyan line) and $\widetilde{R}_2+\vec S_3$ (black dotted curve) coupling $g_2$ decreases monotonically for all the other scenarios due to the negative sign in one-loop beta function ensuring asymptotic freedom of weak interaction. However, while considering one-loop effects only, Planck scale perturbativity is achieved in all the scenarios since the Landau pole in those two above mentioned cases appears beyond the Plank scale, as can  be  seen from Fig. \ref{fig:g2}(a). On the other hand, due to the positive value of the one-loop beta function, the gauge coupling $g_2$ in the three generations of $\vec S_3$ (cyan) and $\widetilde{R}_2+\vec S_3$ (black dashed) models increases monotonically. Now, for two-loop case, all the models acquire additional positive effects that push the RG evolution curves upwards. Therefore, two-loop beta functions of $\vec S_3$ and $\widetilde{R}_2+\vec S_3$ models hit the Landau pole at relatively lower scales, i.e. $10^{19.7}$ GeV (just above the Planck scale) and $10^{14.4}$ GeV (below the GUT scale) respectively, as can be noticed from Fig.~\ref{g23}.


\subsubsection{Beta function of $g_3$: a brief review}

In case of SM, as the scalar and leptons are colour neutral, the one-loop beta function of strong coupling $g_3$ gets contribution only from six quarks which are essentially Dirac fermionic colour triplets under $SU(3)$ gauge group
Thus substituting $T(R_f)=1/2$ and $C_2(G)=3$ in Eq. \eqref{eq:beta_non-ab}, we get:
\begin{equation}
\label{eq:g3sm1l}
\beta(g_3)^{1-loop}_{SM}=\frac{g^3_3}{16\pi^2}\,\bigg[\Big(\frac{4}{3}\times 6\times \frac{1}{2}\Big) - \Big(\frac{11}{3}\times 3\Big)\bigg]=-\,7\,\Big(\frac{g^3_3}{16\pi^2}\Big)~.
\end{equation}
Now, all the leptoquarks are colour triplet complex scalars, i.e. they are in fundamental representation of $SU(3)$ enforcing $T(R_s)=1/2$. However, for the doublet leptoquark we have two such copies of triplets namely $\widetilde{R}_2^{2/3}$ and $\widetilde{R}_2^{1/3}$, whereas there are three scalar triplets for $\vec S_3$ namely 
$S_3^{4/3}$, $S_3^{2/3}$ and $S_3^{1/3}$. Thus the sole contribution of one generation $\widetilde{R}_2^{2/3}$ and $\vec S_3$ in the beta function of $g_3$, as described in the previous subsection for $g_2$, can be written as:
\begin{align}
\Delta\beta(g_3)^{1-loop}_{\widetilde{R}_2}=\frac{g^3_3}{16\pi^2}\,\Big(\frac{1}{3}\times 2\times\frac{1}{2}\Big)=\frac{1}{3}\,\Big(\frac{g^3_3}{16\pi^2}\Big),\\
\Delta\beta(g_3)^{1-loop}_{\vec S_3}=\frac{g^3_3}{16\pi^2}\,\Big(\frac{1}{3}\times 3\times\frac{1}{2}\Big)=\frac{1}{2}\,\Big(\frac{g^3_3}{16\pi^2}\Big).
\end{align}
So, the one-loop beta function of strong coupling $g_3$ for different SM extension with $\widetilde{R}_2$ and $\vec S_3$ are as follows:
\begin{align}
&\beta(g_3)^{1-loop}_{\widetilde R_2,1-gen} = \beta(g_3)^{1-loop}_{SM} + \Delta\beta(g_3)^{1-loop}_{\widetilde R_2}\nn\\
& \qquad \qquad \qquad \quad =-\,\frac{20}{3}\,\Big(\frac{g^3_3}{16\pi^2}\Big),\\
&\beta(g_3)^{1-loop}_{\widetilde R_2,3-gen} = \beta(g_3)^{1-loop}_{SM} + 3\,\Delta\beta(g_3)^{1-loop}_{\widetilde R_2}\nn\\
& \qquad \qquad \qquad \quad=-\,6\,\Big(\frac{g^3_3}{16\pi^2}\Big),\\
&\beta(g_3)^{1-loop}_{\vec S_3,1-gen} = \beta(g_3)^{1-loop}_{SM} + \Delta\beta(g_3)^{1-loop}_{\vec S_3}\nn\\
&\qquad \qquad \qquad \quad=-\,\frac{13}{2}\,\Big(\frac{g^3_3}{16\pi^2}\Big),\\
&\beta(g_3)^{1-loop}_{\vec S_3,3-gen} = \beta(g_3)^{1-loop}_{SM} + 3\,\Delta\beta(g_3)^{1-loop}_{\vec S_3}\nn\\
&\qquad \qquad \qquad \quad=-\,\frac{11}{2}\,\Big(\frac{g^3_3}{16\pi^2}\Big),\\
&\beta(g_3)^{1-loop}_{\widetilde R_2+\vec S_3,1-gen} = \beta(g_3)^{1-loop}_{SM} + \Delta\beta(g_3)^{1-loop}_{\widetilde R_2}\nn\\
&\qquad \qquad \qquad \quad+ \Delta\beta(g_3)^{1-loop}_{\vec S_3}=-\,\frac{37}{6}\,\Big(\frac{g^3_3}{16\pi^2}\Big),\\
\label{eq:g3r2s331l}
&\beta(g_3)^{1-loop}_{\widetilde R_2+\vec S_3,3-gen} = \beta(g_3)^{1-loop}_{SM} + 3\,\Delta\beta(g_3)^{1-loop}_{\widetilde R_2}\nn\\
&\qquad \qquad \qquad \quad+ 3\,\Delta\beta(g_3)^{1-loop}_{\vec S_3}=-\,\frac{9}{2}\,\Big(\frac{g^3_3}{16\pi^2}\Big).
\end{align}
The two-loop beta functions of $g_3$ for all these models are listed in \autoref{sec:g32}.

\subsubsection{Scale variation of $g_3$}

\begin{figure*}[h!]
	\begin{center}
		\subfigure[1-loop]{\includegraphics[width=0.48\linewidth,angle=-0]{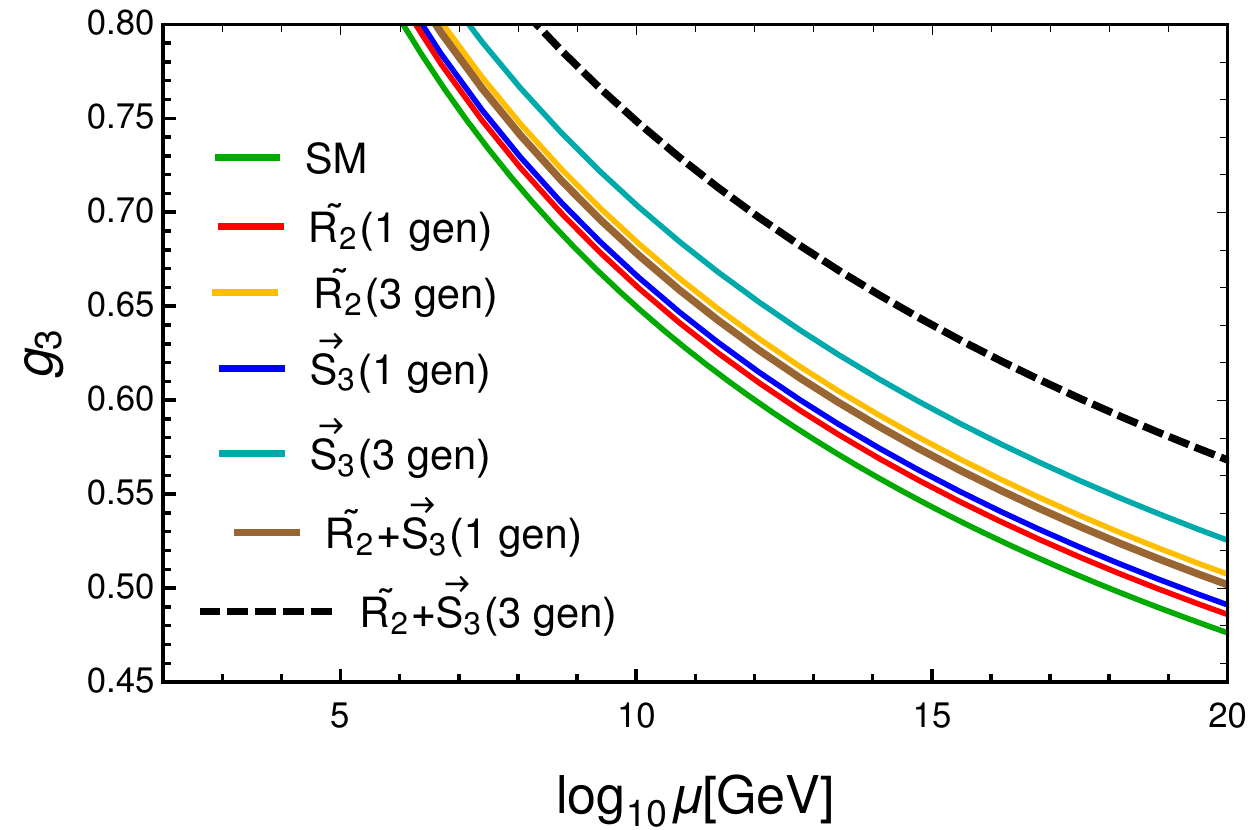}\label{g31}}
		\subfigure[]{\includegraphics[width=0.48\linewidth,angle=-0]{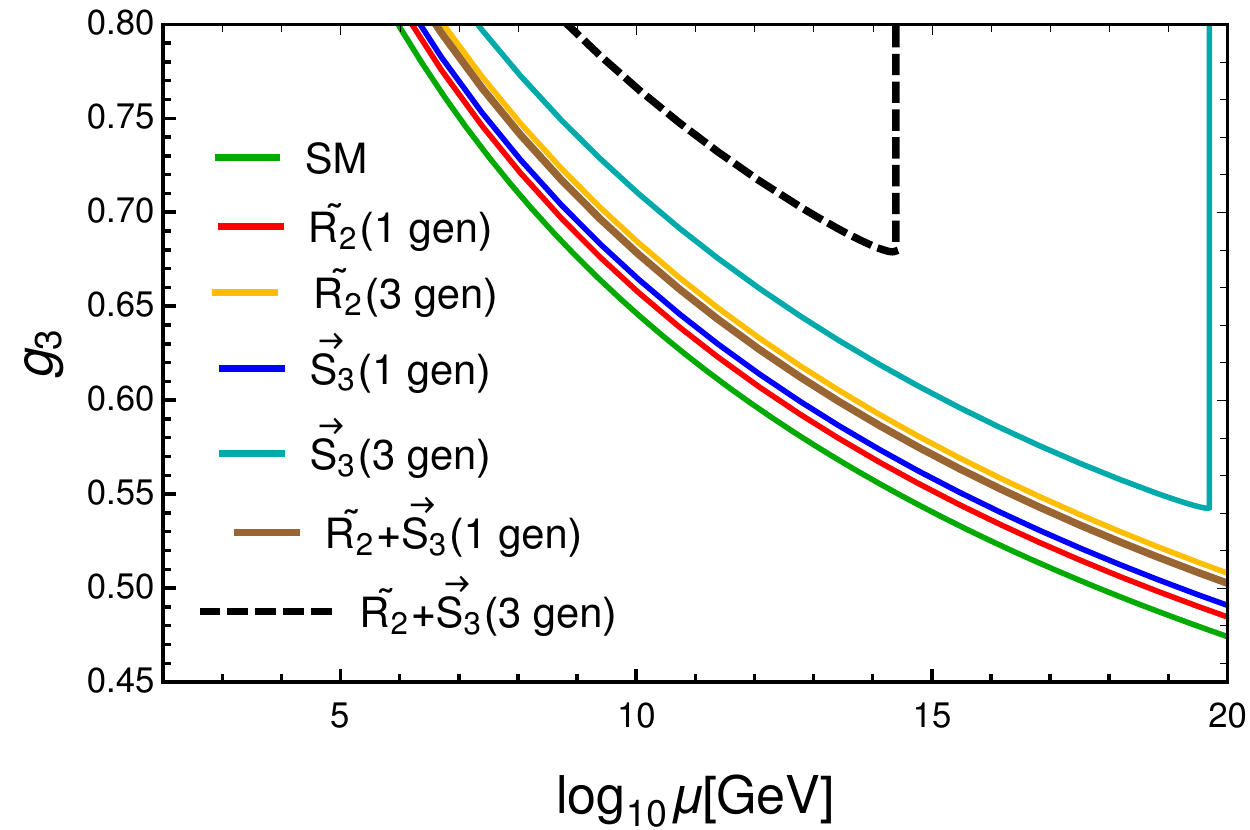}\label{g32}}
		\caption{Running of gauge coupling  $g_3$ with the energy scale $\mu$ for different leptoquark models at one-loop and two-loop order. The green curve represents SM; the red and yellow ones signify extension of SM with one and three generations of $\widetilde{R}_2$ respectively; addition of one and three generations of $\vec S_3$ to SM are indicated by the blue and cyan lines respectively; the brown and dashed black curves illustrate SM extension with one and three generations of both $\widetilde{R}_2$ and $\vec S_3$ respectively. The initial values for SM parameters are listed in Tab. \ref{tab:SMint} along with $Y_\phi=0.1$.}\label{fig:g3}
	\end{center}
\end{figure*}


The variations of strong gauge coupling $g_3$ at one-loop and two-loop with energy scale $\mu$ for different models are depicted in Fig. \ref{fig:g3}. While the left panel signifies the one-loop results, the right panel indicates the full two-loop contributions. The same colour code, mentioned in last section, has also been followed here. The relative positions of the different curves in this plot are mainly determined by coefficients of $(\frac{g_3^3}{16\pi^2})$ in the one-loop beta functions, given by Eqs. \eqref{eq:g3sm1l} - \eqref{eq:g3r2s331l}. This coefficient for SM (green) is $-7$ which gets enhanced to $-20/3$, $-13/2$ and $-37/6$ for $\widetilde{R}_2$ (red), $\vec S_3$ (blue) and the combined scenario (brown) for one generation respectively. For three generations cases this factor gets even more contributions to become $-6$, $-11/2$ and $-9/2$ respectively for $\widetilde{R}_2$ (yellow) , $\vec S_3$ (cyan) and the combined scenario (black dashed). As the one-loop beta function of $g_3$ for all the models remains negative, $g_3$ decreases gradually with increase in energy showing asymptotic freedom. As can be noticed from Fig. \ref{g31} All the models show Planck scale perturbativity at one loop order. At two loop order all of these curves shift upwards due to additional positive contributions as shown in \autoref{sec:g32}.  All the models except two do not exhibit any unusual behaviour. But, as $g_2$ hits Landau pole at $10^{19.7}$ GeV and $10^{14.4}$ GeV for three generations of $\vec S_3$ (cyan) and $\widetilde{R}_2+\vec S_3$ (black dashed) models respectively, $g_3$ shows sudden divergence for these two models at the mentioned energy scales (see Fig. \ref{g32}).

\subsubsection{Beta function of $g_1$: a brief review}

The one-loop beta function for $U(1)_\y$ gauge coupling $g_1$ is given by:
\begin{equation}
\beta(g_1)^{1-loop}=\frac{3}{5}\,\Big(\frac{g_1^3}{16\pi^2}\Big) \,\bigg[\frac{2}{3}\sum_f\y_f^2+\frac{1}{3}\sum_s\y_s^2\bigg],
\end{equation}
where, $\y_{f,s}$ signify the hypercharge of the Weyl fermions and the scalars respectively
\footnote{To relate electromagnetic charge $\cQ$ with hypercharge $\y$, we have followed the convention: $\cQ=\T_3+\y$.}
\textsuperscript{,}\footnote{One can easily compare the above formula with the one-loop beta function for the electromagnetic coupling $\e$ given by: $\displaystyle\beta(\e)^{1-loop}=\frac{\e^3}{16\pi^2} \,\bigg[\frac{4}{3}\sum_f\cQ_f^2+\frac{1}{3}\sum_s\cQ_s^2\bigg]$ where $\cQ_{f,s}$ are the electromagnetic charges of Dirac fermions and scalars.}. 
The $3/5$ factor arises because of $SU(5)$ normalization of the coupling $g_1$. On the other hand, since $U(1)_\y$ gauge boson interacts to left and right handed fermions with different hypercharges, one has to sum over contributions from all the Weyl fermions and hence $2/3$ factor appears for the fermionic effects instead of $4/3$. In SM, there are eighteen left handed quarks (six flavours, three colours) with hypercharge $1/6$, nine right handed up-type quarks (three generations, three colours) with hypercharge $2/3$, nine right handed down-type quarks (three generations, three colours) with hypercharge $-1/3$, six left handed leptons with $\y_f=1/2$ and three right handed charged leptons with $\y_f=-1$. Additionally, there are two scalars ($H^+$ and $H^0$, the components of scalar doublet $H$), each with $\y_s=1/2$. Thus the one-loop beta function for $g_1$ in SM becomes:
\begin{align}
\label{eq:g1sm1l}
&\beta(g_1)^{1-loop}_{SM}=\frac{3}{5}\,\Big(\frac{g_1^3}{16\pi^2}\Big) \,\bigg[\,\frac{2}{3}\times\Big(\frac{1}{2}+4+1+\frac{3}{2}+3\Big)\nn\\
& \qquad \qquad \qquad +\frac{1}{3}\times\frac{1}{2}\bigg]=\frac{41}{10}\,\Big(\frac{g_1^3}{16\pi^2}\Big).
\end{align}

Now, for SM plus one generation of $\widetilde{R}_2$, contribution from six scalars (two flavours, three colours) with hypercharge $1/6$ needs to be added to the SM contribution. Similarly, effects of nine scalars (three flavours, three colours) with $\y_s=1/3$ must be considered while dealing with SM extension by one generation of $\vec S_3$. Thus the sole contribution from one generation of $\widetilde{R}_2$ and $\vec S_3$ to the one-loop beta function of $g_1$ can be calculated as:
\begin{align}
&\Delta\beta(g_1)^{1-loop}_{\widetilde{R}_2}=\frac{3}{5}\,\Big(\frac{g_1^3}{16\pi^2}\Big)\,\bigg[\frac{1}{3}\times 6 \times \frac{1}{36}\bigg]=\frac{1}{30}\,\Big(\frac{g_1^3}{16\pi^2}\Big),\\
&\Delta\beta(g_1)^{1-loop}_{\vec S_3}=\frac{3}{5}\,\Big(\frac{g_1^3}{16\pi^2}\Big)\,\bigg[\frac{1}{3}\times 9 \times \frac{1}{9}\bigg]=\frac{1}{5}\,\Big(\frac{g_1^3}{16\pi^2}\Big),
\end{align}
and hence, the one-loop beta function of $g_1$ for different models we considered becomes:
\begin{align}
&\beta(g_1)^{1-loop}_{\widetilde R_2,1-gen} = \beta(g_1)^{1-loop}_{SM} + \Delta\beta(g_1)^{1-loop}_{\widetilde R_2}\nn\\
&\qquad \qquad \qquad \quad=\frac{62}{15}\,\Big(\frac{g^3_1}{16\pi^2}\Big),\\
&\beta(g_1)^{1-loop}_{\widetilde R_2,3-gen} = \beta(g_1)^{1-loop}_{SM} + 3\,\Delta\beta(g_1)^{1-loop}_{\widetilde R_2}\nn\\
& \qquad \qquad \qquad \quad=\frac{21}{5}\,\Big(\frac{g^3_1}{16\pi^2}\Big),\\
&\beta(g_1)^{1-loop}_{\vec S_3,1-gen} = \beta(g_1)^{1-loop}_{SM} + \Delta\beta(g_1)^{1-loop}_{\vec S_3}\nn \\
&\qquad \qquad \qquad \quad=\frac{43}{10}\,\Big(\frac{g^3_1}{16\pi^2}\Big),\\
&\beta(g_1)^{1-loop}_{\vec S_3,3-gen} = \beta(g_1)^{1-loop}_{SM} + 3\,\Delta\beta(g_1)^{1-loop}_{\vec S_3}\nn \\
&\qquad \qquad \qquad \quad=\frac{47}{10}\,\Big(\frac{g^3_1}{16\pi^2}\Big),\\
&\beta(g_1)^{1-loop}_{\widetilde R_2+\vec S_3,1-gen} = \beta(g_1)^{1-loop}_{SM} + \Delta\beta(g_1)^{1-loop}_{\widetilde R_2}\nn\\
&\qquad \qquad \qquad \quad+ \Delta\beta(g_1)^{1-loop}_{\vec S_3}=\frac{13}{3}\,\Big(\frac{g^3_1}{16\pi^2}\Big),\\
\label{eq:g1r2s331l}
&\beta(g_1)^{1-loop}_{\widetilde R_2+\vec S_3,3-gen} = \beta(g_1)^{1-loop}_{SM} + 3\,\Delta\beta(g_1)^{1-loop}_{\widetilde R_2}\nn\\
&\qquad \qquad \qquad \quad+ 3\,\Delta\beta(g_1)^{1-loop}_{\vec S_3}=\frac{24}{5}\,\Big(\frac{g^3_1}{16\pi^2}\Big).
\end{align}
The two-loop beta functions of $g_1$ for all these models are listed in \autoref{sec:g12}.

\begin{figure*}[h!]
	\begin{center}
		\subfigure[1-loop]{\includegraphics[width=0.48\linewidth,angle=-0]{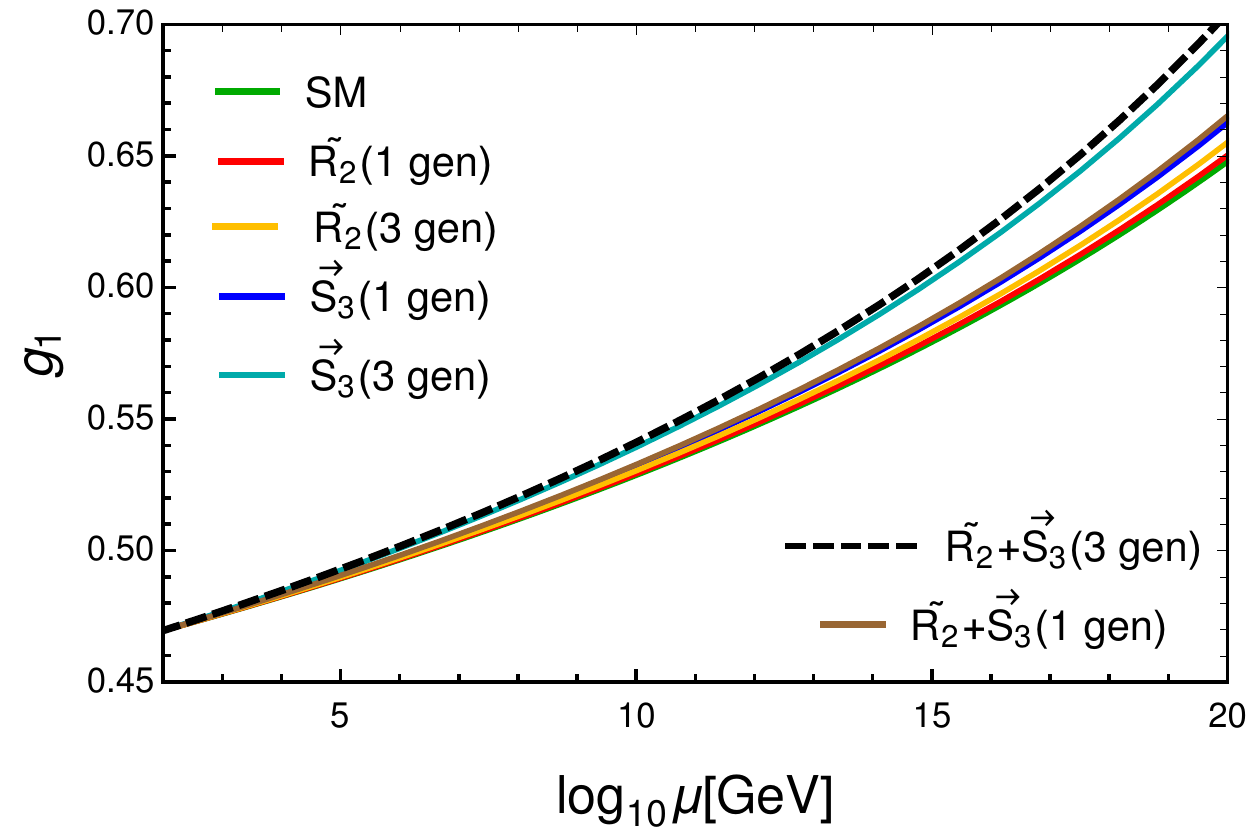}\label{g11}}
		\subfigure[2-loop]{\includegraphics[width=0.48\linewidth,angle=-0]{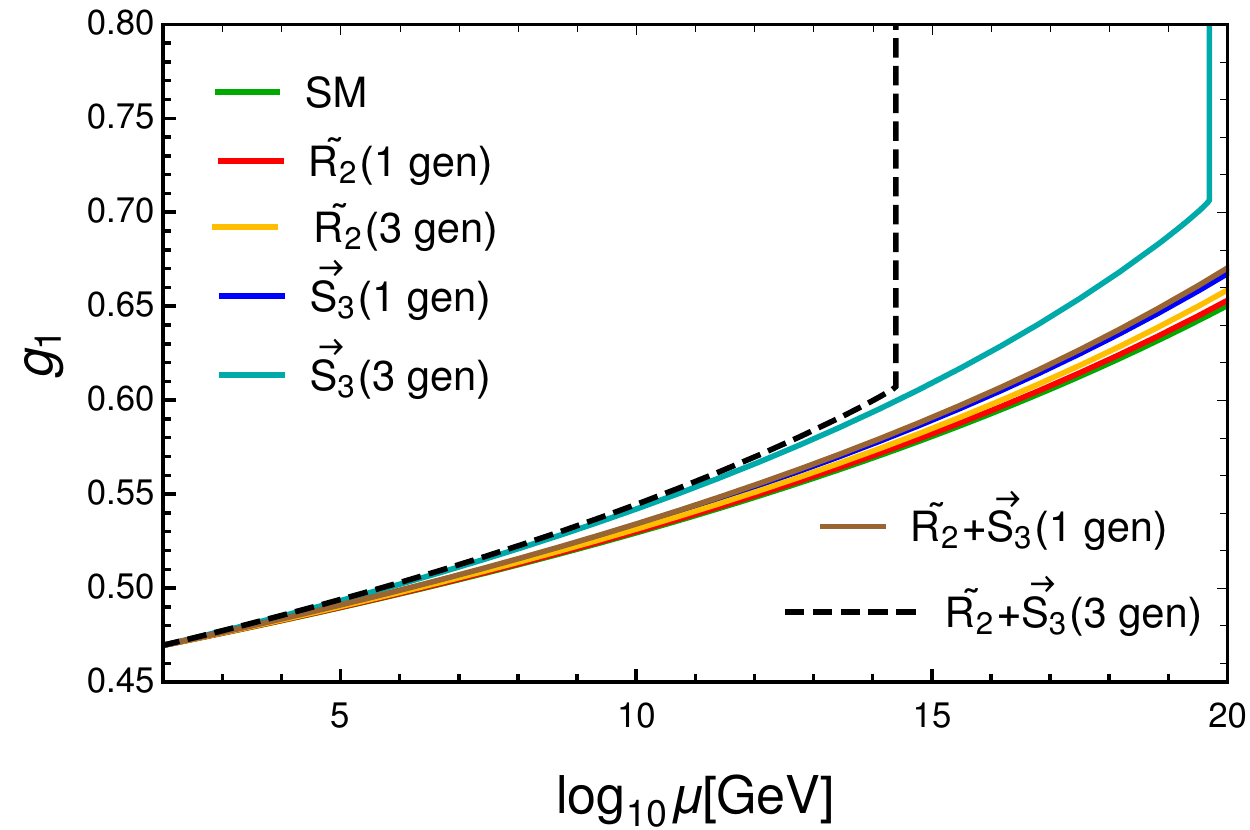}\label{g12}}
		\caption{ Running of gauge coupling $g_1$ with the energy scale $\mu$ for different leptoquark models at one-loop and two-loop order. The green curve represents SM; the red and yellow ones signify extension of SM with one and three generations of $\widetilde{R}_2$ respectively; addition of one and three generations of $\vec S_3$ to SM are indicated by the blue and cyan lines respectively; the brown and dashed black curves illustrate SM extension with one and three generations of both $\widetilde{R}_2$ and $\vec S_3$ respectively. The initial values for SM parameters are listed in Tab. \ref{tab:SMint} along with $Y_\phi=0.1$.}\label{fig:g1}
	\end{center}
\end{figure*}

\subsubsection{Scale variation of $g_1$}

The variations of $g_1$ with the energy scale $\mu$ are displayed in the left panel of Fig. \ref{fig:g1}. The left panel illustrates the one-loop effects and the right panel demonstrates the two-loop effects. The colour codes are same as mentioned before. For this case also the positions of different curves in the above mentioned figure are mainly controlled by the one-loop beta functions, given by Eqs. \eqref{eq:g1sm1l} - \eqref{eq:g1r2s331l}. The coefficient of $(g_1^3/16\pi^2)$ in SM scenario is $41/10$ which gets enhanced to $62/15$ and $21/5$ respectively for one (red) and three (yellow) generations of $\widetilde{R}_2$. For $\vec S_3$ with one (blue) and three (cyan) generations this factor increases to $43/10$ and $47/10$ respectively. For the combined scenario with one (brown) and three generations (black dashed), this prefactor of $(g_1^3/16\pi^2)$ in one-loop beta function of $g_1$ becomes $13/3$ and $24/5$ respectively. Since the one-loop beta functions for all the models are positive, $g_1$ increases moderately with energy. There is also no divergence for one-loop running of $g_1$ in any of the models till Planck scale, which can be verified from Fig. \ref{g11}. Furthermore, two-loop beta function of $g_1$ gets additional positive contributions, presented in \autoref{sec:g12}, that moves all the curves of Fig. \ref{g11} in slightly upward direction resulting in Fig. \ref{g12}. Though all the other scenarios behave smoothly while taking into account two-loop corrections, $g_1$ for three generations of $\vec S_3$ (cyan) and $\widetilde{R}_2+\vec S_3$ (black dashed) models goes to infinity abruptly at $10^{19.7}$ GeV and $10^{14.4}$ GeV respectively due to the divergence of $g_2$.

Thus, we find that the running of gauge couplings at two-loop order for different leptoquark models are predominantly regulated by the corresponding one-loop beta functions, which entirely rely on the properties of the gauge group and the number of different type of particles existing in the model. The two-loop corrections insert additional positive contributions to the running of the gauge couplings. The Yukawa couplings of SM as well as of leptoquarks affect the RG evolution of gauge couplings at two-loop order only, and therefore with the changes of Yukawa couplings of leptoquarks, we do not observe any significant changes. However, it is interesting to notice that Higgs-leptoquark quartic couplings do not appear explicitly in the two-loop beta functions of the gauge couplings at all. It is worth mentioning again that the demand of Planck scale perturbativity rules out the three generations of $\widetilde{R}_2+\vec S_3$ scenario due to the appearance of divergences at much lower scale in two-loop running of the gauge coupling $g_2$. On the other hand, model with three generations of $\vec S_3$ is marginally allowed from Planck scale stability since the gauge coupling $g_2$ hits Landau pole at slightly higher energy scale. These divergences force the other gauge couplings as well as the Yukawa couplings of top quark and leptoquarks (see \autoref{sec:topY} and \autoref{sec:lqY}) for these models to diverge at two-loop level.

\subsection{Higgs-leptoquark quartic couplings}
Now, we step forward to investigate the perturbative bounds on  Higgs-leptoquark quartic couplings.

\subsubsection{Perturbativity of $\widetilde R_2$}
\label{perbR2}
\begin{figure*}[h!]
	\begin{center}
		\mbox{\subfigure[$Y_\phi =0.1$]{\includegraphics[width=0.5\linewidth,angle=-0]{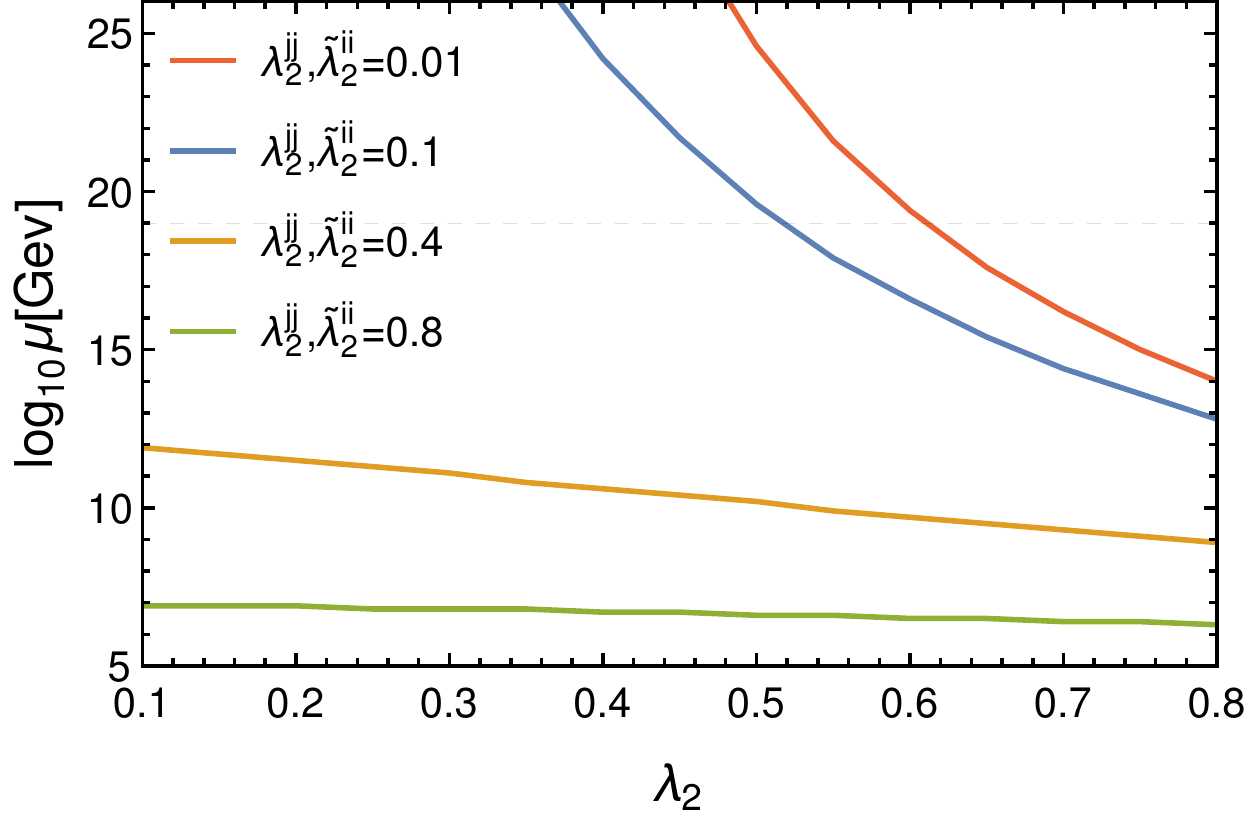}\label{f5a}}
			\subfigure[$Y_\phi =1.0$]{\includegraphics[width=0.5\linewidth,angle=-0]{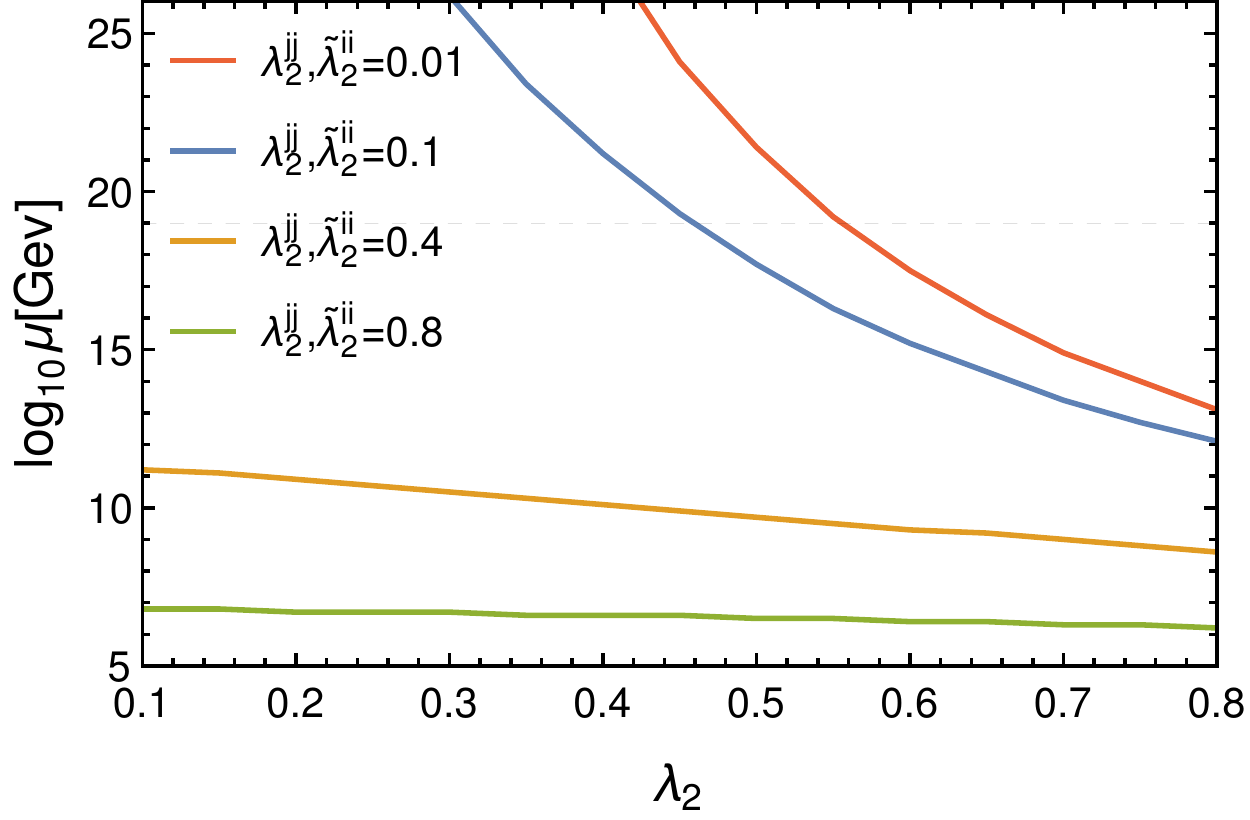}\label{f6a}}}
		\mbox{\subfigure[$Y_\phi =0.1$]{\includegraphics[width=0.5\linewidth,angle=-0]{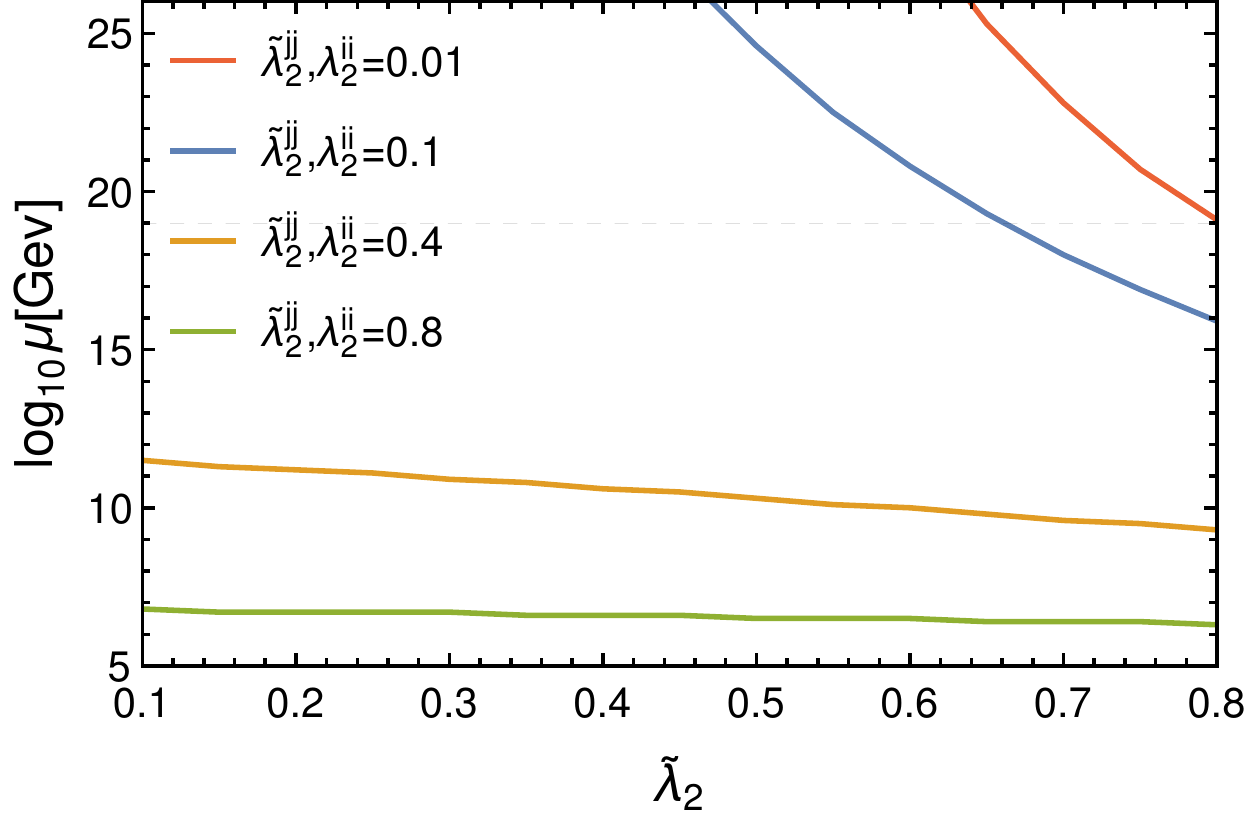}\label{f7a}}
			\subfigure[$Y_\phi =1.0$]{\includegraphics[width=0.5\linewidth,angle=-0]{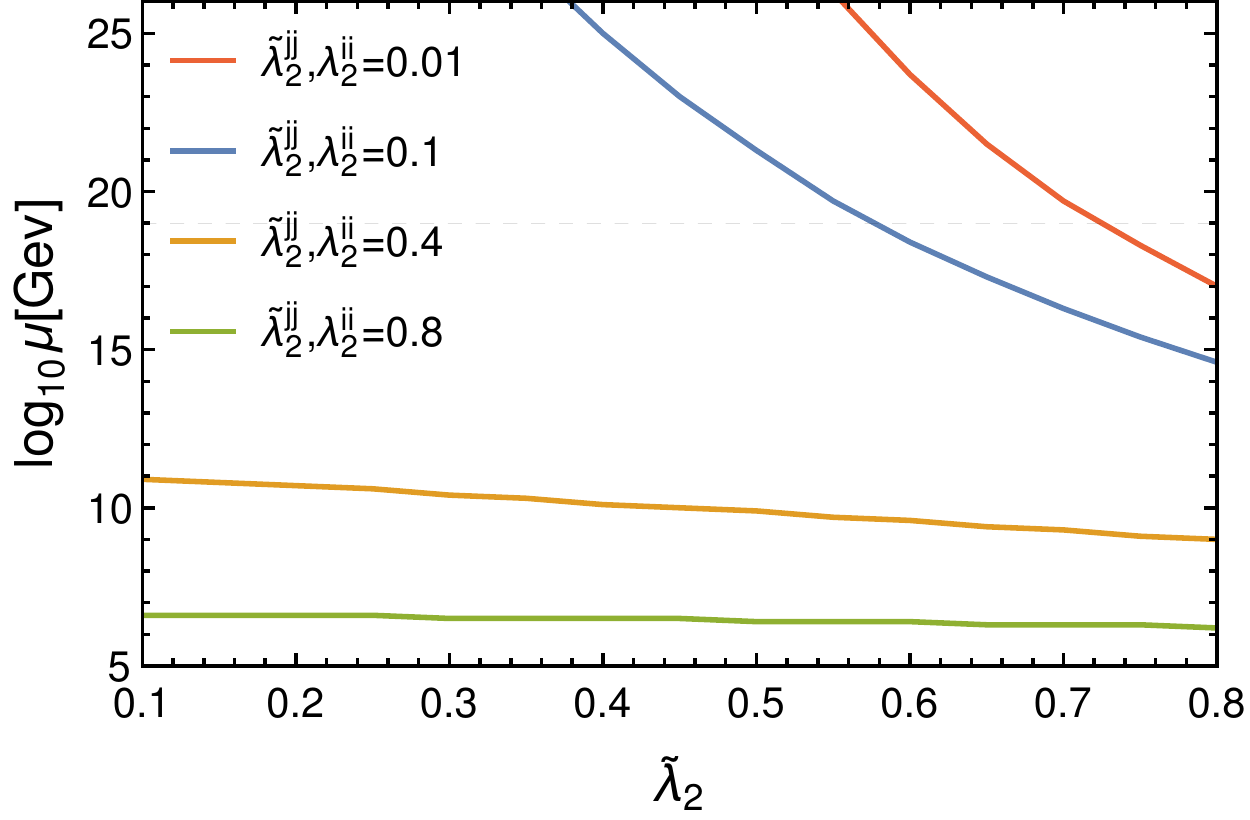}\label{f8a}}}
		\caption{Variation of leptoquark-Higgs quartic coupling $\lambda_2$ and $\wl_2$ with perturbative scale for doublet leptoquark $\widetilde R_2$ with three generations. For plots in first row, $\lambda_2$ variation is considered for any one generation of leptoquark and other generations are defined as $\lambda_2^{jj}$. The other quartic couplings are designated by $\wl_2^{jj}$ for all three generations of leptoquark. The variations are taken for four different EW values of quartic couplings i.e 0.01, 0.1, 0.4 and 0.8 which are depicted by red, blue, orange and green curves respectively. Similarly, for the plots in second row, $\wl_2$ describes the variation of any particular generation and remaining generations are denoted by $\wl^{jj}$. The other quartic coupling terms $\lambda_2^{ii}$ are defined for all three generations of leptoquark. The variations are considered for lower and higher values of $Y_\phi$ i.e 0.1 (left) and 1.0 (right). }\label{fig:r2l2}
	\end{center}
\end{figure*}
In this section, we study the RG evolution of Higgs-lepton
quark quartic couplings of leptoquark $\widetilde{R}_2$, i.e. $\lambda_{2}$ and $\wl_{2}$. As already mentioned, these terms should always remain below $4\pi$ to maintain the perturbativity of the theory. The one-loop beta functions for these two parameters are given below:
\begin{align}
\label{eq:l2r211}
\beta(\lambda_{2})&^{1-loop}_{\widetilde{R}_2,1-gen}=\frac{1}{16\pi^2}\bigg[4\lambda_{2}^2+2\widetilde\lambda_{2}^2+\frac{3}{10}\Big(\frac{1}{10}g_1^4-g_1^2 g_2^2\nonumber\\
&\qquad +\frac{15}{2}g_2^4\Big)-\lambda_{2}\Big(g_1^2+9g_2^2+8g_3^2\Big)+12\lambda_{h}\Big(\lambda_{2}\nn \\ & \qquad +\frac{1}{3}\widetilde\lambda_{2}\Big)+6\lambda_{2}\Tr\Big(\X_\uq+\X_\dq+\frac{1}{3}\X_\cl+\frac{1}{3}\X_2\Big) \nn\\
&\qquad   -4\Tr \Big(\X_2\X_\dq+\widetilde{\X}_\cl\widetilde{\X}_2\Big)\bigg],
\end{align}
\begin{align}
\label{eq:wl2r211}
\beta(\widetilde\lambda_{2})&^{1-loop}_{\widetilde{R}_2,1-gen}=\frac{1}{16\pi^2}\bigg[\frac{3}{5}g_1^2 g_2^2+4\Tr\Big(\wX_2\wX_\cl\Big)+ \wl_2\Big\{8\lambda_{2}\nn\\
&\qquad+4\wl_2-g_1^2-9g_2^2-8g_3^2+4\lambda_{h}+6\Tr\Big(\X_{\uq}+\X_{\dq}\nonumber\\
&\qquad+\frac{1}{3}\X_\cl+\frac{1}{3}\X_2\Big)\Big\}\bigg],
\end{align}
For the three generation case $\lambda_{2}$ and $\wl_2$ become two $3\times3$ matrices whose $ij$-th element indicates the quartic coupling of $i$-th and $j$-th generations of $\widetilde{R}_2$ with two Higgs fields. However, as mentioned earlier, we  restrict our parameter space with no mixing among the generations of leptoquarks at the initial scale; therefore, $\lambda_{2}$ and $\wl_2$ become two diagonal matrices. The one-loop beta functions for these two parameters are simply given by:
\begin{align}
\label{eq:l2r213}
\beta(\lambda_{2}^{ii})^{1-loop}_{\widetilde{R}_2,3-gen}=\bigg[\beta(\lambda_{2})^{1-loop}_{\widetilde{R}_2,1-gen}\bigg]_{\is}\nn\\ \beta(\wl_{2}^{ii})^{1-loop}_{\widetilde{R}_2,3-gen}=\bigg[\beta(\wl_{2})^{1-loop}_{\widetilde{R}_2,1-gen}\bigg]_{\is}
\end{align}
The full two-loop beta functions for these two parameters with both one and three generations are presented in \autoref{sec:H-LQcouplR2}.

Now, we study the variation of quartic coupling among the leptoquark and Higgs with perturbative scale i.e the scale at which any of the coupling diverges. The variations of the quartic couplings $\lambda_2$ and $\wl_2$ for three generations of doublet leptoquark are explained in Fig.\ref{fig:r2l2}. In the first two plots, Fig. \ref{f5a} and  \ref{f6a},  $\lambda_2$ corresponds to quartic coupling term for one particular generation of leptoquark while $\lambda_2^{jj}$ denotes the remaining generations of $\lambda_2$ and all the generations of other quartic coupling term $\wl_2$ are designated as $\wl_2^{ii}$. Similarly, for $\wl_2$ variation in Fig. \ref{f7a} and  \ref{f8a}, $\wl_2$ corresponds to any particular generation of leptoquark while the remaining generations are denoted by $\wl_2^{jj}$ and the other quartic coupling terms $\lambda_2^{ii}$ signify $\lambda_{2}$ for all three generations. The plots in left panel indicate relatively low value of Yukawa, i.e. $Y_\phi=0.1$ whereas the same in right panel illustrate the variation of the mentioned couplings for higher value of Yukawa, i.e. $Y_\phi=1.0$.

In the first two plots, the initial value of $\lambda_2$ is varied from 0.1 to 0.8 keeping the values for other quartic couplings at EW scale to be 0.01, 0.1, 0.4 and 0.8 which are depicted by red, blue, orange and green curves respectively. As can be observed from Eqs. \eqref{eq:l2r211} and \eqref{eq:l2r213} that one-loop beta function of $\lambda_{2}$ receives enhanced contributions from positive valued $\wl_{2}$ and hence $\lambda_{2}$ reaches non-perturbativity quickly for larger values of $\wl_{2}$. It should be noticed from Fig: \ref{f5a} that, for $(\lambda_2^{jj},\wl_2^{ii})$=0.01 and 0.1 at the EW scale, the theory remains perturbative till Planck scale for  $\lambda_2\leq0.62$ and 0.52 respectively with $Y_\phi=0.1$. As we increase the EW values to 0.4 and 0.8, the positive contribution from quartic couplings makes the theory non-perturbative at $\sim 10^{12}$ GeV, $10^{7}$ GeV for lower initial values of $\lambda_2$. For higher EW values of $\lambda_{2}$, this perturbative scale decreases slowly. The variation of $\lambda_2$ with perturbative scale for $Y_\phi=1.0$, as displayed in Fig. \ref{f6a}, looks quite similar to the previous case. However, as can be seen from Eqs. \eqref{eq:l2r211} and \eqref{eq:l2r213}, the one-loop beta function of $\lambda_{2}$ obtains positive contributions from $2\lambda_{2} \Tr \,\X_2$ term (since $Y_\dq$ and $Y_\cl$ are negligible) and therefore, $\lambda_{2}$ becomes non-perturbative at slightly lower energy scale than previous case. In this case, $\lambda_2$ is bounded above to 0.56 and 0.47 for EW values of other quartic couplings $\lambda_2^{jj},\wl_2^{ii}$ to be 0.01 and 0.1 respectively. Further increases in EW values to 0.4 and 0.8 make the theory non-perturbative around $10^{11.2}$ GeV and $10^{6.9}$ GeV respectively for lower initial values of $\lambda_{2}$, and the scale diminishes gently with higher initial values of $\lambda_{2}$.  It is worth mentioning that the non-perturbativity of $\lambda_{2}$ and $\wl_2$, attained with three generations of $\widetilde{R}_2$, is not a result of any Landau pole, which  is also apparent from the different positioning of the non-perturbative scales compared to that of the gauge couplings. 

In a similar fashion, $\wl_{2}$ is altered gradually from 0.1 to 0.8 in the last two plots fixing the values of other quartic couplings to be 0.01, 0.1, 0.4 and 0.8 which are depicted by red, blue, orange and green curves respectively. In this case $\lambda_{2}$ provides positive effect in the running of $\wl_2$, see Eqs. \eqref{eq:wl2r211} and \eqref{eq:l2r213}; therefore, $\wl_2$ moves to non-perturbative region in a faster way for higher values of $\lambda_{2}$. On the other hand, $Y_2$ also contributes positively trough the term $2\wl_2\,\Tr \X_2$ and hence, $\wl_2$ hits non-perturbativity at slightly lower energy scale for higher Yukawa coupling. Form Fig. \ref{f7a} we see that the demand of Planck scale perturbativity constrains $\wl_2$ to be smaller than 0.8 and 0.66 if the EW values of other quartic couplings $(\wl_2^{jj}, \lambda_2^{ii})$ are set to be 0.01 and 0.1 respectively with $Y_\phi=0.1$. With higher values of $\wl_2^{jj}, \lambda_2^{ii}$ at EW scale, i.e. 0.4 and 0.8, the model becomes non-perturbative at much lower energy than Planck scale. From Fig. \ref{f8a}, in comparison with Fig. \ref{f7a}, we observe that $\wl_2$ is restricted to slightly lower values, i.e. 0.73 and 0.59, if we begin with 0.01 and 0.1 respectively for the EW values of other quartic couplings along with $Y_\phi=1.0$. The statement with higher initial values of the quartic couplings remains valid in this scenario too.  

\subsubsection{Perturbativity of \texorpdfstring{$\vec S_3$}{} }
\label{perbS3}

In this section, we scrutinize the RG evolution of Higgs-leptoquark quartic couplings for $\vec S_3$, namely $\lambda_{3}$ and $\wl_{3}$. These two parameters also should be bounded above by $4\pi$. The one-loop beta functions for these two parameters in one generation case are given below:
\begin{align}
\label{eq:l3s31}
\beta(\lambda_{3})&^{1-loop}_{\vec S_3,1-gen}=\frac{1}{4\pi^2}\bigg[\lambda_{3}^2+\frac{1}{4}\,\widetilde{\lambda}_3^2+\frac{3}{4}\,\Big(\frac{1}{25}g_1^4-\frac{2}{5}g_1^2g_2^2+2g_2^4\Big)\nn\\
&-\frac{1}{4}\,\lambda_{3}\Big(\frac{13}{10}g_1^2+\frac{33}{2}g_2^2+8g_3^2\Big)+3\lambda_{h}\Big(\lambda_{3}+\frac{1}{3}\widetilde{\lambda}_3\Big)\nonumber\\
&+\frac{3}{2}\,\lambda_{3}\Tr\Big(\X_\uq+\X_\dq+\frac{1}{3}\X_\cl+\frac{1}{3}\X_3\Big)-\Tr\,\Big(\widetilde{\X}_3\widetilde{\X}_\cl\nn\\
&+\X_3\widetilde{\X}_\dq^T\Big)\bigg],
\end{align}
\begin{align}
\label{eq:wl3s31}
\beta(\widetilde\lambda_{3})&^{1-loop}_{\vec S_3,1-gen}=\frac{1}{4\pi^2}\bigg[\widetilde\lambda_{3}^2+2\,\lambda_{3}\widetilde{\lambda}_3+\lambda_{h}\widetilde{\lambda}_3+\frac{3}{5}\,g_1^2g_2^2\nn\\
&-\frac{1}{4}\,\widetilde\lambda_{3}\Big(\frac{13}{10}g_1^2+\frac{33}{2}g_2^2+8g_3^2\Big)+\frac{3}{2}\,\widetilde\lambda_{3}\Tr\Big(\X_\uq+\X_\dq\nonumber\\
&+\frac{1}{3}\X_\cl+\frac{1}{3}\X_3\Big)+\Tr\,\Big(\widetilde{\X}_3\widetilde{\X}_\cl+\X_3\widetilde{\X}_\dq^T-\X_3\widetilde{\X}_\uq^T\Big)\bigg]
\end{align}
Like the doublet leptoquark case, for three generations scenario, $\lambda_{3}$ and $\wl_{3}$ become two $3\times3$ matrices whose  $ij$-th element indicates the quartic coupling of $i$-th and $j$-th generations of $\vec S_3$ with two Higgs fields. Nevertheless, as mentioned earlier, we have restricted our parameter space with no mixing among the generations of leptoquarks at the initial scale making $\lambda_{3}$ and $\wl_3$ to be two diagonal matrices. The one-loop beta functions for these two parameters are simply given by:
\begin{align}
\label{eq:l3s33}
\beta(\lambda_{3}^{ii})^{1-loop}_{\vec S_3,3-gen}=\bigg[\beta(\lambda_{3})^{1-loop}_{\vec S_3,1-gen}\bigg]_{\is} \nn \\ \beta(\wl_{3}^{ii})^{1-loop}_{\vec S_3,3-gen}=\bigg[\beta(\wl_{3})^{1-loop}_{\vec S_3,1-gen}\bigg]_{\is}
\end{align}
The full two-loop beta functions for $\vec S_3$ with both one and three generations are presented in \autoref{sec:H-LQcouplS3}.

\begin{figure*}[h!]
	\begin{center}
		\mbox{\subfigure[$Y_\phi =0.1$]{\includegraphics[width=0.5\linewidth,angle=-0]{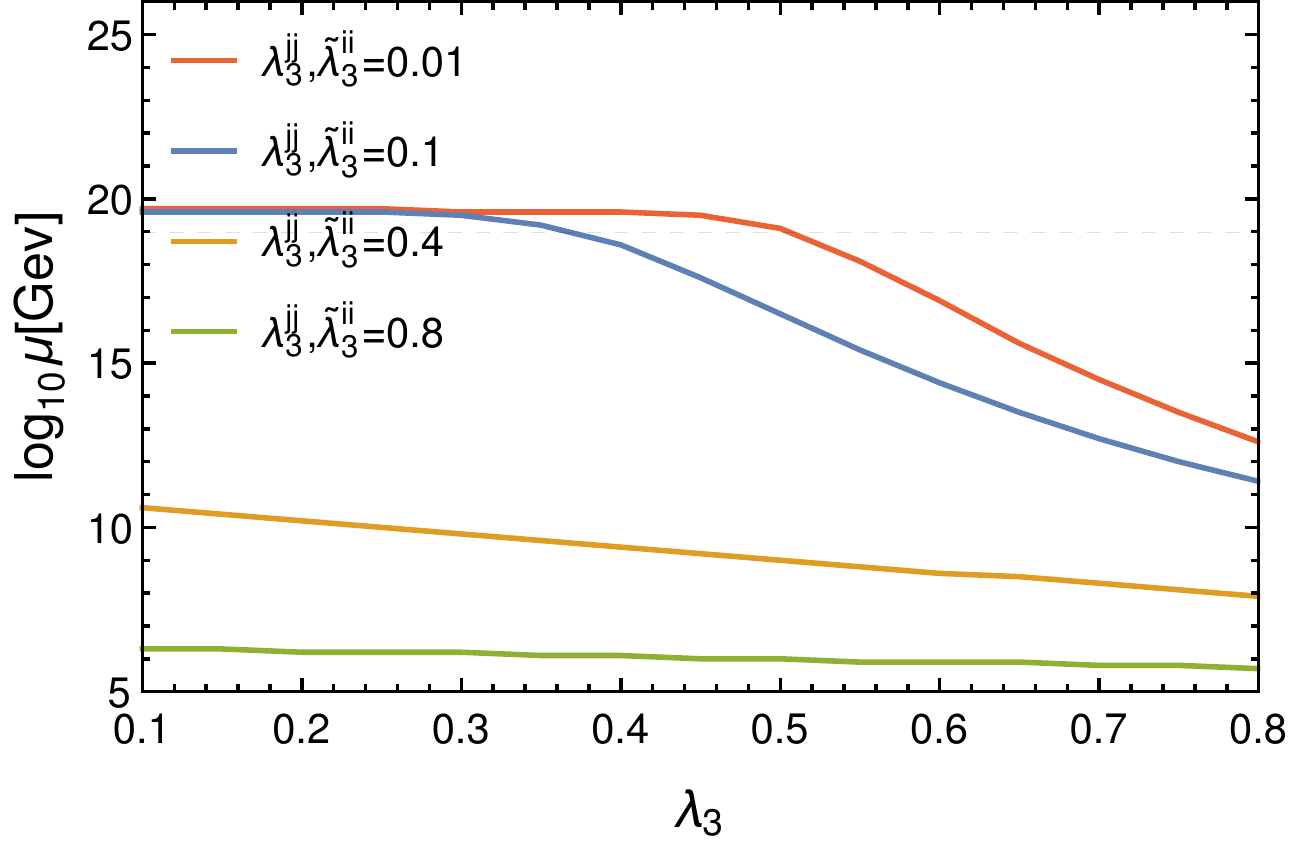}\label{f9a}}
			\subfigure[$Y_\phi =0.8$]{\includegraphics[width=0.5\linewidth,angle=-0]{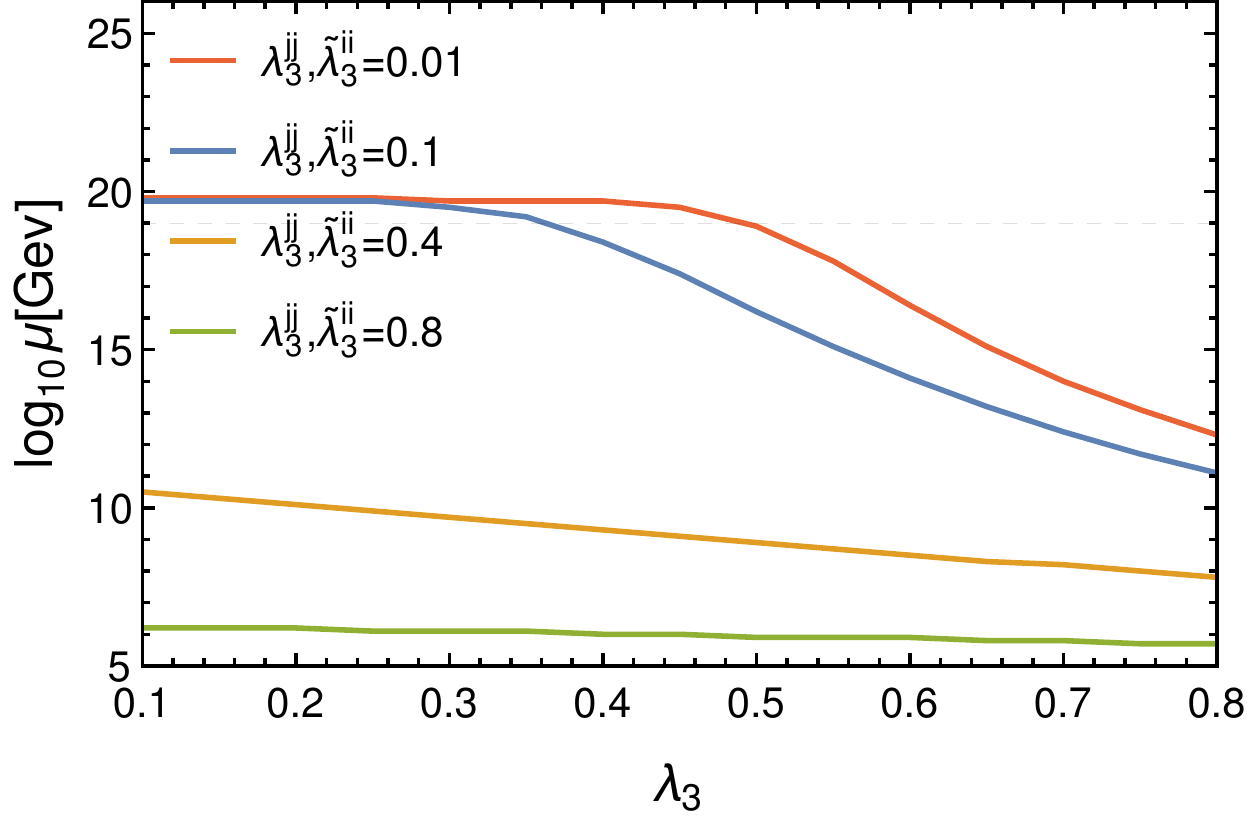}\label{f10a}}}
		\mbox{\subfigure[$Y_\phi =0.1$]{\includegraphics[width=0.5\linewidth,angle=-0]{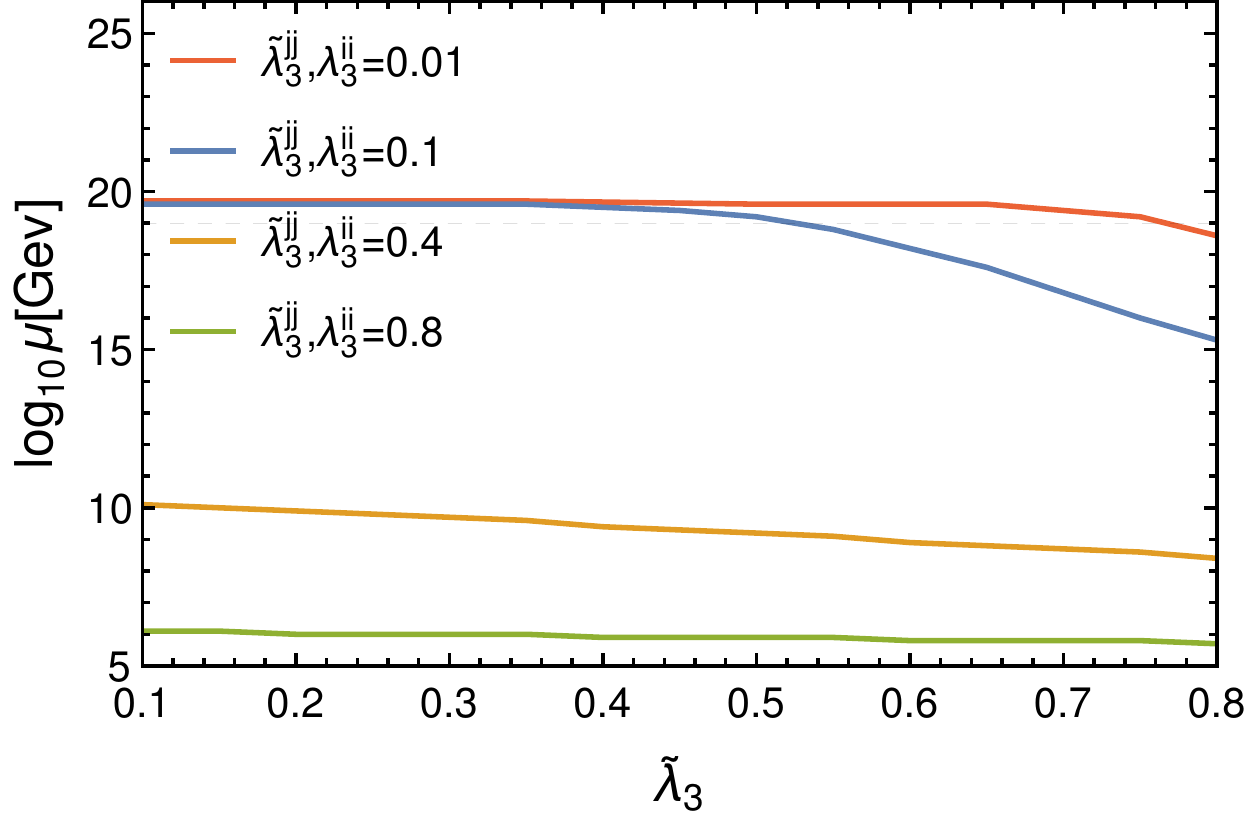}\label{f11a}}
			\subfigure[$Y_\phi =0.8$]{\includegraphics[width=0.5\linewidth,angle=-0]{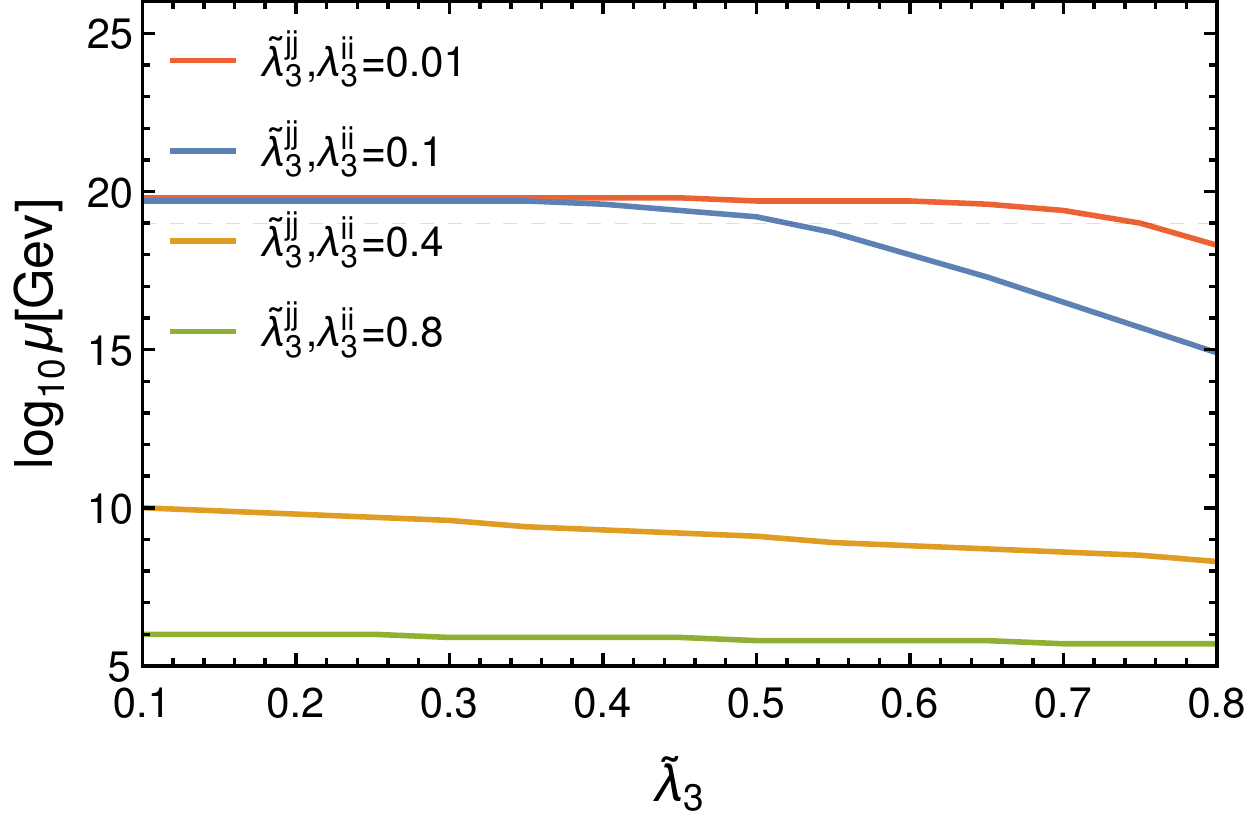}\label{f12a}}}
		\caption{Variation of leptoquark-Higgs quartic coupling $\lambda_3$ and $\wl_3$ for triplet leptoquark $\vec S_3$ with the perturbative scale. For the plots in first row, $\lambda_3$ variation is considered for any particular generation of leptoquark and the same for remaining generations are denoted by $\lambda_3^{jj}$. The other leptoquark-Higgs quartic couplings $\wl_3^{ii}$ include all three leptoquark generations. Similarly, for the plots in second row, the variation of quartic coupling $\wl_3$ for any particular leptoquark generation is depicted while symbolizing the same for the remaining generations by $\wl_3^{jj}$. The other quartic coupling $\lambda_3^{ii}$ includes all three generations of leptoquark. The variations are considered for four different initial values, i.e. 0.01, 0.1, 0.4 and 0.8, at EW scale which are described by red, blue, orange and green curves respectively. Here, two different values for $Y_\phi$ have been considered which are 0.1 and 0.8.}\label{fig:s3l3}
	\end{center}
\end{figure*}

Now, we consider the variation of quartic coupling between the leptoquark $\vec S_3 $ and Higgs with perturbative scale and it has been illustrated in Fig.\ref{fig:s3l3} for three generations case. In the first two plots, Fig. \ref{f9a} and  \ref{f10a}, $\lambda_3$ corresponds to quartic coupling term for one particular generation of leptoquark while $\lambda_3^{jj}$ denote the remaining generations of $\lambda_3$ and all the generations of other quartic coupling term $\wl_3$ are designated as $\wl_3^{ii}$. Similarly, for $\wl_3$ variation in Fig. \ref{f11a} and  \ref{f12a}, $\wl_3$ corresponds to any particular generation of leptoquark while the remaining generations are denoted by $\wl_3^{jj}$ and the other quartic coupling terms $\lambda_3^{ii}$ signify $\lambda_{3}$ for all three generations. The plots in left panel indicate relatively low value of Yukawa, i.e. $Y_\phi=0.1$ whereas the same in right panel illustrate the variation of the mentioned couplings for higher value of Yukawa, i.e. $Y_\phi=0.8$.

\begin{figure*}[h!]
	\begin{center}
		\mbox{\subfigure[$Y_\phi =0.1$]{\includegraphics[width=0.5\linewidth,angle=-0]{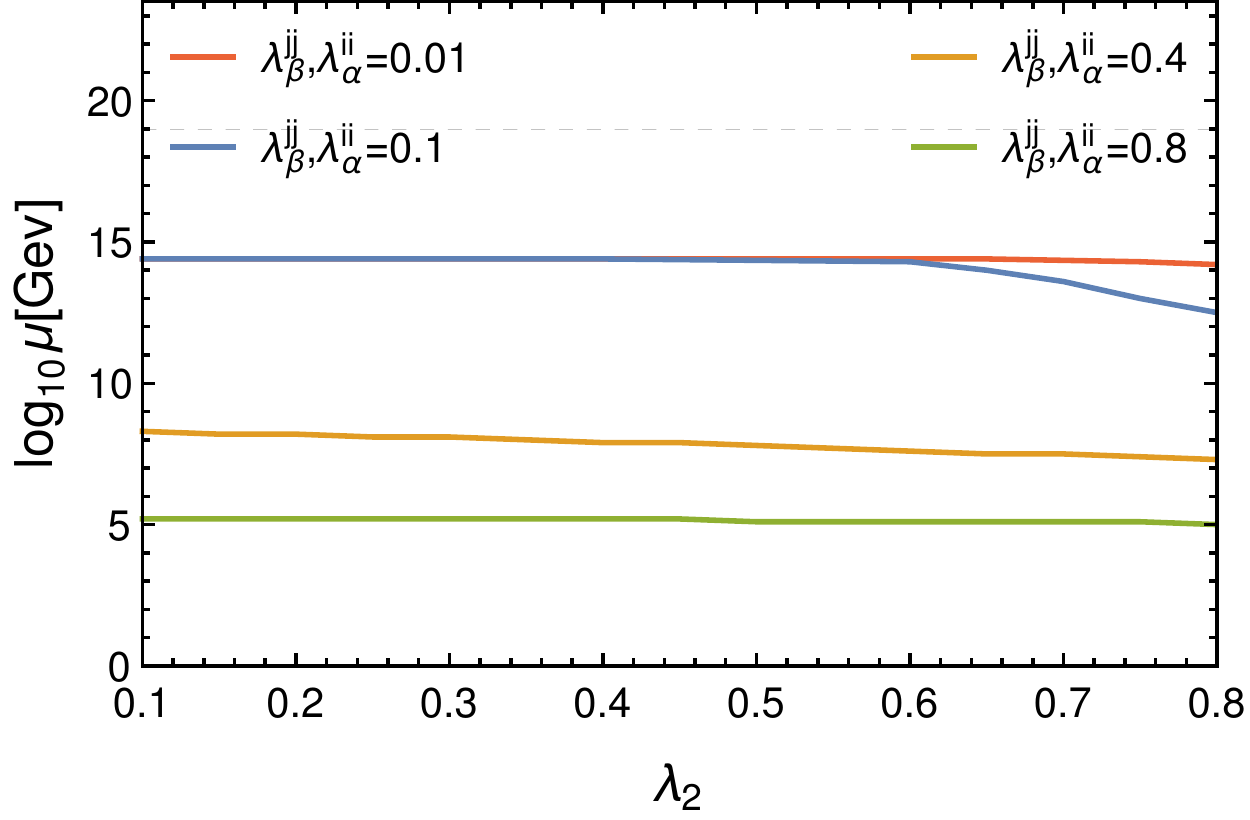}\label{f13a}}
			\subfigure[$Y_\phi =0.1$]{\includegraphics[width=0.5\linewidth,angle=-0]{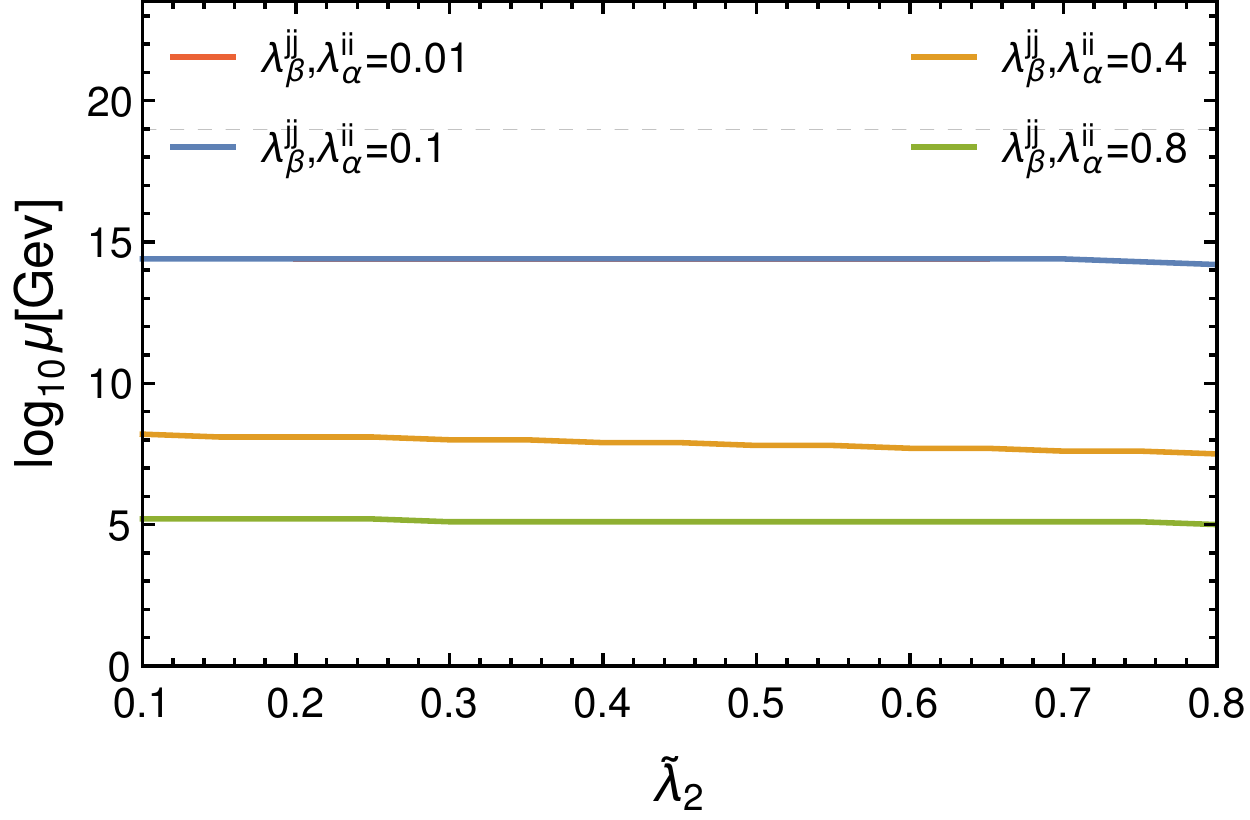}\label{f14a}}}
		\mbox{\subfigure[$Y_\phi =0.1$]{\includegraphics[width=0.5\linewidth,angle=-0]{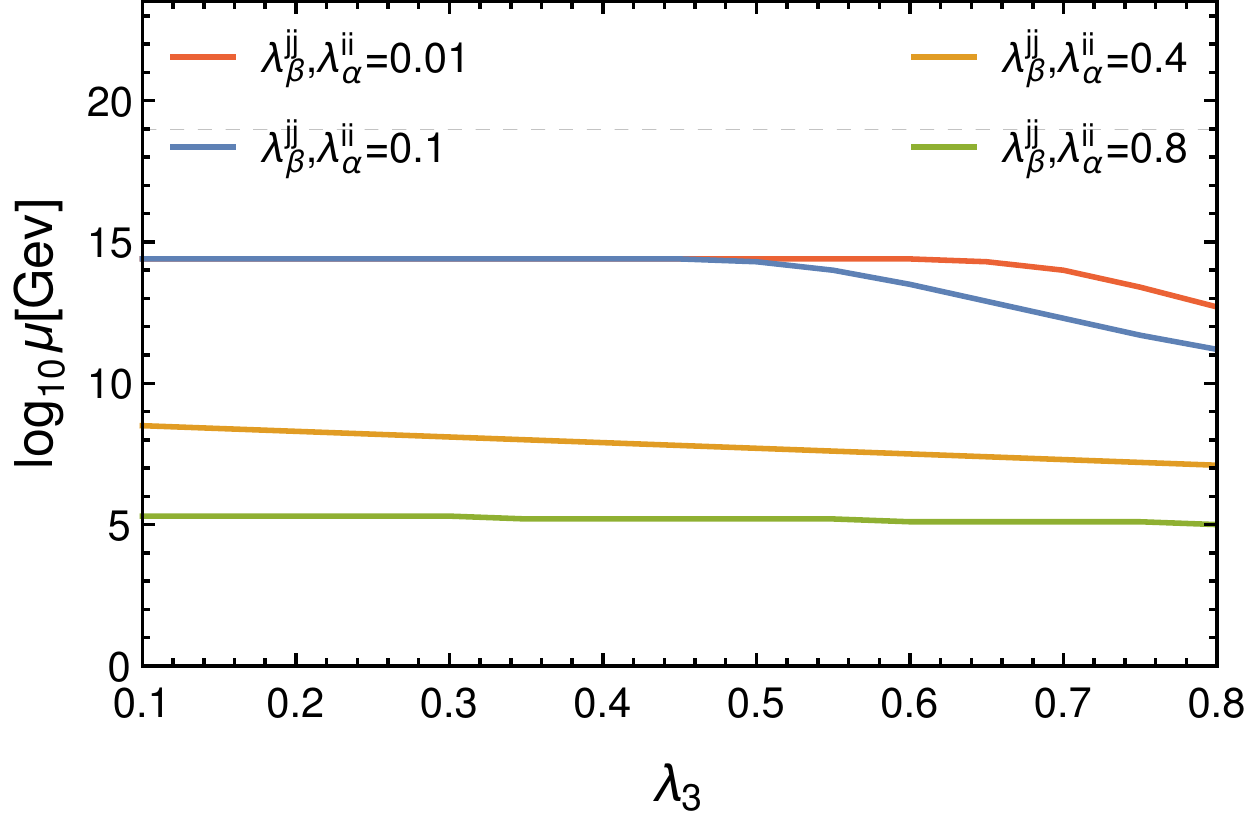}\label{f15a}}
			\subfigure[$Y_\phi =0.1$]{\includegraphics[width=0.5\linewidth,angle=-0]{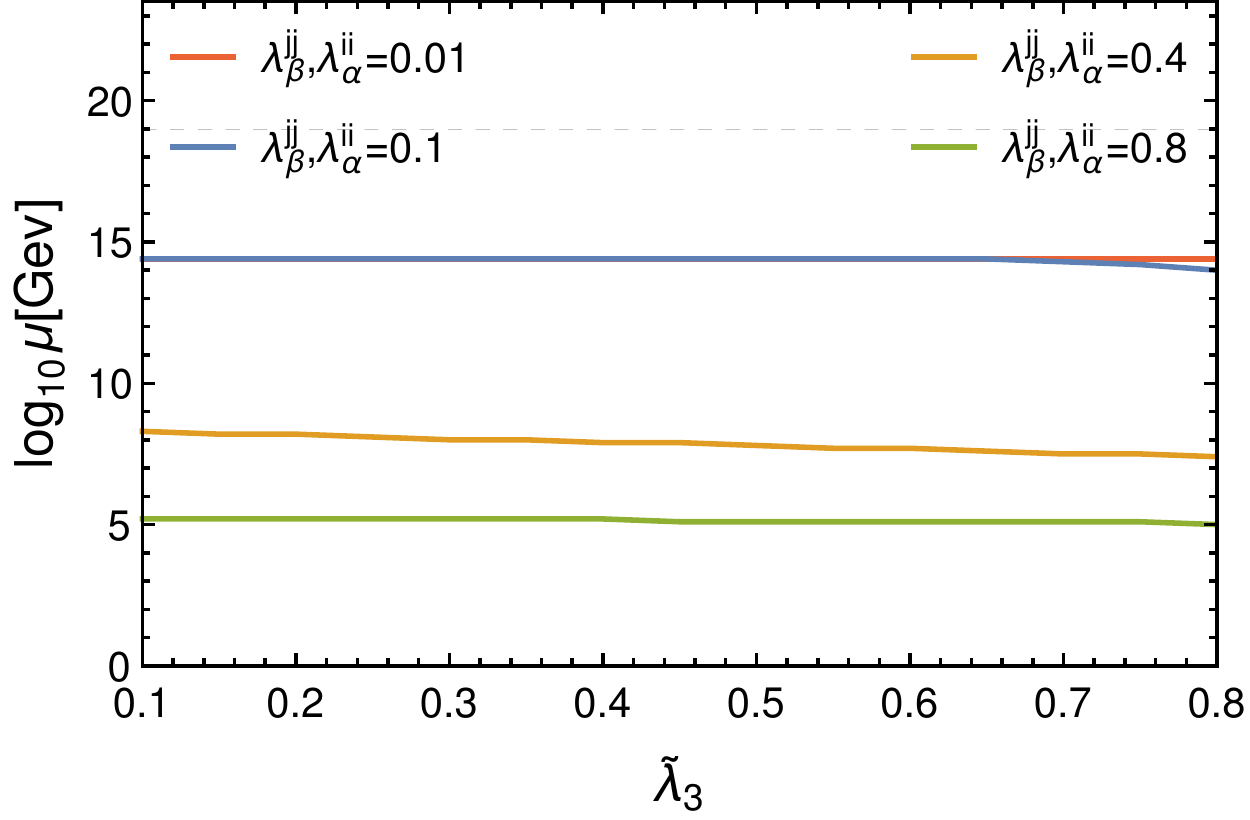}\label{f16a}}}
		\caption{Variation of quartic coupling $\lambda_2, \wl_2, \lambda_3$ and $\wl_3$ with perturbative scale for three generations of $\widetilde R_2 + \vec S_3$. Here, $\lambda_2$ variation is shown for any one generation of $\widetilde R_2$ and remaining generations for $\lambda_2$ term of $\vec R_2$ are defined  by $\lambda_\beta^{jj}$. The $\lambda_\alpha^{ii}$ term corresponds to three generations of $\widetilde R_2$ for $\wl_2$ term and three generations of $\vec S_3$ for $\lambda_3$ and $\wl_3$ term. Again the $\wl_2$ variation is depicted for any one generation of $\widetilde R_2$ and remaining generations of $\widetilde R_2$ are given by $\lambda_\beta^{jj}$ for $\wl_2$ terms. In this case, the $\lambda_\alpha^{ii}$ term corresponds to three generations of $\widetilde R_2$ for $\lambda_2$ terms and three generations of $\vec S_3$ for $\lambda_3$ and $\wl_3$ terms. Similar notation has been followed for the variation of $\lambda_3$ and $\wl_3$. The EW  scale values for the quartic couplings other than the coupling whose variation is considered are set to four different values i.e 0.01, 0.1, 0.4 and 0.8 that are illustrated by red, blue, orange and green curves respectively taking $Y_\phi$=0.1. Here, $Y_\phi$ signifies both $Y_2$ and $Y_3$ with three generations.}\label{fig:s3r2l23}
	\end{center}
\end{figure*}
In the first two plots, Fig. \ref{f9a} and \ref{f10a}, we have gradually varied the initial values for $\lambda_{3}$ from 0.1 to 0.8 keeping the EW values for other quartic couplings to be 0.01, 0.1, 0.4 and 0.8 respectively, which are presented by red, blue, orange and green lines. The similar things for $\wl_{3}$ are presented in Fig. \ref{f11a} and \ref{f12a}. As we have already shown in the earlier sections, all the other couplings for three generations of triplet leptoquark diverge at $10^{19.7}$ GeV due to the gauge coupling $g_2$. The couplings $\lambda_{3}$ and $\wl_{3}$ are also not different from that behaviour. Therefore, unlike $\widetilde{R}_2$ case, here $\lambda_{3}$ and $\wl_3$ diverge at $10^{19.7}$ GeV for any smaller initial values of $\lambda_{3}$ and $\wl_{3}$ at EW scale with any value of Yukawa coupling $Y_\phi$. Now, as can be noticed from Eqs. \eqref{eq:l3s31} and \eqref{eq:l3s33}, $\wl_{3}$ contributes positively in the one-loop beta function of $\lambda_{3}$ and hence  $\lambda_{3}$ reaches non-perturbativity at early stage with higher values of $\wl_{3}$. On the other hand, due to positive effect of $\X_3$ at one-loop order, all the lines shift slightly downward with higher values of Yukawa couplings but the shifts are almost unnoticeable. Both of the above statements are true for running of $\wl_{3}$ also. For $\lambda_{3}$, Planck scale perturbativity is achieved till 0.51 and 0.37 with other quartic coupling at EW scale being 0.01 and 0.1 respectively for both the Yukawa coupling. However, for higher values of other quartic couplings at EW scale, i.e. 0.4 and 0.8, $\lambda_3$ diverges at much lower scale, like  $10^{10.8}$ GeV and $10^{6.4}$ GeV, with its lower initial values, and this decreases with enhancement in beginning value of $\lambda_3$. Likewise, the quartic coupling $\widetilde \lambda_3$ is constrained to 0.76 and 0.52 for Planck scale perturbativity with EW values of other quartic couplings to be 0.01 and 0.1 respectively. For higher EW values of quartic couplings the theory becomes non-perturbative at much lower scales as previously discussed.

\subsubsection{Perturbativity of \texorpdfstring{$\widetilde{R}_2+\vec S_3$}{} with 3-gen}

Now, we move to the combine combined scenario of $\widetilde{R}_2$ and $\vec S_3$ with three generations. The one-loop beta functions for all the Higgs-leptoquark quartic couplings in this case can easily be written as:
\begin{align}
&\beta(\lambda_{2}^{ii})^{1-loop}_{\widetilde{R}_2+\vec S_3,3-gen}=\beta(\lambda_{2}^{ii})^{1-loop}_{\widetilde{R}_2,3-gen},\nn\\
&\beta(\wl_{2}^{ii})^{1-loop}_{\widetilde{R}_2+\vec S_3,3-gen}=\beta(\wl_{2}^{ii})^{1-loop}_{\widetilde{R}_2,3-gen},\nonumber\\
&\beta(\lambda_{3}^{ii})^{1-loop}_{\widetilde{R}_2+\vec S_3,3-gen}=\beta(\lambda_{3}^{ii})^{1-loop}_{\vec S_3,3-gen},\nn\\
&\beta(\wl_{3}^{ii})^{1-loop}_{\widetilde{R}_2+\vec S_3,3-gen}=\beta(\wl_{3}^{ii})^{1-loop}_{\vec S_3,3-gen}.
\end{align}
The full two-loop beta functions of all the Higgs-leptoquark quartic couplings in this 
scenario are listed in \autoref{sec:H-LQcouplR2S3}.

\begin{figure*}[h!]
	\begin{center}
		\mbox{\subfigure[$Y_\phi =0.1$]{\includegraphics[width=0.5\linewidth,angle=-0]{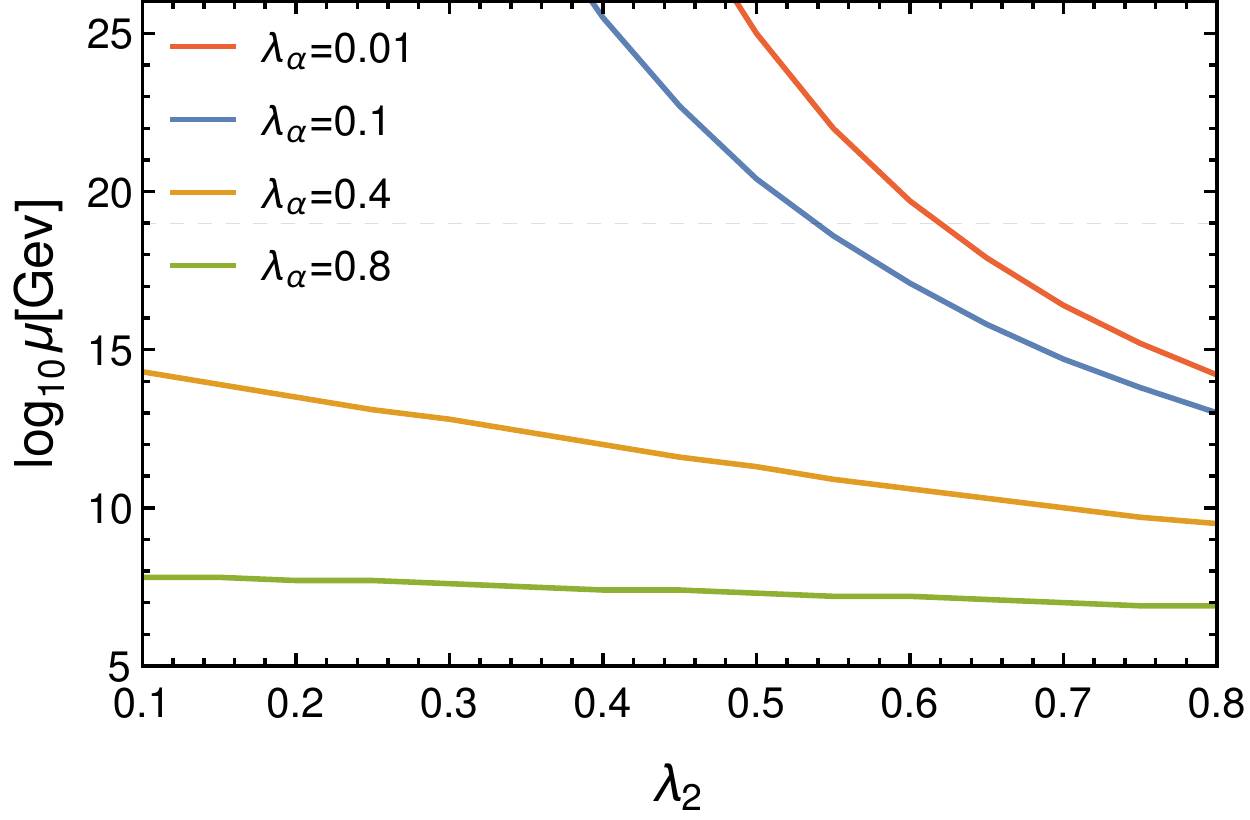}\label{f17a}}
			\subfigure[$Y_\phi =0.8$]{\includegraphics[width=0.5\linewidth,angle=-0]{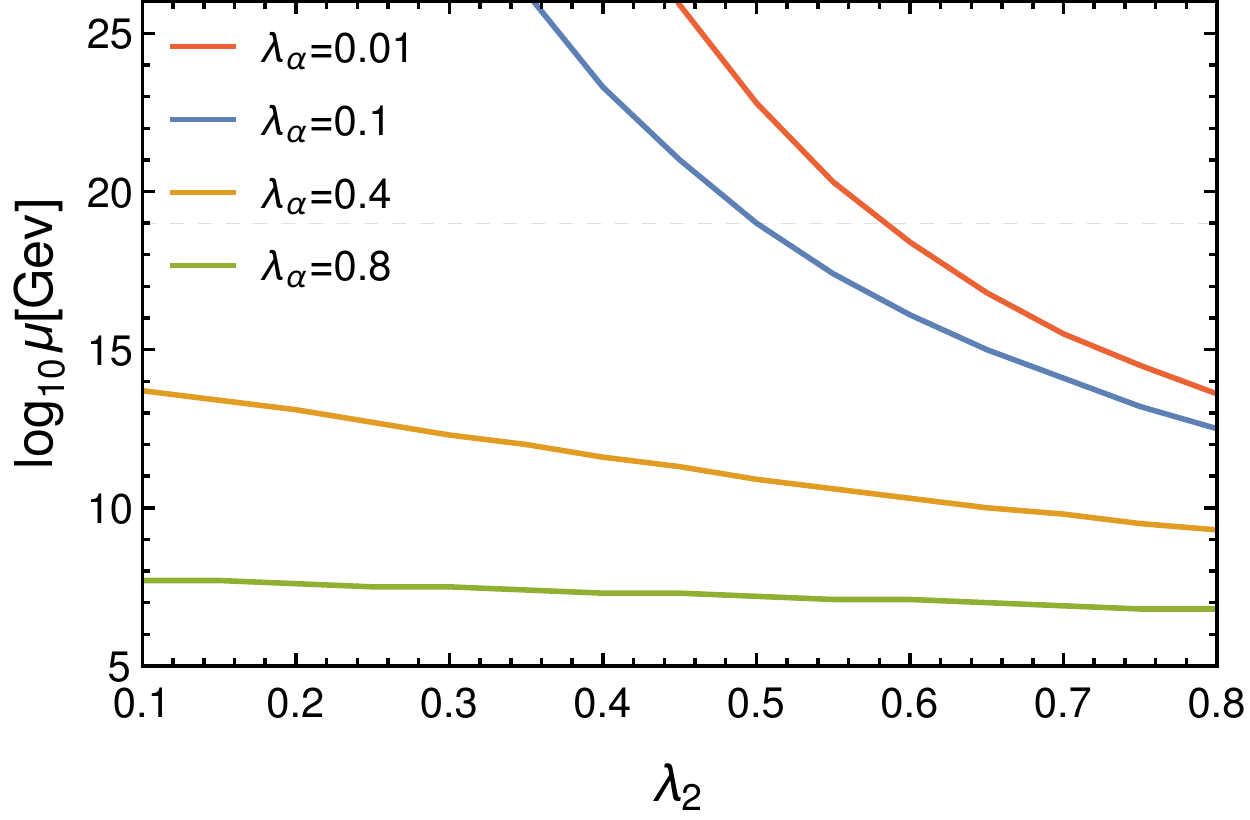}\label{f18a}}}
		\mbox{\subfigure[$Y_\phi =0.1$]{\includegraphics[width=0.5\linewidth,angle=-0]{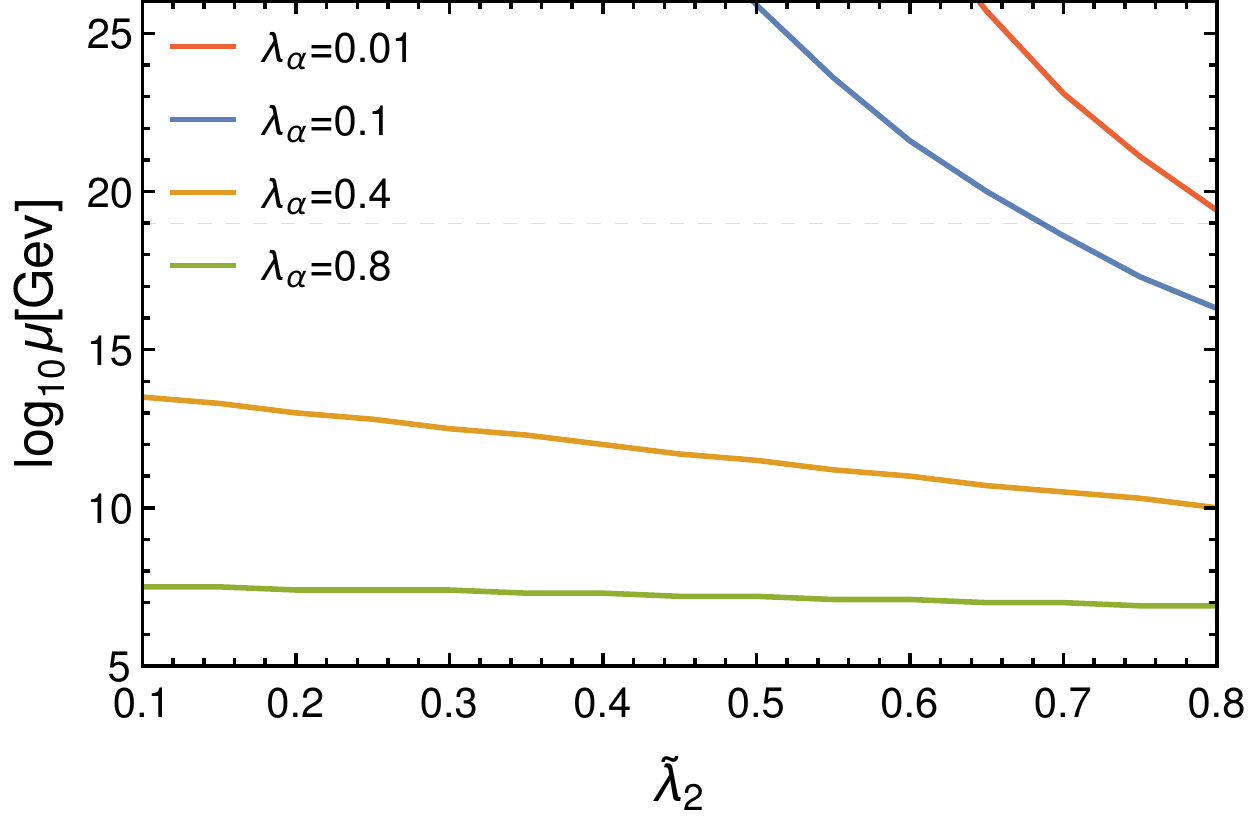}\label{f19a}}
			\subfigure[$Y_\phi =0.8$]{\includegraphics[width=0.5\linewidth,angle=-0]{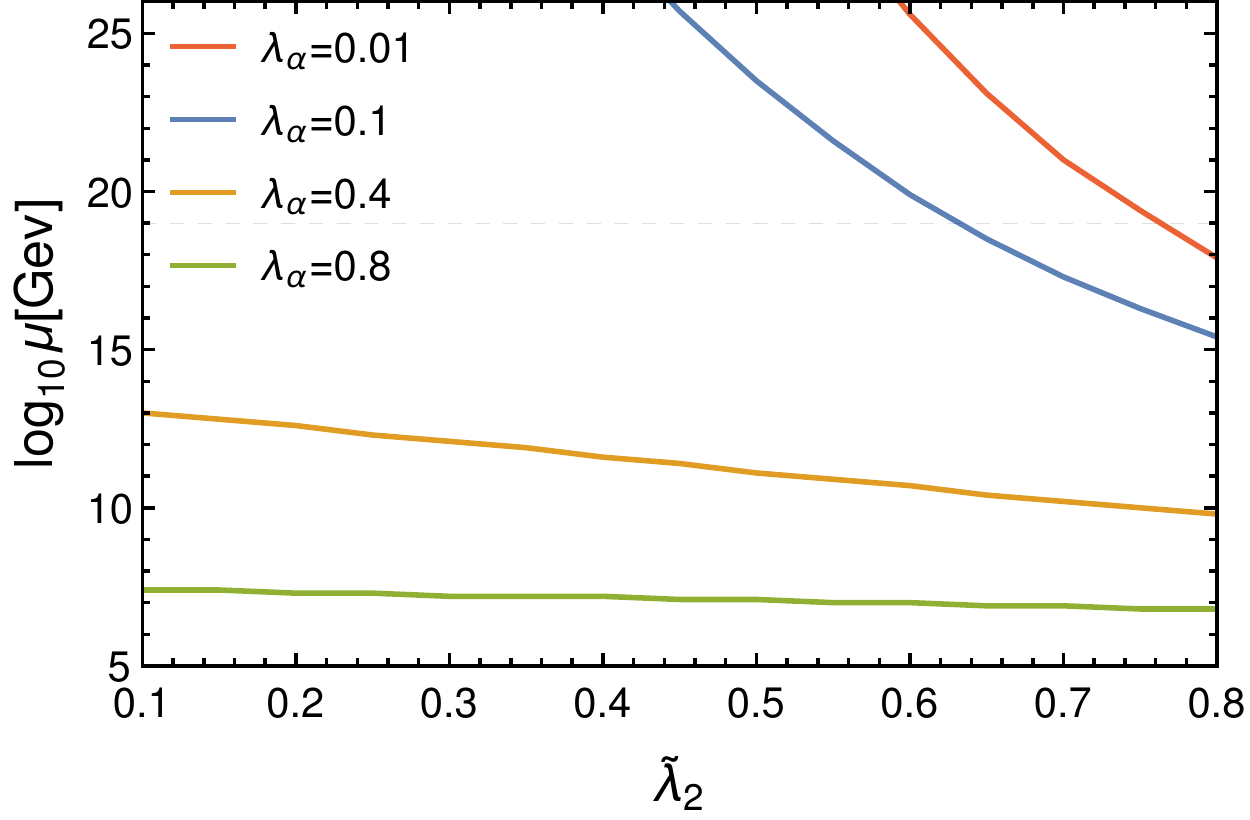}\label{f20a}}}
		\caption{Variation of quartic couplings $\lambda_2$ and $\wl_2$ with the perturbative scale for one generation of $\widetilde R_2 + \vec S_3$. Here, For $\lambda_2$ variation, $\lambda_\alpha$ corresponds to \{$\wl_2, \lambda_3$, $\wl_3$\}, and if we consider the behaviour of $\wl_2$ then $\lambda_\alpha$ includes $\lambda_2, \lambda_3$ and $\wl_3$. All quartic couplings other than for which variation is considered are assigned four different values at the EW scale i.e 0.01, 0.1, 0.4 and 0.8 and these are delineated by red, blue, orange and green curves respectively. Two different values of $Y_\phi$ have been i.e 0.1 and 0.8 with $Y_\phi$ representing the Yukawa couplings for both the leptoquarks.}\label{fig:l2r2s31}
	\end{center}
\end{figure*}

\begin{figure*}[h!]
	\begin{center}
		\mbox{\subfigure[$Y_\phi =0.1$]{\includegraphics[width=0.5\linewidth,angle=-0]{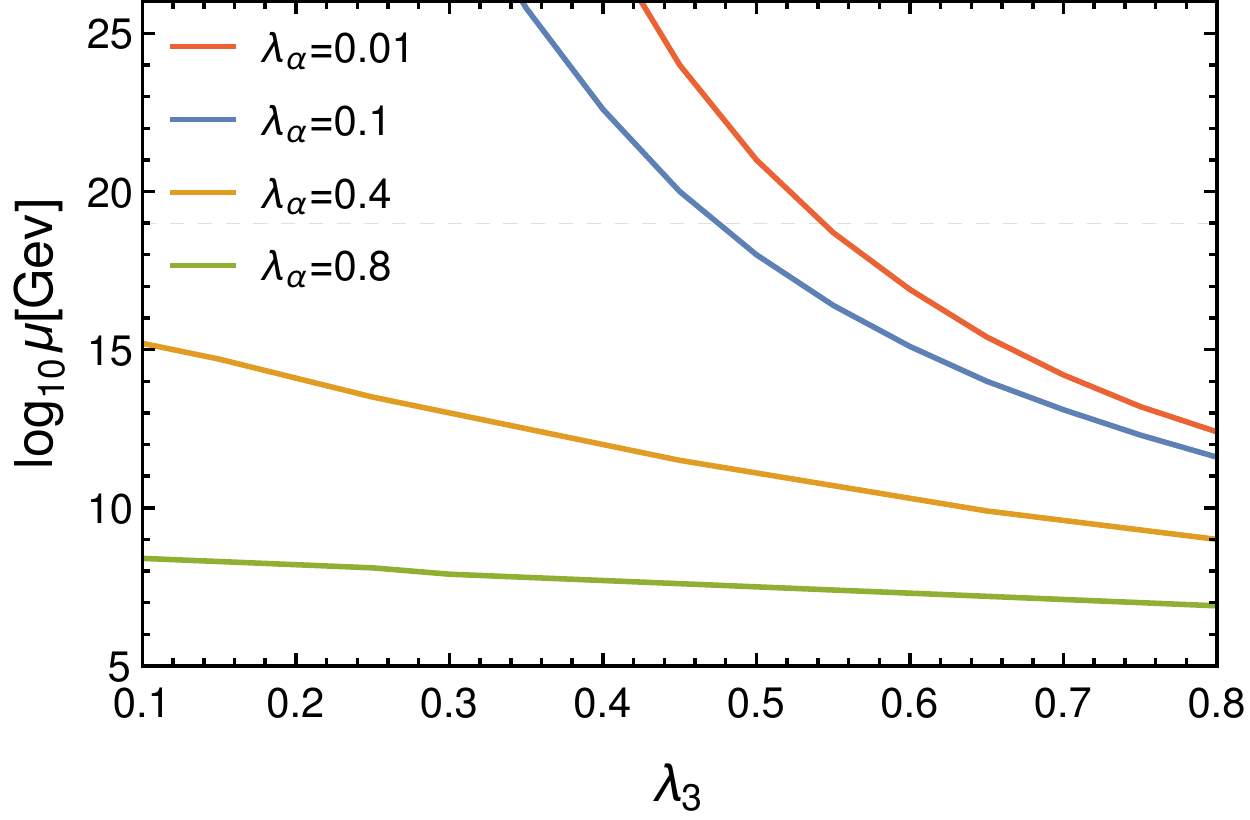}\label{f21a}}
			\subfigure[$Y_\phi =0.8$]{\includegraphics[width=0.5\linewidth,angle=-0]{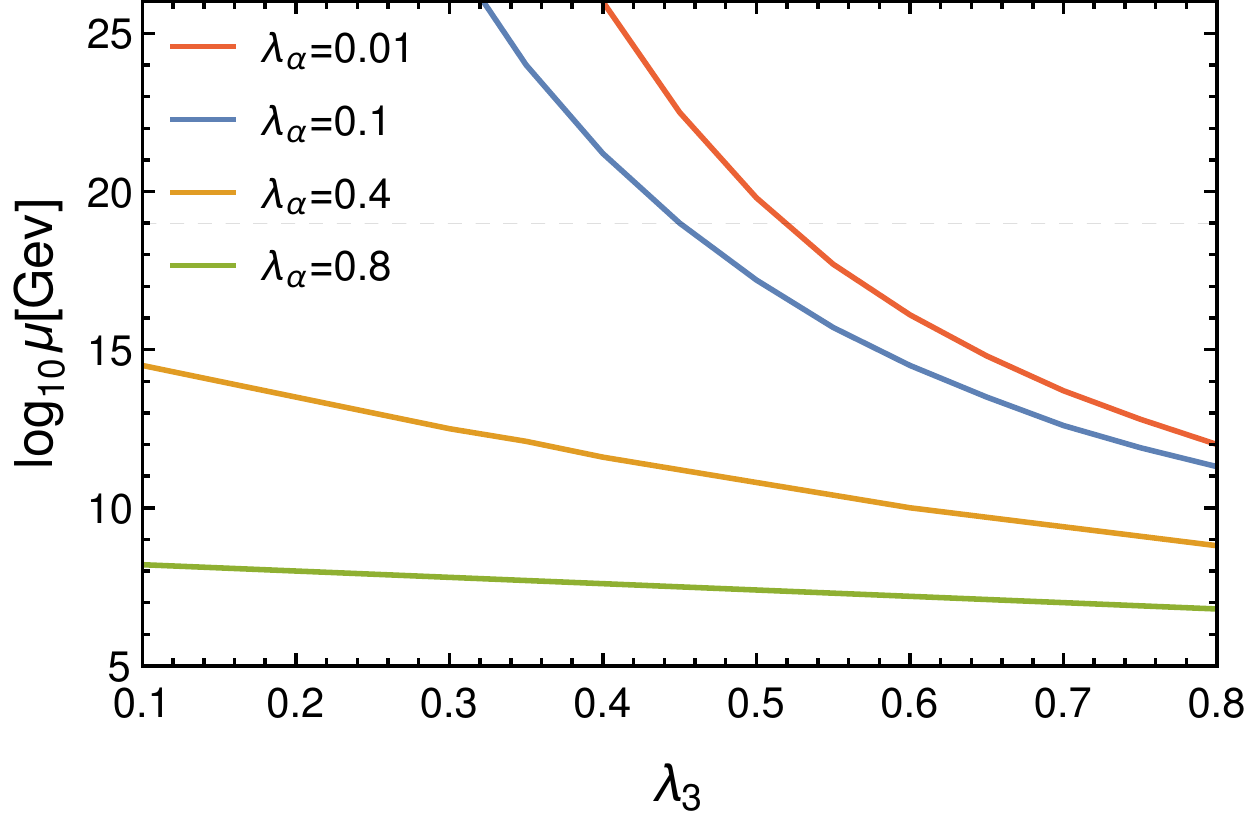}\label{f22a}}}
		\mbox{\subfigure[$Y_\phi =0.1$]{\includegraphics[width=0.5\linewidth,angle=-0]{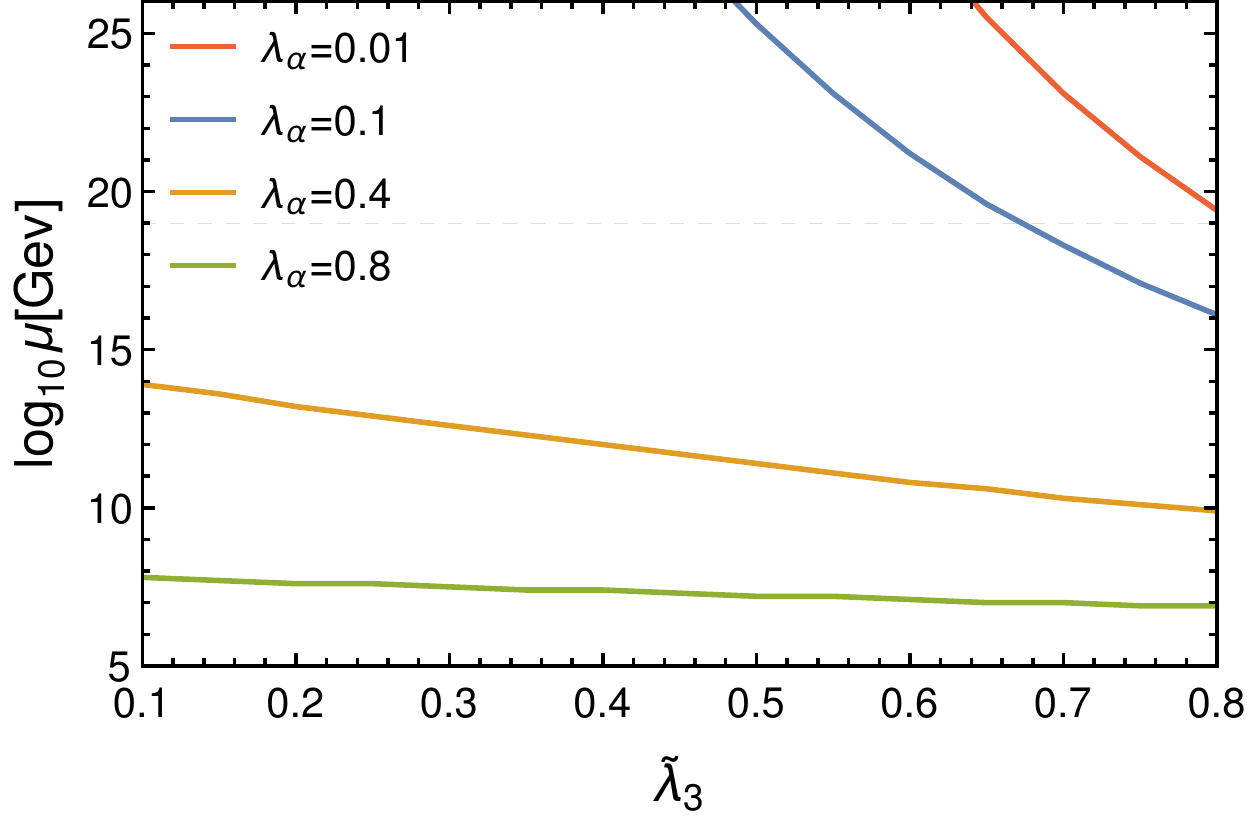}\label{f23a}}
			\subfigure[$Y_\phi =0.8$]{\includegraphics[width=0.5\linewidth,angle=-0]{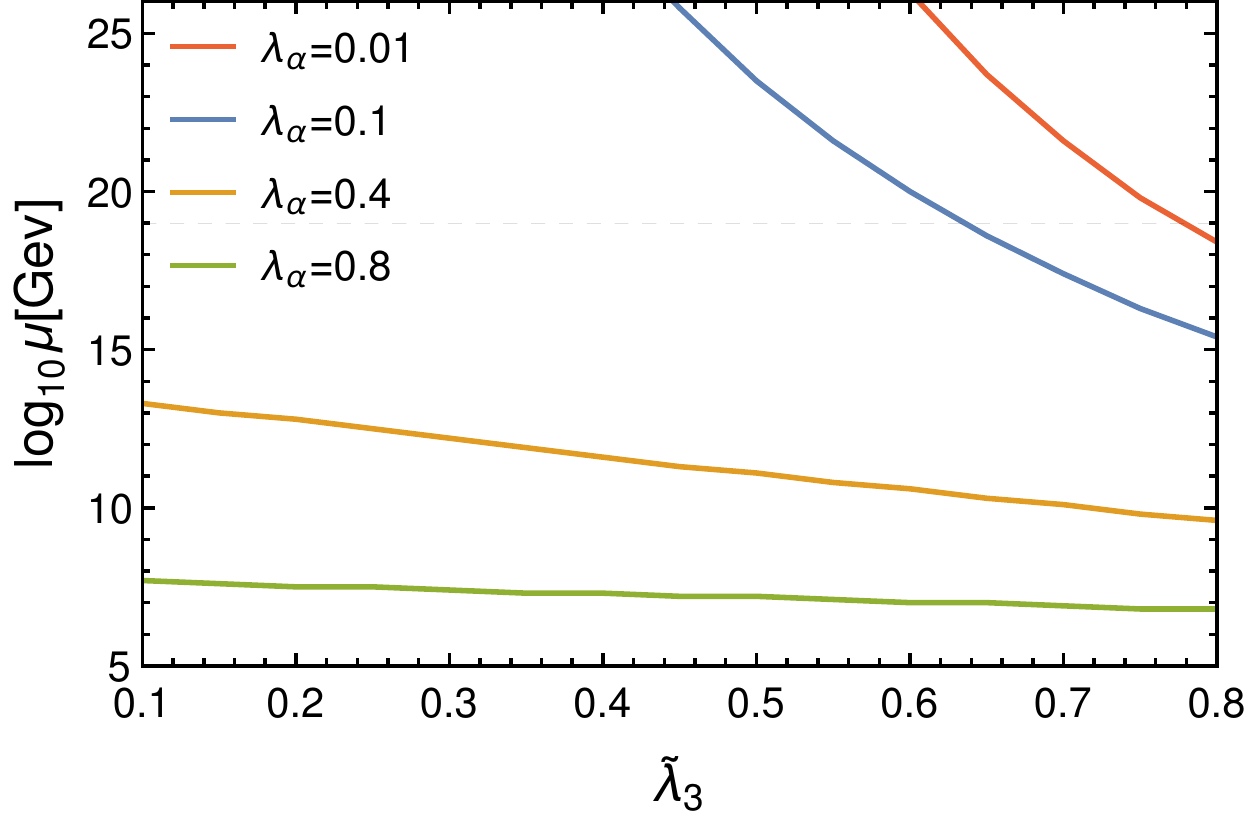}\label{f24a}}}
		\caption{Variation of triplet leptoquark-Higgs quartic coupling $\lambda_3$ and $\wl_3$ for one generation of $\widetilde R_2 + \vec S_3$. Here, for $\lambda_3$ variation, $\lambda_\alpha \in \{\lambda_2, \wl_2, \wl_3\}$ and for $\wl_3$, $\lambda_\alpha \in \{\lambda_2, \wl_2, \lambda_3\}$. We consider four different values of $\lambda_\alpha$ at the EW scale i.e 0.01, 0.1, 0.4 and 0.8 which are explained by red, blue, orange and green curves respectively. The black dotted line parallel to x-axis denotes the Planck scale.}\label{fig:l3r2s31}
	\end{center}
\end{figure*}

For three generations of $\vec S_3 + \widetilde R_2$, we have already seen that all the gauge couplings diverge below Planck scale, i.e at $10^{14.4}$ GeV, mainly due to typical behaviour of $g_2$ at two loop order. This affects the running of quartic couplings too. We study the variation of these couplings with perturbative scale in Fig. \ref{fig:s3r2l23} assuming $Y_\phi=0.1$. The adjustments in these plots with larger $Y_\phi$ are not very significant and hence we do not present them. While examining the variation of $\lambda_{2}$ for any particular generation, the remaining generations of $\lambda_{2}$ are denoted as $\lambda_\beta^{jj}$ whereas the other quartic couplings like $\wl_2$, $\lambda_{3}$ and $\wl_{3}$ with all the generations are designated as $\lambda_\alpha^{ii}$. The same notation has been followed for all the other quartic couplings too. The colour codes have been discussed previously. It can be noticed from Figs. \ref{f13a} - \ref{f16a} that even for lower initial values of $\lambda_{\alpha}^{ii}$ and $\lambda_{\beta}^{jj}$, like 0.01 and 0.1, the quartic couplings go to non-perturbative region at $10^{14.4}$ GeV due to appearance of Landau pole in $g_2$. For higher values of the parameters at EW scale, non-perturbativity is reached even at much lower scale. Thus, the demand of Planck scale perturbativity rules out the three generations scenario of $\vec S_3 + \widetilde R_2$ model for any values of the leptoquark-Higgs quartic couplings. So, we have to consider the one generation scenario of $\vec S_3 + \widetilde R_2$ model.

\subsubsection{Perturbativity of \texorpdfstring{$\widetilde{R}_2+\vec S_3$}{} with 1-gen}

In this section, we look into the perturbativity of Higgs-leptoquark quartic couplings for combined scenario of $\widetilde{R}_2$ and $\vec S_3$ with one generation. The one-loop beta functions for all these parameters in this case can easily be written as:
\begin{align}
&\beta(\lambda_{2})^{1-loop}_{\widetilde{R}_2+\vec S_3,1-gen}=\beta(\lambda_{2})^{1-loop}_{\widetilde{R}_2,1-gen}, \nn\\
&\beta(\wl_{2})^{1-loop}_{\widetilde{R}_2+\vec S_3,1-gen}=\beta(\wl_{2})^{1-loop}_{\widetilde{R}_2,1-gen},\nonumber\\
&\beta(\lambda_{3})^{1-loop}_{\widetilde{R}_2+\vec S_3,1-gen}=\beta(\lambda_{3})^{1-loop}_{\vec S_3,1-gen},\nn\\ &\beta(\wl_{3})^{1-loop}_{\widetilde{R}_2+\vec S_3,1-gen}=\beta(\wl_{3})^{1-loop}_{\vec S_3,1-gen}.
\end{align}
The full two-loop beta functions of all the Higgs-leptoquark quartic couplings in this 
scenario are listed in \autoref{sec:H-LQcouplR2S3}.

Now, we study the variation of leptoquark-Higgs quartic couplings $\lambda_2, \wl_2$, $\lambda_3$ and $\wl_3$ with the perturbative scale for one generation of $\widetilde R_2 + \vec S_3$ model. The results for variation of $\lambda_{2}$ and $\wl_{2}$ are presented in Fig. \ref{fig:l2r2s31}. When considering the variation of $\lambda_2$, we denote all the other leptoquark-Higgs quartic couplings, namely $\wl_2, \lambda_3$ and $\wl_3$, as $\lambda_\alpha$. By the same token, while examining the behaviour of $\wl_2$ the other leptoquark-Higgs quartic couplings, viz. $\lambda_2, \lambda_3$ and $\wl_3$, are taken as $\lambda_\alpha$. The colour codes have already been mentioned in earlier sections. As can be noticed from Fig. \ref{f17a} and \ref{f18a}, the initial value of $\lambda_2$ is restricted to 0.62 and 0.54 from Planck scale perturbativity for EW values of other quartic couplings being 0.01 and 0.1 respectively with $Y_\phi=0.1$, whereas with $Y_\phi=0.8$, these upper bounds roll down to 0.59 and 0.51 respectively. For higher values of other quartic couplings at the EW scale like 0.4 and 0.8, theory becomes non-perturbative around $10^{14.1}$ GeV and $10^{7.9}$ GeV with $Y_\phi=0.1$ which differ slightly (about 0.2 GeV) in $Y_\phi=0.8$ case even if the initial value of $\lambda_{2}$ is taken to be very small. Similarly, for $\wl_2$, Planck scale perturbativity with $Y_\phi=0.1$ is achieved till $\wl_2 \leq$ 0.82 and 0.68, which diminish to 0.76 and 0.63 respectively with $Y_\phi=0.8$, while taking the initial values for other quartic couplings as 0.01 and 0.1 at EW scale. Again, for higher EW values of $\lambda_\alpha$, like 0.4 and 0.8, the theory becomes non-perturbative at much lower scales as described in Figs. \ref{f19a} and \ref{f20a}. The reason for all these typical behaviours are already discussed in the previous section \ref{perbR2}.

Correspondingly, the changes in $\lambda_{3}$ and $\wl_{3}$ with perturbative scale are displayed in Fig. \ref{fig:l3r2s31}. Here, for  $\lambda_3$ variation, we symbolize $\{\lambda_2, \wl_2, \wl_3\}$ as $\lambda_\alpha$ whereas for $\wl_3$ variation, we assume $\lambda_\alpha  \in \{\lambda_2, \wl_2, \lambda_3\}$. The colour codes have already been discussed previously. Here, Plank scale perturbativity with $Y_\phi=0.1$ restricts $\lambda_3$ to 0.55 and 0.47 (see Fig. \ref{f21a}), which change to 0.53 and 0.45 respectively with $Y_\phi=0.8$ (Fig. \ref{f22a}), for $\lambda_\alpha$=0.01, 0.1 at the EW scale. Similarly, from Figs. \ref{f23a} and \ref{f24a} one can observe that for $Y_\phi=0.1$, $\wl_{3}$ should be bounded above till 0.83 and 0.67, which reduce to 0.78 and 0.62 respectively for $Y_\phi=0.8$, in order to respect Plank scale perturbativity with $\lambda_\alpha$=0.01, 0.1 at the EW scale. For higher initial values of $\lambda_\alpha$, like 0.4 and 0.8, the theory becomes non-perturbative at very low scale like $\sim$ $10^{14-15}$ GeV and $10^{8-9}$ GeV even with very small EW value of $\lambda_{3}$ and $\wl_3$ at both the Yukawa couplings, and the scale decreases gradually with increase in initial values of these two parameters. The reason for all these typical behaviours are already discussed in the previous section \ref{perbS3}. It is worth reminding that there is no Landau pole of any gauge coupling in this model and the non-perturbativity, discussed here, appears because of the Higgs-leptoquark quartic couplings growing beyond $4\pi$ during the RG evolution.

\subsubsection{Effects of self-quartic couplings of leptoquarks}

Up to this point, we do not consider self-quartic couplings of the leptoquark for simplicity. In this subsection, we discuss the effects of such couplings on perturbativity of the model. We find that introduction of these couplings does not affect the running of gauge couplings much; however, it brings in non-negligible positive contribution to the running of Higgs-leptoquark quartic couplings up to two-loop order. Therefore, Higgs-leptoquark quartic couplings attain the non-perturbative limit earlier compared to the scenario with self-quartic couplings of leptoquarks being neglected. For instance, one can add the self-interaction term of $\omega_2 (\widetilde{R}_2^\dagger\widetilde{R}_2)(\widetilde{R}_2^\dagger\widetilde{R}_2)$ to Lagrangian given by Eq. \ref{eq:lag2}. With values 0.47 and 0.64 at EW scale, $\lambda_2$ goes to non-perturbative region at Planck scale in this case for other quartic couplings $\lambda_{2}^{jj}$, $\widetilde{\lambda}_2^{ii}$ and the newly introduced self-quartic coupling of $\widetilde R_2$ (three generations, without any generation mixing) being 0.01 and 0.1 respectively assuming $Y_\phi=1.0$. Before the introduction of self-quartic coupling of $\widetilde R_2$, the values of $\lambda_{2}$ for which non-perturbativity was achieved at Planck scale under the same values of other quartic couplings were 0.47 and 0.31 respectively (see Fig. \ref{f6a}). With the same value of $Y_\phi$ and other quartic couplings, $\widetilde \lambda_{2}$ maintains Planck scale perturbativity until values at EW scale being 0.64 and 0.41 which were 0.73 and 0.58 respectively (see Fig. \ref{f8a}) before the introduction of self-quartic coupling of $\widetilde R_2$(three generations, without any generation mixing). On the other hand, for $\vec S_3$ with three  generations, the positive effects of self-quartic couplings of leptoquarks are even stronger. As an example,
we add self-quartic term\footnote{There could be another term like $\Tr\big[(S_3^{ad})^\dagger S_3^{ad}(S_3^{ad})^\dagger S_3^{ad}\big]$.} of $\Tr\big[(S_3^{ad})^\dagger S_3^{ad}\big]^2$ to the Lagrangian given by Eq. \ref{eq:lag3}. With $Y_\phi=0.8$, the parameters $\lambda_{3}$ and $\widetilde\lambda_{3}$ now cannot achieve Planck scale perturbativity for small value of other quartic couplings like 0.01 at EW scale. Before the consideration of self-quartic coupling of leptoquark, $\lambda_{3}$ and $\widetilde\lambda_{3}$ were achieving Planck scale perturbativity with other quartic couplings being 0.1 at EW scale (see Fig. \ref{f10a} and \ref{f12a}).

\section{Vacuum stability}
\label{sec:stab}
There exists two approaches in literatures regarding the stability analysis. The first one is the running of Higgs quartic coupling $\lambda_h$ using  beta-functions, and  the other method the Coleman-Weinberg effective potential approach ~\cite{Coleman:1973jx}.

At first, we scrutinize the running of self-quartic coupling for Higgs boson, i.e. $\lambda_{h}$, which in turn would indicate the change in stability of Higgs vacuum. This parameter is also expected to be below $4\pi$ at all energy scale to respect the perturbativity. However, for the purpose of this section, we focus on stability of vacuum which suggests that $\lambda_{h}$ should be a positive quantity at all the energy scale. The one and two-loop beta functions for $\lambda_{h}$ under SM are given by:
\begin{align}
&\beta(\lambda_{h})^{1-loop}_{SM}=\frac{3}{8\pi^2}\bigg[\lambda_h^2+\frac{3}{200}\,g_1^4+\frac{3}{16}\Big(\frac{g_1^2}{5}+g_2^2-4\lambda_h\Big)^2\nn\\
&+2\,\lambda_h\,\Tr\Big(\X_\uq +\X_\dq+\frac{1}{3}\X_\cl\Big)-\Tr\Big(\X_{\uq}^2+\X_{\dq}^2+\frac{1}{3}\X_\cl^2\Big)\bigg],
\end{align}
\begin{align}  
\beta(\lambda_{h})&^{2-loop}_{SM}  =  \beta(\lambda_{h})^{1-loop}_{SM}+\frac{1}{(16\pi^2)^2}\bigg[\Big(
-\frac{3411}{2000} g_{1}^{6} \nn \\
& -\frac{1677}{400} g_{1}^{4} g_{2}^{2} -\frac{289}{80} g_{1}^{2} g_{2}^{4} +\frac{305}{16} g_{2}^{6} +\frac{1887}{200} g_{1}^{4} \lambda_h \nonumber\\
&+\frac{117}{20} g_{1}^{2} g_{2}^{2} \lambda_h -\frac{73}{8} g_{2}^{4} \lambda_h +\frac{108}{5} g_{1}^{2} \lambda_h^{2} +108 g_{2}^{2} \lambda_h^{2} \nn \\
&-312 \lambda_h^{3}\Big)+\Big\{\frac{9}{20}\,\Big(g_1^4+6g_1^2g_2^2-5g_2^4\Big)+\frac{5}{2}\,\lambda_h\,\Big(g1^2\nonumber \\ 
& +9g_2^2+32g_3^2\Big)-144\lambda_h^2\Big\}\Tr\,\X_\dq -\Big\{\frac{3}{4}\Big(3g_1^4-\frac{22}{5}g_1^2g_2^2\nn\\
&+g_2^4\Big)-\frac{15}{2}\,\lambda_{h}\,\Big(g_1^2+g_2^2\Big)+48\lambda_{h}^2\Big\}\Tr\,\X_\cl -\Big\{\frac{9}{10}\,\nonumber \\ 
&\Big(\frac{19}{10}g_1^4-7g_1^2g_2^2+\frac{5}{2}g_2^4\Big)-\lambda_{h}\Big(\frac{17}{2}g_1^2+\frac{45}{2}g_2^2+80g_3^2\Big)\nonumber \\ 
&+144\lambda_{h}^2\Big\} \Tr\,\X_\uq -\Tr\,\Big\{\lambda_{h}\Big(\X_\cl^2+3\X_\uq^2+3\X_\dq^2+42\widetilde\X_\uq\widetilde\X_\dq\Big)\nn\\
&-\frac{4}{5}\,g_1^2\,\Big(\X_\dq^2-2\X_\uq^2-3\X_\cl^2\Big)+32\,g_3^2\Big(\X_\uq^2+\X_\dq^2\Big)\Big\}\nonumber \\ 
&+10\,\Tr\,\Big\{\X_\cl^3+3\Big(\X_\uq^3+\X_\dq^3\Big)-\frac{3}{5}\widetilde\X_\dq\widetilde\X_\uq\Big(\widetilde\X_\uq-\widetilde\X_\dq\Big)\Big\}\bigg].
\end{align}
It is well known that, in case of SM, $\lambda_h$ enters into negative valued region between $10^{9}$ GeV and $10^{10}$ GeV  energy scale \cite{Buttazzo:2013uya,EliasMiro:2011aa} at two-loop order. At this point it is worth mentioning that in case of $\lambda_{h}$ two-loop contributions affect the running significantly. The addition of right-handed neutrinos pulls the stability scale further down with more negative contributions\cite{exwfermion,Coriano:2015sea,DelleRose:2015bms,Jangid:2020dqh,Garg:2017iva}. In contrast, the presence of scalar leptoquarks is expected to push the stability scale further by adding positive contributions to these beta functions.

\subsection{Vacuum stability of $\widetilde R_2$}\label{sec:staR2}
\begin{figure*}[h!]
	\begin{center}
		\mbox{\subfigure[2-loop $Y_\phi =1.0$]{\includegraphics[width=0.5\linewidth,angle=-0]{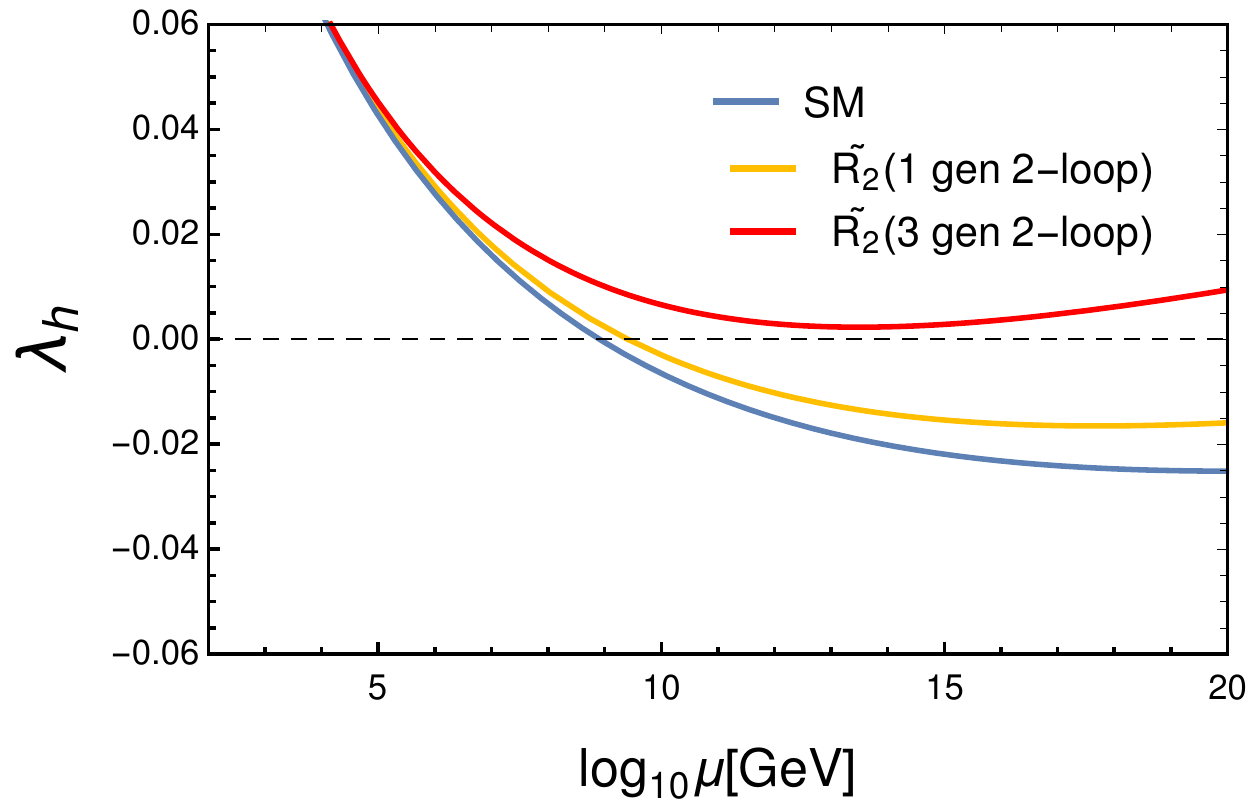}\label{f28a}}
			\subfigure[2-loop $Y_\phi =1.36$]{\includegraphics[width=0.5\linewidth,angle=-0]{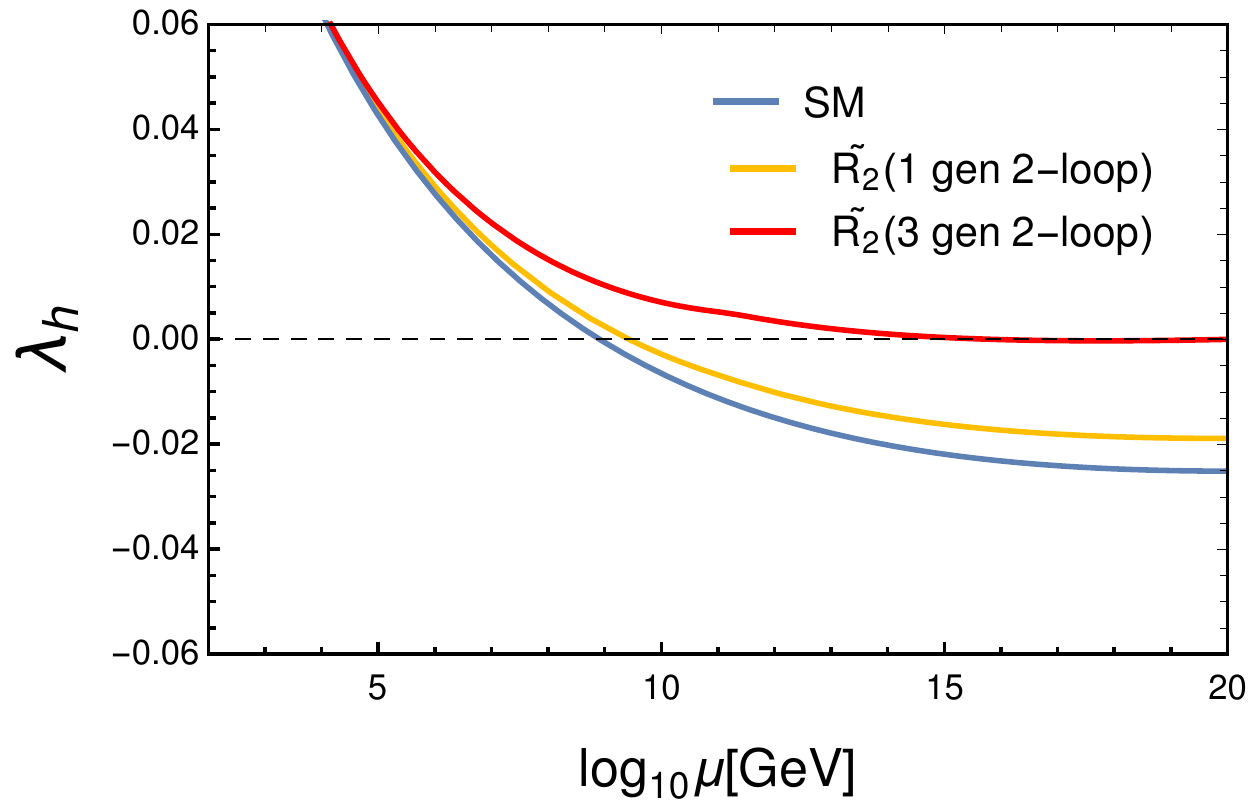}\label{f27a}}}
		\caption{Running of SM Higgs quartic coupling with scale for stability analysis. Stability scale is defined as the scale after which $\lambda_h<0$. Here, blue, yellow and red curves describes the running of $\lambda_h$ for SM, one generation and three generations of $\widetilde R_2$. With three generations of $\widetilde R_2$, the positive contribution from gauge couplings and other quartic couplings is large enough and is compensated by the negative contribution of $Y_2$. The crucial value of $Y_2$ is 1.36, $\lambda_h$ will go negative if higher values of $Y_2$ are considered and stability is lost. On the other hand, for one generation of $\widetilde{R}_2$, $\lambda_{h}$ always goes to negative value before Planck scale.}\label{fig:r2lh}
	\end{center}
\end{figure*}
At first, we look into the effects of doublet leptoquark $\widetilde R_2$. The one and two-loop beta functions for $\lambda_{h}$ in this case are given by:
\begin{align}
\label{eq:lhr21}
&\beta(\lambda_{h})^{1-loop}_{\widetilde{R}_2,1-gen}= \beta(\lambda_{h})^{1-loop}_{SM}+\Delta\beta(\lambda_{h})^{1-loop}_{\widetilde{R}_2}\nn\\ &\beta(\lambda_{h})^{1-loop}_{\widetilde{R}_2,3-gen}= \beta(\lambda_{h})^{1-loop}_{SM}+\sum_{i=1}^{3}\Big[\Delta\beta(\lambda_{h})^{1-loop}_{\widetilde{R}_2}\Big]_{\is}\nonumber\\
&\text{with}\quad\Delta\beta(\lambda_{h})^{1-loop}_{\widetilde{R}_2}=\frac{3}{8\pi^2}\Big(\lambda_2^2+\lambda_2\,\widetilde\lambda_2+\frac{1}{2}\widetilde\lambda_2^2\Big),
\end{align}
\begin{align}
\label{eq:lhr22}
&\beta(\lambda_{h})^{2-loop}_{\widetilde{R}_2,1-gen}= \beta(\lambda_{h})^{2-loop}_{SM}+\Delta\beta(\lambda_{h})^{2-loop}_{\widetilde{R}_2}\nn \\ &\beta(\lambda_{h})^{2-loop}_{\widetilde{R}_2,3-gen}= \beta(\lambda_{h})^{2-loop}_{SM}+\sum_{i=1}^{3}\Big[\Delta\beta(\lambda_{h})^{2-loop}_{\widetilde{R}_2}\Big]_{\is}\nonumber\\
\text {with} &\quad \Delta\beta(\lambda_{h})^{2-loop}_{\widetilde{R}_2,1-gen}  = \Delta\beta(\lambda_{h})^{1-loop}_{\widetilde{R}_2}+\frac{3}{(16\pi^2)^2}\bigg[\nn\\
&-\frac{7}{8}\Big(\frac{1}{125}g_1^6+\frac{1}{75}g_1^4g_2^2+\frac{1}{5}g_1^2g_2^4+g_2^6\Big)+\frac{15}{2}\Big(\frac{1}{75}\,g_1^4\nonumber\\
&+\,g_2^4\,\Big)\,\Big(\lambda_2+\frac{1}{2}\,\widetilde{\lambda}_2+\frac{11}{30}\,\lambda_{h}\Big)+\frac{1}{2}\,g_1^2g_2^2\,\widetilde\lambda_2-8\,\widetilde\lambda_2^2\,\Big(\lambda_2\nn\\
&+\frac{1}{2}\widetilde\lambda_2+\frac{3}{8}\,g_2^2+\frac{1}{4}\lambda_{h}\Big)+12\,\Big(\frac{1}{45}\,g_1^2+g_2^2+\frac{16}{9}\,g_3^2\nonumber\\
&-\frac{5}{3}\lambda_{h}-\frac{2}{3}\lambda_2-\frac{1}{3}\,\widetilde\lambda_2\Big)\,\Big(\lambda_2^2+\lambda_2\widetilde\lambda_2+\frac{1}{2}\widetilde\lambda_2^2\Big)-6\,\lambda_{h}\,\nn\\
&\Tr\Big(\X_2\,\X_\dq+\frac{1}{2}\widetilde{\X}_2\,\widetilde{\X}_\cl\Big)+4\,\Tr\,\Big(\X_2\,\X_\dq^2+Y_2\,\X_\cl Y_2^\dagger\X_\dq\nonumber\\
&+\frac{1}{2}\widetilde{\X}_2\,\widetilde{\X}_\cl^2\Big)\bigg].
\end{align}	
In the last section, we observe that there is not much room for the Higgs-leptoquark quartic couplings to be varied randomly from the perspective of Planck scale perturbativity. However, the Yukawa couplings for leptoquarks do not attain such serious constraints. Therefore, we address issue of vacuum stability from the effects of leptoquark Yukawa coupling. But it should be noticed from Eqs. \eqref{eq:lhr21} and  \eqref{eq:lhr22} that the contributions of $Y_2$ appear at two-loop level only. The effects of $Y_2$ in the running of $\lambda_h$ for $\widetilde R_2$ with both one generation and three generations cases have been portrayed in Fig. \ref{fig:r2lh}. Here the blue, yellow and red curves explain the running of $\lambda_h$ for SM, one  generation of $\widetilde R_2$ and three generations of $\widetilde R_2$ respectively. For all the analyses we assume every Higgs-leptoquark quartic coupling to be 0.01.

As already mentioned, the stability scale, after which $\lambda_{h}$ turns negative, for SM is just above $10^9$ GeV at two-loop level. But, while considering the RG evolution of $\lambda_{h}$ in $\widetilde{R}_2$ case, the gauge couplings and other quartic couplings contributes positively whereas the Yukawa coupling of leptoquark inserts negative contributions at two-loop level, as can be seen from Eqs. \eqref{eq:lhr21} and \eqref{eq:lhr22}. Again, since the additional contribution in beta function of $\lambda_{h}$ for three generations case is the sum of all individual generations and the gauge couplings at any particular scale for three generations case are higher than the same at one generation case, three generations scenario obtain more positive contributions than the one generation case. It should also be noticed that though there are two negative and three positive terms containing $\X_2$ in Eq. \eqref{eq:lhr22}, the positive terms are quadratic in $\X_\dq$ and $\X_\cl$ and therefore smaller than the negative terms which are linear in $\X_\dq$ and $\X_\cl$. Thus the yellow curve representing leptoquark $R_2$ with one generation stays above the blue line depicting SM and the red curve signifying $R_2$ with three generations lies at further above region. However, due to the negative contributions of the Yukawa couplings of leptoquark the red and yellow line move downward with enhancement in $Y_\phi$. In Figs. \ref{f28a} and \ref{f27a}, we depict the variations of $\lambda_{h}$ with energy scale taking the initial values for $Y_\phi$ to 1.0 and 1.36 respectively. As can be observed, for both the cases the vacuum of leptoquark model $\widetilde{R}_2$ with one generation remains stable up to $\sim 10^{9.5}$ GeV, slightly higher than the SM estimates. However, it is interesting to perceive in the left panel that the red curve remains in the positive region of $\lambda_{h}$ for all the energies indicating stability of the vacuum all the way till Planck scale with three generations of $\widetilde R_2$ and $Y_\phi=1.0$. Once we start with initial value of $Y_\phi$ to be 1.36, we observe in the right panel that the red curve touches the $\lambda_{h}=0$ line, and thus for higher values of $Y_\phi$ the Planck scale stability will be lost. One can also notice that  for this particular value of $Y_\phi$ the red curve touches the $\lambda_{h}=0$ line at $\sim 10^{14.5}$ GeV and remains very flat till the Planck scale. One can also find that this value 1.36 of $Y_\phi$ is relatively higher than the required Yukawa coupling $Y_N$ in inert doublet+type III seesaw or inverse type III seesaw to maintain the Planck scale stability \cite{Bandyopadhyay:2020djh}. On the contrary, it should be noted that leptoquark $\widetilde{R}_2$ with one generation does not show Planck scale stability even with very low Yukawa. Now, it is worth mentioning that with change in Higgs-leptoquark couplings from 0.01 to 0.1, we don't find any significant changes in the behaviour of $\lambda_{h}$. Though very high values of $\lambda_2$ and $\wl_2$ might shift the red curve in upward direction, but these higher values are disfavoured from Planck scale perturbativity of $\lambda_2$ and $\wl_2$. Consideration of self-quartic coupling of leptoquark introduces positive contributions indicating need of higher initial value of $Y_\phi$ to push $\lambda_{h}$ to the negative region. However, for $\widetilde{R}_2$ with three generations, we do not find much difference in the critical value of $Y_\phi$.

\subsection{Vacuum stability of \texorpdfstring{$\vec S_3$}{}}

\begin{figure*}[h!]
	\begin{center}
		\mbox{\subfigure[2-loop $Y_\phi =1.29$]{\includegraphics[width=0.5\linewidth,angle=-0]{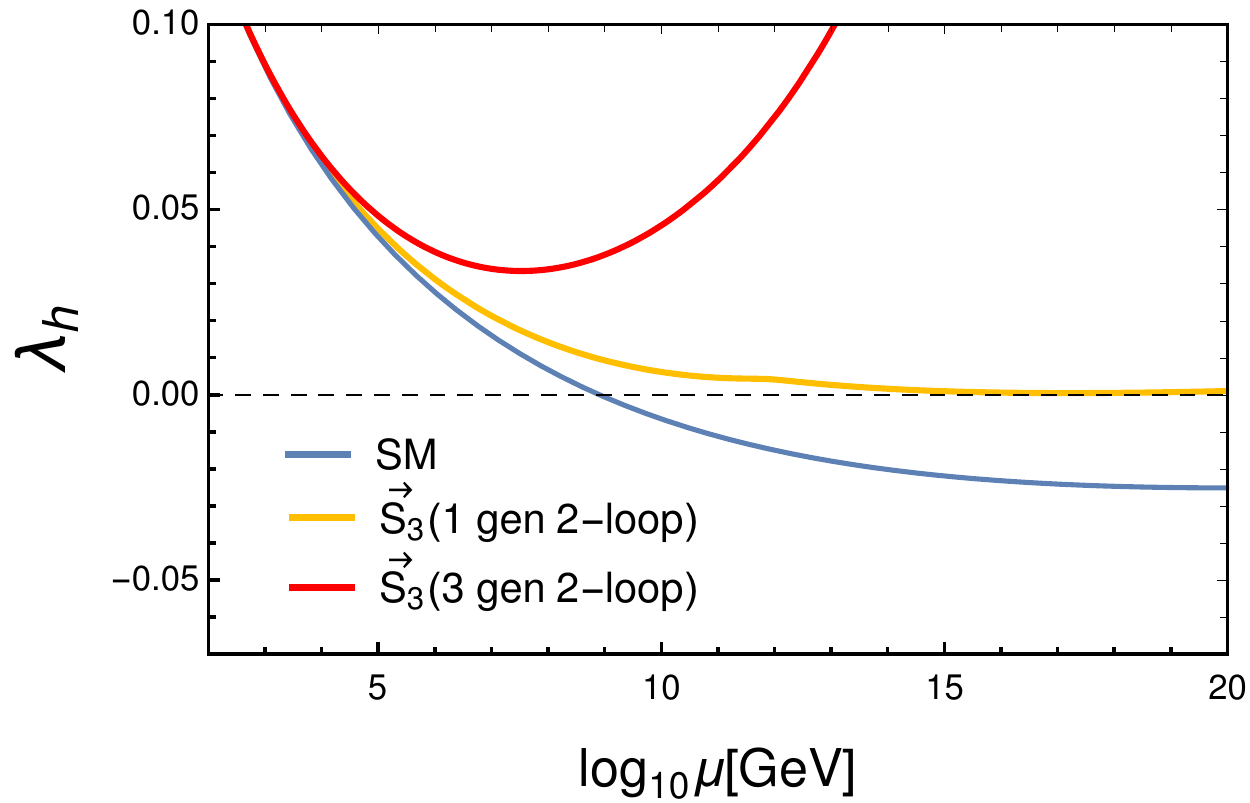}\label{f32a}}
			\subfigure[2-loop $Y_\phi =3.9$]{\includegraphics[width=0.5\linewidth,angle=-0]{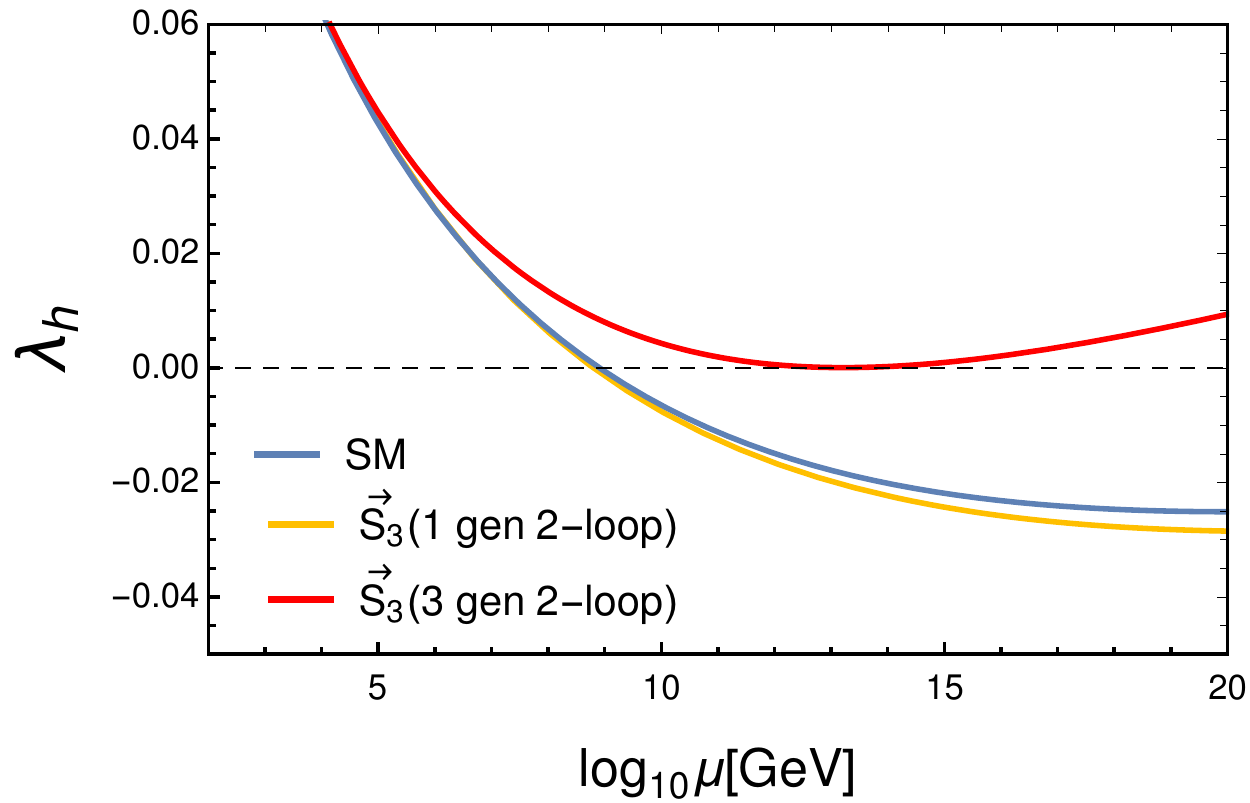}\label{f31a}}	}
		\caption{Running of Higgs quartic coupling with scale  for triplet leptoquark $\vec S_3$ at two-loop. Here,  $\lambda_h$ running for SM, one generation and three generations of $\vec S_3$ is delineated by blue, yellow and red curves respectively. If $Y_\phi$ is assumed to be greater than 1.29, the one generation model of $\vec S_3$ loses stability at two-loop order, though the three generations scenario remains stable. However, if we consider $Y_\phi>3.9$, three generations of $\vec S_3$ model also leaves the stable region at two-loop order. }\label{fig:s3lh}
	\end{center}
\end{figure*}

Now, we discuss the stability of Higgs vacuum for $\vec S_3$ scenario. The one and two-loop beta functions of $\lambda_h$ in this case are as follows:
\begin{align}
\label{eq:lhs31}
&\beta(\lambda_{h})^{1-loop}_{\vec S_3,1-gen}= \beta(\lambda_{h})^{1-loop}_{SM}+\Delta\beta(\lambda_{h})^{1-loop}_{\vec S_3}\nn\\
& \beta(\lambda_{h})^{1-loop}_{\vec S_3,3-gen}= \beta(\lambda_{h})^{1-loop}_{SM}+\sum_{i=1}^{3}\Big[\Delta\beta(\lambda_{h})^{1-loop}_{\vec S_3}\Big]_{\is}\nonumber\\
& \text{with}\quad\Delta\beta(\lambda_{h})^{1-loop}_{\vec S_3}=\frac{9}{16\pi^2}\Big(\lambda_3^2+\lambda_3\,\widetilde\lambda_3+\frac{5}{12}\,\widetilde\lambda_3^2\Big),
\end{align}
\begin{align}
\label{eq:lhs32}
&\beta(\lambda_{h})^{2-loop}_{\vec S_3,1-gen}= \beta(\lambda_{h})^{2-loop}_{SM}+\Delta\beta(\lambda_{h})^{2-loop}_{\vec S_3} \nn\\
&\beta(\lambda_{h})^{2-loop}_{\vec S_3,3-gen}= \beta(\lambda_{h})^{2-loop}_{SM}+\sum_{i=1}^{3}\Big[\Delta\beta(\lambda_{h})^{2-loop}_{\vec S_3}\Big]_{\is}\nonumber\\
\text {with} &\quad \Delta\beta(\lambda_{h})^{2-loop}_{\widetilde{R}_2,1-gen}  =\Delta\beta(\lambda_{h})^{1-loop}_{\vec S_3}+\frac{3}{(16\pi^2)^2}\bigg[-\frac{7}{2}\,\nn \\
&\Big(\frac{3}{250}\,g_1^6 +\frac{1}{50}\,g1^4g_2^2+\frac{1}{5}\,g_1^2g_2^4+g_2^6\Big)+2\widetilde\lambda_3\,g_1^2g_2^2\nn\\
&+30\,\Big(\frac{1}{50}\,g_1^4+g_2^4\Big)\Big(\lambda_3+\frac{1}{2}\widetilde\lambda_3+\frac{11}{30}\lambda_h\Big)-8\,\widetilde\lambda_3^2\,\Big(\lambda_3\nn\\
&+\frac{1}{2}\widetilde\lambda_3+\frac{3}{8}g_2^2+\frac{1}{4}\lambda_{h}\Big)+48\,\Big(\frac{1}{30}\,g_1^2+g_2^2+\frac{2}{3}\,g_3^2\nonumber\\
&-\frac{1}{4}\lambda_3-\frac{1}{8}\widetilde\lambda_3-\frac{5}{8}\lambda_{h}-\frac{1}{8}\Tr\,\X_3\Big)\Big(\lambda_3^2+\lambda_3\widetilde\lambda_3+\frac{5}{12}\,\widetilde\lambda_3^2\Big)\nn\\
&-\frac{9}{2}\lambda_{h}\Tr\Big(\widetilde\X_3\widetilde{\X}_\cl+\X_3\widetilde{\X}_\dq^T+\X_3\widetilde{\X}_\uq^T\Big)+3\,\Tr\Big\{\widetilde{\X}_3\widetilde{\X}_\cl^2\nonumber\\
&+\X_3(\widetilde{\X}_\dq^{T})^2+\X_3(\widetilde{\X}_\uq^{T})^2+\frac{4}{3}\,Y_3\widetilde{\X}_\cl Y_3^\dagger\big(\widetilde{\X}_\dq^T+\frac{1}{2}\widetilde{\X}_\uq^T\big)\Big\}\bigg].
\end{align}

The running of Higgs quartic coupling for triplet leptoquark $\vec S_3$ has been portrayed in Fig. \ref{fig:s3lh} taking the EW value of $\lambda_3$ and $\wl_{3}$ to be 0.01. Here, blue, yellow and red curves denote the RG evolution of $\lambda_h$ for SM, one generation of $\vec S_3$ and three generations of $\vec S_3$ respectively at two-loop order. As discussed in the last section \ref{sec:staR2}, gauge couplings and Higgs-leptoquark quartic couplings contribute positively in the running of $\lambda_{h}$ while the leptoquark Yukawa coupling brings in negative effects (see Eqs. \eqref{eq:lhs31} and \eqref{eq:lhs32}). Furthermore, since $\widetilde{R}_2$ lies in fundamental representation of $SU(2)_L$, while $\vec S_3$ stays in adjoint representation the positive effects in case of $\vec S_3$ are very large compared to the same for $\widetilde{R}_2$; therefore large Yukawa coupling for $\vec S_3$ will be needed to make the vacuum unstable. We depict the results for $Y_\phi$ being 1.29 and 3.9 in Figs. \ref{f32a} and \ref{f31a} respectively. In the left panel, we see that both the cases of $\vec S_3$ with one generation and three generations show Planck scale stability for  $Y_\phi=1.29$, but the yellow curve touches the $\lambda_{h}=0$ line implying further increment in $Y_\phi$ will make the theory unstable before the Planck is reached. It is also interesting to notice that the yellow curve touches the $\lambda_{h}=0$ line at $\sim 10^{15}$ GeV and remains very flat till the Planck scale like the $\widetilde{R}_2$ scenario with three generations. In the right panel, one can observe that the higher value of $Y_\phi$, i.e. 3.9, has forced the red and yellow curves to move downward pushing the one generation of $\vec S_3$ to unstable region. However, the red curve touches the $\lambda_{h}=0$ line at this Yukawa coupling, and it indicates that $Y_\phi\leq3.9$ in order to preserve Planck scale stability with three generations of $\vec S_3$. It is worth mentioning that here the red curve just kisses the $\lambda_{h}=0$ line at a lower energy scale $\sim 10^{13.5}$ GeV and then the positive contributions make it grow faster in the positive direction unlike the previous cases. However, to ensure perturbativity of the model, $Y_\phi\le\sqrt {4\pi}\approx3.54$. Therefore, combining vacuum stability and perturbativity, one should consider $\sqrt{4\pi}$ as the upper limit of $Y_\phi$ for three generations scenario of $\vec S_3$. Like $\widetilde{R}_2$, in this case also, the behaviour of these plots do not show any notable alteration if $\lambda_3$ and $\wl_{3}$ are increased to 0.1 from 0.01. Inclusion of leptoquark self-quartic coupling inserts huge positive effect for  $\vec S_3$. Due to this, with three generations of $\vec S_3$, the critical initial value of 3.9 for $Y_\phi$ now goes beyond 5. However, since $\sqrt{4\pi}\le 5$, the upper bound on $Y_\phi$ remains $\sqrt{4\pi}$ while considering combined constraint from vacuum stability and perturbativity.

\subsection{Vacuum stability of \texorpdfstring{$\widetilde{R}_2+\vec S_3$}{} with 3-gen}
\begin{figure}[h!]
	\begin{center}
		\includegraphics[width=0.9\linewidth,angle=-0]{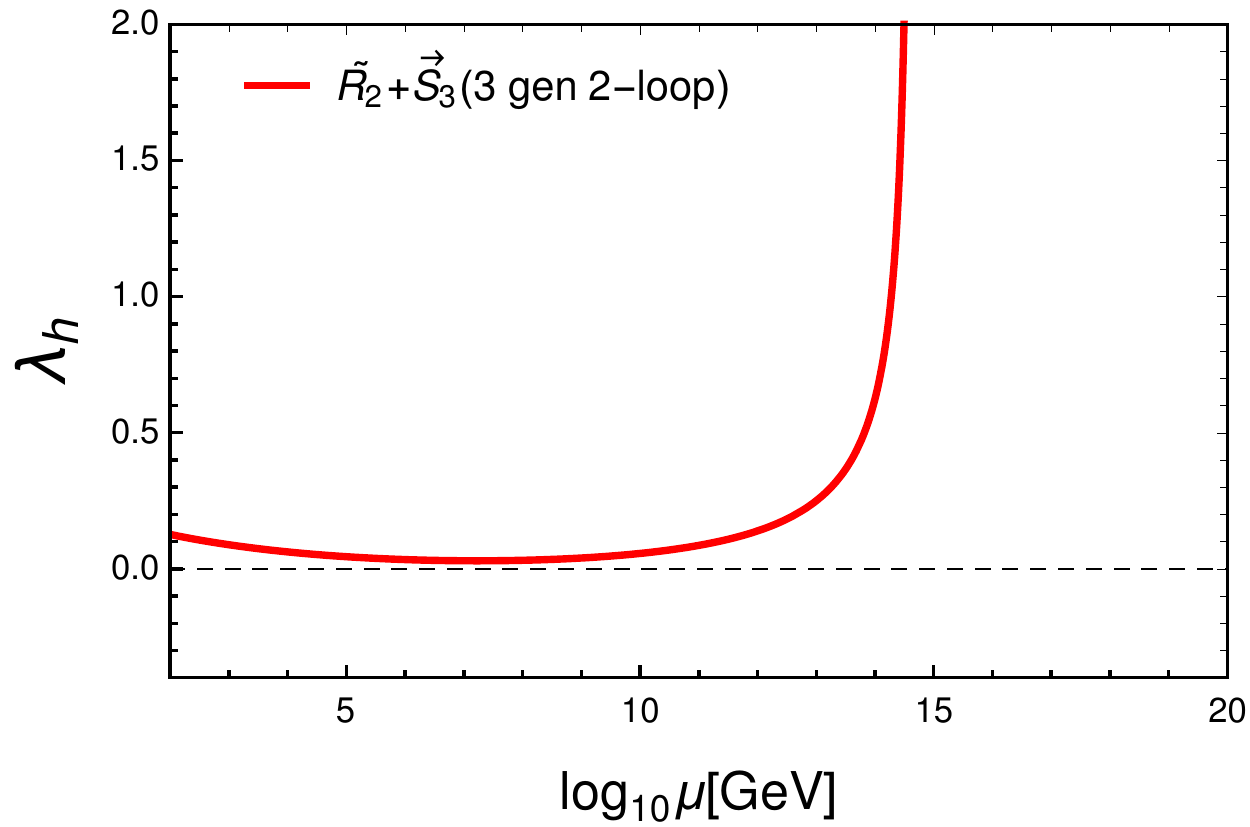}
		\caption{Running of Higgs quartic coupling $\lambda_h$ for three generations of $\widetilde R_2 + \vec S_3$ at two-loop. Here all the leptoquark Yukawa couplings are assumed to be 1.0 and all the Higgs-leptoquark couplings are taken to be 0.01. In this model $\lambda_h$ diverges at an energy scale ($\sim$ $10^{14.4}$ GeV) far below the Planck scale  at two-loop order.}\label{fig:r2s3lh}
	\end{center}
\end{figure}
The one-loop and two-loop beta functions for $\lambda_{h}$ with three generations of $\widetilde{R}_2+\vec S_3$ can be written as:
\begin{align}
\beta(\lambda_{h})^{1-loop}_{\widetilde{R}_2+\vec S_3,3-gen}= \sum_{i=1}^3\Big[\beta(\lambda_{h})^{1-loop}_{\widetilde{R}_2+\vec S_3,1-gen}\Big]_{\is}\nn \\ \beta(\lambda_{h})^{2-loop}_{\widetilde{R}_2+\vec S_3,3-gen}= \sum_{i=1}^3\Big[\beta(\lambda_{h})^{2-loop}_{\widetilde{R}_2+\vec S_3,1-gen}\Big]_{\is}
\end{align}  
The result at two-loop order for this scenario with all the leptoquark Yukawa couplings being 1.0 and all the Higgs-leptoquark couplings being 0.01 are shown in Fig. \ref{fig:r2s3lh}. We have already seen that in this model all the parameters blows up at the energy scale $10^{14.4}$ GeV. The parameter $\lambda_{h}$ is also no different from them. With any value of Higgs-leptoquark coupling or Yukawa coupling less than one, this divergence is unavoidable for this model. It is also noteworthy that $\lambda_{h}$ grows into non-perturbative region before the emergence of instability in this model. Therefore, we will discuss the behaviour of $\lambda_h$ for one generation of $\widetilde R_2 + \vec S_3$.

%
%
%
%

\subsection{Vacuum stability of \texorpdfstring{$\widetilde{R}_2+\vec S_3$}{} with 1-gen}

\begin{figure}[h!]
	\begin{center}
		\includegraphics[width=0.9\linewidth,angle=-0]{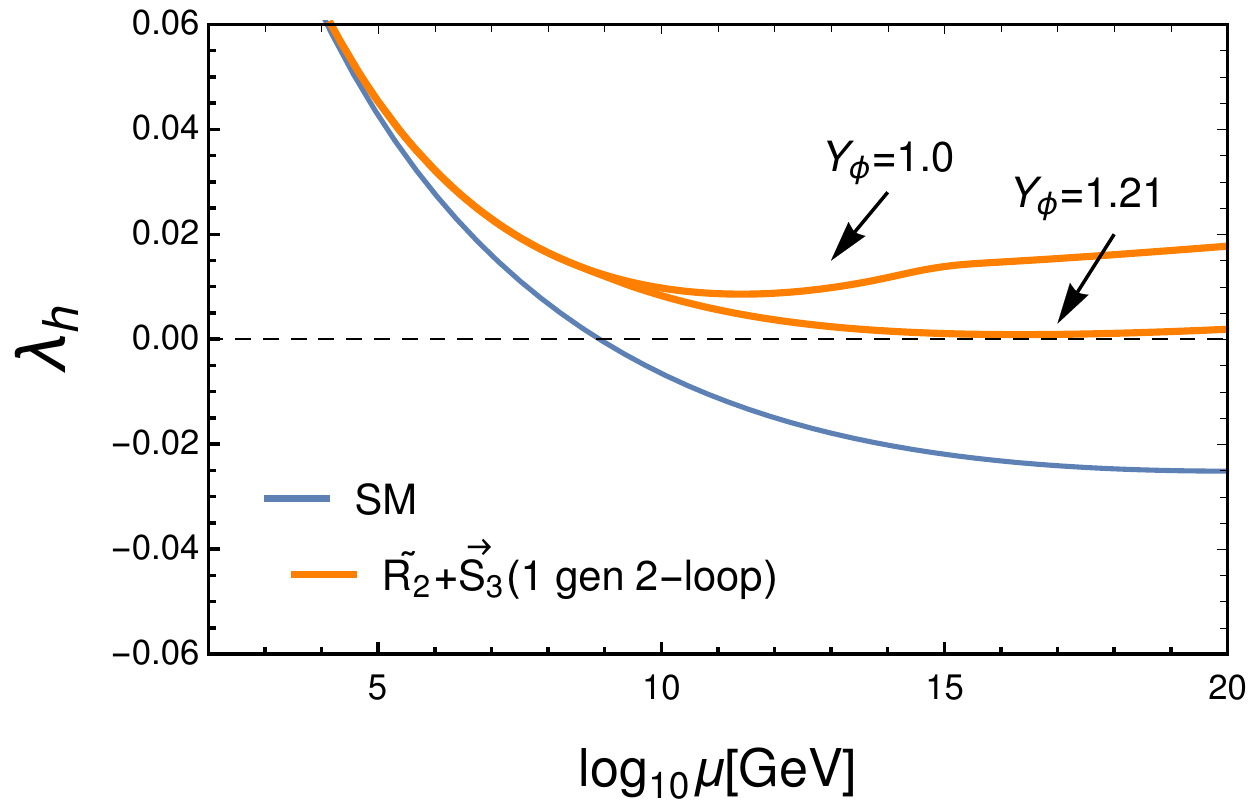}
		\caption{Higgs quartic coupling running with scale is given for one generation of $\widetilde R_2 + \vec S_3$ . Here, $\lambda_h$ running for SM and one 0 of $\widetilde R_2 + \vec S_3$  are explained by blue and yellow curves respectively. In order to maintain Planck scale stability, the upper bound on $Y_\phi$ for this model is 1.21. Here, $Y_\phi$ includes the leptoquark Yukawa couplings for both $\widetilde R_2$ and $\vec S_3$. }\label{fig:r2s31g}
	\end{center}
\end{figure}

The one and two-loop beta functions for $\lambda_{h}$ for combined scenario of $\widetilde{R}_2+\vec S_3$ with one generation can simply be expressed as:
\begin{align}
&\beta(\lambda_{h})^{1-loop}_{\widetilde{R}_2+\vec S_3,1-gen}= \beta(\lambda_{h})^{1-loop}_{SM}+\Delta\beta(\lambda_{h})^{1-loop}_{\widetilde{R}_2}\nn\\
&\qquad \qquad \qquad \qquad+\Delta\beta(\lambda_{h})^{1-loop}_{\vec S_3},\nonumber\\
&\beta(\lambda_{h})^{2-loop}_{\widetilde{R}_2+\vec S_3,1-gen}= \beta(\lambda_{h})^{2-loop}_{SM}+\Delta\beta(\lambda_{h})^{2-loop}_{\widetilde{R}_2}\nn\\
&\qquad \qquad \qquad \qquad+\Delta\beta(\lambda_{h})^{2-loop}_{\vec S_3}.
\end{align} 
The two-loop result for this case with all the leptoquark-Higgs coupling being 0.01 is portrayed in Fig: \ref{fig:r2s31g}, where the blue curve represents SM and the yellow line signifies this particular model. As can be noticed, with $Y_\phi=1.0$, $\lambda_{h}$ entirely stays in the positive region whereas for $Y_\phi\geq1.21$ this model no longer remains stable. With $Y_\phi=1.21$, the orange curve touches the $\lambda_{h}=0$ line at a relatively higher scale, $\sim 10^{16}$ GeV, and remains mostly flat till Planck Scale. Here, $Y_\phi$ includes the leptoquark Yukawa couplings for both $\widetilde R_2$ and $\vec S_3$. The result remains almost same with all the leptoquark-Higgs coupling being 0.1 also.

%

\subsection{Bounds from Effective potential stability constraints:}

\begin{table*}[h!]
	\renewcommand{\arraystretch}{1.3}
	\begin{center}
		\begin{tabular}{|c|c|c|c|c|c|c|}\hline
			Particles & $i$ & $F$ & $n_i$ & $c_i$ & $\kappa_i$ & $\kappa'_i$ \\ \hline
			\multirow{6}{*}{SM}& $W^\pm$ & 0 & 6 & 5/6 & $g_2^2/4$ & 0\\
			& $Z$ & 0 & 3 & 5/6 & $(g_1^2+g_2^2)/4$ & 0\\
			& $t$ & 1 & 12 & 3/2 & $Y_t^2$ & 0\\
			& $h$ & 0 & 1 & 3/2 & $\lambda_h$ & $m^2$\\
			& $G^\pm$ & 0 & 2 & 3/2 & $\lambda_h$ & $m^2$\\
			& $G^0$ & 0 & 1 & 3/2 & $\lambda_h$ & $m^2$\\ \hline
			\multirow{2}{*}{$\widetilde R_2$} & 	$\widetilde R_2^{2/3}$ & 0 &18 (6) & 3/2 & $\lambda_2/2$ & $m_2^2$\\
			& $\widetilde R_2^{1/3}$& 0 & 18 (6) & 3/2 & $(\lambda_2+\wl_2)/2$ & $m_2^2$\\ \hline
			\multirow{3}{*}{$\vec S_3$} &$ S_3^{4/3}$& 0 & 18 (6) & 3/2 & $\lambda_3/2$ & $m_3^2$\\
			& $ S_3^{2/3}$ & 0 & 18 (6) & 3/2 & $(\lambda_3+\wl_3)/2$ & $m_3^2$\\
			& $ S_3^{1/3}$ & 0 & 18 (6) & 3/2 & $(2 \lambda_3+\wl_3)/4$ & $m_3^2$\\
			\hline
		\end{tabular}
	\end{center}
	\caption{Different particles and the corresponding coefficients which contribute to the Coleman-Weinberg effective potential cf.~Eq.~\eqref{qc}. Here, the number of degrees of freedom for three generations of leptoquarks, i.e. 18, is shown outside the parentheses while the same with one generation of leptoquark, i.e. 6, is listed inside the brackets.}	\label{table:1}
\end{table*}

Now, to study the stability, we follow the Coleman-Weinberg effective potential approach ~\cite{Coleman:1973jx} where the one-loop contributions from all the particles at zero temperature with vanishing moments are included in effective coupling $\lambda_{eff}$. The effective potential for high field values in the h-direction can be defined as
\begin{align}
V_{\rm eff}(h,\mu) \ \simeq \ \lambda_{\rm eff}(h,\mu)\frac{h^4}{4},\quad {\rm with}~h\gg v \, ,
\label{leff}
\end{align} 
The possibility of a minima in the leptoquark direction can lead to charge and color breaking minima, which is physically unwanted. However, such possibilities have little to do in our case. Firstly, unlike the Higgs field, the bare mass term for leptoquark is chosen sufficiently large and positive, ensuring positive sign of the effective leptoquark mass term i.e. for $\widetilde{R}_2$, $m^2_2 + \lambda_{2}\frac{v^2}{2}, m^2_2 + (\lambda_{2}+ \tilde{\lambda}_2\frac{v^2}{2} > 0$, which gives $<\widetilde{R}_2>=0$, for both with and without self leptoquark couplingas at the tree-level. The possibility of non-zero vev at loop-level in the presence of the self-quartic coupling and with the negative Higgs-leptoquark quartic coupling, though possible, but for the choice of large positve bare leptoquark mass term, which is of the order of TeV, are diminished in our case. Point to be noted that the possibility of the resultant  negative mass, gives rise to the unphysicial solution.

Such observations are also been made in the context of 2HDM that if $v_1\gg v_2$, where $v_{1,2}$ are the two VEVs corresponding to the two Higgs doublets $\Phi_{1,2}$, the potential along the $\Phi_2$ direction remains almost flat and hence it is instructive to show the variation of the potential perpendicular to it, i.e, along $\Phi_1$ [147,111, 118]. Even at the one-loop $\phi_2$ direction cannot have any deeper minima as compared to the $\phi_1$ direction. Similarly, in out leptoquark case, as the tree-level vev in the leptoquark direction is zero, the possibility of a deeper minima in the that direction also cease to exist.

The total potential including tree-level potential as well as one-loop contributions from SM particles and leptoquarks can be defined as;
\bea
V= V_0 + V_1^{SM} + V_1^{\widetilde R_2 / \vec S_3 / \widetilde R_2 + \vec S_3},\label{eff}
\eea
where $V_0$ is the tree-level potential of the model and $V_1$ is the one-loop effective potential which includes the contributions from SM particles as well as the leptoquarks and can be expressed as:
{\small
\begin{align}\label{qc}
\hspace*{-1mm}V_1(h, \mu) \ = \ \frac{1}{64\pi^2}\sum_{i} (-1)^F n_i \mathcal M_i^4(h) \Bigg[\log\frac{\mathcal M_i^2(h)}{\mu^2}-c_i\Bigg].
\end{align}}
Here, the summation includes all the particles which couple to Higgs field $h$ at tree level, $n_i$ denotes the number of degrees of freedom for those particles, $c_i$ is a constant taking value $\frac{5}{6}$ for gauge bosons and $\frac{3}{2}$ for fermions and scalars, and the quantity $F$ is another constant which becomes 0 for bosons and 1 for fermions. The entity $\mathcal M_i$ which is given by:
\begin{align}
\mathcal M_i^2(h) \ = \ \kappa_i h^2-\kappa'_i,
\label{eq:4.5}
\end{align}
signify field dependent masses for the particles in the model with $\kappa$ and $\kappa^\prime$ being two constants. All the particles, relevant for this paper, are listed in Tab.\ref{table:1} along with all the corresponding constants. For the numerical analysis we have considered $h=\mu$  since potential remains invariant at this scale \cite{Casas:1994us}.

The full effective potential in \eqref{eff} can be redefined in terms of an effective quartic coupling $\lambda_{eff}$, as in \eqref{leff} using one-loop potential \eqref{qc} as follows;
{\allowdisplaybreaks  \begin{align} \label{totalL}
	\lambda_{\rm eff}\left(h,\mu\right) & \ \simeq \  \underbrace{\lambda_h\left(\mu\right)}_{\text{tree-level}}+\underbrace{\frac{1}{16\pi^2}\sum_{\substack{i=W^\pm, Z, t, \\ h, G^\pm, G^0}} n_i\kappa_i^2 \Big[\log\frac{\kappa_i h^2}{\mu^2}-c_i\Big]}_{\text{Contribution from SM}} \nn \\
	   & \qquad  +\underbrace{\frac{1}{16\pi^2}\sum_{}n_i\kappa_i^2 \Big[\log\frac{\kappa_i h^2}{\mu^2}-c_i\Big]}_{\text{ Contribution from \texorpdfstring{$\widetilde R_2$}{}/\texorpdfstring{$\vec S_3$}{}/\texorpdfstring{$\widetilde R_2 +\vec S_3$}{}  }},
	\end{align}}


Now, let us consider that there are two minima of the Higgs potential and we reside at the first one. If the second minimum is higher than the first one, the tunnelling from first minimum to the second one will be impossible which in turn would indicate that the first minimum lies in the {\tt stable region}, denoted by $\lambda_{\rm eff}>0$. But if the height of second minimum is lower than that of the first one, there would be a finite probability for the system to tunnel to the second one. In this scenario, if the tunnelling lifetime becomes greater than the age of the universe, we term the first minimum as {\tt metastable region}.

\begin{figure*}[h!]
	\begin{center}
		\mbox{\subfigure[1-gen]{\includegraphics[width=0.5\linewidth,angle=-0]{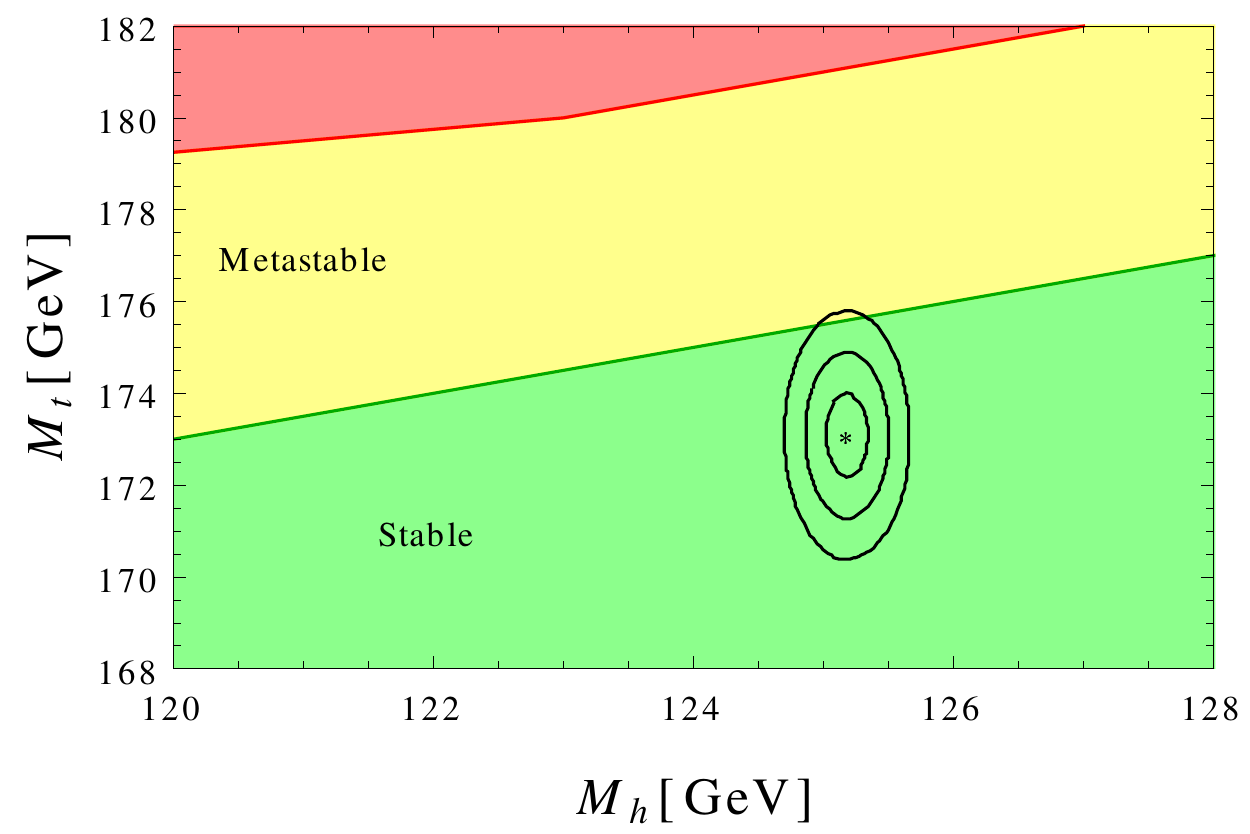}\label{f36a}}
			\subfigure[3-gen]{\includegraphics[width=0.5\linewidth,angle=-0]{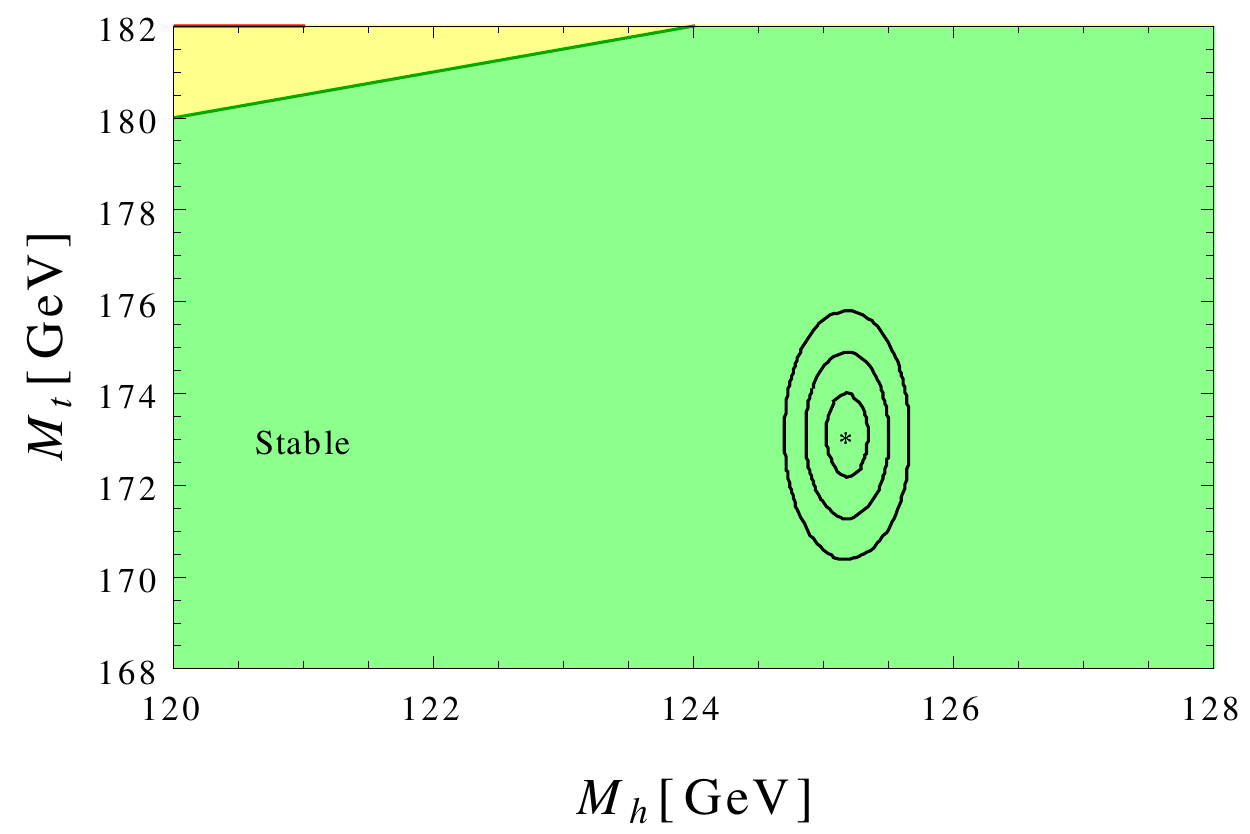}\label{f37a}}}
		\caption{Phase space diagram with Higgs mass $M_h$ vs top mass $M_t$ in GeV for $\widetilde{R}_2$. Green, yellow and red colours correspond to stable, metastable and unstable regions respectively. The black dotted circles denote 1$\sigma$, 2$\sigma$ and 3$\sigma$ contours and black dot denotes the current Higgs mass and top mass value.}\label{fig:r2eff}
	\end{center}
\end{figure*}

\begin{figure*}[h!]
	\begin{center}
		\mbox{\subfigure[1-gen]{\includegraphics[width=0.5\linewidth,angle=-0]{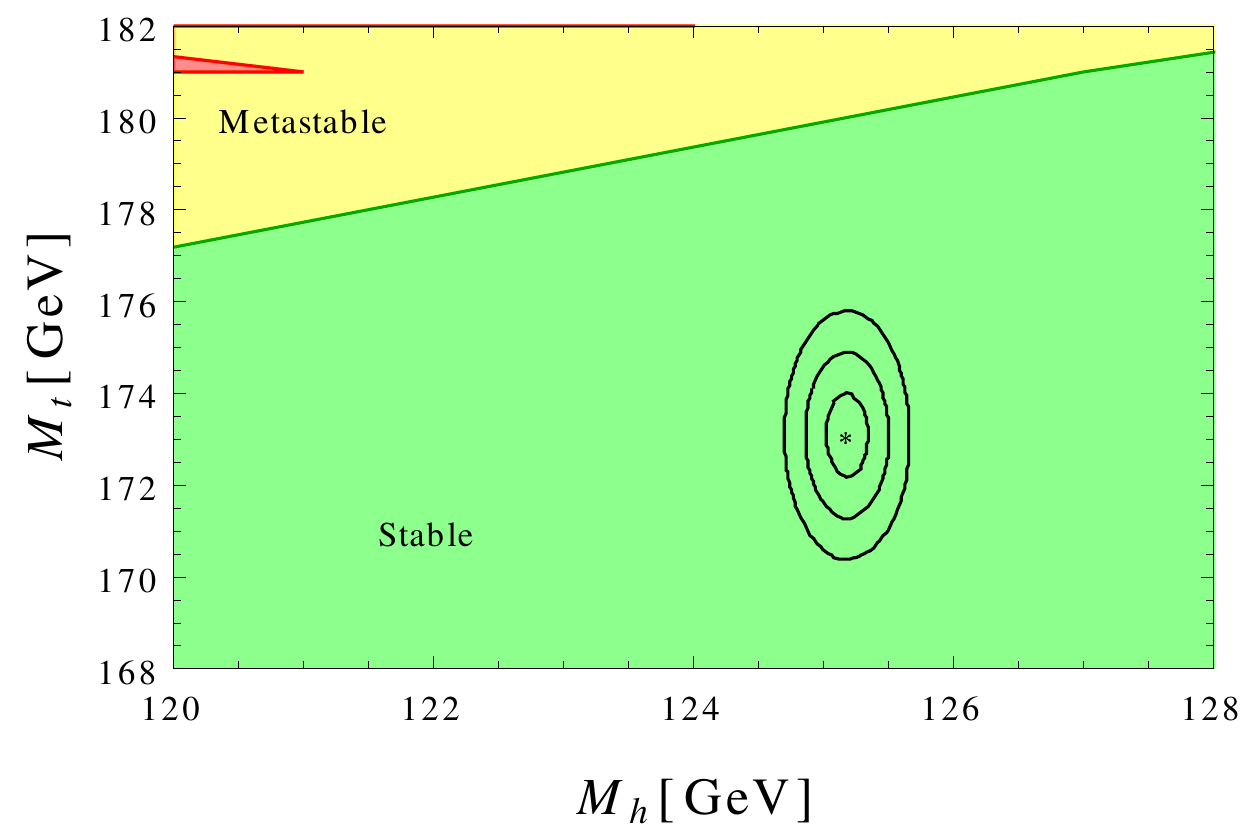}\label{f38a}}
			\subfigure[3-gen]{\includegraphics[width=0.5\linewidth,angle=-0]{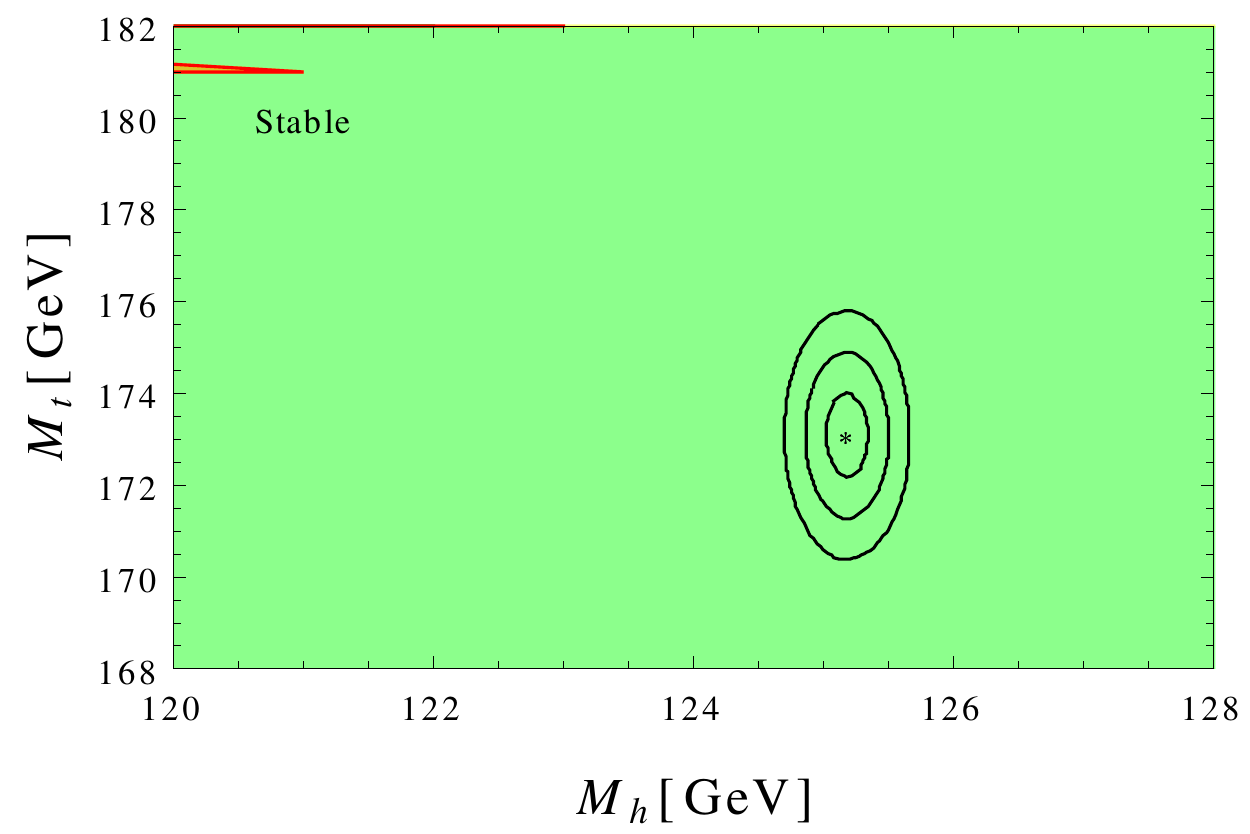}\label{f39a}}}
		\caption{Phase diagram for $\vec S_3$ with $M_h$ in GeV vs $M_t$ in GeV. The stable, metastable and unstable regions are delineated by green, yellow and red colours respectively. Black dot denotes the current values of Higgs mass and top mass in GeV and black circles are 1$\sigma$, 2$\sigma$ and 3$\sigma$ contours. }\label{fig:s3eff}
	\end{center}
\end{figure*}

\begin{figure}[h!]
	\begin{center}
		\mbox{\includegraphics[width=0.9\linewidth,angle=-0]{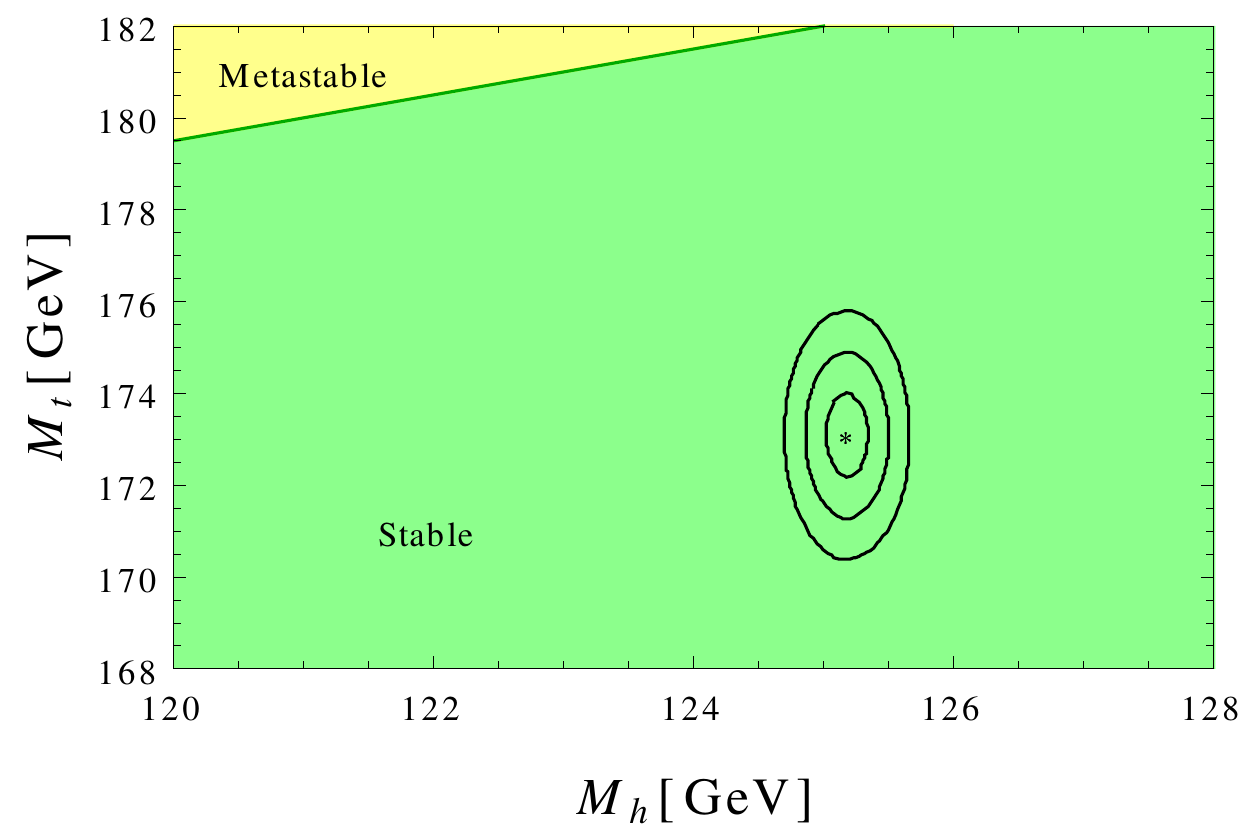}\label{f40a}}
		\caption{Higgs mass $M_h$ vs top quark mass $M_t$ plot in GeV for one generation of $\widetilde{R}_2$ and $\vec S_3$. The green and yellow colours are used to present stable and metastable regions. The Black circles are 1$\sigma$, 2$\sigma$ and 3$\sigma$ contours with the current experimental values of Higgs mass and top quark mass are denoted by dot at centre. }\label{fig:r2s3eff}
	\end{center}
\end{figure}


The tunnelling probability in this scenario is given by:
\bea
\rm P \ = \rm \ T_0^4 {\mu}^4 e^{\frac{-8 {\pi}^2}{3 \lambda_{\rm eff}(\mu)}} \, ,
\label{eq:P}
\eea
where, $\mu$ is the scale at which the probability is maximum, i.e. $\frac{\partial P}{\partial \mu}=0$, and $T_0$ is the age of the universe. Using condition $\frac{\partial P}{\partial \mu}=0$ along with $\beta_{\lambda}=0$, we can get the expression of $\lambda_{\rm eff}$ at different scales:
\bea
\lambda_{\rm eff}(\mu) \ = \ \frac{\lambda_{\rm eff}(v)}{1-\frac{3}{2\pi^2}\log\left(\frac{v}{\mu}\right)\lambda_{\rm eff}(v)} \, .
\label{eq:lamb}
\eea
Now if we set $P=1$, $T_0=10^{10}$ years and $\mu=v$ where $v \simeq 246$ GeV is the EW vev in Eq: \eqref{eq:P} then $\lambda_{eff}(v)$ comes out to be 0.0623. But, if we consider $P<1$ with $T_0=10^{10}$ years, then it will be equivalent of demanding that tunnelling probability from first vacuum to the deeper one is greater than $T_0$ and we will obtain the condition for {\tt metastability} as \cite{Isidori:2001bm}:
\bea \label{meta}
0 \ > \ \lambda_{\rm eff}(\mu) \ \gtrsim \ \frac{-0.065}{1-0.01 \log\left(\frac{v}{\mu}\right)}.
\eea

Lastly, if the tunnelling probability from first minimum to the deeper one is lesser than the age of universe, i.e $\lambda_{\rm eff}<0$, then the first minimum will be named as the {\tt unstable region}. We know that the SM vacuum lies in the {\tt metastable region}. But the presence of leptoquarks will exert extra effects in $\lambda_{\rm eff}$ which will alter the metastability of the Higgs vacuum. Different regions regarding stability, metastability and instability for  $\widetilde R_2$,  $\vec S_3$ and one generation of $\widetilde R_2 + \vec S_3$ have been presented in Figs. \ref{fig:r2eff}, \ref{fig:s3eff} and \ref{fig:r2s3eff}. We refrain ourselves from three generations of $\widetilde R_2 + \vec S_3$ since it attains serious constraints from Planck scale perturbativity and stability. We have plotted Higgs mass $M_h$ (in GeV) vs top mass $M_t$ (in GeV) in those above mentioned figures along with
the stable, metastable and unstable regions coloured by green, yellow and red respectively. The black circles defines  1$\sigma$, 2$\sigma$ and 3$\sigma$ contours with a dot at the centre denoting the current Higgs mass and top mass values \cite{Buttazzo:2013uya,Masina:2012tz,10.1093/ptep/ptaa104}. In Figs. \ref{f36a} and \ref{f37a}, the results for one generation and three generations of $\widetilde{R}_2$ have been illustrated. For this analysis $M_h$ is varied between 119 GeV and 135 GeV, whereas $M_t$ has been altered from 165 GeV to 185 GeV with fixing $\lambda_h=0.1264$ and $Y_\uq^{33}=0.9369$ at the EW scale.  The other quartic couplings $\lambda_2$ and $\widetilde{\lambda}_2$ are varied from 0.1 to 0.8. As can be seen, for one generation of $\widetilde R_2$, only the 3$\sigma$ contour hits the metastability while the three generations scenario resides entirely inside the stable region as the positive effects of gauge couplings and  quartic couplings are very large. Again, the positive contributions form gauge couplings in triplet leptoquark case are even higher than the $\widetilde R_2$ scenario. Therefore, we get the complete stable region with both one generation and three generations of $\vec S_3$, shown in Figs. \ref{f38a} and \ref{f39a}. The positive gauge coupling contributions are more high for $\widetilde R_2 +\vec S_3$ case and hence we get completely stable region for this case also, see Fig. \ref{fig:r2s3eff}.

\section{Phenomenology}
\label{sec:ph}
In this section, we discuss different experimental bounds on the parameter space of scalar leptoquarks and compare them with the theoretical bounds arising from the demand of perturbativity and stability of the theory till Planck scale. There are both direct and indirect bounds on leptoquarks. While the indirect limits are obtained using effective four-fermion interactions induced by leptoquarks at various low energy experiments, the direct ones are drawn from the cross-section involving their production (if any) at high energy colliders. B-anomalies in semi-leptonic B decays, lepton flavour non-universality, lepton flavour violating decays, anomalous magnetic moment of muon, rare kaon decays are few low energy phenomena constraining leptoquarks. A comprehensive list containing all the indirect bounds on leptoquarks can be found in the ``Indirect Limits for Leptoquarks'' section of Ref. \cite{10.1093/ptep/ptaa104}. However, most of the indirect limits involve bounds on product of one diagonal and one off-diagonal Yukawa coupling of the leptoquarks with quarks and leptons \cite{Carpentier:2010ue,Davidson:2018rqt,Mandal:2019gff}. Since, this coupling has been considered diagonal in our analysis, those indirect limits are automatically satisfied. 
\begin{figure}[h!]
	\includegraphics[width=0.95\linewidth]{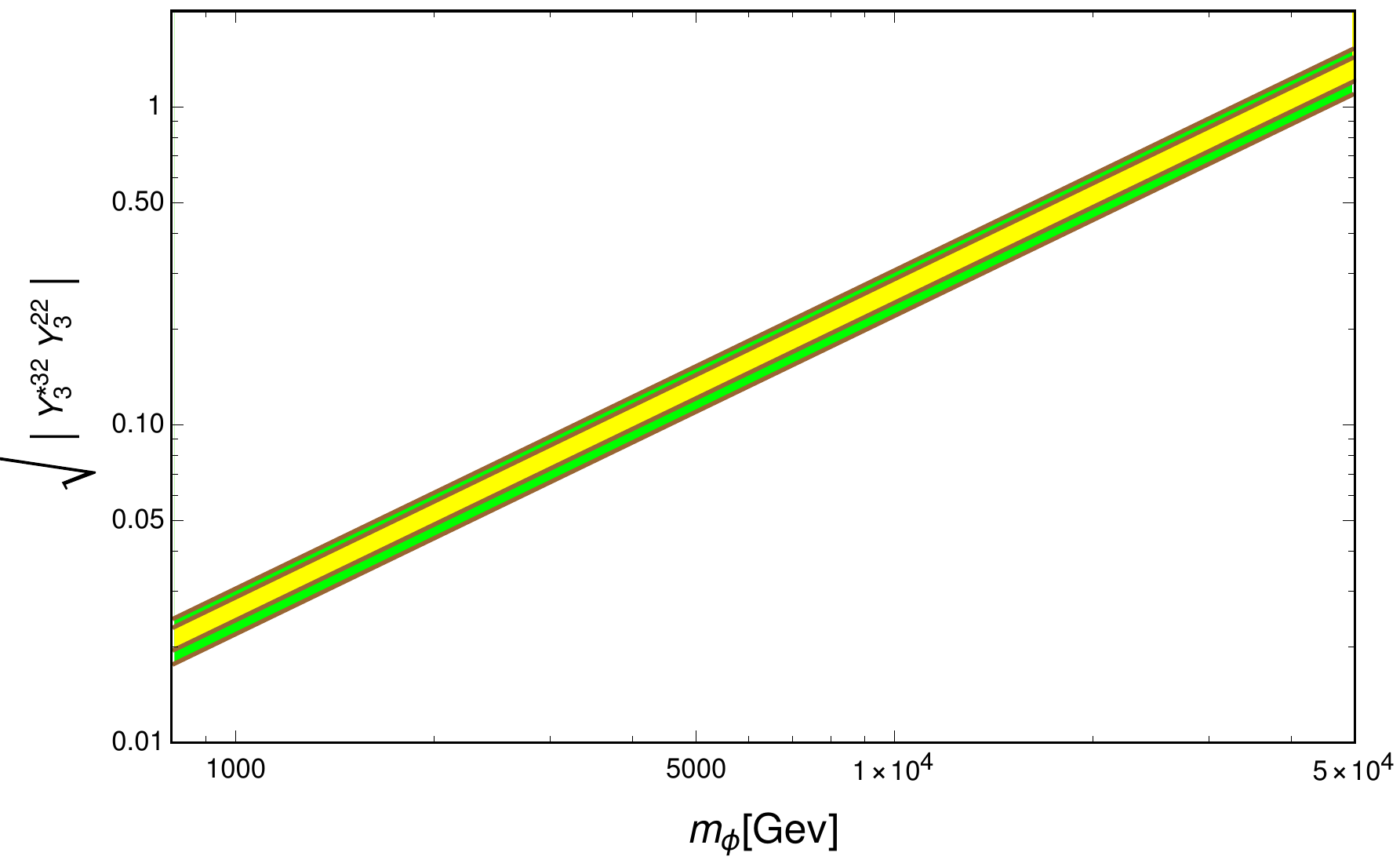}
	\caption{Constraint on Yukawa coupling of $\vec S_3$ as a function of its mass describing $R_{K^{(*)}_{}}$ anomalies \cite{Lee:2021jdr}. The yellow and green colours indicate $1\sigma$ and $2\sigma$ allowed regions.}
	\label{fig:rd}
\end{figure}
\begin{figure}[h!]
	\includegraphics[width=0.95\linewidth]{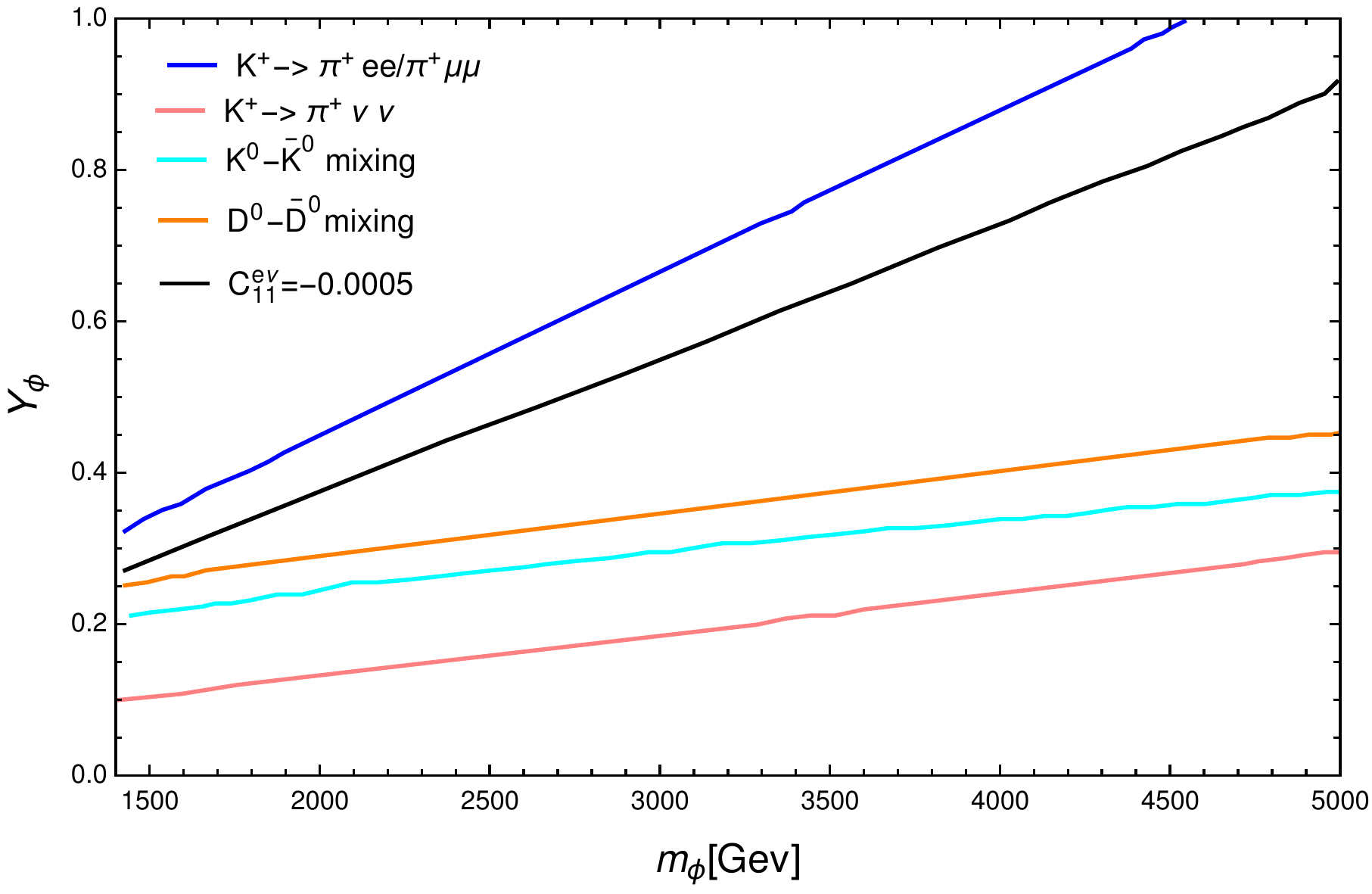}
	\caption{Flavour constraint on Yukawa coupling of first generation $\vec S_3$ as a function of its mass \cite{Crivellin:2021egp}. The blue and red lines indicate the bounds from the ratio $\mathcal B (K^+\to \pi^+ ee)/\mathcal B (K^+\to \pi^+ \mu\mu)$ and the branching fraction  $\mathcal B (K^+\to \pi^+ \nu\nu)$. The cyan and orange lines indicate constraints from neutral kaon and D-meson mixing respectively. The black line signifies leptoquark ($\vec S_3$) contribution of -0.0005 to the Wilson coefficient $(C_{11}^{e\nu})$ involved in the ratio $\mathcal B(\pi\to \mu\nu)/\mathcal B(\pi\to e\nu)$.}
	\label{fig:flavlq}
\end{figure}
On the other hand, it is well known that leptoquarks coupling to multiple generations of quarks and leptons are capable of inducing flavour changing neutral currents. For example, non-chiral leptoquarks, that can interact with both left- and right-handed leptons, obtain stringent constraints from muon $g-2$ \cite{Cheung:2001ip} and the ratio of partial decay rates $(\pi\to e\nu)/(\pi\to \mu\nu)$ \cite{Shanker:1982nd}, if they are allowed to interact with multiple generations of quarks and fermions. In our analysis, we neither do force any leptoquark to couple to different generations of quarks and leptons, nor we work with any non-chiral leptoquark\footnote{Both $\widetilde{R}_2$ and $\vec S_3$ are chiral leptoquarks since they couple to left-handed leptons only.}. Therefore, the constraints arising from flavour changing neutral currents will be much weaker in our scenarios. It is interesting to mention that the possibilities of larger Yukawa couplings of leptoquarks, i.e. $\mathcal{O}(1)$, are not completely ruled out by the low energy observables \cite{Angelescu:2021lln,Angelescu:2018tyl,Crivellin:2017zlb,ColuccioLeskow:2016dox,Crivellin:2020mjs,Bandyopadhyay:2021pld}.

Now, it is impossible to find any single scalar leptoquark solution to all the flavour anomalies and therefore combination of different scalar leptoquarks are essential to take various flavour anomalies into account. For example, leptoquarks $S_1$ and $R_2$ can explain the observed anomalies in $R_{D^{(*)}_{}}$ whereas leptoquark $\vec S_3$ can account for $R_{K^{(*)}_{}}$ anomalies \cite{Angelescu:2021lln}. So, in order to describe both the B-anomalies, one should consider $S_1-\vec S_3$ or $R_2-\vec S_3$ pairs\footnote{Leptoquark $\widetilde{R}_2$ cannot explain any of these two anomalies.}. In Fig. \ref{fig:rd}, we depict the constraints on the parameter space of $\vec S_3$ describing $R_{K^{(*)}_{}}$ anomalies where the yellow and green regions indicate $1\sigma$ and $2\sigma$ allowed ranges \cite{Lee:2021jdr}.
	Again, To generate tiny neutrino masses through loops within the framework of leptoquark models, one has to combine $S_1$ or $\vec S_3$ with $\widetilde R_2$ \cite{Dorsner:2019itg,Babu:2020hun}. Moreover, though non-chiral leptoquarks $S_1$ and $R_2$ can accommodate muon and electron $(g-2)$, the masses of the leptoquarks required for illustrating the experimental values are $\approx 100$ TeV considering the Yukawa couplings under perturbative limit \cite{Dorsner:2019itg}. Therefore, one should consider combinations of $S_1$ \& $\vec S_3$, $\widetilde{S}_1$ \& $\vec S_3$ or $R_2$ and $\widetilde{R}_2$ mixing through Higgs field \cite{Dorsner:2019itg}. However, imposition of various flavour physics constraints along with LHC bounds and $\mu\to e \gamma$ result suggests that none of these scenarios can accommodate for both muon and electron $(g-2)$. Therefore, to get a complete picture regarding various low-energy observables, study of bounds on the parameter space of different leptoquarks is indispensable.

We have already mentioned that we have considered diagonal Yukawa couplings only whereas most of the indirect bounds involve off-diagonal elements also. For instance, Fig. \ref{fig:rd} shows bound on $\sqrt{|Y_3^{*32} Y_3^{22}|}$ as a function of mass for $\vec S_3$ to explain $R_{K^{(*)}_{}}$ anomalies. Now, the upper limits on diagonal Yukawa couplings, derived from the demand of Planck scale stability and perturbativity, are not expected to alter much with the introduction of small off-diagonal couplings. However, these small off-diagonal couplings along with large diagonal elements can now be used to explain various flavour anomalies respecting different indirect bounds. Again, there arises some additional flavour constraints on the parameter space of first generation scalar triplet leptoquark $(\vec S_3)$ \cite{Crivellin:2021egp}, which have been depicted in Fig. \ref{fig:flavlq}; but such bounds do not appear for $\widetilde{R}_2$. Moreover, different low-energy bounds on the Yukawa couplings of $\widetilde{R}_2-\vec S_3$ model are described in Ref. \cite{Babu:2020hun}. However, we are mostly interested in the constraints from the collider perspective.

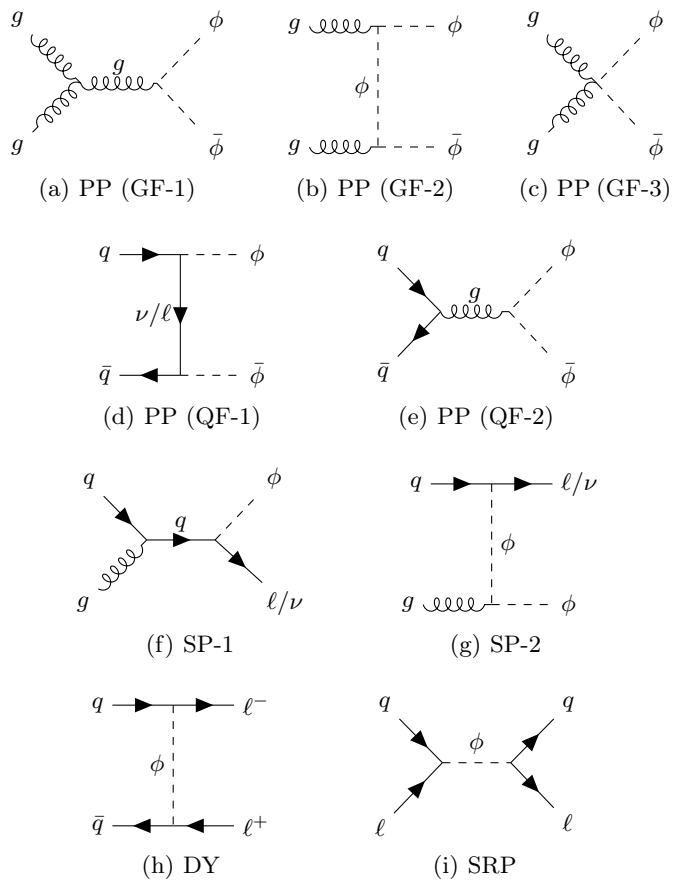
\begin{figure}[h!]
	\centering
	\subfigure[PP (GF-1)]{\begin{tikzpicture}
		\begin{feynman}
		\vertex (n2);
		\vertex [above left=0.9cm of n2] (z1){\( g \)} ;
		\vertex [below left=0.9cm of n2] (s1){\( g \)};
		\vertex [right=1cm of n2] (n3);
		\vertex [above right=0.8cm of n3] (z4){\( \phi \)};
		\vertex [below right=0.8cm of n3] (s4){{\( \bar{\phi} \)}};
		\diagram{(z1)--[gluon](n2)--[gluon,edge label=\( g\)](n3),
			(n2)--[gluon](s1),
			(z4)--[scalar](n3)--[scalar](s4)};
		\end{feynman}
		\end{tikzpicture}}
	\hfill
	\subfigure[PP (GF-2)]{\begin{tikzpicture}
		\begin{feynman}
		\vertex (n2);
		\vertex [left=0.9cm of n2] (n1){\( g \)} ;
		\vertex [right=0.8cm of n2] (n3){\( \phi \)};
		\vertex [below=1.6cm of n2] (s2);
		\vertex [left=0.9cm of s2] (s1){\( g \)};
		\vertex [right=0.8cm of s2] (s3){\( \bar{\phi} \)};
		\diagram{(n1)--[gluon](n2)--[scalar](n3),
			(n2)--[scalar, edge label'=\( \phi \)](s2),
			(s2)--[scalar](s3),
			(s1)--[gluon](s2)};
		\end{feynman}
		\end{tikzpicture}}
	\hfill
	\subfigure[PP (GF-3)]{\begin{tikzpicture}
		\begin{feynman}
		\vertex (n2);
		\vertex [above left=0.9cm of n2] (z1){\( g \)} ;
		\vertex [below left=0.9cm of n2] (s1){\( g \)};
		\vertex [above right=0.8cm of n2] (z4){\( \phi \)};
		\vertex [below right=0.8cm of n2] (s4){{\( \bar{\phi} \)}};
		\diagram{(z1)--[gluon](n2)--[gluon](s1),
			(z4)--[scalar](n2)--[scalar](s4)};
		\end{feynman}
		\end{tikzpicture}}
	
	\subfigure[PP (QF-1)]{\begin{tikzpicture}
		\begin{feynman}
		\vertex (n2);
		\vertex [left=0.8cm of n2] (n1){\( q \)} ;
		\vertex [right=0.8cm of n2] (n3){\( \phi \)};
		\vertex [below=1.6cm of n2] (s2);
		\vertex [left=0.8cm of s2] (s1){\( \bar{q} \)};
		\vertex [right=0.8cm of s2] (s3){\( \bar{\phi} \)};
		\diagram{(n1)--[fermion](n2)--[scalar](n3),
			(n2)--[fermion, edge label'=\( \nu / \ell \)](s2),
			(s2)--[scalar](s3),
			(s2)--[fermion](s1)};
		\end{feynman}
		\end{tikzpicture}}
	\hfil
	\subfigure[PP (QF-2)]{\begin{tikzpicture}
		\begin{feynman}
		\vertex (n2);
		\vertex [above left=0.8cm of n2] (z1){\( q \)} ;
		\vertex [below left=0.8cm of n2] (s1){\( \bar{q} \)};
		\vertex [right=0.9cm of n2] (n3);
		\vertex [above right=0.8cm of n3] (z4){\( \phi \)};
		\vertex [below right=0.8cm of n3] (s4){{\( \bar{\phi} \)}};
		\diagram{(z1)--[fermion](n2)--[gluon,edge label=\( g \)](n3),
			(n2)--[fermion](s1),
			(z4)--[scalar](n3)--[scalar](s4)};
		\end{feynman}
		\end{tikzpicture}}
	
	\subfigure[SP-1]{\begin{tikzpicture}
		\begin{feynman}
		\vertex (a1);
		\vertex [above left=0.8cm of a1] (a0){$q$};
		\vertex [right=0.9cm of a1] (a2);
		\vertex [above right=0.8 cm of a2] (a3){$\phi$};
		\vertex [below left=0.9cm of a1] (b0){$g$};
		\vertex [below right=0.8cm of a2] (b3){$\ell/\nu$};
		\diagram {(a0)--[fermion](a1)--[fermion,edge label=$q$](a2)--[scalar](a3),
			(b0)--[gluon](a1),(a2)--[fermion](b3)};
		\end{feynman}
		\end{tikzpicture}}
	\hfil
	\subfigure[SP-2]{
		\begin{tikzpicture}		
		\begin{feynman}
		\vertex (a1);
		\vertex [above=0.8cm of a1] (z1);
		\vertex [left=8mm of z1] (z0) {$q$};
		\vertex [right=8mm of z1] (z2) {$\ell/\nu$};
		\vertex [below=0.8cm of a1] (b1);
		\vertex [left=9mm of b1] (b0){$g$};
		\vertex [right=8mm of b1] (b2) {$\phi$};
		\diagram {(z0)--[fermion](z1)--[scalar, edge label=$\phi$](b1)--[scalar](b2),
			(b0)--[gluon](b1),
			(z1)--[fermion](z2)};
		\end{feynman}
		\end{tikzpicture}}
	
	\subfigure[DY]{\begin{tikzpicture}
		\begin{feynman}
		\vertex (n2);
		\vertex [left=0.8cm of n2] (n1){\( q \)} ;
		\vertex [right=0.8cm of n2] (n3){\( \ell^- \)};
		\vertex [below=1.6cm of n2] (s2);
		\vertex [left=0.8cm of s2] (s1){\( \bar{q} \)};
		\vertex [right=0.8cm of s2] (s3){\( \ell^+ \)};
		\diagram{(n1)--[fermion](n2)--[fermion](n3),
			(n2)--[scalar, edge label'=\(\phi \)](s2),
			(s2)--[anti fermion](s3),
			(s2)--[fermion](s1)};
		\end{feynman}
		\end{tikzpicture}}
	\hfil
	\subfigure[SRP]{\begin{tikzpicture}
		\begin{feynman}
		\vertex (a1);
		\vertex [above left=0.8cm of a1] (a0){$q$};
		\vertex [right=0.9cm of a1] (a2);
		\vertex [above right=0.8 cm of a2] (a3){$q$};
		\vertex [below left=0.9cm of a1] (b0){$\ell$};
		\vertex [below right=0.8cm of a2] (b3){$\ell$};
		\diagram {(a0)--[fermion](a1)--[scalar,edge label=$\phi$](a2)--[fermion](a3),
			(b0)--[fermion](a1),(a2)--[fermion](b3)};
		\end{feynman}
		\end{tikzpicture}}
	\caption{Leading order Feynman diagrams involving direct bounds on leptoquarks. The first two rows correspond to Leptoquark pair production (PP) at LHC while the third and fourth rows signify single production (SP) of leptoquark associated with a quark, leptoquark contribution to Drell-Yan like di-lepton process and single resonant production of leptoquark (SRP). Regarding pair production, the first three diagrams indicate gluon fusion (GF), while the last two illustrate quark fusion (QF). The photon and Z mediated diagrams have been ignored due to very small contribution.}
	\label{FDLQ}
\end{figure}
\begin{figure*}[h!]
	\subfigure[$\phi$ coupling to $e$ and $u$]{\includegraphics[width=0.49\linewidth]{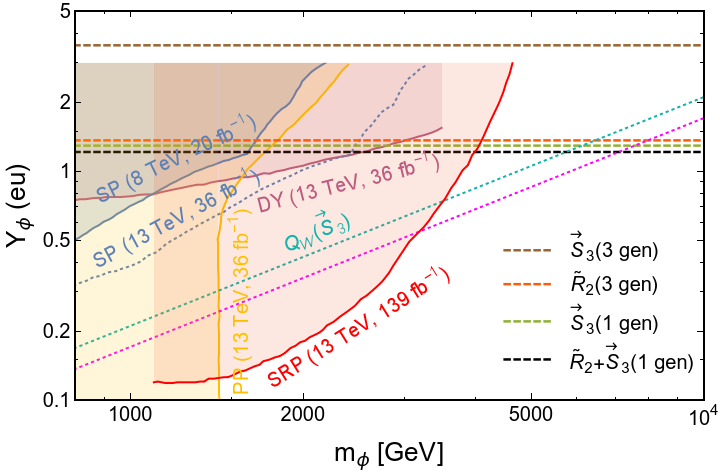}}\hfill
	\subfigure[$\phi$ coupling to $e$ and $d$]{\includegraphics[width=0.49\linewidth]{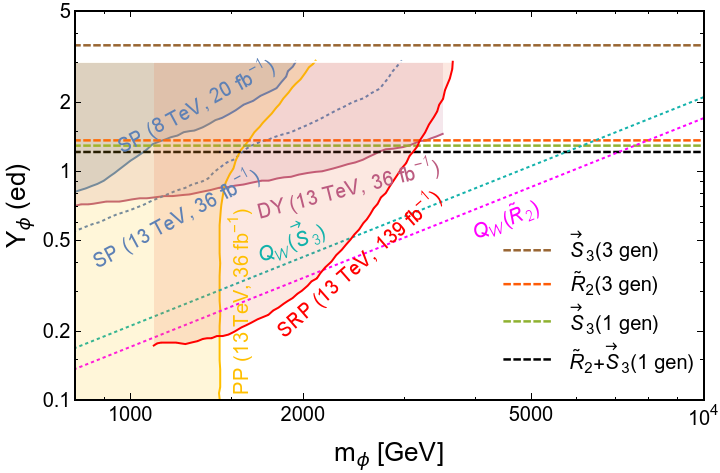}}
	
	\subfigure[$\phi$ coupling to $\nu$ and $u$]{\includegraphics[width=0.49\linewidth]{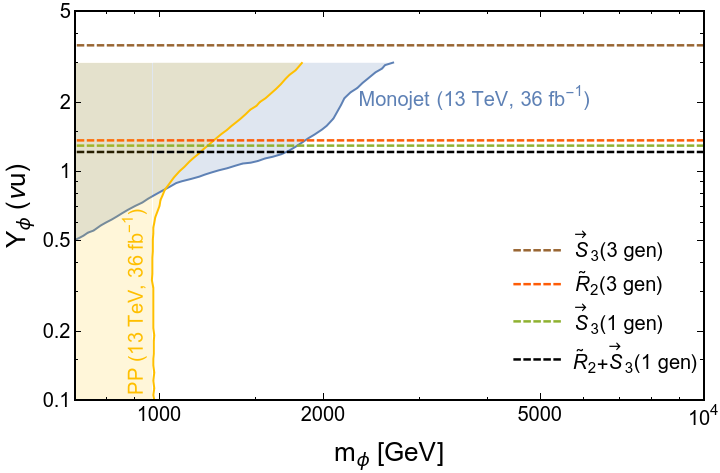}}\hfill
	\subfigure[$\phi$ coupling to $\nu$ and $d$]{\includegraphics[width=0.48\linewidth]{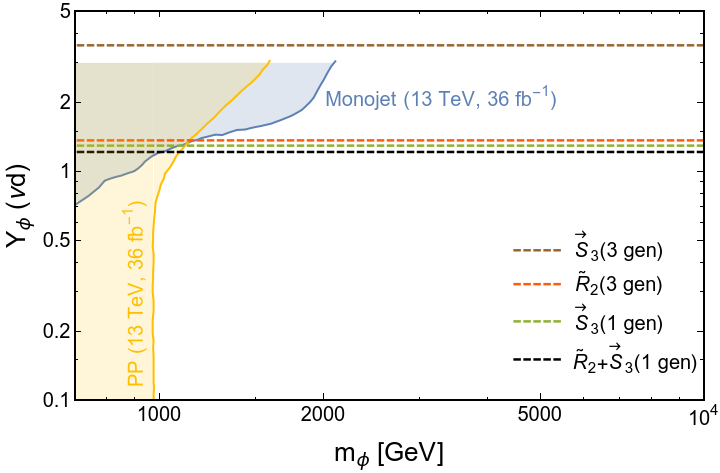}}
	\caption{Bounds on parameter space of scalar leptoquark coupling to first generation of quarks and leptons \cite{Schmaltz:2018nls, Buonocore:2020erb}. The shaded regions are disallowed by direct detection. The limits from pair production involving charged leptons are recast from \cite{CMS:2018sxp}, while the same involving neutrinos recast from \cite{CMS:2018qqq}. These bounds are shown in yellow colour. The limits emerging from single production (SP), Drell-Yan (DY) and single resonant production (SRP), shown by bluish, maroonish purple and reddish portions, are based on Refs. \cite{CMS:2018ipm,CMS:2015xzc, Buonocore:2020erb}. On the other hand, the mono-jet limits (bluish) are drawn from Ref. \cite{CMS:2017zts}. The dotted lines with magenta and seagreen colours represent constraints from weak hypercharge measurements involving $\widetilde{R}_2$ and $\vec S_3$ respectively. Finally, the dashed lines indicate theoretical upper bounds on the Yukawa coupling appearing from Planck scale stability up to two loop order; the brown and red lines represent the limits for three generations of $\vec S_3$ and $\widetilde{R}_2$, whereas the green and black lines portray the same for one generation of $\vec S_3$ and $\widetilde{R}_2+\vec S_3$.}
	\label{limit1}
\end{figure*}
\begin{figure*}[h!]
	\subfigure[$\phi$ coupling to $\mu$ and $c$]{\includegraphics[width=0.49\linewidth]{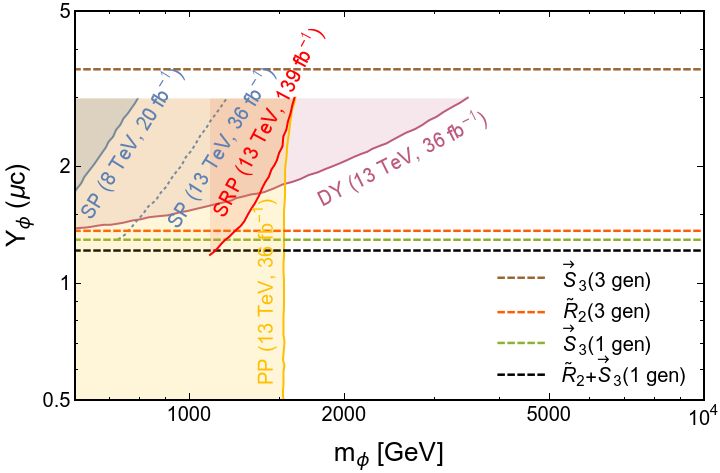}}\hfill
	\subfigure[$\phi$ coupling to $\mu$ and $s$]{\includegraphics[width=0.49\linewidth]{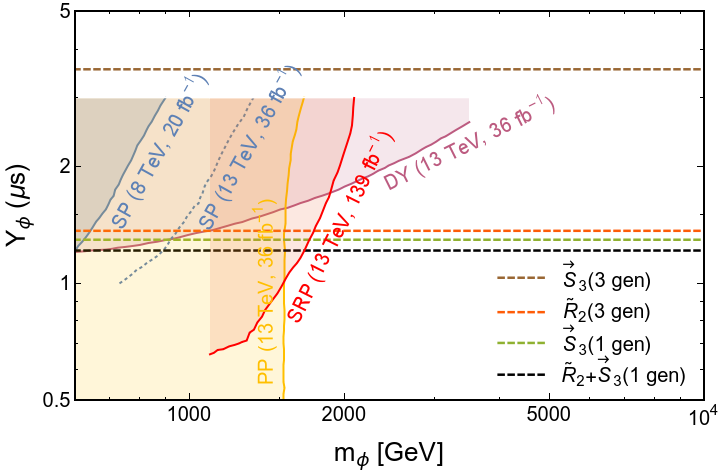}}
	
	\subfigure[$\phi$ coupling to $\nu$ and $c$]{\includegraphics[width=0.49\linewidth]{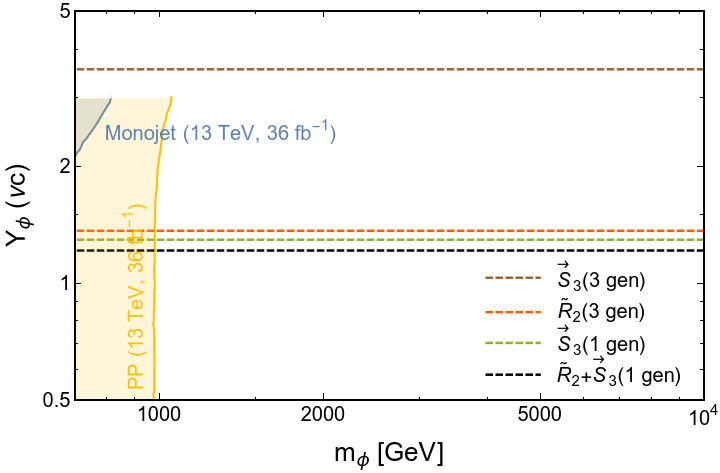}}\hfill
	\subfigure[$\phi$ coupling to $\nu$ and $s$]{\includegraphics[width=0.49\linewidth]{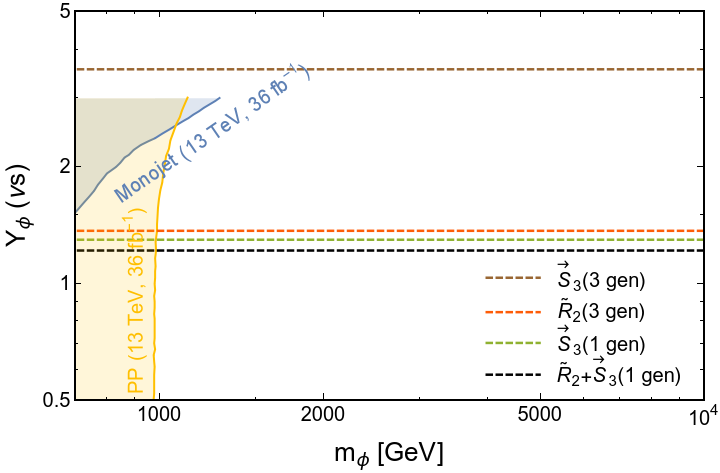}}
	\caption{Bounds on parameter space of scalar leptoquark coupling to second generation of quarks and leptons \cite{Schmaltz:2018nls, Buonocore:2020erb}. The shaded regions are disallowed by direct detection. The limits from pair production involving charged leptons are recast from \cite{CMS:2018sgp}, while the same involving neutrinos recast from \cite{CMS:2018qqq}. These bounds are shown in yellow colour.  The limits emerging from single production (SP), Drell-Yan (DY) and single resonant production (SRP), shown by bluish, maroonish purple and reddish portions, are based on Refs. \cite{CMS:2018ipm,CMS:2015xzc, Buonocore:2020erb}. On the other hand, the mono-jet limits (bluish) are drawn from Ref. \cite{CMS:2017zts}. Finally, the dashed lines indicate theoretical upper bounds on the Yukawa coupling appearing from Planck scale stability up to two loop order; the colour codes are already mentioned in Fig. \ref{limit1}. }
	\label{limit2}
\end{figure*}
\begin{figure*}[h!]
	\subfigure[$\phi$ coupling to $\tau$ and $b$]{\includegraphics[width=0.49\linewidth]{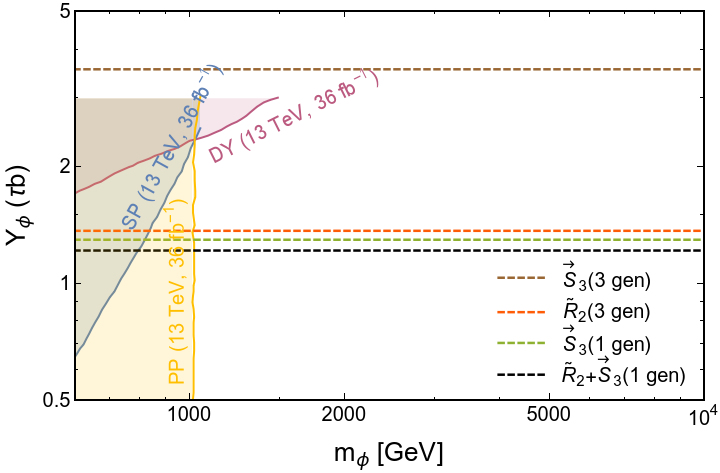}}\hfill
	\subfigure[$\phi$ coupling to $\nu$ and $b$]{\includegraphics[width=0.49\linewidth]{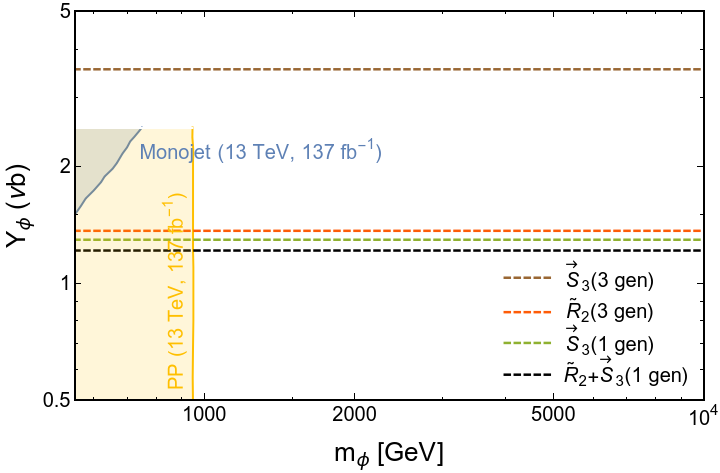}}
	\caption{Bounds on parameter space of scalar leptoquark coupling to third generation of quarks and leptons \cite{Schmaltz:2018nls,CMS:2020wzx}. The shaded regions are disallowed by direct detection. The limit from pair production involving charged leptons is recast from \cite{CMS:2018eud} which is shown in yellow colour. The limits emerging from single production (SP) and Drell-Yan (DY), shown by bluish and maroonish purple portions, are based on Refs. \cite{CMS:2018hjx,ATLAS:2017eiz}. Finally, the dashed lines indicate theoretical upper bounds on the Yukawa coupling appearing from Planck scale stability up to two loop order; the colour codes are already mentioned in Fig. \ref{limit1}. }
	\label{limit3}
\end{figure*}
While discussing the direct bounds on leptoquarks, we consider pair production (PP), single production (SP) associated with a quark, Drell-Yan processes (DY) and single resonant production of leptoquark (SRP). At $pp$ collider, like LHC, pair production of leptoquarks can occur through gluon fusion (GF) as well as via quark fusion (QF) whose corresponding Feynman diagrams are shown in first and second rows of Fig. \ref{FDLQ}. On the other hand, the Feynman diagrams for single production of leptoquark, contribution to Drell-Yan like dilepton processes and SRP are presented in third and fourth rows of Fig. \ref{FDLQ}. Regarding the coupling of leptoquarks to charged leptons, we get opposite sign di-lepton (OSD) signature for DY processes as shown in Fig~\ref{FDLQ}(h), whereas PP and SP provide di-jet plus OSD and mono-jet plus OSD finalstates at the detector \cite{Bandyopadhyay:2018syt, Bandyopadhyay:2021pld}. Conversely, for leptoquarks coupling to neutrinos, we have di-jet plus missing energy and mono-jet plus missing energy signatures only. The full data set collected at HERA in $ep$ collision excluded first generation of leptoquark with mass up to 800 GeV at 95\% C.L. for coupling to be 0.3 \cite{Collaboration:2011qaa}. In more recent study they have modified $Y_\phi/m_\phi$ limits for first generation of leptoquarks \cite{Abramowicz:2019uti}. The CMS collaboration at the LHC also searched for single production of leptoquarks which probe the high coupling region of leptoquarks \cite{CMS:2015xzc,CMS:2020gru}.


We depict different direct constraints on the parameter space of scalar leptoquarks in Figs. \ref{limit1}, Fig. \ref{limit2} and Fig. \ref{limit3}. These bounds can be recasted for different models of scalar leptoquarks depending on the cross-sections and the corresponding decay branching fractions leading to the finalstates. Fig. \ref{limit1} summarizes the bounds for first generation of leptoquark, Fig. \ref{limit2} and Fig. \ref{limit3} portray the same for second and third generations of leptoquarks, respectively. All the plots presented in Fig. \ref{limit1} and Fig. \ref{limit2} are taken from Ref. \cite{Schmaltz:2018nls, Buonocore:2020erb}, which uses Refs. \cite{CMS:2018sxp,CMS:2018sgp,CMS:2018qqq} for PP, Ref. \cite{CMS:2018ipm} for SP, Ref. \cite{CMS:2015xzc} for DY, LHC Run II data for SRP and  Ref. \cite{CMS:2017zts} for mono-jet signature with first and second generations of leptons to restrict the parameter space for leptoquark-quark-lepton coupling below 3.0 with mass of leptoquark below 3 TeV. Conversely, Fig. \ref{limit3}(a) describing constraints on $\phi\tau b$ coupling is taken from Ref. \cite{Schmaltz:2018nls} that uses Refs. \cite{CMS:2018eud,CMS:2018hjx,ATLAS:2017eiz} for their analysis and Fig. \ref{limit3}(b) illustrating limits on $\phi\nu b$ coupling is taken from Ref. \cite{CMS:2020wzx}. For the finalstates involving charged leptons, the yellow, blueish, maroonish purple and reddish portions indicate the prohibited region from PP, SP, DY and SRP processes. On the contrary, for the finalstates involving missing energy, the yellow and bluish regions signify PP and mono-jet signals. 

We impose the theoretical bounds obtained from the perturbative unitarity and the stability at the two-loop for the dimensionless couplings in $m_{\phi}-Y_{\phi}$ plane for $\widetilde{R}_2$, $\vec S_3$ and $\widetilde{R}_2+\vec S_3$, respectively.  The brown $(Y_\phi=3.90)$ and the red $(Y_\phi=1.36)$ dashed lines depict the theoretical upper limits on the Yukawa couplings of leptoquarks for three generations of $\vec S_3$ and $\widetilde{R}_2$, respectively, considering Planck scale stability at two-loop level. The same for one generation of $S_3$ and $\widetilde{R}_2+\vec S_3$ are presented by the green $(Y_\phi=1.29)$ and the black $(Y_\phi=1.21)$ dashed lines.

At this point it is worth mentioning that we do not present the bounds on $\widetilde{R}_2$ with one generation and $\widetilde{R}_2+\vec S_3$ with three generations in these plots. Actually, as described earlier, $\widetilde{R}_2$ with one generation cannot achieve Planck scale stability for any small value of $Y_\phi$  at two-loop order. On the other hand, though $\widetilde{R}_2+\vec S_3$ with three generations shows stability for $Y_\phi\leq 1.0$, it looses perturbativity at at an energy scale ($\sim$ $10^{14.4}$ GeV) far below the Planck scale  at two-loop order.

For the first generation leptoquark coupling to charged lepton, there exists another bound from measurement of weak charge of proton and nuclei \cite{Schmaltz:2018nls}. This quantity is measured through atomic parity violation and parity violating electron scattering \cite{10.1093/ptep/ptaa104,Qweak:2018tjf}. For $\widetilde R_2$ and $\vec S_3$ these measurements translate into $Y_\phi\leq0.17 \Big(\frac{m_\phi}{1 \rm{TeV}}\Big)$ and $Y_\phi\leq0.21 \Big(\frac{m_\phi}{1 \rm{TeV}}\Big)$, respectively, which are shown by the dotted lines in magenta and seagreen colours, respectively. Since, $\widetilde{R}_2$ couples to the down type quarks only, while $\vec S_3$ interacts with both up-type and down-type quarks, we find the magenta line in Fig. \ref{limit1}(b) only, whereas the seagreen line exits in both Fig. \ref{limit1}(a) and Fig. \ref{limit1}(b). Since, the nuclei do not contain other generations of quarks as valance quarks, this kind of limit does not appear for other generations of leptoquarks. 

From these results it is evident that the theoretical limits coming from Planck scale stability and perturbative unitarity up to two-loop order might put stronger constraints on the parameter space of leptoquarks with higher mass range specially for second and third generations of the leptoquarks. On the other hand, bounds on the Higgs-leptoquark quartic coupling are not very well studied in literature. In our analysis we find that this coupling being larger than $\sim0.2$ disturbs the perturbativity of the theory till Planck scale\footnote{To be more specific, for three generations of $\widetilde R_2$ one needs $(\lambda_{2},\widetilde \lambda_{2})\leq0.22$ and for three generations of $\vec S_3$ we require $(\lambda_{3},\widetilde \lambda_{3})\leq0.18$ in order to confirm Planck scale perturbativity.}. 


\section{Conclusion}
\label{sec:concl}

In this paper, we have studied the scalar doublet leptoquark $\widetilde{R}_2$, the scalar triplet leptoquark $\vec S_3$ and their combination with one generation as well as three generations in light of the perturbativity and the stability of the Higgs vacuum. The extra contribution in the running of the gauge couplings at one-loop mainly depends on the number of the leptoquark components present in the model, which is determined by the gauge structure of it. Though at two-loop, they depend on the leptoquark Yukawa couplings but they do not depend on the Higgs-leptoquark couplings explicitly. With the two-loop effects, the gauge coupling $g_2$ for the leptoquark $\vec S_3$ and the combined scenario of  $\widetilde{R}_2$ and $\vec S_3$ with three generations diverges at $10^{19.7}$ GeV and $10^{14.4}$ GeV, respectively, which forces the other couplings to hit singularity at those scales. But at one-loop, all the leptoquark models considered in this paper achieve Planck scale perturbativity with gauge couplings. It is also noteworthy that no Landau pole emerges in the running of gauge couplings for two generations of these leptoquarks. The Higgs-leptoquark quartic couplings acquire sever constraints from Planck scale perturbativity. With larger EW values of these couplings (like 0.3) the theories become non-perturbative at much lower energy scales than the Planck scale. These constraints do not change much due to alteration in the leptoquark Yukawa couplings. For three generations scenario with $\widetilde R_2$ and $\vec S_3$ combined, the Higgs-leptoquark quartic couplings diverge much below the Planck scale. On the other hand the leptoquark Yukawa couplings get upper bound from the Planck scale perturbativity and stability of the Higgs vacuum. In the running of $\lambda_{h}$, the gauge couplings exert positive contributions, whereas the Yukawa couplings of leptoquarks introduce negative effects.
For three generations of $\widetilde{R}_2$ with the Higgs-leptoquark quartic couplings being 0.1, the Yukawa coupling should be smaller than 1.36 for the theory maintaining stability till Planck scale. This number becomes 1.29, 3.9\footnote{The upper bound on $Y_\phi$ would be $\sqrt{4\pi}$ considering perturbative unitarity and Planck scale stability for three generations of $\vec S_3$.} and 1.21 for one generation of $\vec S_3$, three generations of $\vec S_3$ and one generation of $\widetilde{R}_2+\vec S_3$ respectively. Finally, regarding the Coleman-Weinberg effective potential approach, the presence of any of these leptoquarks with any number of generations pushes the metastable vacuum of SM to the stable region although the $3\sigma$ contour of $\widetilde{R}_2$ with one generation marginally touches the metastable region. The phenomenological bounds obtained from mainly the collider experiments are also drawn along with out theoretical bounds. We see that the Planck scale perturbativity and stability puts some theoretical additional restrictions to the parameter space of the leptoquarks on top of the experimental bounds.

\subsection*{Acknowledgements}
The authors acknowledge SERB CORE Grant CRG/2018/ 004971 and MATRICS Grant MTR/2020/000668 for the financial support. SJ thanks DST/INSPIRES/03/2018/ 001207 for the financial support towards finishing this work. This work has also been supported in part by MCIN/ AEI/10.13039/501100011033 Grant No. PID2020-114473 GB-I00,
and Grant PROMETEO/2021/071 (Generalitat Valenciana). The authors also thank Alexander Bednyakov for some useful comments.

\appendix
\section{Two-loop beta functions of $g_3$}
\label{sec:g32}
\vspace*{+2mm}
Using \texttt{SARAH}, we generate the beta function of $g_3$ for different models till two-loops which are given below:
\vspace*{+2mm}
{
\begin{align}
&\beta(g_3)^{2-loop}_{SM}=-\,7\Big(\frac{g_{3}^{3}}{16\pi^2}\Big)+\frac{g_{3}^{3}}{(16\pi^2)^2}\bigg[\frac{11}{10}\,g_1^2+\frac{9}{2}\,g_2^2-26\,g_3^2\nn\\
&\qquad \qquad \qquad -2\,\Tr\,\big(\X_\uq+\X_\dq\big)\bigg],\\[2mm]
&\beta(g_3)^{2-loop}_{\widetilde{R}_2,1-gen}=-\,\frac{20}{3}\Big(\frac{g_{3}^{3}}{16\pi^2}\Big)+\frac{g_{3}^{3}}{(16\pi^2)^2}\bigg[\frac{7}{6}\,g_1^2+\frac{15}{2}\,g_2^2\nn\\
&\qquad \qquad \quad \quad -\frac{56}{3}\,g_3^2-2\,\Tr\,\Big(\X_\uq+\X_\dq+\frac{1}{2}\X_2\Big)\bigg],\\[2mm]
&\beta(g_3)^{2-loop}_{\widetilde{R}_2,3-gen}=-\,6\Big(\frac{g_{3}^{3}}{16\pi^2}\Big)+\frac{g_{3}^{3}}{(16\pi^2)^2}\bigg[\frac{13}{10}\,g_1^2+\frac{27}{2}\,g_2^2\nn\\
&\qquad \qquad \quad -4\,g_3^2-2\,\Tr\,\Big(\X_\uq+\X_\dq+\frac{1}{2}\sum_{i=1}^3\X_{2,i}\Big)\bigg],\\[2mm]
&\beta(g_3)^{2-loop}_{\vec S_3,1-gen}= -\,\frac{13}{2}\Big(\frac{g_{3}^{3}}{16\pi^2}\Big)+ 
\frac{g_3^3}{(16\pi^2)^2}\bigg[ \frac{3}{2}\, g_{1}^{2} +\frac{33}{2}\, g_{2}^{2} \nn\\
&\qquad \qquad \qquad -15\,g_{3}^{2}-2\,\Tr\Big(\X_\uq+\X_\dq
+\frac{3}{4}\,\X_{3}\Big)\bigg], \,   \\[2mm]
&\beta(g_3)^{2-loop}_{\vec S_3,3-gen}= -\,\frac{11}{2}\Big(\frac{g_{3}^{3}}{16\pi^2}\Big)+ 
\frac{g_3^3}{(16\pi^2)^2}\bigg[ \frac{23}{10}\, g_{1}^{2} +\frac{81}{2}\, g_{2}^{2} \nn\\
&\qquad \quad \quad \quad +7g_{3}^{2}-2\,\Tr\Big(\X_\uq+\X_\dq
+\frac{3}{4} \sum_{i=1}^{3}\X_{3,i}\Big)\bigg], \,   \\[2mm]
&\beta(g_3)^{2-loop}_{\widetilde{R}_2+\vec S_3,1-gen}= -\,\frac{37}{6}\Big(\frac{g_{3}^{3}}{16\pi^2}\Big)+ 
\frac{g_3^3}{(16\pi^2)^2}\bigg[ \frac{47}{30}\, g_{1}^{2} \nn\\ &   +\frac{39}{2}\, g_{2}^{2}-\frac{23}{3}\,g_{3}^{2}-2\,\Tr\Big(\X_\uq+\X_\dq
+\frac{1}{2}\,\X_2+\frac{3}{4}\X_{3}\Big)\bigg], \,   \\[2mm]
&\beta(g_3)^{2-loop}_{\widetilde{R}_2+\vec S_3,3-gen}= -\,\frac{9}{2}\Big(\frac{g_{3}^{3}}{16\pi^2}\Big)+ 
\frac{g_3^3}{(16\pi^2)^2}\bigg[ \frac{5}{2}\, g_{1}^{2} +\frac{99}{2}\, g_{2}^{2} \nn\\
&\quad   +29\,g_{3}^{2}-2\,\Tr\Big(\X_\uq+\X_\dq
+\frac{1}{2}\sum_{i=1}^{3}\X_{2,i}+\frac{3}{4}\sum_{i=1}^{3}\X_{3,i}\Big)\bigg].
\end{align}}

\section{Two-loop beta functions of $g_1$}
\label{sec:g12}
\vspace*{+2mm}
Now, with the help of \texttt{SARAH}, we show the two-loops beta function of $g_1$ for all the models as following:
\vspace*{+2mm}
{ 
\begin{align}
&\beta(g_1)^{2-loop}_{SM}=\frac{41}{10}\,\Big(\frac{g^3_1}{16\pi^2}\Big)+ \frac{g^3_1}{(16\pi^2)^2}\,\bigg[\frac{199}{50}\,g_1^2+\frac{27}{10}\,g_2^2\nn\\
&\qquad \qquad \quad +\frac{44}{5}\,g_3^2-\frac{3}{2}\Tr\,\Big(\X_\cl+\frac{17}{15}\X_\uq+\frac{1}{3}\X_\dq\Big)\bigg],\\[2mm]
&\beta(g_1)^{2-loop}_{\widetilde{R}_2,1-gen}=\frac{62}{15}\,\Big(\frac{g^3_1}{16\pi^2}\Big)+ \frac{g^3_1}{(16\pi^2)^2}\,\bigg[\frac{299}{75}\,g_1^2+3\,g_2^2\nn\\
&\quad  +\frac{28}{3}\,g_3^2-\frac{3}{2}\Tr\,\Big(\X_\cl+\frac{17}{15}\X_\uq+\frac{1}{3}\X_\dq+\frac{13}{15}\X_{2}\Big)\bigg],\\[2mm]
&\beta(g_1)^{2-loop}_{\widetilde{R}_2,3-gen}=\frac{21}{5}\,\Big(\frac{g^3_1}{16\pi^2}\Big)+ \frac{g^3_1}{(16\pi^2)^2}\,\bigg[4\,g_1^2+\frac{18}{5}\,g_2^2\nn\\
&\quad +\frac{52}{5}\,g_3^2-\frac{3}{2}\Tr\,\Big(\X_\cl+\frac{17}{15}\X_\uq+\frac{1}{3}\X_\dq+\frac{13}{15}\sum_{i=1}^{3}\X_{2,i}\Big)\bigg],\\[2mm]
&\beta(g_1)^{2-loop}_{\vec S_3,1-gen}=\frac{43}{10}\,\Big(\frac{g^3_1}{16\pi^2}\Big)+ \frac{g^3_1}{(16\pi^2)^2}\,\bigg[\frac{207}{50}\,g_1^2+\frac{15}{2}\,g_2^2\nn\\
&\qquad \quad +12\,g_3^2-\frac{3}{2}\Tr\,\Big(\X_\cl+\frac{17}{15}\X_\uq+\frac{1}{3}\X_\dq+\X_3\Big)\bigg],\\[2mm]
&\beta(g_1)^{2-loop}_{\vec S_3,3-gen}=\frac{47}{10}\,\Big(\frac{g^3_1}{16\pi^2}\Big)+ \frac{g^3_1}{(16\pi^2)^2}\,\bigg[\frac{223}{50}\,g_1^2+\frac{171}{10}\,g_2^2\nn\\
&\;\;\; +\frac{92}{5}\,g_3^2-\frac{3}{2}\Tr\,\Big(\X_\cl+\frac{17}{15}\X_\uq+\frac{1}{3}\X_\dq+\sum_{i=1}^3\X_{3,i}\Big)\bigg],\\[2mm]
&\beta(g_1)^{2-loop}_{\widetilde{R}_2+\vec S_3,1-gen}=\frac{13}{3}\,\Big(\frac{g^3_1}{16\pi^2}\Big)+ \frac{g^3_1}{(16\pi^2)^2}\,\bigg[\frac{311}{75}\,g_1^2\nn\\
&\qquad \qquad \qquad +\frac{39}{5}\,g_2^2+\frac{188}{15}\,g_3^2-\frac{3}{2}\Tr\,\Big(\X_\cl+\frac{17}{15}\X_\uq\nn\\
&\qquad \qquad \qquad+\frac{1}{3}\X_\dq+\frac{13}{15}\X_{2}+\X_{3}\Big)\bigg],\\[2mm]
&\beta(g_1)^{2-loop}_{\widetilde{R}_2+\vec S_3,3-gen}=\frac{24}{5}\,\Big(\frac{g^3_1}{16\pi^2}\Big)+ \frac{g^3_1}{(16\pi^2)^2}\,\bigg[\frac{112}{25}\,g_1^2\nn\\
&\qquad \qquad \qquad+18\,g_2^2+20\,g_3^2-\frac{3}{2}\Tr\,\Big(\X_\cl+\frac{17}{15}\X_\uq\nn\\
&\qquad \qquad \qquad+\frac{1}{3}\X_\dq+\frac{13}{15}\sum_{i=1}^{3}\X_{2,i}+\sum_{i=1}^{3}\X_{3,i}\Big)\bigg].
\end{align}}

\section{Running of Top Yukawa Coupling}
\label{sec:topY}
\vspace*{+2mm}
Top Yukawa coupling plays very important role in the stability of Higgs vacuum. So, it is important to study the RG evolution of this parameter. As already mentioned in Eq.  \eqref{eq:pbnd}, the absolute value for top Yukawa coupling at any energy scale must be less than $\sqrt{4\pi}$ in order to ensure the perturbativity of the model. It is worth mentioning that although the Yukawa couplings for leptons and other quarks also vary with the scale, their initial value at the EW scale are so small that they usually never cross the perturbativity bound unless some other parameter hits the divergence. Therefore, we restrict our discussion for the top Yukawa coupling only
Now, to investigate the running of top Yukawa coupling, we study the RG evolution of $Y_\uq$ (Yukawa matrix for up-type quarks) whose (3,3) component would provide us the desired result. The one-loop and two-loop beta functions of $Y_\uq$ under SM are as follows:
\vspace*{+2mm}
{
\begin{align}
\label{eq:yusm}
&\beta(Y_\uq)^{1-loop}_{SM}
=\frac{Y_\uq}{16\pi^2}\bigg[\frac{3}{2}\,\Big(\widetilde{\X}_\uq-\widetilde{\X}_\dq\Big)-\text{I}_3\,\Big(\frac{17}{20}g_1^2+\frac{9}{4}g_2^2\nn\\
&\qquad \qquad \qquad  +8g_3^2\Big)+3\,\text{I}_3\Tr\Big(\X_\uq+\X_\dq+\frac{1}{3}\X_\cl\Big)\bigg],
\end{align}}
{
\begin{align}
&\beta(Y_\uq)^{2-loop}_{SM}=\beta(Y_\uq)^{1-loop}_{SM}+\frac{Y_\uq}{(16\pi^2)^2}\bigg[\Big(\frac{11}{4}\,\widetilde{\X}_\dq^2-\widetilde{\X}_\uq\widetilde{\X}_\dq\nn\\
&\quad\quad\quad +\frac{3}{2}\widetilde{\X}_\uq^2-\frac{1}{4}\widetilde{\X}_\dq\widetilde{\X}_\uq\Big)
+\widetilde{\X}_\uq\Big\{\frac{223}{80}g_1^2+\frac{135}{16}g_2^2\nonumber\\
&\quad\quad\quad+16g_3^2-12\lambda_{h}-\frac{27}{4}\,\Tr\big(\X_\uq+\X_\dq+\frac{1}{3}\,\X_\cl\big)\Big\}\nn\\
&\quad\quad\quad-\widetilde{\X}_\dq\Big\{\frac{43}{80}g_1^2-\frac{9}{16}g_2^2+16g_3^2-\frac{15}{4}\,\Tr\big(\X_\uq\nn\\
&\quad\quad\quad+\X_\dq+\frac{1}{3}\,\X_\cl\big)\Big\}+\text{I}_3\,\Big(6\lambda_{h}^2+\frac{1187}{600}g_1^4-\frac{23}{4}g_2^4\nonumber\\
&\quad\quad\quad-108g_3^4-\frac{9}{20}g_1^2 g_2^2+\frac{19}{15}g_1^2 g_3^2+9g_2^2 g_3^2\Big)\nn\\
&\quad\quad\quad+\frac{5}{8}\,\I\,\Tr\,\Big\{\Big(g_1^2+9g_2^2+32g_3^2\Big)\X_\dq+\Big(\frac{17}{5}g_1^2\nonumber\\
&\quad\quad\quad+9g_2^2+32g_3^2\Big)\X_\uq+3\Big(g_1^2+g_2^2\Big)\X_\cl\Big\}\nn\\
&\quad \quad \quad-\frac{27}{4}\,\I\,\Tr\,\Big(\X_\uq^2+\X_\dq^2-\frac{2}{9}\,\widetilde{\X}_\uq\widetilde{\X}_\dq+\frac{1}{3}\X_\cl^2\Big)\bigg].
\end{align}}

\vspace*{+2mm}
The above two expressions are matrix equations with $\I$ indicating $3\times3$ identity matrix in flavour-space of up-type quarks.

Now, for leptoquark $\widetilde{R}_2$ (with both one generation and three generations),
one-loop beta function of $Y_\uq$ does not get any additional contribution at one-loop level, i.e. $\Delta\beta(Y_\uq)^{1-loop}_{\widetilde{R}_2}=0$. Hence, it  remains same like that of SM:  
\vspace*{+2mm}
{\begin{align}
\label{eq:yur2}
\beta(Y_\uq)^{1-loop}_{\widetilde{R}_2,1-gen}=\beta(Y_\uq)^{1-loop}_{\widetilde{R}_2,3-gen}=\beta(Y_\uq)^{1-loop}_{SM}~.
\end{align}}

\vspace*{+2mm}
Nevertheless, there exist some non-vanishing two-loop contributions to it, and hence the full two-loop beta functions of $Y_\uq$ for leptoquark $\widetilde{R}_2$ with one generation as well as three generations can be expressed as
\footnote{It should be noted that while taking $\displaystyle\sum_{i=1}^3\Delta\beta$ for any parameter, one has to change $\X_\gamma$ to $\displaystyle\sum_{i=1}^3\X_{\gamma,i}$ and $f(\lambda_\gamma, \widetilde{\lambda}_\gamma)$ to $\displaystyle\sum_{i=1}^3f(\lambda_\gamma^{ii}, \widetilde{\lambda}_\gamma^{ii})$ along with making all the other additive terms thrice of the one generation case.}: 
\vspace*{+2mm}
{
\begin{align}
&\beta(Y_\uq)^{2-loop}_{\widetilde{R}_2,1-gen}=\beta(Y_\uq)^{2-loop}_{SM}+\Delta\beta(Y_\uq)^{2-loop}_{\widetilde{R}_2} \; \nn \\ &\beta(Y_\uq)^{2-loop}_{\widetilde{R}_2,3-gen}=\beta(Y_\uq)^{2-loop}_{SM}+\sum_{i=1}^3\Big[\Delta\beta(Y_\uq)^{2-loop}_{\widetilde{R}_2}\Big]_{\is}\nonumber\\
&\text{where,}\quad\Delta\beta(Y_\uq)^{2-loop}_{\widetilde{R}_2}=\frac{Y_\uq}{(16\pi^2)^2}\bigg[\frac{7}{4}Y_\dq^\dagger\X_2 Y_\dq+\I\,\Big\{\Big(\frac{2}{45}g_1^4\nn\\
&\qquad \qquad \qquad  +\frac{3}{2}g_2^4+\frac{44}{9}g_3^4\Big)+3\,\Big(\lambda_2^2+\lambda_2\widetilde\lambda_2+\widetilde\lambda_2^2\Big)\nonumber\\
&\qquad \qquad \qquad -\frac{9}{2}\,\Tr\,\Big(\X_2\X_\dq+\frac{1}{2}\widetilde{\X}_2\widetilde{\X}_\cl\Big)\Big\}\bigg]~.
\end{align}}

\vspace*{+2mm}
In case of leptoquark $\vec S_3$, the correction to one-loop beta function of $Y_\uq$ contains one term only, and hence, it looks like:
\vspace*{+2mm}
{\begin{align}
\label{eq:yus3}
&\beta(Y_\uq)^{1-loop}_{\vec S_3,1-gen}=\beta(Y_\uq)^{1-loop}_{SM}+\Delta\beta(Y_\uq)^{1-loop}_{\vec S_3}\;\nn\\
&\beta(Y_\uq)^{1-loop}_{\vec S_3,3-gen}=\beta(Y_\uq)^{1-loop}_{SM}+\sum_{i=1}^3\Big[\Delta\beta(Y_\uq)^{1-loop}_{\vec S_3}\Big]_{\is}\nonumber\\
&\text{with}\quad\Delta\beta(Y_\uq)^{1-loop}_{\vec S_3}=\frac{3}{64\pi^2}\,Y_\uq\X_3^T~.
\end{align}}

\vspace*{+2mm}
The full two-loop beta function for $Y_\uq$ in this scenario becomes as follows:
\vspace*{+2mm}
{\begin{align}
&\beta(Y_\uq)^{2-loop}_{\vec S_3,1-gen}=\beta(Y_\uq)^{2-loop}_{SM}+\Delta\beta(Y_\uq)^{2-loop}_{\vec S_3}\;\nn\\
&\beta(Y_\uq)^{2-loop}_{\vec S_3,3-gen}=\beta(Y_\uq)^{2-loop}_{SM}+\sum_{i=1}^3\Big[\Delta\beta(Y_\uq)^{2-loop}_{\vec S_3}\Big]_{\is} \text{   with}\nonumber\\
&\Delta\beta(Y_\uq)^{2-loop}_{\vec S_3}=\Delta\beta(Y_\uq)^{1-loop}_{\vec S_3}+\frac{Y_\uq}{(16\pi^2)^2}\bigg[\I \Big\{\frac{4}{15}g_1^4+6g_2^4\nn\\
&\qquad  
+\frac{22}{3}g_3^4+\frac{9}{2}\Big(\lambda_{3}^2+\lambda_{3}\widetilde{\lambda}_3+\frac{3}{4}\widetilde{\lambda}_3^2\Big) -\frac{27}{8}\Tr\Big(\widetilde\X_3 \widetilde\X_\cl\nonumber\\
&\qquad +\X_3\widetilde{X}_\dq^T+\X_3\widetilde{X}_\uq^T\Big)\Big\}+\Big(-3\lambda_{3}-\frac{9}{2}\widetilde{\lambda}_3+\frac{43}{20}g_1^2\nn\\
&\qquad +\frac{45}{4}g_2^2+\frac{11}{2}g_3^2 -\frac{9}{8}\,\Tr\,\X_3\Big)\X_3^T-\frac{3}{16}Y_3^*\widetilde{\X}_\cl^T Y_3^T\nonumber\\
&\qquad +\frac{3}{2}\X_3^T \widetilde{\X}_\dq-\frac{3}{8}\X_3^T \widetilde{\X}_\uq-\frac{27}{32}(\X_3^T)^2\bigg]~.
\end{align}}

Now, in the combined scenario of $\widetilde{R}_2$ and $\vec S_3$, apart from the individual contributions of $\widetilde{R}_2$ and $\vec S_3$ to the running of $Y_\uq$, there emerges another at two-loop which contains effects of $Y_2$ and $Y_3$ simultaneously. Therefore, the beta function for $Y_\uq$ up to two-loop order in this case can be expressed as:
{\small\begin{align}
&\beta(Y_\uq)^{2-loop}_{\widetilde{R}_2+\vec S_3,1-gen}=\beta(Y_\uq)^{2-loop}_{SM}+\Delta\beta(Y_\uq)^{2-loop}_{\widetilde{R}_2}\nn\\
&\qquad \qquad +\Delta\beta(Y_\uq)^{2-loop}_{\vec S_3}-\frac{9}{4096\pi^2} Y_\uq Y_3^* \wX_2^T Y_3^T~,
\end{align}}

{\begin{align}
&\beta(Y_\uq)^{2-loop}_{\widetilde{R}_2+\vec S_3,3-gen}=\beta(Y_\uq)^{2-loop}_{SM}+\sum_{i=1}^3\Big[\Delta\beta(Y_\uq)^{2-loop}_{\widetilde{R}_2}\Big]_{\is}\nn\\
& +\sum_{i=1}^3\Big[\Delta\beta(Y_\uq)^{2-loop}_{\vec S_3}\Big]_{\is}-\frac{9\,Y_\uq}{4096\pi^2} \sum_{(i,l)=1}^3 Y_{3,i}^* \wX_{2,l}^T Y_{3,i}^T~.
\end{align}}

At this point it is important to mention that for our choice of leptoquark couplings in three generation cases $Y_{\gamma,i}Y_{\gamma^\prime,j}=0$ for $i\neq j$ where $(\gamma,\gamma^\prime)\in\{2,3\}$.

\begin{figure*}[h!]
	\begin{center}
		\mbox{\subfigure[$Y_\phi=0.4$]{\includegraphics[width=0.5\linewidth,angle=-0]{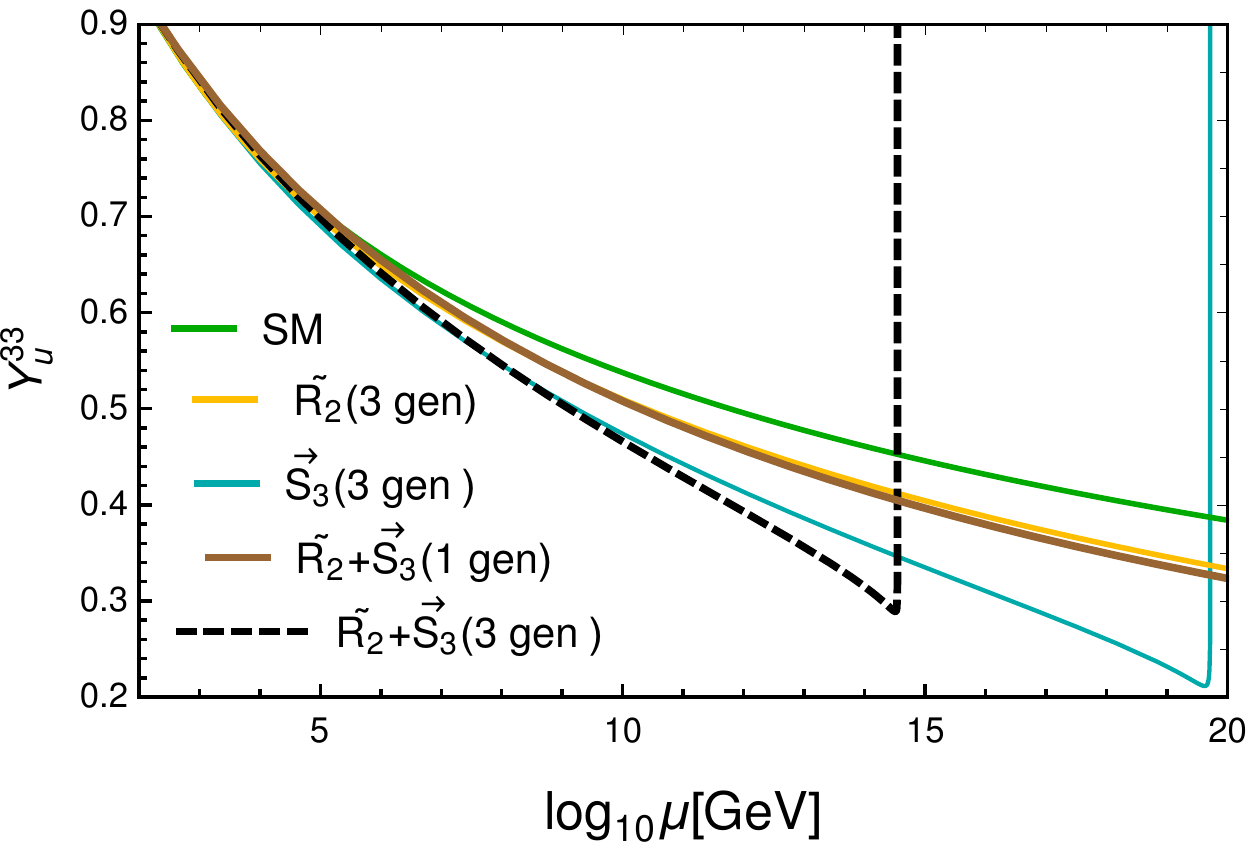}\label{f3a}}
			\subfigure[$Y_\phi=1.0$]{\includegraphics[width=0.5\linewidth,angle=-0]{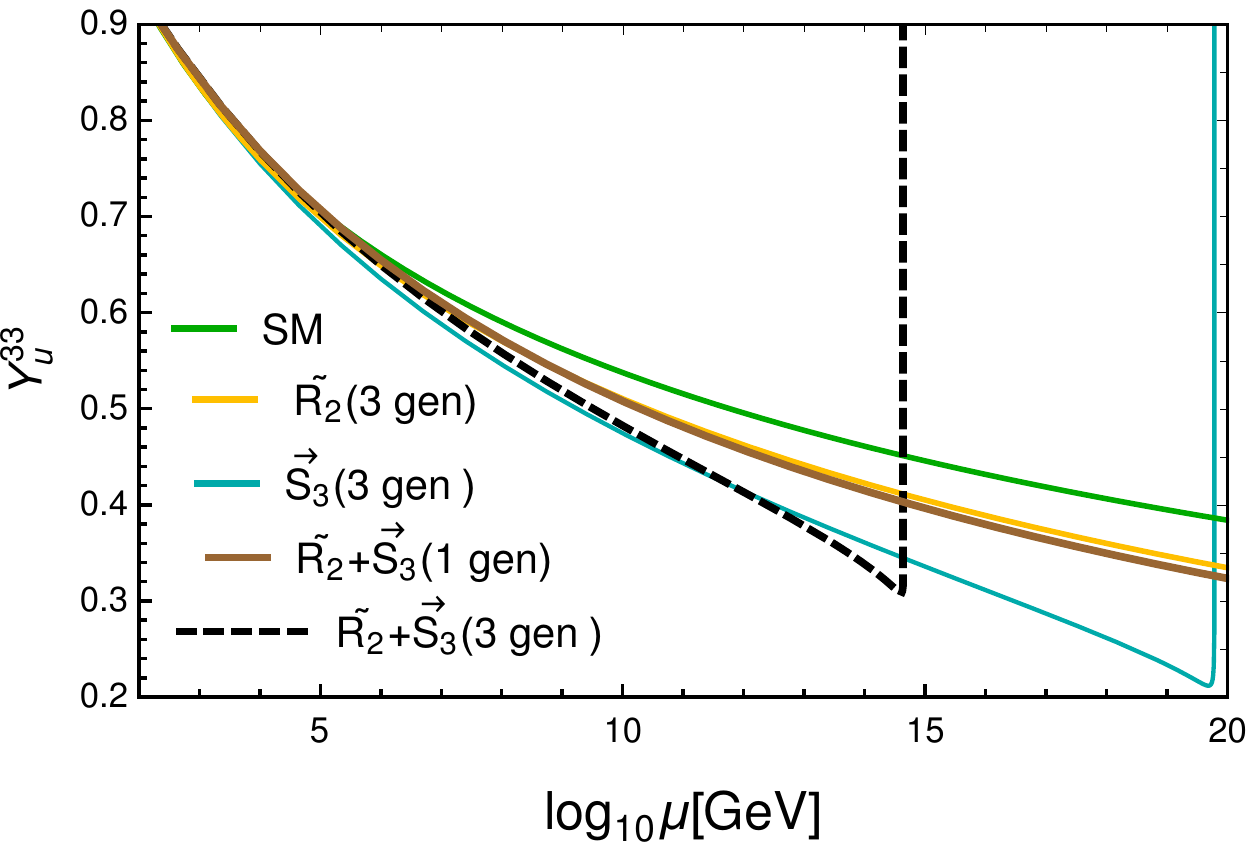}\label{f4a}}}
		\caption{Variation of top quark Yukawa with scale. }\label{fig:yt}
	\end{center}
\end{figure*}

\vspace*{+2mm}
We depict the results for variation of top Yukawa coupling with the energy scale at two-loop level under different models in Fig. \ref{fig:yt} where the left and right panels signify the leptoquark coupling with quarks and leptons ($Y_\phi$) to be 0.4 and 1.0 respectively. While the green curve explains the SM scenario, the yellow and blue lines illustrates $\widetilde{R}_2$ and $\vec S_3$ leptoquarks with three generations. As expected the SM value of top Yukawa coupling decreases with energy. With the inclusion of leptoquarks this coupling shifts further down and for the case of $\vec S_3$ it achieves divergence at $10^{19.7}$ GeV. Since $\widetilde{R}_2$ and $\vec S_3$ with one generation do not show any abnormal behaviour, we do not present them here. On the other hand, the brown (solid) and black (dashed) curves represent the combined models of $\widetilde{R}_2$ and $\vec S_3$ with one and three generations respectively. As anticipated, the case with three generations of both the leptoquarks stays at the bottom of all the other lines for lower values of $Y_\phi$ (like $Y_\phi=0.4$, shown by left panel of Fig. \ref{fig:yt}), although only this curve gets noticeable effect if $Y_\phi$ is increased to some sufficiently higher value (like $Y_\phi=1.0$, as exhibited in the right panel of Fig. \ref{fig:yt}). Like all the other couplings for this scenario, it also diverges at $10^{14.4}$ GeV. The relative positions of the curves in the above mentioned plot primarily depend on the negative contributions from the gauge couplings at one-loop level, see Eqs \eqref{eq:yusm} - \eqref{eq:yus3}. With increase in number of leptoquark components, the values of gauge couplings get enhanced at any particular energy scale which in turn will push the top Yukawa coupling downward.

\onecolumn

\section{Running of Leptoquark Yukawa Couplings}
\label{sec:lqY}
Now, let us discuss the evolution of Leptoquark Yukawa couplings $Y_2$ and $Y_3$. As mentioned in Eq. \eqref{eq:pbnd}, these parameters also should have an upper bound of $\sqrt{4\pi}$ at any energy scale $\mu$. For $\widetilde R_2$ scenario, the one-loop and two-loop beta functions for $Y_2$ are given by:
\begin{align}
\beta(Y_2)^{1-loop}_{\widetilde{R}_2,1-gen}=\frac{1}{16\pi^2}\bigg[\X_\dq Y_2+\frac{5}{2}\X_2Y_2+\frac{1}{2}Y_2\wX_\cl-\Big(\frac{13}{20}g_1^2+\frac{9}{4}g_2^2+4g_3^2-\Tr \X_2\Big)Y_2\bigg],
\end{align}
\begin{align}
\beta(Y_2)&^{2-loop}_{\widetilde{R}_2,1-gen}=\beta(Y_2)^{1-loop}_{\widetilde{R}_2,1-gen}+\frac{1}{(16\pi^2)^2}\bigg[Y_2 \wX_\cl\Big\{\frac{117}{80}g_1^2+\frac{33}{16}g_2^2-2\lambda_2+2\widetilde\lambda_2-\frac{9}{4}\Tr\Big(\X_\uq+\X_\dq+\frac{1}{3}\X_\cl\Big)\Big\}\nonumber\\
&+\X_\dq Y_2\Big\{\frac{61}{120}g_1^2+\frac{51}{8}g_2^2-\frac{16}{3}g_3^2-4\lambda_2-2\widetilde\lambda_2-\frac{9}{2}\Tr\Big(\X_\uq+\X_\dq+\frac{1}{3}\X_\cl\Big)\Big\}+\frac{1}{2}\X_2^2 Y_2-\frac{3}{4}\X_2\X_\dq Y_2\nonumber\\
&-\frac{1}{4}Y_2 \widetilde\X_\cl\widetilde{\X}_2 -\frac{1}{4}Y_2\widetilde\X_\cl^2+2\X_\dq Y_2 \widetilde{X}_\cl-\frac{1}{4}\X_\dq^2Y_2-\frac{1}{4}Y_\dq\widetilde{\X}_\uq Y_\dq^\dagger Y_2+\X_2Y_2\Big(\frac{107}{48}g_1^2+\frac{201}{16}g_2^2+\frac{73}{3}g_3^2-\frac{15}{4}\Tr \X_2\Big)\nonumber\\
&+Y_2\Big\{\lambda_2^2+\lambda_2 \widetilde\lambda_2+\widetilde\lambda_2^2+\frac{3961}{1800}g_1^4-\frac{17}{4}g_2^4-\frac{173}{9}g_3^4+\frac{9}{20}g_1^2 g_2^2+\frac{17}{15}g_1^2g_3^2-9g_2^2g_3^2\nonumber\\
&+\Big(\frac{13}{24}g_1^2+\frac{15}{8}g_2^2+\frac{10}{3}g_3^2\Big)\Tr \X_2-\frac{3}{2}\Tr\Big(\X_2 \X_\dq+\frac{1}{2}\widetilde{X}_\cl \widetilde{\X}_2+\frac{5}{2}\X_2^2\Big)\Big\}\bigg].
\end{align}

Now, for three generations of $\widetilde R_2$, we have three Yukawa matrices of leptoquark ($Y_{2,i}$) corresponding to three different generations of quarks and leptons. The running of each of these Yukawa matrices at one-loop can be expressed as:
\begin{align}
\label{eq:r2Y2i}
\beta(Y_{2,i})^{1-loop}_{\widetilde{R}_2,3-gen}=\Big[\beta(Y_{2})^{1-loop}_{\widetilde{R}_2,1-gen}\Big]_{\is}+\frac{1}{16\pi^2}\sum_{j\neq i}\bigg[\frac{3}{2}Y_{2,i}\,\wX_{2,j}+ \X_{2,j}\,Y_{2,i}+Y_{2,j}\,\Tr\Big(Y_{2,i} Y_{2,j}^\dagger\Big)\bigg].
\end{align}
At this point we remind the reader again that $\big[\beta\big]_{\is}$ for any parameter indicates beta function of that parameter with the replacement of $f(Y_\gamma,\,\X_\gamma,\,\wX_\gamma,\,\lambda_\gamma,\,\widetilde{\lambda}_\gamma)$ to $f(Y_{\gamma,i}\,,\,\X_{\gamma,i}\,,\,\wX_{\gamma,i}\,,\,\lambda_\gamma^{ii}\,,\,\widetilde{\lambda}_\gamma^{ii})$  with $\gamma\in\{2,3\}$ and $i$ representing the generation. It is interesting to notice that there appear some additional terms with inter-generation interactions. The beta function for $i$-th generation of $Y_{2}$ at two-loop order is given by:
{\small
\begin{align}
\beta(Y_{2,i})&^{2-loop}_{\widetilde{R}_2,3-gen}=\bigg\{\beta(Y_{2,i})^{1-loop}_{\widetilde{R}_2,3-gen}-\Big[\beta(Y_{2})^{1-loop}_{\widetilde{R}_2,1-gen}\Big]_{\is}\bigg\}+
\Big[\beta(Y_{2})^{2-loop}_{\widetilde{R}_2,1-gen}\Big]_{\is}+\frac{Y_{2,i}}{(16\pi^2)^2}\,\Big(\frac{7}{90}g_1^4+3g_2^4+\frac{32}{9}g_3^4\Big)\nonumber\\
&+\frac{1}{(16\pi^2)^2}\sum_{j\neq i}\bigg[-\frac{3}{2}\,Y_{2,i}\Tr\Big(\X_{2,i}\,\X_{2,j}+\frac{3}{2}\,\wX_{2,i}\,\wX_{2,j}\Big)+\Big(\frac{47}{80}g_1^2+\frac{99}{16}g_2^2+17 g_3^2\Big)Y_{2,i}\,\wX_{2,j}\nonumber\\
&+\Big(\frac{197}{120}g_1^2+\frac{51}{8}g_2^2+\frac{22}{3}g_3^2\Big)\X_{2,j}\,Y_{2,i}-\frac{3}{4}\X_{2,i}\X_{2,j}Y_{2,i}-\frac{3}{4}Y_{2,i}Y_{2,j}^\dagger\X_\dq Y_{2,j}+\frac{5}{4}Y_{2,i}Y_{2,j}^\dagger\X_{2,i} Y_{2,j}\nonumber\\
&-\frac{3}{4}Y_{2,i} \wX_{2,j}\wX_{2,i}-\frac{3}{4}Y_{2,i} \wX_{2,j}^2-\frac{1}{4}Y_{2,j}\wX_\cl Y_{2,j}^\dagger Y_{2,i}+\frac{5}{4}Y_{2,j}\wX_{2,i}Y_{2,j}^\dagger Y_{2,i}+2\,\X_{2,j}Y_{2,i}\wX_{2,j}-\frac{3}{4}\X_{2,j}^2 Y_{2,i}\nonumber\\
&+\sum_{k\slashed\in \{i,j\}}\Big(2Y_{2,j}Y_{2,k}^\dagger Y_{2,i}Y_{2,j}^\dagger Y_{2,k}-\frac{3}{4}Y_{2,j}\wX_{2,k} Y_{2,j}^\dagger Y_{2,i}-\frac{3}{4}Y_{2,i}Y_{2,j}^\dagger \X_{2,k}  Y_{2,j}\Big)+\Big\{\Big(\frac{13}{24}g_1^2+\frac{15}{8}g_2^2+\frac{10}{3}g_3^2\Big)Y_{2,j}\nonumber\\
&-\frac{3}{2}Y_{2,j}\wX_{2,i}-\frac{9}{4}\wX_{2,i}Y_{2,j}\Big\}\Tr\Big(Y_{2,i}Y_{2,j}^\dagger\Big)+\frac{15}{4}Y_{2,i}Y_{2,j}^\dagger Y_{2,i} \Tr\Big(Y_{2,j}Y_{2,i}^\dagger\Big)-\Big(\frac{9}{4}Y_{2,i}\,\wX_{2,j}+\frac{3}{2}\X_{2,j} Y_{2,i}\Big)\Tr\wX_{2,j}\nonumber\\
&-\frac{3}{2}\,\sum_{k\slashed\in \{i,j\}}\bigg\{\Big(\frac{3}{2}\,Y_{2,i}\,Y_{2,j}^\dagger\,Y_{2,k}+Y_{2,k}\,Y_{2,j}^\dagger\,Y_{2,i}\Big)\Tr\Big(Y_{2,j}\,Y_{2,k}^\dagger\Big)+Y_{2,j}\,\Tr\Big(\frac{3}{2}Y_{2,i}\wX_{2,k}Y_{2,j}^\dagger+\X_{2,k} Y_{2,i}Y_{2,j}^\dagger\Big)\bigg\}\nonumber\\
&-\frac{3}{2}\,Y_{2,j}\,\Tr\Big(\X_\dq\,Y_{2,i}Y_{2,j}^\dagger+\frac{1}{2}Y_{2,i}\wX_\cl Y_{2,j}^\dagger+\frac{5}{2} \X_{2,i} Y_{2,i} Y_{2,j}^\dagger+\frac{5}{2}Y_{2,i}\wX_{2,j}Y_{2,j}^\dagger\Big)\bigg].
\end{align}}
\normalfont
The term within the curly bracket is added to the above expression in order to incorporate the extra contribution at one-loop order coming from cross-generation interaction, as shown in Eq. \eqref{eq:r2Y2i}. The rest of the terms arise from two-loop contributions.


In a similar fashion, the one-loop and two-loop beta functions for $Y_3$ in the case of leptoquark $\vec S_3$ can be expressed as the following:
\begin{equation}
\beta(Y_3)^{1-loop}_{\vec S_3,1-gen}=\frac{1}{32\pi^2}\bigg[Y_3\widetilde\X_\cl+6Y_3\widetilde\X_3+\widetilde{\X}_\dq^TY_3+\widetilde{\X}_\uq^TY_3-\Big(g_1^2+9g_2^2+8g_3^2-2\Tr \X_3\Big)Y_3\bigg],
\end{equation}
\begin{align}
\beta(Y_3)&^{2-loop}_{\vec S_3,1-gen}=\beta(Y_2)^{1-loop}_{\vec S_3,1-gen}+\frac{1}{(16\pi^2)^2}\bigg[Y_3\widetilde{\X}_\cl\Big\{\frac{69}{80}g_1^2+\frac{33}{16}g_2^2-2\lambda_{3}+\widetilde{\lambda}_3-\frac{9}{4}\Tr\Big(\X_\uq+\X_\dq+\frac{1}{3}\X_\cl\Big)\Big\}\nonumber\\
&+\widetilde\X_\dq^T Y_3\Big\{\frac{7}{240}g_1^2+\frac{33}{16}g_2^2-\frac{8}{3}g_3^2-2\lambda_{3}+\widetilde{\lambda}_3-\frac{9}{4}\Tr\Big(\X_\uq+\X_\dq+\frac{1}{3}\X_\cl\Big)\Big\}+\widetilde\X_\uq^T Y_3\Big\{\frac{427}{240}g_1^2+\frac{33}{16}g_2^2-\frac{8}{3}g_3^2\nonumber\\
&-2\lambda_{3}-3\widetilde\lambda_{3}-\frac{9}{4}\Tr\Big(\X_\uq+\X_\dq+\frac{1}{3}\X_\cl\Big)\Big\}+\X_3 Y_3\Big(\frac{23}{10}g_1^2+48 g_2^2+31 g_3^2-\frac{9}{2}\Tr \X_3\Big)-\frac{1}{4}Y_3\widetilde\X_\cl^2-\frac{3}{16}Y_3\widetilde{\X}_\cl\widetilde{\X}_3\nonumber\\
&+\frac{13}{16}Y_3 \widetilde{\X}_3^2-\frac{9}{16}\X_3\widetilde{\X}_\dq Y_3-\frac{9}{16}\X_3\widetilde{\X}_\uq Y_3+2\widetilde{X}_\dq Y_3 \widetilde{X}_\cl-\frac{1}{4}\widetilde\X_\dq^2 Y_3-\frac{1}{4}\widetilde\X_\uq^2 Y_3+Y_3\Big\{\lambda_{3}^2+\lambda_{3}\widetilde\lambda_{3}+\frac{3}{4}\widetilde\lambda_{3}^2\nonumber\\
&+\frac{599}{400}g_1^4-\frac{173}{16}g_2^4-\frac{55}{3}g_3^4-\frac{23}{40}g_1^2 g_2^2-31 g_2^2 g_3^2-\frac{1}{15}g_1^2 g_3^2+\frac{5}{12}\Big(g_1^2+9g_2^2+8g_3^2\Big)\Tr \X_3\nonumber\\
&-\frac{3}{4}\Tr\Big(\widetilde{\X}_3\widetilde{\X}_\cl+\X_3\widetilde{\X}_\dq^T+\X_3\widetilde{\X}_\uq^T+6\X_3^2\Big)\Big\}\bigg].
\end{align}
For three generations case of $\vec S_3$, the one-loop beta of $i$-th generation leptoquark Yukawa coupling takes the form:
\begin{align}
\beta(Y_{3,i})^{1-loop}_{\vec{S}_3,3-gen}=\Big[\beta(Y_{2})^{1-loop}_{\vec{S}_3,1-gen}\Big]_{\is}+\frac{9}{64\pi^2}\sum_{j\neq i}\bigg[Y_{3,i}\wX_{3,j}+\frac{1}{3}\X_{3,j}Y_{3,i}+\frac{4}{9}Y_{3,j}\Tr\Big(Y_{3,i} Y_{3,j}^\dagger\Big)\bigg].
\end{align}
Like the case of $\widetilde R_2$, here also some new terms appear due to inter-generation interactions at one-loop level. Encompassing these additional terms, the two-loop beta function can be written as:
{\small\begin{align}
\beta(Y&_{3,i})^{2-loop}_{\vec{S}_3,3-gen}=\bigg\{\beta(Y_{3,i})^{2-loop}_{\vec{S}_3,3-gen}-\Big[\beta(Y_{3})^{1-loop}_{\vec{S}_3,1-gen}\Big]_{\is}\bigg\}+
\Big[\beta(Y_{3})^{2-loop}_{\vec{S}_3,1-gen}\Big]_{\is}\nonumber\\
&+\frac{Y_{3,i}}{(16\pi^2)^2}\Big(\frac{49}{150}g_1^4+21g_2^4+\frac{16}{3}g_3^4\Big)+\frac{1}{(16\pi^2)^2}\sum_{j\neq i}\bigg[-\frac{9}{8}Y_{3,i}\Tr\Big(\X_{3,i}\,\X_{3,j}+3\wX_{3,i}\,\wX_{3,j}\Big)\nonumber\\
&+Y_{3,i}\wX_{3,j}\Big(\frac{3}{5}g_1^2+36g_2^2+\frac{51}{2}g_3^2\Big)+\X_{3,j} Y_{3,i}\Big(\frac{17}{10}g_1^2+12g_2^2+\frac{11}{2}g_3^2\Big)-\frac{27}{32}\X_{3,i}\X_{3,j}Y_{3,i}\nonumber\\
&+\frac{53}{32}Y_{3,i}Y_{3,j}^\dagger \X_{3,i}Y_{3,j}-\frac{27}{32}Y_{3,i}\,\wX_{3,j}\wX_{3,i}-\frac{27}{32}Y_{3,i}\,\wX_{3,j}^2 -\frac{27}{32}Y_{3,i}\,Y_{3,j}^\dagger\X_{3,k}Y_{3,j}-\frac{9}{16}Y_{3,i}\,Y_{3,j}^\dagger\wX_{\dq}^T \,Y_{3,j}\nonumber\\
&-\frac{9}{16}Y_{3,i}\,Y_{3,j}^\dagger\wX_{\uq}^T \,Y_{3,j}-\frac{3}{16}Y_{3,j}\wX_{\cl}Y_{3,j}^\dagger Y_{3,i}+\frac{53}{32}Y_{3,j}\wX_{3,i}Y_{3,j}^\dagger Y_{3,i}+\frac{5}{2}\X_{3,j}Y_{3,i}\wX_{3,j}-\frac{27}{32}\X_{3,j}^2Y_{3,i}\nonumber\\
&+\Big\{\Big(\frac{5}{12}g_1^2+\frac{15}{4}g_2^2+\frac{10}{3}g_3^2\Big)Y_{3,j}-\frac{9}{8}Y_{3,j} \wX_{3,i}-\frac{27}{8}\X_{3,i}Y_{3,j}\Big\}\Tr\Big(Y_{3,i}Y_{3,j}^\dagger\Big)-\frac{9}{2}Y_{3,i}Y_{3,j}^\dagger Y_{3,i}\Tr\Big(Y_{3,j}Y_{3,i}^\dagger\Big)\nonumber\\
&-\Big(\frac{27}{8}Y_{3,i}\,\wX_{3,j}+\frac{9}{8}\,\X_{3,j}Y_{3,i}\Big)\Tr\X_{3,j}-\frac{3}{4}Y_{3,j}\Tr\Big(\wX_{\cl}Y_{3,j}^\dagger Y_{3,i}\Big)-\frac{9}{2}Y_{3,j}\Tr\Big(\X_{3,i} Y_{3,i}Y_{3,j}^\dagger\Big)-\frac{9}{2} Y_{3,j}\Tr\Big(Y_{3,i} \wX_{3,j}Y_{3,j}^\dagger\Big)\nonumber\\
&-\frac{3}{4}Y_{3,j}\Tr\Big(Y_{3,i}Y_{3,j}^\dagger\wX_\dq^T\Big)-\frac{3}{4}Y_{3,j}\Tr\Big(Y_{3,i}Y_{3,j}^\dagger\wX_\uq^T\Big) +\sum_{k\slashed\in \{i,j\}}\bigg\{\frac{5}{2}Y_{3,j}Y_{3,k}^\dagger Y_{3,i}Y_{3,j}^\dagger Y_{3,k}-\frac{27}{32}Y_{3,j}\wX_{3,k}Y_{3,j}^\dagger Y_{3,i}\nonumber\\
&-\Big(\frac{27}{8}Y_{3,i}Y_{3,j}^\dagger Y_{3,k}+\frac{9}{8}Y_{3,k}Y_{3,j}^\dagger Y_{3,i}\Big)\Tr\Big(Y_{3,j}Y_{3,k}^\dagger\Big)-\frac{27}{8}Y_{3,j}\Tr\Big(Y_{3,i}\,\wX_{3,k}Y_{3,j}^\dagger\Big)-\frac{9}{8}Y_{3,j}\Tr\Big(Y_{3,i}\,Y_{3,j}^\dagger\X_{3,k}\Big)\bigg\}\bigg].
\end{align}}

Now, for the combined scenario of $\widetilde R_2$ and $\vec S_3$, the above expressions get modified and extra contributions from interactions of doublet and triplet leptoquarks emerges at both one and two-loop level. Thus, they can be written in the following way:
\begin{align}
\beta(Y_2)&^{2-loop}_{\widetilde{R}_2+\vec S_3,1-gen}=\frac{9}{64\pi^2}Y_2 \widetilde\X_3+\beta(Y_2)^{2-loop}_{\widetilde{R}_2,1-gen}+\frac{1}{(16\pi^2)^2}\bigg[Y_2\widetilde{\X}_3\Big(\frac{21}{20}g_1^2+\frac{135}{4}g_2^2+\frac{51}{2}g_3^2-\frac{27}{8}\Tr \X_3\Big)\nonumber\\
&-\frac{9}{8}Y_2\Big(\widetilde\X_3\widetilde\X_2+\frac{3}{4}\widetilde\X_3^2+\frac{1}{2}Y_3^\dagger\widetilde\X_{\dq}^T Y_3+\frac{1}{2}Y_3^\dagger\widetilde\X_{\uq}^T Y_3\Big)+3Y_\dq Y_3^*Y_2^TY_\dq^* Y_3-\frac{3}{8}Y_\dq\X_3^T Y_\dq^\dagger Y_2\nonumber\\
&+Y_2\Big\{\frac{7}{30}g_1^4+6g_2^4+\frac{8}{3}g_3^4-\frac{27}{8}\Tr\Big(\wX_2\wX_3\Big)\Big\}\bigg].
\end{align}
\vspace*{-3mm}
\begin{align}
\beta(Y_3)&^{2-loop}_{\widetilde{R}_2+\vec S_3,1-gen}=\frac{3}{32\pi^2}Y_3 \widetilde\X_2+\beta(Y_3)^{2-loop}_{\widetilde{R}_2,1-gen}+\frac{1}{(16\pi^2)^2}\bigg[-\frac{3}{4}Y_3\Big(\wX_2^2+\frac{3}{4}\wX_2 \wX_3+Y_2^\dagger\X_\dq Y_2\Big)\nonumber\\
&-\frac{1}{4}Y_\dq^T \X_2^TY_\dq^* Y_3+2Y_\dq^TY_2^*Y_3^TY_\dq^\dagger Y_2-Y_3\wX_2\Big(\frac{1}{80}g_1^2-\frac{99}{16}g_2^2-17g_3^2+\frac{9}{4}\Tr\X_2\Big)\nonumber\\
&+Y_3\Big\{\frac{49}{1800}g_1^4+\frac{21}{8}g_2^4+\frac{16}{9}g_3^4-\frac{9}{4}\Tr\Big(\wX_2\wX_3\Big)\Big\}\bigg].
\end{align}
\begin{align}
\beta(Y_{2,i})&^{2-loop}_{\widetilde{R}_2+\vec S_3,3-gen}=\frac{9}{64\pi^2}\sum_{l=1}^3Y_{2,i}\,\wX_{3,l}+\beta(Y_{2,i})^{2-loop}_{\widetilde{R}_2,3-gen}+\frac{Y_{2,i}}{(16\pi^2)^2}\Big(\frac{7}{10}g_1^4+18g_2^4+8g_3^4\Big)\nonumber\\
&+\frac{1}{(16\pi^2)^2}\sum_{l=1}^3\bigg[-\frac{27}{8}Y_{2,i}\Tr\Big(\wX_{2,i}\,\wX_{3,l}\Big)+Y_{2,i}\,\widetilde{\X}_{3,l}\Big(\frac{21}{20}g_1^2+\frac{135}{4}g_2^2+\frac{51}{2}g_3^2\Big)\nonumber\\
&-\frac{9}{8}Y_{2,i}\Big(\wX_{3,l}\wX_{2,i}+\frac{1}{2}Y_{3,l}^\dagger\wX_\dq^T Y_{3,l}+\frac{1}{2}Y_{3,l}^\dagger\wX_\uq^T Y_{3,l}+\frac{3}{4}\sum_{k=1}^3Y_{3,l}^\dagger\X_{3,k}Y_{3,l}\Big)+\frac{9}{8}\sum_{j\neq i}Y_{2,j}\wX_{3,l}Y_{2,j}^\dagger Y_{2,i}\nonumber\\
&+3Y_\dq Y_{3,l}^*Y_{2,i}^T Y_\dq^* Y_{3,l}-\frac{3}{8}Y_\dq \X_{3,l}^T Y_\dq^\dagger Y_{2,i}-\frac{27}{8}\sum_{k=1}^3 Y_{2,i}Y_{3,l}^\dagger Y_{3,k}\Tr\Big(Y_{3,l} Y_{3,k}^\dagger\Big)\bigg].
\end{align}
\vspace*{-3mm}
\begin{align}
\beta(Y_{3,i})&^{2-loop}_{\widetilde{R}_2+\vec S_3,3-gen}=\frac{3}{32\pi^2}\sum_{l=1}^3 Y_{3,i}\,\wX_{2,l}+\beta(Y_{2,i})^{2-loop}_{\widetilde{R}_2,3-gen}+\frac{Y_{3,i}}{(16\pi^2)^2}\Big(\frac{49}{600}g_1^4+\frac{63}{8}g_2^4+\frac{16}{3}g_3^4\Big)\nonumber\\
&+\frac{1}{(16\pi^2)^2}\sum_{l=1}^3\bigg[-\frac{9}{4} Y_{3,i}\Tr\Big(\wX_{2,l}\wX_{3,i}\Big)+Y_{3,i}\wX_{2,l}\Big(-\frac{1}{80}g_1^2+\frac{99}{16}g_2^2+17 g_3^2\Big)\nonumber\\
&-\frac{3}{4}Y_{3,i}\Big(Y_{2,l}^\dagger\X_\dq Y_{2,l}+\frac{3}{4}\wX_{2,l} \wX_{3,i}+\sum_{k=1}^3Y_{2,l}^\dagger\X_{2,k} Y_{2,l}\Big)-\frac{1}{4}Y_\dq^T \X_{2,l}^TY_\dq^* Y_{3,i}+2Y_\dq^TY_{2,l}^*Y_{3,i}^TY_\dq^\dagger Y_{2,l}\nonumber\\
&-\frac{9}{4}\sum_{k=1}^3Y_{3,i}Y_{2,l}^\dagger Y_{2,k}\Tr\Big(Y_{2,l}Y_{2,k}^\dagger\Big)-\frac{9}{4}\sum_{j\neq i}Y_{3,j}\Tr\Big(\wX_{2,l}Y_{3,j}^\dagger Y_{3,i}\Big)\bigg].
\end{align}
In all of the above four expressions, the first term indicates the extra contribution at one-loop order due to presence of both doublet and triplet leptoquarks.

\begin{figure}[h!]
	\begin{center}
		\subfigure[$Y_\phi=0.4$]{\includegraphics[width=0.5\linewidth,angle=-0]{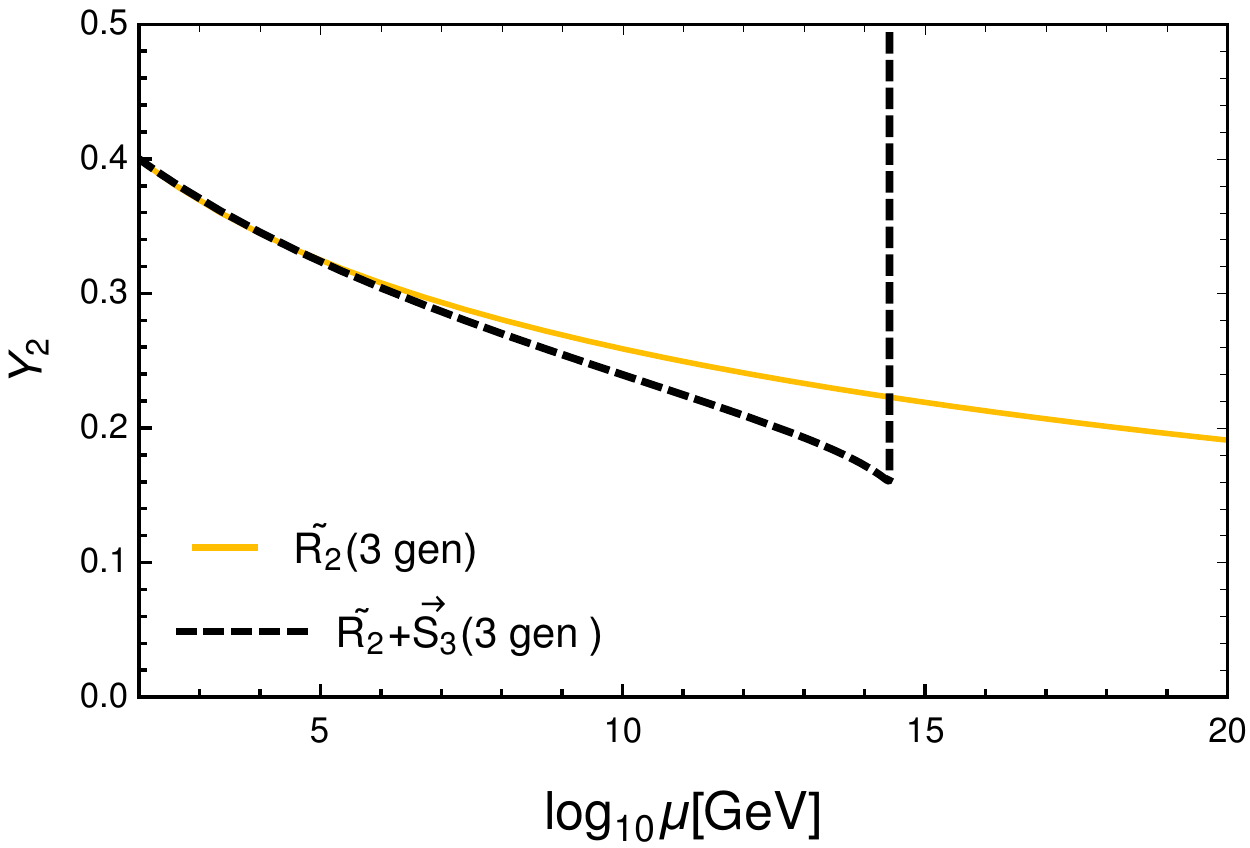}\label{y2a}}\hfil
			\subfigure[$Y_\phi=1.0$]{\includegraphics[width=0.5\linewidth,angle=-0]{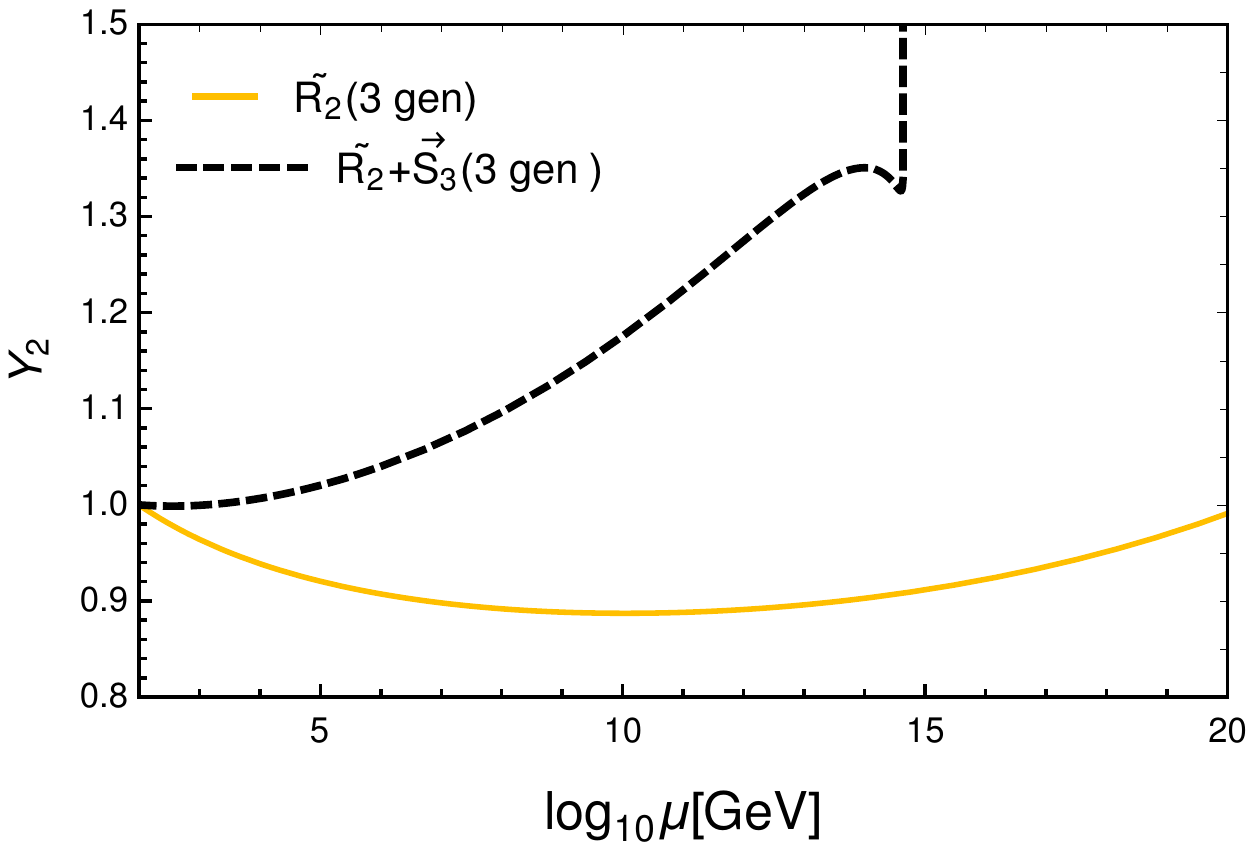}\label{y2b}}
			
		\subfigure[$Y_\phi=0.4$]{\includegraphics[width=0.5\linewidth,angle=-0]{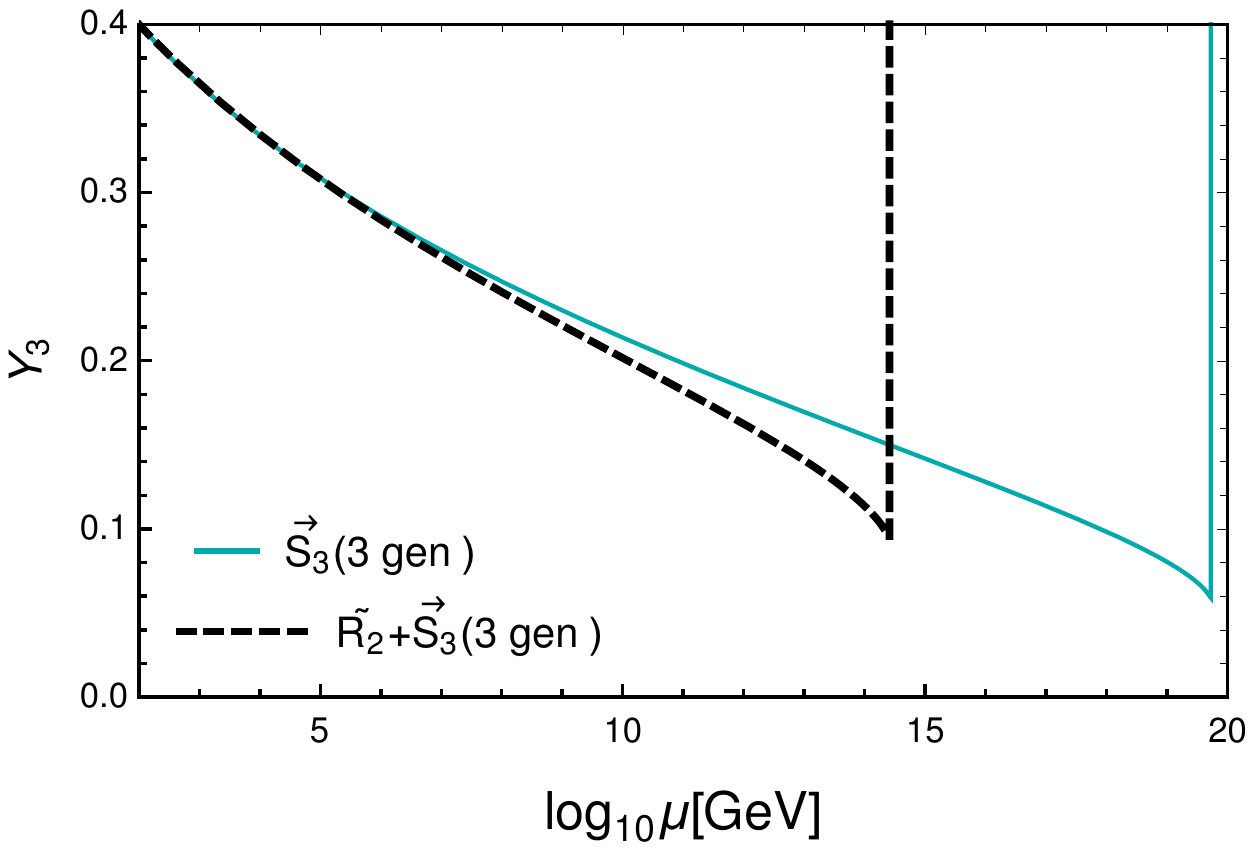}\label{y3a}}\hfil
			\subfigure[$Y_\phi=1.0$]{\includegraphics[width=0.5\linewidth,angle=-0]{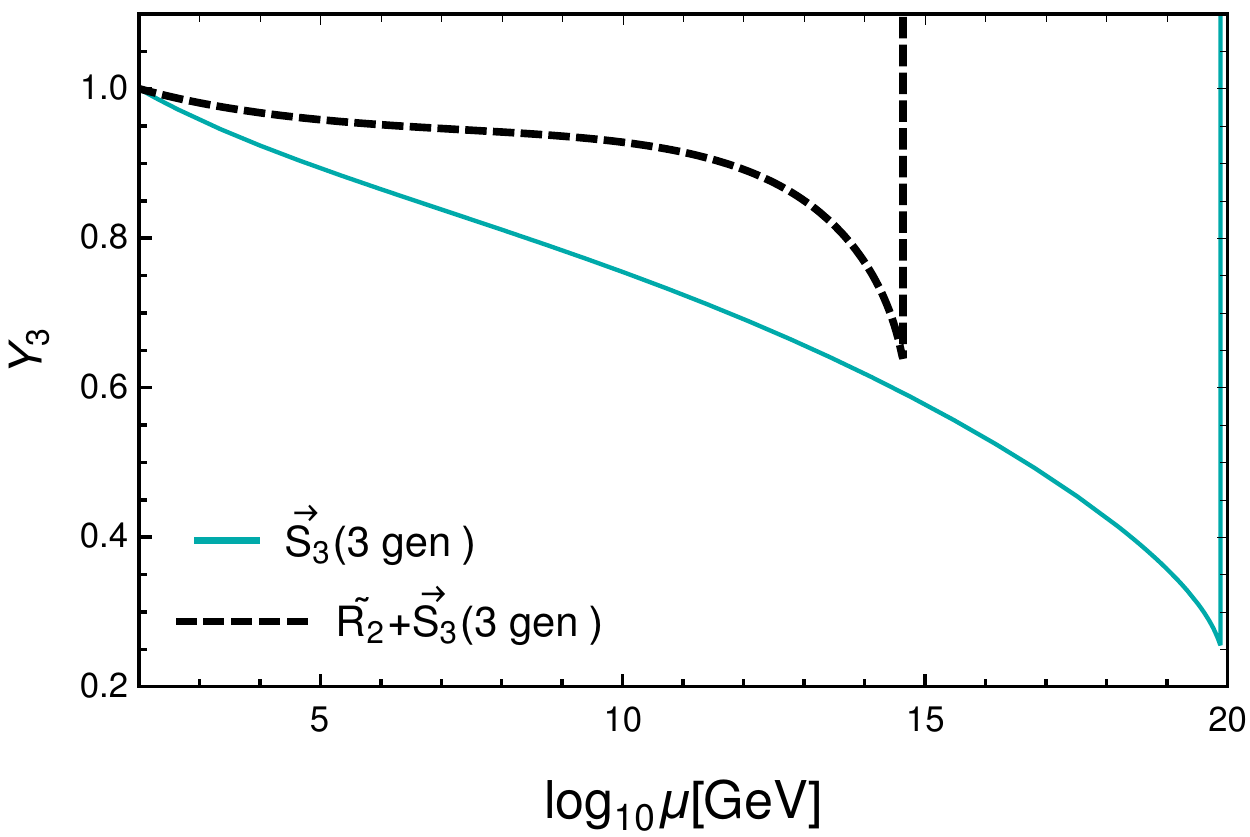}\label{y3b}}
		\caption{Variation of leptoquark Yukawa with scale. }\label{ylq}
	\end{center}
\end{figure}

Since one generation cases do not show any irregularities, we depict the variations of $Y_2$ and $Y_3$ in three generations scenarios of the leptoquarks in Fig. \ref{ylq}. While Figs. \ref{y2a} and \ref{y2b} in the first row illustrate the variations of any diagonal element of $Y_2$ starting from 0.4 and 1.0 respectively, the Figs. \ref{y3a} and \ref{y3b} in the second row demonstrate the similar thing for $Y_3$. As can be observed, for low Yukawa, the combined scenarios stay below the individual cases whereas the situation flips for higher Yukawa cases due to large effects from combined terms of $Y_2$ and $Y_3$. As expected, $Y_3$ for three generation of $\vec S_3$ decreases monotonically with energy and hits the divergence at $10^{19.7}$ GeV while both $Y_2$ and $Y_3$ diverge at $10^{14.4}$ GeV for three generation of $\widetilde{R}_2+\vec S_3$ case. But $Y_2$ shows different behaviour for large Yukawa. For $\widetilde{R}_2$ case, as can be noticed from Fig. \ref{y2b}, initially it decreases with scale, then reaches a minimum and gradually starts increasing. For the $\widetilde{R}_2+\vec S_3$ case,  it grows with energy from the beginning, then reaches a maximum and starts to fall off; but suddenly it blows up at $10^{14.4}$ GeV.

\section{Two-loop beta functions of Higgs-leptoquark quartic couplings for $\widetilde{R}_2$}
\label{sec:H-LQcouplR2}
\begin{align}
\beta(\lambda_{2})&^{2-loop}_{\widetilde{R}_2,1-gen}=\beta(\lambda_{2})^{1-loop}_{\widetilde{R}_2,1-gen}+\frac{1}{(16\pi^2)^2}\bigg[-\frac{1091}{3000}g_1^6+\frac{263}{8}g_2^6+\frac{1031}{600}g_1^4 g_2^2-\frac{1}{5}g_1^4g_3^2-\frac{29}{120}g_1^2g_2^4-15g_2^4g_3^2\nonumber\\&+2g_1^2 g_2^2 g_3^2+\lambda_{h}\Big(\frac{3}{10}g_1^4-g_1^2g_2^2+\frac{45}{4}g_2^4\Big)-16\Big(\lambda_{2}^3+\frac{1}{2}\lambda_{2}^2\,\widetilde{\lambda}_2+\frac{5}{4}\lambda_{2}\, \widetilde{\lambda}_2^2+\widetilde{\lambda}_2^3\Big)+\lambda_{2}\Big(\frac{929}{200}g_1^4+\frac{53}{8}g_2^4\nonumber\\
&-\frac{254}{9}g_3^4+\frac{21}{20}g_1^2 g_2^2+\frac{2}{9}g_1^2g_3^2+10 g_2^2 g_3^2+\frac{72}{5}g_1^2\lambda_h+72g_2^2\lambda_{h}-60\lambda_{h}^2\Big)+\widetilde\lambda_{2}\Big(\frac{7}{15}g_1^4+15g_2^4+\frac{40}{3}g_3^4\nonumber\\
&-\frac{22}{15}g_1^2 g_2^2+\frac{8}{3}g_2^2 g_3^2+\frac{24}{5}g_1^2\lambda_{h}+36 g_2^2\lambda_{h}-16\lambda_{h}^2\Big)+\lambda_{2}^2\Big(\frac{2}{3}g_1^2+6g_2^2+\frac{16}{3}g_3^2-72\lambda_{h}\Big)\nonumber\\
&+\widetilde{\lambda}_2^2\Big(-\frac{4}{15}g_1^2+6g_2^2+\frac{8}{3}g_3^2-28\lambda_{h}\Big)-\lambda_{2}\widetilde\lambda_{2}\Big(12g_2^2+32\lambda_{h}\Big)-\Big\{\frac{73}{100}g_1^4+\frac{3}{4}g_2^4-\frac{13}{10}g_1^2 g_2^2+4\lambda_{2}^2+2\widetilde\lambda_{2}^2\nonumber\\
&-\lambda_2\Big(\frac{13}{12}g_1^2+\frac{15}{4}g_2^2+\frac{20}{3}g_3^2\Big)\Big\}\Tr \X_2-\Big\{12\lambda_{2}^2+6\widetilde\lambda_{2}^2-\frac{1}{20}g_1^4+\frac{9}{4}g_2^4+32g_3^4+\frac{9}{10}g_1^2 g_2^2+24\lambda_{h}\widetilde\lambda_{2}\nonumber\\
&-5\lambda_{2}\Big(\frac{1}{4}g_1^2+\frac{9}{4}g_2^2+8g_3^2-\frac{72}{5}\lambda_{h}\Big)\Big\}\Tr \X_\dq-\Big\{\frac{1}{4}g_1^4+\frac{3}{4}g_2^4+\frac{11}{10}g_1^2 g_2^2+4\lambda_{2}^2+2\widetilde\lambda_{2}^2+8\lambda_{h}\widetilde\lambda_{2}\nonumber\\
&-\frac{15}{4}\lambda_{2}\Big(g_1^2+g_2^2-\frac{32}{5}\lambda_{h}\Big)\Big\}\Tr\X_\cl-\Big\{12\lambda_{2}^2+6\widetilde\lambda_2^2+\frac{19}{100}g_1^4+\frac{9}{4}g_2^4+32 g_3^4+\frac{21}{10}g_1^2 g_2^2+24\lambda_{h}\widetilde\lambda_{2}\nonumber\\
&-\lambda_{2}\Big(\frac{17}{4}g_1^2+\frac{45}{4}g_2^2+40g_3^2-72\lambda_{h}\Big)\Big\}\Tr\X_\uq-\Tr\Big\{\Big(\frac{8}{15}g_1^2 +\frac{32}{3}g_3^2\Big)\X_2\X_\dq+\frac{16}{5} g_1^2\widetilde\X_\cl\widetilde\X_2\nonumber\\
&
+\Big(21\lambda_{2}+24\widetilde\lambda_{2}\Big)\wX_\uq \wX_\dq+\lambda_{2} \Big(\frac{15}{2}\X_2^2+4\X_2\X_\dq-2\wX_\cl\wX_2+\frac{9}{2}\X_\cl^2+\frac{27}{2}\X_\uq^2+\frac{27}{2}\X_\dq^2\Big)\Big\}\nonumber\\
&+\Tr\Big(26\X_2^2\X_\dq+20\wX_2^2\wX_\cl+14\X_2\X_\dq^2+12\wX_2\wX_\cl^2-2\X_2 Y_\dq \wX_\uq Y_d^\dagger+14Y_2\wX_\cl Y^\dagger\X_\dq\Big)\bigg].
\end{align}

\begin{align}
\beta(\widetilde\lambda_{2})&^{2-loop}_{\widetilde{R}_2,1-gen}=\beta(\widetilde\lambda_{2})^{1-loop}_{\widetilde{R}_2,1-gen}+\frac{1}{(16\pi^2)^2}\bigg[27g_2^6-\frac{583}{150}g_1^4 g_2^2-\frac{49}{15}g_1^2g_2^4-4g_1^2g_2^2g_3^2+2g_1^2 g_2^2 \lambda_{h}+\frac{2}{5}g_1^2g_2^2 \lambda_{2}\nonumber\\
&+4\Big(\wl_2^3-8\lambda_{2}\wl_2^2-8\lambda_{2}^2\wl_2\Big)+\wl_2\Big(\frac{2227}{600}g_1^4-\frac{187}{8}g_2^4-\frac{494}{9}g_3^4+\frac{287}{60}g_1^2g_2^2+\frac{2}{9}g_1^2g_3^2+\frac{14}{3}g_2^2g_3^2\nonumber\\
&+\frac{24}{5}g_1^2\lambda_{h}-28\lambda_{h}^2\Big)+2\wl_2^2\Big(\frac{14}{15}g_1^2+9g_2^2+\frac{8}{3}g_3^2-20\lambda_{h}\Big)+4\lambda_{2}\wl_2\Big(\frac{1}{3}g_1^2+9g_2^2+\frac{8}{3}g_3^2-20\lambda_{h}\Big)\nonumber\\
&+\Big\{\frac{13}{5}g_1^2g_2^2+\wl_2\Big(\frac{13}{12}g_1^2+\frac{15}{4}g_2^2+\frac{20}{3}g_3^2-8\lambda_{2}-4\wl_2\Big)\Big\}\Tr\X_2+\Big\{-\frac{9}{5}g_1^2g_2^2+\wl_2\Big(\frac{5}{4}g_1^2+\frac{45}{4}g_2^2+40g_3^2\nonumber\\
&-24\lambda_{h}-24\lambda_{2}-12\wl_2\Big)\Big\}\Tr\X_\dq+\Big\{-\frac{21}{5}g_1^2g_2^2+\wl_2\Big(\frac{17}{4}g_1^2+\frac{45}{4}g_2^2+40g_3^2-24\lambda_{h}-24\lambda_{2}-12\wl_2\Big)\Big\}\Tr\X_\uq\nonumber\\
&+\Big\{-\frac{11}{5}g_1^2g_2^2+\wl_2\Big(\frac{15}{4}g_1^2+\frac{15}{4}g_2^2-8\lambda_{h}-8\lambda_{2}-4\wl_2\Big)\Big\}\Tr\X_\cl-\Tr\Big\{8\Big(\lambda_{2}+\frac{7}{4}\wl_2-\frac{2}{5}g_1^2\Big)\wX_2\wX_\cl\nonumber\\
&+\frac{27}{2}\wl_2\Big(\frac{5}{9}\X_2^2+\X_{\uq}^2+\X_{\dq}^2+\frac{1}{3}\X_{\cl}^2+\frac{8}{27}\X_2\X_\dq-2\wX_\uq\wX_\dq\Big)\Big\}-16\Tr\Big\{\wX_2^2\wX_\cl+\frac{1}{2}\wX_2\wX_\cl^2+\frac{3}{4}Y_2\wX_\cl Y_2^\dagger \X_\dq\Big\}\bigg].
\end{align}
\begin{align}
\beta(\lambda_{2}^{ii})&^{2-loop}_{\widetilde{R}_2,3-gen}=\bigg[\beta(\lambda_{2})^{2-loop}_{\widetilde{R}_2,1-gen}\bigg]_{\is}+\frac{1}{(16\pi^2)^2}\bigg\{-\frac{7}{750}g_1^6+\frac{7}{150}g_1^4g_2^2+\frac{7}{10}g_1^2g_2^4-\frac{21}{2}g_2^6\nonumber\\
&+\lambda_{2}^{ii}\,\Big(\frac{11}{90} g_1^4+\frac{33}{2} g_2^4
+\frac{88}{9} g_3^4\Big)\bigg\}+\frac{1}{(16\pi^2)^2}\sum_{j\neq i}\bigg[-6\lambda_{2}^{ii}\Big\{\Big(\lambda_{2}^{jj}\Big)^2+\lambda_{2}^{jj}\,\widetilde\lambda_{2}^{jj}+\Big(\widetilde\lambda_{2}^{jj}\Big)^2\Big\}-6\widetilde\lambda_{2}^{ii}\Big(\widetilde\lambda_{2}^{jj}\Big)^2\nonumber\\
&+\Big(\lambda_{2}^{jj}+\frac{1}{2}\widetilde\lambda_{2}^{jj}\Big)\Big(\frac{1}{30}g_1^4+\frac{45}{2}g_2^4+\frac{80}{3}g_3^4\Big)-\frac{1}{2}\widetilde\lambda_{2}^{jj}g_1^2 g_2^2-\frac{9}{2}\lambda_2^{ii}\,\Tr\Big\{2\,\X_\dq\X_{2,j}+\wX_\cl\wX_{2,j}+\frac{2}{3}\X_{2,i}\,\X_{2,j}\nonumber\\
&+\wX_{2,i}\,\wX_{2,j}\Big\}+4\Tr\bigg\{\X_\dq \Big(\X_{2,i} \X_{2,j}+\X_{2,j} \X_{2,i}+\frac{3}{2} Y_{2,i} \wX_{2,j} Y_{2,i}^\dagger+3 Y_{2,j} \wX_{2,i} Y_{2,j}^\dagger\Big)\bigg\}\nonumber\\
&+6\Tr\bigg\{\wX_\cl\Big( \wX_{2,i} \wX_{2,j}+\wX_{2,j} \wX_{2,i}+\frac{2}{3}\,Y_{2,i}^\dagger \X_{2,j} Y_{2,i}+\frac{2}{3} Y_{2,j}^\dagger \X_{2,i} Y_{2,j}\Big)\bigg\}\bigg].
\end{align}
\begin{align}
\beta(\wl_{2}^{ii})&^{2-loop}_{\widetilde{R}_2,3-gen}=\bigg[\beta(\wl_{2})^{2-loop}_{\widetilde{R}_2,1-gen}\bigg]_{\is}+\frac{1}{(16\pi^2)^2}\bigg\{\wl_{2}^{ii}\,\Big(\frac{11}{90} g_1^4+\frac{33}{2} g_2^4+\frac{88}{9} g_3^4\Big)-\frac{7}{75}g_1^4 g_2^2 -\frac{7}{5}g_1^2 g_2^4\bigg\}\nonumber\\
&+\frac{1}{(16\pi^2)^2}\sum_{j\neq i}\bigg[-6\wl_{2}^{ii}\Big\{\Big(\lambda_{2}^{jj}\Big)^2+\lambda_{2}^{jj}\,\wl_{2}^{jj}+\Big(\wl_{2}^{jj}\Big)^2\Big\}+\wl_{2}^{jj}g_1^2 g_2^2-\frac{9}{2}\,\wl_{2}^{ii}\,\Tr\Big\{2\,\X_\dq\X_{2,j}+\wX_\cl \wX_{2,j}\nonumber\\
&+\frac{2}{3}\X_{2,i} \X_{2,j}+\wX_{2,i}\wX_{2,j}\Big\}-6\Tr\Big\{\wX_\cl\Big(\wX_{2,i}\wX_{2,j}+\wX_{2,j}\wX_{2,i}+\frac{2}{3} Y_{2,i}^\dagger\X_{2,j}Y_{2,i}\Big)\Big\}\bigg].
\end{align}

\section{Two-loop beta functions of Higgs-leptoquark quartic couplings for \texorpdfstring{$\vec S_3$}{}}
\label{sec:H-LQcouplS3}
\begin{align}
\beta(\lambda_{3})&^{2-loop}_{\vec S_3,1-gen}=\beta(\lambda_{3})^{1-loop}_{\vec S_3}+\frac{1}{(16\pi^2)^2}\bigg[-\frac{789}{500}g_1^6+\frac{689}{100}g_1^4g_2^2+\frac{176}{15}g_1^2g_2^4+\frac{49}{6}g_2^6-\frac{4}{5}g_1^4g_3^2+8g_1^2g_2^2g_3^2\nonumber\\
&-40g_2^4g_3^2+\frac{6}{5}g_1^4\lambda_{h}-4g_1^2g_2^2\lambda_{h}+60g_2^4\lambda_{h}+\lambda_{3}\Big(\frac{25067}{3600}g_1^4+\frac{199}{120}g_1^2g_2^2+\frac{12661}{48}g_2^4+\frac{8}{9}g_1^2g_3^2+\frac{80}{3}g_2^2g_3^2\nonumber\\
&-\frac{112}{9}g_3^4+\frac{72}{5}g_1^2\lambda_h+72g_2^2\lambda_{h}-60\lambda_{h}^2\Big)+\widetilde\lambda_{3}\,\Big(\frac{17}{20}g_1^4-\frac{26}{3}g_1^2g_2^2+\frac{505}{4}g_2^4+\frac{8}{3}g_2^2g_3^2+20g_3^4+\frac{24}{5}g_1^2\lambda_{h}\nonumber\\
&+36g_2^2\lambda_{h}-16\lambda_{h}^2\Big)+\lambda_{3}^2\,\Big(\frac{13}{15}g_1^2+11g_2^2+\frac{16}{3}g_3^2-72\lambda_{h}\Big)+\widetilde\lambda_{3}^2\,\Big(-\frac{23}{60}g_1^2+\frac{11}{4}g_2^2+\frac{4}{3}g_3^2-18\lambda_{h}\Big)\nonumber\\
&+\lambda_{3}\widetilde\lambda_{3}\,\Big(-12g_2^2-32\lambda_{h}\Big)-\Big(19\lambda_{3}^3+11\lambda_{3}^2\widetilde\lambda_{3}+\frac{57}{4}\lambda_{3}\widetilde\lambda_{3}^2+\frac{19}{2}\widetilde\lambda_{3}^3\Big)+\Big\{\frac{1}{5}g_1^4-\frac{18}{5}g_1^2g_2^2-6g_2^4-32g_3^4\nonumber\\
&+\lambda_{3}\,\Big(\frac{5}{4}g_1^2+\frac{45}{4}g_2^2+40g_3^2-72\lambda_{h}\Big)-24\widetilde\lambda_{3}\lambda_{h}-12\lambda_{3}^2-3\widetilde\lambda_{3}^2\Big\}\Tr\,\X_{\dq}+\Big\{-\frac{19}{25}g_1^4-\frac{42}{5}g_1^2g_2^2-6g_2^4\nonumber\\
&-32g_3^4+\lambda_{3}\,\Big(\frac{17}{4}g_1^2+\frac{45}{4}g_2^2+40g_3^2-72\lambda_{h}\Big)-24\widetilde\lambda_{3}\lambda_{h}-12\lambda_{3}^2-3\widetilde\lambda_{3}^2\Big\}\Tr\,\X_{\uq}+\Big\{-\frac{2}{5}g_1^4-\frac{4}{5}g_1^2g_2^2+g_2^4\nonumber\\
&+\lambda_{3}\,\Big(\frac{5}{6}g_1^2+\frac{15}{2}g_2^2+\frac{20}{3}g_3^2\Big)-4\lambda_{3}^2-\widetilde\lambda_{3}^2\Big\}\Tr\,\X_{3}+\Big\{-2g_2^4-g_1^4-\frac{22}{5}g_1^2g_2^2+\frac{3}{4}\lambda_{3}\big(5g_1^2+5g_2^2-32\lambda_{h}\big)\nonumber\\
&-8\widetilde\lambda_{3}\lambda_{h}-4\lambda_3^2-\widetilde\lambda_3^2\Big\}\Tr\X_\cl-\Tr\Big\{\frac{27}{2}\lambda_{3}\widetilde\X_\dq^2+21\lambda_{3}\widetilde{\X}_\uq\widetilde{\X}_\dq+24\widetilde\lambda_{3}\widetilde{\X}_\uq\widetilde{\X}_\dq\Big\}-\frac{9}{2}\lambda_{3}\Tr\X_\cl^2-9\lambda_{3}\Tr\X_3^2\nonumber\\
&+\Tr\Big(\widetilde\X_3\widetilde\X_\cl\Big)\Big(-\frac{1}{4}\lambda_{3}-\frac{14}{5}g_1^2+2g_2^2\Big)-\Tr\Big(\X_3\widetilde\X_\dq^T\Big)\Big(\frac{1}{4}\lambda_{3}+\frac{2}{15}g_1^2-2g_2^2+\frac{32}{3}g_3^2\Big)\nonumber\\
&-\Tr\Big(\X_3\widetilde\X_\uq^T\Big)\Big(\frac{33}{4}\lambda_{3}-4\widetilde\lambda_{3}\Big)+\Tr\Big(-\frac{27}{2}\lambda_{3}\X_\uq^2+12\widetilde\X_\cl^2\widetilde\X_3+25\wX_3^2\wX_\cl+12 Y_3^\dagger\wX_\dq^T Y_3\wX_\cl+2 Y_3^\dagger\wX_\uq^T Y_3\wX_\cl\nonumber\\
&+27\X_3^2\wX_\dq^T+6\X_3^2\wX_\uq^T+12\wX_\dq^T\X_3\wX_\dq^T-2\X_3\wX_\dq^T\wX_\uq^T-2\X_3\wX_\uq^T\wX_\dq^T+4\wX_\uq^T\X_3\wX_\uq^T\Big)\bigg].
\end{align}
\begin{align}
\beta(\widetilde\lambda_{3})&^{2-loop}_{\vec S_3,1-gen}=\beta(\widetilde\lambda_{3})^{1-loop}_{\vec S_3}+\frac{1}{(16\pi^2)^2}\bigg[-\frac{427}{25}g_1^4g_2^2-\frac{547}{15}g_1^2g_2^4-16g_1^2g_2^2g_3^2+8g_1^2g_2^2\Big(\lambda_{h}+\frac{1}{5}\lambda_{3}\Big)\nonumber\\
&+\widetilde\lambda_{3}\Big(\frac{18947}{3600}g_1^4+\frac{541}{48}g_2^4-\frac{472}{9}g_3^4+\frac{2567}{120}g_1^2g_2^2+\frac{8}{9}g_1^2g_3^2+\frac{64}{3}g_2^2g_3^2+\frac{24}{5}g_1^2\lambda_{h}-28\lambda_{h}^2\Big)+\widetilde\lambda_{3}^2\Big(\frac{31}{15}g_1^2-\frac{13}{4}\widetilde\lambda_{3}\Big)\nonumber\\
&+\lambda_{3}\widetilde\lambda_{3}\Big(\frac{26}{15}g_1^2-35\lambda_{3}-35\widetilde\lambda_{3}\Big)+\wl_{3}\Big(\lambda_{3}+\frac{1}{2}\wl_{3}\Big)\Big(46g_2^2+\frac{32}{3}g_3^2-80\lambda_{h}\Big)+\bigg\{\frac{36}{5}g_1^2g_2^2+\wl_{3}\Big(\frac{5}{4}g_1^2+\frac{45}{4}g_2^2\nonumber\\
&+40g_3^2-24\lambda_{h}-24\lambda_{3}-12\wl_{3}\Big)\bigg\}\Tr\X_\dq+\bigg\{\frac{44}{5}g_1^2g_2^2+\wl_{3}\Big(\frac{15}{4}g_1^2+\frac{15}{4}g_2^2-8\lambda_{h}-8\lambda_{3}-4\wl_{3}\Big)\bigg\}\Tr\X_\cl\nonumber\\
&+\bigg\{\frac{8}{5}g_1^2g_2^2+\wl_{3}\Big(\frac{5}{6}g_1^2+\frac{15}{2}g_2^2+\frac{20}{3}g_3^2-8\lambda_{3}-4\wl_{3}\Big)\bigg\}\Tr \X_{3}+\bigg\{\frac{84}{5}g_1^2g_2^2+\wl_{3}\Big(\frac{17}{4}g_1^2+\frac{45}{4}g_2^2+40g_3^2\nonumber\\
&-24\lambda_{h}-24\lambda_{3}-12\wl_{3}\Big)\bigg\}\Tr \X_\uq+\wl_{3}\Tr\Big\{-\frac{27}{2}\X_\uq^2-\frac{27}{2}\X_\dq^2+27\wX_\dq\wX_\uq-\frac{9}{2}\X_\cl^2-9\X_3^2\Big\}\nonumber\\
&+\Big\{\frac{14}{5}g_1^2-2g_2^2-8\lambda_{3}-\frac{49}{4}\,\wl_{3}\Big\}\Tr\Big(\wX_\cl \wX_3\Big)+\Big\{\frac{2}{15}g_1^2-2g_2^2+\frac{32}{3}g_3^2-8\lambda_{3}-\frac{49}{4}\,\wl_{3}\Big\}\Tr\Big(\X_3\wX_\dq^T\Big)\nonumber\\
&-\Big\{\frac{14}{15}g_1^2-2g_2^2+\frac{32}{3}g_3^2-8\lambda_{3}+\frac{17}{4}\,\wl_{3}\Big\}\Tr\Big(\X_3\wX_\uq^T\Big)-\Tr\Big(8\wX_3\wX_\cl^2+23\wX_3^2\wX_\cl+12\wX_\cl Y_3^\dagger \wX_\dq Y_3\nonumber\\
&+21\X_3^2\wX_\dq^T-21\X_3^2\wX_\uq^T+8\wX_\dq^T\X_3\wX_\dq^T-8\wX_\uq^T\X_3\wX_\uq^T\Big)\bigg].
\end{align}
\begin{align}
\beta(\lambda_{3}^{ii})&^{2-loop}_{\vec S_2,3-gen}=\bigg[\beta(\lambda_{3})^{2-loop}_{\vec S_3,1-gen}\bigg]_{\is}+\frac{1}{(16\pi^2)^2}\bigg\{+\lambda_{3}^{ii}\Big(\frac{143}{150}g_1^4+121 g_2^4+\frac{44}{3}g_3^4\Big)-\frac{28}{125}g_1^6-112g_2^6\nonumber\\
&+\frac{28}{25}g_1^4 g_2^2+\frac{56}{5}g_1^2g_2^4\bigg\}+\frac{1}{(16\pi^2)^2}\sum_{j\neq i}\bigg[-9\lambda_{3}^{ii}\Big\{\Big(\lambda_{3}^{jj}\Big)^2+\lambda_{3}^{jj}\,\wl_{3}^{jj}+\frac{3}{4} \Big(\wl_{3}^{jj}\Big)^2\Big\}-6\wl_{3}^{ii}\Big(\wl_{3}^{jj}\Big)^2\nonumber\\
&-8\wl_{3}^{jj}g_1^2 g_2^2+\Big(\lambda_{3}^{jj}+\frac{1}{2}\wl_{3}^{jj}\Big)\Big(\frac{4}{5}g_1^4+240 g_2^4+40g_3^4\Big)-\frac{27}{4}\lambda_{3}^{ii}\Tr\Big\{\wX_\cl\wX_{3,j}+\frac{1}{3}\X_{3,i}\X_{3,j}+\wX_{3,i}\wX_{3,j}\nonumber\\
&+\X_{3,j}\wX_\uq^T+\X_{3,j}\wX_\dq^T\Big\}+9\Tr\Big\{\wX_\cl \Big(\wX_{3,i}\wX_{3,j}+\wX_{3,j}\wX_{3,i}+\frac{1}{3} Y_{3,i}^\dagger\X_{3,j} Y_{3,i}+\frac{4}{9}Y_{3,j}^\dagger\X_{3,i} Y_{3,j}\Big)\Big\}\nonumber\\
&+3\Tr\Big\{\wX_{\dq}^T\Big(\X_{3,i}\X_{3,j}+\X_{3,j}\X_{3,i}+3Y_{3,i}\wX_{3,j}^T Y_{3,i}^\dagger\Big)+4\Big(\wX_\dq^T+\frac{1}{2}\wX_\uq^T\Big)Y_{3,j}\wX_{3,i} Y_{3,j}^\dagger\Big\}\bigg].
\end{align}
\begin{align}
\beta(\wl_{3}^{ii})&^{2-loop}_{\vec S_2,3-gen}=\bigg[\beta(\wl_{3})^{2-loop}_{\vec S_3,1-gen}\bigg]_{\is}+\frac{1}{(16\pi^2)^2}\bigg\{\wl_{3}^{ii}\Big(\frac{143}{150}g_1^4+121 g_2^4+\frac{44}{3}g_3^4\Big)-\frac{56}{25}g_1^4 g_2^2-\frac{112}{5}g_1^2 g_2^4\bigg\}\nonumber\\
&+\frac{1}{(16\pi^2)^2}\sum_{j\neq i}\bigg[-9\wl_{3}^{ii}\Big\{\Big(\lambda_{3}^{jj}\Big)^2+\lambda_{3}^{jj}\,\wl_{3}^{jj}-\frac{7}{12} \Big(\wl_{3}^{jj}\Big)^2\Big\}+16\wl_{3}^{jj}g_1^2 g_2^2-\frac{27}{4}\wl_{3}^{ii}\,\Tr\Big\{\wX_\cl\wX_{3,j}+\frac{1}{3}\X_{3,i}\X_{3,j}\nonumber\\
&+\wX_{3,i}\wX_{3,j}+\X_{3,j}\wX_\dq^T+\X_{3,j}\wX_\uq^T\Big\}-9\Tr\Big\{\wX_\cl\Big(\wX_{3,i}\wX_{3,j}+\wX_{3,j}\wX_{3,i}+\frac{1}{3} Y_{3,i}^\dagger\X_{3,j} Y_{3,i}+\frac{2}{9}Y_{3,j}^\dagger\X_{3,i} Y_{3,j}\Big)\Big\}\nonumber\\
&+3\Tr\Big\{\Big(\wX_\uq^T-\wX_\dq^T\Big)\Big(\X_{3,i}\X_{3,j}+\X_{3,j}\X_{3,i}+3Y_{3,i}\wX_{3,j}Y_{3,i}^\dagger+2Y_{3,j}\wX_{3,i}Y_{3,j}^\dagger\Big)\Big\}\bigg].
\end{align}

\section{Two-loop beta functions of Higgs-leptoquark quartic couplings for \texorpdfstring{$\widetilde{R}_2+\vec S_3$}{}}
\label{sec:H-LQcouplR2S3}

\begin{align}
\beta(\lambda_{2})&^{2-loop}_{\widetilde{R}_2+\vec S_3,1-gen}=\beta(\lambda_{2})^{2-loop}_{\widetilde{R}_2,1-gen}+\frac{1}{(16\pi^2)^2}\bigg[-9\lambda_{2}\Big(\lambda_{3}^2+\lambda_{3}\wl_{3}+\frac{3}{4}\,\wl_{3}^2\Big)-6\wl_{2}\wl_{3}^2-2\wl_{3} g_1^2 g_2^2\nonumber\\
&+\Big(\lambda_{3}+\frac{1}{2}\wl_{3}\Big)\Big(\frac{1}{5}g_1^4+90g_2^4+40g_3^4\Big)-\frac{7}{5}\,\Big(\frac{1}{50}g_1^6-\frac{1}{10}g_1^4 g_2^2-g_1^2g_2^4+15g_2^6\Big)+11\lambda_{2}\Big(\frac{1}{30} g_1^4+3g_2^4+\frac{2}{3} g_3^4\Big)\nonumber\\
&-\frac{27}{4}\lambda_{2}\Tr\Big(\wX_3\wX_2+\wX_3\wX_\cl+\X_3\wX_\dq^T+\X_3\wX_\uq^T\Big)+9\Tr\Big\{Y_2\wX_3Y_2^\dagger\X_\dq+\wX_2\wX_3\wX_\cl+\wX_3\wX_2\wX_\cl+\frac{4}{3}\wX_2 Y_3^\dagger\wX_\dq^T Y_3\nonumber\\
&+\frac{2}{3}\wX_2 Y_3^\dagger\wX_\uq^T Y_3+\frac{4}{9}Y_2Y_3^\dagger Y_\dq^TY_2^*Y_3^T Y_\dq^\dagger+\frac{1}{3}\X_{3}Y_\dq^T \X_2^T Y_\dq^*\Big\}\bigg],
\end{align}
\begin{align}
\beta(\wl_{2})&^{2-loop}_{\widetilde{R}_2+\vec S_3,1-gen}=\beta(\wl_{2})^{2-loop}_{\widetilde{R}_2,1-gen}+\frac{1}{(16\pi^2)^2}\bigg[-9\wl_{2}\Big(\lambda_{3}^2+\lambda_{3}\wl_{3}+\frac{7}{12}\,\wl_{3}^2\Big)+4\wl_{3}\,g_1^2g_2^2-\frac{7}{25}g_1^4 g_2^2-\frac{14}{5}g_1^2 g_2^4\nonumber\\
&+11\wl_{2}\Big(\frac{1}{30} g_1^4+3g_2^4+\frac{2}{3} g_3^4\Big)-\frac{27}{4}\wl_{2}\Tr\Big(\wX_3\wX_2+\wX_3\wX_\cl+\X_3\wX_\dq^T+\X_3\wX_\uq^T\Big)-9\Tr\Big\{\wX_2\wX_3\wX_\cl+\wX_{3}\wX_{2}\wX_{\cl}\nonumber\\
&+\frac{2}{9}Y_2Y_3^\dagger Y_\dq^TY_2^*Y_3^T Y_\dq^\dagger+\frac{2}{3}\wX_2 Y_3^\dagger\wX_\dq^T Y_3-\frac{2}{3}\wX_2 Y_3^\dagger\wX_\uq^T Y_3\Big\}\bigg],
\end{align}
\begin{align}
\beta(\lambda_{3})&^{2-loop}_{\widetilde{R}_2+\vec S_3,1-gen}=\beta(\lambda_{3})^{2-loop}_{\vec S_3,1-gen}+\frac{1}{(16\pi^2)^2}\bigg[-6\Big(\lambda_{2}^2\lambda_{3}+\lambda_{2}\wl_{2}\lambda_{3}+\wl_{2}^2\,\lambda_{3}+\wl_{2}^2\,\wl_{3}\Big)-\frac{7}{5}\,\Big(\frac{1}{75}g_1^6-\frac{1}{15}g_1^4 g_2^2\nonumber\\
&-g_1^2g_2^4+10g_2^6\Big)-2\wl_{2}g_1^2g_2^2+\Big(\lambda_{2}+\frac{1}{2}\wl_{2}\Big)\Big(\frac{2}{15}g_1^4+60g_2^4+\frac{80}{3}g_3^4\Big)+11\lambda_{3}\Big(\frac{13}{1800}g_1^4+\frac{11}{8}g_2^4+\frac{4}{9}g_3^4\Big)\nonumber\\
&-\frac{9}{2}\lambda_{3}\Tr\Big\{2\X_{2}\X_{\dq}+\wX_{2}\wX_{3}+\wX_{2}\wX_{\cl}\Big\}+6\Tr\Big\{2 Y_2\wX_3Y_2^\dagger\X_\dq+\wX_{2}\wX_{3}\wX_\cl+\wX_{3}\wX_{2}\wX_\cl+\frac{2}{3}Y_2Y_3^\dagger Y_\dq^TY_2^*Y_3^T Y_\dq^\dagger\nonumber\\
&+\wX_2 Y_3^\dagger\wX_\dq^T Y_3+\frac{2}{3}\X_3 Y_\dq^T \X_2^T Y_\dq^*\Big\}\bigg],
\end{align}
\begin{align}
\beta(\wl_{3})&^{2-loop}_{\widetilde{R}_2+\vec S_3,1-gen}=\beta(\wl_{3})^{2-loop}_{\vec S_3,1-gen}+\frac{1}{(16\pi^2)^2}\bigg[-6\wl_3\Big(\lambda_{2}^2+\lambda_{2}\wl_{2}-\wl_{2}^2\Big)+4\wl_{2}\,g_1^2g_2^2-\frac{14}{75}g_1^4g_2^2-\frac{14}{5}g_1^2 g_2^4\nonumber\\
&+11\wl_{3}\Big(\frac{13}{1800}g_1^4+\frac{11}{8}g_2^4+\frac{4}{9}g_3^4\Big)-\frac{9}{2}\wl_{3}\Tr\Big\{2\X_2\X_\dq+\wX_{2}\wX_3+\wX_{2}\wX_\cl\Big\}-6\Tr\Big\{\wX_2\wX_3\wX_\cl+\wX_3\wX_2\wX_\cl\nonumber\\
&+\frac{2}{3}Y_2Y_3^\dagger Y_\dq^TY_2^*Y_3^T Y_\dq^\dagger+\wX_2 Y_3^\dagger\wX_\dq^T Y_3-\wX_2 Y_3^\dagger\wX_\uq^T Y_3+\frac{2}{3}\X_3 Y_\dq^T \X_2^T Y_\dq^*\Big\}\bigg].
\end{align}

\begin{align}
\beta(\lambda_{2}^{ii})&^{2-loop}_{\widetilde{R}_2+\vec S_3,3-gen}=\beta(\lambda_{2}^{ii})^{2-loop}_{\widetilde{R}_2,3-gen}+\frac{1}{(16\pi^2)^2}\Big\{11\lambda_{2}^{ii}\Big(\frac{1}{10}g_1^4+9g_2^4+2g_3^4\Big)-\frac{21}{250}g_1^6+\frac{21}{50}g_1^4g_2^2\nonumber\\
&+\frac{21}{5}g_1^2g_2^4-63g_2^6\Big\}+\frac{1}{(16\pi^2)^2}\sum_{l=1}^3\bigg[-6\wl_2^{ii}\Big(\wl_{3}^{ll}\Big)^2-2\wl_{3}^{ll}g_1^2g_2^2-9\lambda_{2}^{ii}\Big\{\Big(\lambda_{3}^{ll}\Big)^2+\lambda_{3}^{ll}\,\wl_{3}^{ll}+\frac{3}{4}\Big(\wl_{3}^{ll}\Big)^2\Big\}\nonumber\\
&+\Big(\lambda_{3}^{ll}+\frac{1}{2}\wl_3^{ll}\Big)\Big(\frac{1}{5}g_1^4+90g_2^4+40g_3^4\Big)-\frac{27}{4}\lambda_{2}^{ii}\,\Tr\Big(\wX_{2,i}\wX_{3,l}+\X_{3,l}\wX_\dq^T+\X_{3,l}\wX_\uq^T+\wX_{3,l}\wX_\cl\Big)\nonumber\\
&+9\Tr\Big\{
Y_{2,i}\wX_{3,l}Y_{2,i}^\dagger\X_\dq+\wX_{2,i}\wX_{3,l}\wX_\cl+\frac{4}{9}Y_{2,i}Y_{3,l}^\dagger Y_\dq^T Y_{2,i}^* Y_{3,l}^T Y_\dq^\dagger+\frac{4}{3}\wX_{2,i} Y_{3,l}^\dagger \wX_\dq^T Y_{3,l}+\frac{2}{3}\wX_{2,i} Y_{3,l}^\dagger \wX_\uq^T Y_{3,l}\nonumber\\
&+Y_{2,i}\wX_{3,l} Y_{2,i}^\dagger \X_\dq+\wX_{2,l}\wX_{3,i}\wX_\cl+\frac{1}{3}\X_{3,l} Y_\dq^T\X_{2,i}^T Y_\dq^*\Big\}\bigg],
\end{align}
\begin{align}
\beta(\wl_{2}^{ii})&^{2-loop}_{\widetilde{R}_2+\vec S_3,3-gen}=\beta(\wl_{2}^{ii})^{2-loop}_{\widetilde{R}_2,3-gen}+\frac{1}{(16\pi^2)^2}\Big\{11\wl_{2}^{ii}\Big(\frac{1}{10}g_1^4+9 g_2^4+2 g_3^4\Big)-\frac{21}{25}g_1^4 g_2^2-\frac{42}{5}g_1^2 g_2^4\Big\}\nonumber\\
&+\frac{1}{(16\pi^2)^2}\sum_{l=1}^{3}\bigg[9\wl_{2}^{ii}\Big\{\frac{7}{12}\Big(\wl_{3}^{ll}\Big)^2-\lambda_{3}^{ll}\,\wl_{3}^{ll}-\Big(\lambda_{3}^{ll}\Big)^2\Big\}+4\wl_{3}^{ll}g_1^2 g_2^2-\frac{27}{4}\wl_{2}^{ii}\,\Tr\Big\{\wX_{2,i}\wX_{3,l}+\wX_{\cl}\wX_{3,l}\nonumber\\
&+\X_{3,l}\wX_\uq^T+\X_{3,l}\wX_\dq^T\Big\}-9\Tr\Big\{\wX_{2,i}\wX_{3,l}\wX_\cl+\wX_{2,i}\wX_\cl\wX_{3,l}+\frac{2}{9}Y_{2,i}Y_{3,l}^\dagger Y_\dq^T Y_{2,i}^* Y_{3,l}^T Y_\dq^\dagger+\frac{2}{3}\wX_{2,i} Y_{3,l}^\dagger\wX_\dq^T Y_{3,l}\nonumber\\
&-\frac{2}{3}\wX_{2,i} Y_{3,l}^\dagger\wX_\uq^T Y_{3,l}\Big\}\bigg],
\end{align}
\begin{align}
\beta(\lambda_{3}^{ii})&^{2-loop}_{\widetilde{R}_2+\vec S_3,3-gen}=\beta(\lambda_{3}^{ii})^{2-loop}_{\vec S_3,3-gen}+\frac{1}{(16\pi^2)^2}\Big\{11\lambda_{3}^{ii}\Big(\frac{13}{600}g_1^4+\frac{33}{8}g_2^4+\frac{4}{3}g_3^4\Big)-\frac{7}{125}g_1^6+\frac{7}{25}g_1^4g_2^2\nonumber\\
&+\frac{21}{5}g_1^2g_2^4-42g_2^6\Big\}+\frac{1}{(16\pi^2)^2}\sum_{l=1}^{3}\bigg[-6\wl_{3}^{ii}\Big(\wl_{2}^{ll}\Big)^2-2\wl_{2}^{ll}g_1^2g_2^2-6\lambda_{3}^{ii}\Big\{\Big(\wl_{2}^{ll}\Big)^2+\lambda_{2}^{ll}\,\wl_{2}^{ll}+\Big(\lambda_{2}^{ll}\Big)^2\Big\}\nonumber\\
&+\Big(\lambda_{2}^{ll}+\frac{1}{2}\wl_{2}^{ll}\Big)\Big(\frac{2}{15}g_1^4+60 g_2^4+\frac{80}{3}g_3^4\Big)-\frac{9}{2}\lambda_{3}^{ii}\Tr\Big\{2\X_{2,l}\X_\dq+\wX_{3,i}\wX_{2,l}+\wX_\cl \wX_{2,l}\Big\}\nonumber\\
&+6\Tr\Big\{2Y_{2,l}\wX_{3,i}Y_{2,l}^\dagger\X_\dq+\wX_{2,l}\wX_{3,i}\wX_\cl+\frac{2}{3}Y_{2,l}Y_{3,i}^\dagger Y_\dq^T Y_{2,l}^* Y_{3,i}^T Y_\dq^\dagger+\wX_{2,l} Y_{3,i}^\dagger\wX_\dq^T Y_{3,i}+\wX_{2,l} \wX_\cl\wX_{3,i}\nonumber\\
&+\frac{2}{3}\X_{3,i} Y_\dq^T \X_{2,l}^T Y_\dq^*\Big\}\bigg],
\end{align}
\begin{align}
\beta(\wl_{3}^{ii})&^{2-loop}_{\widetilde{R}_2+\vec S_3,3-gen}=\beta(\wl_{3}^{ii})^{2-loop}_{\vec S_3,3-gen}+\frac{1}{(16\pi^2)^2}\Big\{-\frac{14}{25}g_1^4 g_2^2-\frac{42}{5}g_1^2g_2^4+11\lambda_{3}^{ii}\Big(\frac{13}{600}g_1^4+\frac{33}{8}g_2^4+\frac{4}{3}g_3^4\Big)\Big\}\nonumber\\
&+\frac{1}{(16\pi^2)^2}\sum_{l=1}^{3}\bigg[6\wl_{3}^{ii}\Big\{\Big(\wl_{2}^{ll}\Big)^2-\lambda_{2}^{ll}\,\wl_{2}^{ll}-\Big(\lambda_{2}^{ll}\Big)^2\Big\}+4\wl_{2}^{ll}g_1^2 g_2^2-\frac{9}{2}\wl_{3}^{ii}\Tr\Big\{2\X_{2,l}\X_\dq+\wX_{3,i}\wX_{2,l}+\wX_\cl \wX_{2,l}\Big\}\nonumber\\
&-6\Tr\Big(\wX_{2,l}\wX_{3,i}\wX_\cl+\wX_{3,i}\wX_{2,l}\wX_\cl+\frac{2}{3}Y_{2,l}Y_{3,i}^\dagger Y_\dq^T Y_{2,l}^* Y_{3,i}^TY_\dq^\dagger+\wX_{2,l} Y_{3,i}^\dagger \wX_\dq^T Y_{3,i}-\wX_{2,l} Y_{3,i}^\dagger \wX_\uq^T Y_{3,i}\nonumber\\
&+\frac{2}{3}\X_{3,i}Y_\dq^T\X_{2,l}^T Y_\dq^*\Big)\bigg].
\end{align}

\twocolumn[]
\bibliographystyle{Ref}
\bibliography{References}

\providecommand{\href}[2]{#2}\begingroup\begin{thebibliography}{100}

\bibitem{Aad:2012tfa}
{\scshape ATLAS} collaboration, G.~Aad et~al., \textit{{Observation of a new
  particle in the search for the Standard Model Higgs boson with the ATLAS
  detector at the LHC}},
  \href{https://doi.org/10.1016/j.physletb.2012.08.020}{\textit{Phys. Lett. B}
  {\bfseries 716} }, [\href{https://arxiv.org/abs/1207.7214}{{\ttfamily
  1207.7214}}].

\bibitem{Chatrchyan:2012xdj}
{\scshape CMS} collaboration, S.~Chatrchyan et~al., \textit{{Observation of a
  New Boson at a Mass of 125 GeV with the CMS Experiment at the LHC}},
  \href{https://doi.org/10.1016/j.physletb.2012.08.021}{\textit{Phys. Lett. B}
  {\bfseries 716} }, [\href{https://arxiv.org/abs/1207.7235}{{\ttfamily
  1207.7235}}].

\bibitem{Dorsner:2016wpm}
I.~Dor\v{s}ner, S.~Fajfer, A.~Greljo, J.~Kamenik and N.~Ko\v{s}nik,
  \textit{{Physics of leptoquarks in precision experiments and at particle
  colliders}},
  \href{https://doi.org/10.1016/j.physrep.2016.06.001}{\textit{Phys. Rept.}
  {\bfseries 641} (2016) 1--68},
  [\href{https://arxiv.org/abs/1603.04993}{{\ttfamily 1603.04993}}].

\bibitem{Pati:1973uk}
J.~C. Pati and A.~Salam, \textit{{Unified Lepton-Hadron Symmetry and a Gauge
  Theory of the Basic Interactions}},
  \href{https://doi.org/10.1103/PhysRevD.8.1240}{\textit{Phys. Rev. D}
  {\bfseries 8} (1973) 1240--1251}.

\bibitem{Pati:1974yy}
J.~C. Pati and A.~Salam, \textit{{Lepton Number as the Fourth Color}},
  \href{https://doi.org/10.1103/PhysRevD.10.275}{\textit{Phys. Rev. D}
  {\bfseries 10} (1974) 275--289}. [Erratum: Phys. Rev. D 11, 703--703 (1975)].

\bibitem{Marzocca:2018wcf}
D.~Marzocca, \textit{{Addressing the B-physics anomalies in a fundamental
  Composite Higgs Model}},
  \href{https://doi.org/10.1007/JHEP07(2018)121}{\textit{JHEP} {\bfseries 07}
  (2018) 121}, [\href{https://arxiv.org/abs/1803.10972}{{\ttfamily
  1803.10972}}].

\bibitem{Gherardi:2020qhc}
V.~Gherardi, D.~Marzocca and E.~Venturini, \textit{{Low-energy phenomenology of
  scalar leptoquarks at one-loop accuracy}},
  \href{https://doi.org/10.1007/JHEP01(2021)138}{\textit{JHEP} {\bfseries 01}
  (2021) 138}, [\href{https://arxiv.org/abs/2008.09548}{{\ttfamily
  2008.09548}}].

\bibitem{Crivellin:2017zlb}
A.~Crivellin, D.~M\"uller and T.~Ota, \textit{{Simultaneous explanation of
  R(D$^{(*)}$) and $b\to s\mu^{+}\mu^{-}$ : the last scalar leptoquarks
  standing}}, \href{https://doi.org/10.1007/JHEP09(2017)040}{\textit{JHEP}
  {\bfseries 09} (2017) 040},
  [\href{https://arxiv.org/abs/1703.09226}{{\ttfamily 1703.09226}}].

\bibitem{Crivellin:2019dwb}
A.~Crivellin, D.~M\"uller and F.~Saturnino, \textit{{Flavor Phenomenology of
  the Leptoquark Singlet-Triplet Model}},
  \href{https://doi.org/10.1007/JHEP06(2020)020}{\textit{JHEP} {\bfseries 06}
  (2020) 020}, [\href{https://arxiv.org/abs/1912.04224}{{\ttfamily
  1912.04224}}].

\bibitem{Aydemir:2019ynb}
U.~Aydemir, T.~Mandal and S.~Mitra, \textit{{Addressing the ${\mathbf
  R_{D^{(*)}}}$ anomalies with an ${\mathbf S_1}$ leptoquark from
  $\mathbf{SO(10)}$ grand unification}},
  \href{https://doi.org/10.1103/PhysRevD.101.015011}{\textit{Phys. Rev. D}
  {\bfseries 101} (2020) 015011},
  [\href{https://arxiv.org/abs/1902.08108}{{\ttfamily 1902.08108}}].

\bibitem{Becirevic:2018afm}
D.~Be\v{c}irevi\'c, I.~Dor\v{s}ner, S.~Fajfer, N.~Ko\v{s}nik, D.~A. Faroughy
  and O.~Sumensari, \textit{{Scalar leptoquarks from grand unified theories to
  accommodate the $B$-physics anomalies}},
  \href{https://doi.org/10.1103/PhysRevD.98.055003}{\textit{Phys. Rev. D}
  {\bfseries 98} (2018) 055003},
  [\href{https://arxiv.org/abs/1806.05689}{{\ttfamily 1806.05689}}].

\bibitem{Bigaran:2019bqv}
I.~Bigaran, J.~Gargalionis and R.~R. Volkas, \textit{{A near-minimal leptoquark
  model for reconciling flavour anomalies and generating radiative neutrino
  masses}}, \href{https://doi.org/10.1007/JHEP10(2019)106}{\textit{JHEP}
  {\bfseries 10} (2019) 106},
  [\href{https://arxiv.org/abs/1906.01870}{{\ttfamily 1906.01870}}].

\bibitem{Mandal:2018kau}
T.~Mandal, S.~Mitra and S.~Raz, \textit{{$R_{D^{(*)}}$ motivated
  $\mathcal{S}_1$ leptoquark scenarios: Impact of interference on the exclusion
  limits from LHC data}},
  \href{https://doi.org/10.1103/PhysRevD.99.055028}{\textit{Phys. Rev. D}
  {\bfseries 99} (2019) 055028},
  [\href{https://arxiv.org/abs/1811.03561}{{\ttfamily 1811.03561}}].

\bibitem{Iguro:2020keo}
S.~Iguro, M.~Takeuchi and R.~Watanabe, \textit{{Testing leptoquark/EFT in
  ${\bar{B}} \rightarrow {D^{(*)}}l{\bar{\nu }}$ at the LHC}},
  \href{https://doi.org/10.1140/epjc/s10052-021-09125-5}{\textit{Eur. Phys. J.
  C} {\bfseries 81} (2021) 406},
  [\href{https://arxiv.org/abs/2011.02486}{{\ttfamily 2011.02486}}].

\bibitem{Lee:2021jdr}
H.~M. Lee, \textit{{Leptoquark option for B-meson anomalies and leptonic
  signatures}},
  \href{https://doi.org/10.1103/PhysRevD.104.015007}{\textit{Phys. Rev. D}
  {\bfseries 104} (2021) 015007},
  [\href{https://arxiv.org/abs/2104.02982}{{\ttfamily 2104.02982}}].

\bibitem{Bordone:2020lnb}
M.~Bordone, O.~Cat\`a, T.~Feldmann and R.~Mandal, \textit{{Constraining flavour
  patterns of scalar leptoquarks in the effective field theory}},
  \href{https://doi.org/10.1007/JHEP03(2021)122}{\textit{JHEP} {\bfseries 03}
  (2021) 122}, [\href{https://arxiv.org/abs/2010.03297}{{\ttfamily
  2010.03297}}].

\bibitem{Borschensky:2021hbo}
C.~Borschensky, B.~Fuks, A.~Kulesza and D.~Schwartl\"ander, \textit{{Scalar
  leptoquark pair production at the LHC: precision predictions in the era of
  flavour anomalies}},  \href{https://arxiv.org/abs/2108.11404}{{\ttfamily
  2108.11404}}.

\bibitem{Browder:2021hbl}
T.~E. Browder, N.~G. Deshpande, R.~Mandal and R.~Sinha, \textit{{Impact of
  $B\to K^* \nu \bar \nu$ measurements on beyond the Standard Model theories}},
  \href{https://doi.org/10.1103/PhysRevD.104.053007}{\textit{Phys. Rev. D}
  {\bfseries 104} (2021) 053007},
  [\href{https://arxiv.org/abs/2107.01080}{{\ttfamily 2107.01080}}].

\bibitem{Sheng:2021iss}
J.-H. Sheng, J.~Zhu and Q.-Y. Hu, \textit{{Investigation on the New Physics
  effects of the vector leptoquark on semileptonic $\bar{B}^*\rightarrow V \tau
  ^-\bar{\nu _{\tau }}$ decays}},
  \href{https://doi.org/10.1140/epjc/s10052-021-09322-2}{\textit{Eur. Phys. J.
  C} {\bfseries 81} (2021) 524}.

\bibitem{Cornella:2021sby}
C.~Cornella, D.~A. Faroughy, J.~Fuentes-Martin, G.~Isidori and M.~Neubert,
  \textit{{Reading the footprints of the B-meson flavor anomalies}},
  \href{https://doi.org/10.1007/JHEP08(2021)050}{\textit{JHEP} {\bfseries 08}
  (2021) 050}, [\href{https://arxiv.org/abs/2103.16558}{{\ttfamily
  2103.16558}}].

\bibitem{Crivellin:2021lix}
A.~Crivellin, J.~F. Eguren and J.~Virto, \textit{{Next-to-Leading-Order QCD
  Matching for $\Delta F=2$ Processes in Scalar Leptoquark Models}},
  \href{https://arxiv.org/abs/2109.13600}{{\ttfamily 2109.13600}}.

\bibitem{Angelescu:2018tyl}
A.~Angelescu, D.~Be\v{c}irevi\'c, D.~A. Faroughy and O.~Sumensari,
  \textit{{Closing the window on single leptoquark solutions to the $B$-physics
  anomalies}}, \href{https://doi.org/10.1007/JHEP10(2018)183}{\textit{JHEP}
  {\bfseries 10} (2018) 183},
  [\href{https://arxiv.org/abs/1808.08179}{{\ttfamily 1808.08179}}].

\bibitem{Angelescu:2021lln}
A.~Angelescu, D.~Be\v{c}irevi\'c, D.~A. Faroughy, F.~Jaffredo and O.~Sumensari,
  \textit{{Single leptoquark solutions to the B-physics anomalies}},
  \href{https://doi.org/10.1103/PhysRevD.104.055017}{\textit{Phys. Rev. D}
  {\bfseries 104} (2021) 055017},
  [\href{https://arxiv.org/abs/2103.12504}{{\ttfamily 2103.12504}}].

\bibitem{Arnan:2019olv}
P.~Arnan, D.~Becirevic, F.~Mescia and O.~Sumensari, \textit{{Probing low energy
  scalar leptoquarks by the leptonic $W$ and $Z$ couplings}},
  \href{https://doi.org/10.1007/JHEP02(2019)109}{\textit{JHEP} {\bfseries 02}
  (2019) 109}, [\href{https://arxiv.org/abs/1901.06315}{{\ttfamily
  1901.06315}}].

\bibitem{ColuccioLeskow:2016dox}
E.~Coluccio~Leskow, G.~D'Ambrosio, A.~Crivellin and D.~M\"uller,
  \textit{{$(g-2)_\mu$, lepton flavor violation, and $Z$ decays with
  leptoquarks: Correlations and future prospects}},
  \href{https://doi.org/10.1103/PhysRevD.95.055018}{\textit{Phys. Rev. D}
  {\bfseries 95} (2017) 055018},
  [\href{https://arxiv.org/abs/1612.06858}{{\ttfamily 1612.06858}}].

\bibitem{Crivellin:2020mjs}
A.~Crivellin, C.~Greub, D.~M\"uller and F.~Saturnino, \textit{{Scalar
  Leptoquarks in Leptonic Processes}},
  \href{https://doi.org/10.1007/JHEP02(2021)182}{\textit{JHEP} {\bfseries 02}
  (2021) 182}, [\href{https://arxiv.org/abs/2010.06593}{{\ttfamily
  2010.06593}}].

\bibitem{Saad:2020ihm}
S.~Saad, \textit{{Combined explanations of $(g-2)_{\mu}$, $R_{D^{(*)}}$,
  $R_{K^{(*)}}$ anomalies in a two-loop radiative neutrino mass model}},
  \href{https://doi.org/10.1103/PhysRevD.102.015019}{\textit{Phys. Rev. D}
  {\bfseries 102} (2020) 015019},
  [\href{https://arxiv.org/abs/2005.04352}{{\ttfamily 2005.04352}}].

\bibitem{Saad:2020ucl}
S.~Saad and A.~Thapa, \textit{{Common origin of neutrino masses and
  $R_{D^{(\ast)}}$, $R_{K^{(\ast)}}$ anomalies}},
  \href{https://doi.org/10.1103/PhysRevD.102.015014}{\textit{Phys. Rev. D}
  {\bfseries 102} (2020) 015014},
  [\href{https://arxiv.org/abs/2004.07880}{{\ttfamily 2004.07880}}].

\bibitem{Babu:2020hun}
K.~S. Babu, P.~S.~B. Dev, S.~Jana and A.~Thapa, \textit{{Unified framework for
  $B$-anomalies, muon $g - 2$ and neutrino masses}},
  \href{https://doi.org/10.1007/JHEP03(2021)179}{\textit{JHEP} {\bfseries 03}
  (2021) 179}, [\href{https://arxiv.org/abs/2009.01771}{{\ttfamily
  2009.01771}}].

\bibitem{Chang:2021axw}
W.-F. Chang, \textit{{One colorful resolution to the neutrino mass generation,
  three lepton flavor universality anomalies, and the Cabibbo angle anomaly}},
  \href{https://doi.org/10.1007/JHEP09(2021)043}{\textit{JHEP} {\bfseries 09}
  (2021) 043}, [\href{https://arxiv.org/abs/2105.06917}{{\ttfamily
  2105.06917}}].

\bibitem{Zhang:2021dgl}
D.~Zhang, \textit{{Radiative neutrino masses, lepton flavor mixing and muon g
  \ensuremath{-} 2 in a leptoquark model}},
  \href{https://doi.org/10.1007/JHEP07(2021)069}{\textit{JHEP} {\bfseries 07}
  (2021) 069}, [\href{https://arxiv.org/abs/2105.08670}{{\ttfamily
  2105.08670}}].

\bibitem{Georgi:1974my}
H.~Georgi, \textit{{The State of the Art\textemdash{}Gauge Theories}},
  \href{https://doi.org/10.1063/1.2947450}{\textit{AIP Conf. Proc.} {\bfseries
  23} (1975) 575--582}.

\bibitem{Georgi:1974sy}
H.~Georgi and S.~Glashow, \textit{{Unity of All Elementary Particle Forces}},
  \href{https://doi.org/10.1103/PhysRevLett.32.438}{\textit{Phys. Rev. Lett.}
  {\bfseries 32} (1974) 438--441}.

\bibitem{Dimopoulos:1979es}
S.~Dimopoulos and L.~Susskind, \textit{{Mass Without Scalars}}, .

\bibitem{Farhi:1980xs}
E.~Farhi and L.~Susskind, \textit{{Technicolor}},
  \href{https://doi.org/10.1016/0370-1573(81)90173-3}{\textit{Phys. Rept.}
  {\bfseries 74} (1981) 277}.

\bibitem{Schrempp:1984nj}
B.~Schrempp and F.~Schrempp, \textit{{Light Leptoquarks}},
  \href{https://doi.org/10.1016/0370-2693(85)91450-9}{\textit{Phys. Lett. B}
  {\bfseries 153} (1985) 101--107}.

\bibitem{Wudka:1985ef}
J.~Wudka, \textit{{Composite Leptoquarks}},
  \href{https://doi.org/10.1016/0370-2693(86)90356-4}{\textit{Phys. Lett. B}
  {\bfseries 167} (1986) 337--342}.

\bibitem{Nilles:1983ge}
H.~P. Nilles, \textit{{Supersymmetry, Supergravity and Particle Physics}},
  \href{https://doi.org/10.1016/0370-1573(84)90008-5}{\textit{Phys. Rept.}
  {\bfseries 110} (1984) 1--162}.

\bibitem{Haber:1984rc}
H.~E. Haber and G.~L. Kane, \textit{{The Search for Supersymmetry: Probing
  Physics Beyond the Standard Model}},
  \href{https://doi.org/10.1016/0370-1573(85)90051-1}{\textit{Phys. Rept.}
  {\bfseries 117} (1985) 75--263}.

\bibitem{Assad:2017iib}
N.~Assad, B.~Fornal and B.~Grinstein, \textit{{Baryon Number and Lepton
  Universality Violation in Leptoquark and Diquark Models}},
  \href{https://doi.org/10.1016/j.physletb.2017.12.042}{\textit{Phys. Lett. B}
  {\bfseries 777} (2018) 324--331},
  [\href{https://arxiv.org/abs/1708.06350}{{\ttfamily 1708.06350}}].

\bibitem{Perez:2021ddi}
P.~F. Perez, C.~Murgui and A.~D. Plascencia, \textit{{Leptoquarks and matter
  unification: Flavor anomalies and the muon g-2}},
  \href{https://doi.org/10.1103/PhysRevD.104.035041}{\textit{Phys. Rev. D}
  {\bfseries 104} (2021) 035041},
  [\href{https://arxiv.org/abs/2104.11229}{{\ttfamily 2104.11229}}].

\bibitem{Murgui:2021bdy}
C.~Murgui and M.~B. Wise, \textit{{Scalar leptoquarks, baryon number violation,
  and Pati-Salam symmetry}},
  \href{https://doi.org/10.1103/PhysRevD.104.035017}{\textit{Phys. Rev. D}
  {\bfseries 104} (2021) 035017},
  [\href{https://arxiv.org/abs/2105.14029}{{\ttfamily 2105.14029}}].

\bibitem{Bandyopadhyay:2018syt}
P.~Bandyopadhyay and R.~Mandal, \textit{{Revisiting scalar leptoquark at the
  LHC}}, \href{https://doi.org/10.1140/epjc/s10052-018-5959-x}{\textit{Eur.
  Phys. J. C} {\bfseries 78} (2018) 491},
  [\href{https://arxiv.org/abs/1801.04253}{{\ttfamily 1801.04253}}].

\bibitem{Bhaskar:2020gkk}
A.~Bhaskar, T.~Mandal and S.~Mitra, \textit{{Boosting vector leptoquark
  searches with boosted tops}},
  \href{https://doi.org/10.1103/PhysRevD.101.115015}{\textit{Phys. Rev. D}
  {\bfseries 101} (2020) 115015},
  [\href{https://arxiv.org/abs/2004.01096}{{\ttfamily 2004.01096}}].

\bibitem{Bhaskar:2021pml}
A.~Bhaskar, D.~Das, T.~Mandal, S.~Mitra and C.~Neeraj, \textit{{Precise limits
  on the charge-$2/3$ $U_1$ vector leptoquark}},
  \href{https://arxiv.org/abs/2101.12069}{{\ttfamily 2101.12069}}.

\bibitem{Bhaskar:2021gsy}
A.~Bhaskar, T.~Mandal, S.~Mitra and M.~Sharma, \textit{{Improving
  third-generation leptoquark searches with combined signals and boosted top}},
   \href{https://arxiv.org/abs/2106.07605}{{\ttfamily 2106.07605}}.

\bibitem{DaRold:2021pgn}
L.~Da~Rold, M.~Epele, A.~Medina, N.~I. Mileo and A.~Szynkman,
  \textit{{Enhancement of the double Higgs production via leptoquarks at the
  LHC}},  \href{https://arxiv.org/abs/2105.06309}{{\ttfamily 2105.06309}}.

\bibitem{Hiller:2021pul}
G.~Hiller, D.~Loose and I.~Ni\v{s}and\v{z}i\'c, \textit{{Flavorful leptoquarks
  at the LHC and beyond: spin 1}},
  \href{https://doi.org/10.1007/JHEP06(2021)080}{\textit{JHEP} {\bfseries 06}
  (2021) 080}, [\href{https://arxiv.org/abs/2103.12724}{{\ttfamily
  2103.12724}}].

\bibitem{Haisch:2020xjd}
U.~Haisch and G.~Polesello, \textit{{Resonant third-generation leptoquark
  signatures at the Large Hadron Collider}},
  \href{https://doi.org/10.1007/JHEP05(2021)057}{\textit{JHEP} {\bfseries 05}
  (2021) 057}, [\href{https://arxiv.org/abs/2012.11474}{{\ttfamily
  2012.11474}}].

\bibitem{Chandak:2019iwj}
K.~Chandak, T.~Mandal and S.~Mitra, \textit{{Hunting for scalar leptoquarks
  with boosted tops and light leptons}},
  \href{https://doi.org/10.1103/PhysRevD.100.075019}{\textit{Phys. Rev. D}
  {\bfseries 100} (2019) 075019},
  [\href{https://arxiv.org/abs/1907.11194}{{\ttfamily 1907.11194}}].

\bibitem{Bhaskar:2020kdr}
A.~Bhaskar, D.~Das, B.~De and S.~Mitra, \textit{{Enhancing scalar productions
  with leptoquarks at the LHC}},
  \href{https://doi.org/10.1103/PhysRevD.102.035002}{\textit{Phys. Rev. D}
  {\bfseries 102} (2020) 035002},
  [\href{https://arxiv.org/abs/2002.12571}{{\ttfamily 2002.12571}}].

\bibitem{Alves:2018krf}
A.~Alves, O.~J.~t. Eboli, G.~Grilli Di~Cortona and R.~R. Moreira,
  \textit{{Indirect and monojet constraints on scalar leptoquarks}},
  \href{https://doi.org/10.1103/PhysRevD.99.095005}{\textit{Phys. Rev. D}
  {\bfseries 99} (2019) 095005},
  [\href{https://arxiv.org/abs/1812.08632}{{\ttfamily 1812.08632}}].

\bibitem{Dorsner:2019vgp}
I.~Dor\v{s}ner, S.~Fajfer and M.~Patra, \textit{{A comparative study of the
  $S_1$ and $U_1$ leptoquark effects in the light quark regime}},
  \href{https://doi.org/10.1140/epjc/s10052-020-7754-8}{\textit{Eur. Phys. J.
  C} {\bfseries 80} (2020) 204},
  [\href{https://arxiv.org/abs/1906.05660}{{\ttfamily 1906.05660}}].

\bibitem{Mandal:2018qpg}
S.~Mandal, M.~Mitra and N.~Sinha, \textit{{Probing leptoquarks and heavy
  neutrinos at the LHeC}},
  \href{https://doi.org/10.1103/PhysRevD.98.095004}{\textit{Phys. Rev. D}
  {\bfseries 98} (2018) 095004},
  [\href{https://arxiv.org/abs/1807.06455}{{\ttfamily 1807.06455}}].

\bibitem{Padhan:2019dcp}
R.~Padhan, S.~Mandal, M.~Mitra and N.~Sinha, \textit{{Signatures of
  $\tilde{R}_2$ class of Leptoquarks at the upcoming $ep$ colliders}},
  \href{https://doi.org/10.1103/PhysRevD.101.075037}{\textit{Phys. Rev. D}
  {\bfseries 101} (2020) 075037},
  [\href{https://arxiv.org/abs/1912.07236}{{\ttfamily 1912.07236}}].

\bibitem{Baker:2019sli}
M.~J. Baker, J.~Fuentes-Mart\'\i{}n, G.~Isidori and M.~K\"onig, \textit{{High-
  $p_T$ signatures in vector\textendash{}leptoquark models}},
  \href{https://doi.org/10.1140/epjc/s10052-019-6853-x}{\textit{Eur. Phys. J.
  C} {\bfseries 79} (2019) 334},
  [\href{https://arxiv.org/abs/1901.10480}{{\ttfamily 1901.10480}}].

\bibitem{Nadeau:1993zv}
H.~Nadeau and D.~London, \textit{{Leptoquarks at e gamma colliders}},
  \href{https://doi.org/10.1103/PhysRevD.47.3742}{\textit{Phys. Rev. D}
  {\bfseries 47} (1993) 3742--3749},
  [\href{https://arxiv.org/abs/hep-ph/9303238}{{\ttfamily hep-ph/9303238}}].

\bibitem{Atag:1994hk}
S.~Atag and O.~Cakir, \textit{{Pair production of scalar leptoquarks at TeV
  energy gamma p colliders}},
  \href{https://doi.org/10.1103/PhysRevD.49.5769}{\textit{Phys. Rev. D}
  {\bfseries 49} (1994) 5769--5772}.

\bibitem{Atag:1994np}
S.~Atag, A.~Celikel and S.~Sultansoy, \textit{{Scalar leptoquark production at
  TeV energy gamma p colliders}},
  \href{https://doi.org/10.1016/0370-2693(94)91212-2}{\textit{Phys. Lett. B}
  {\bfseries 326} (1994) 185--189}.

\bibitem{Buchmuller:1986zs}
W.~Buchmuller, R.~Ruckl and D.~Wyler, \textit{{Leptoquarks in Lepton - Quark
  Collisions}},
  \href{https://doi.org/10.1016/0370-2693(87)90637-X}{\textit{Phys. Lett. B}
  {\bfseries 191} (1987) 442--448}. [Erratum: Phys.Lett.B 448, 320--320
  (1999)].

\bibitem{Hewett:1987yg}
J.~Hewett and S.~Pakvasa, \textit{{Leptoquark Production in Hadron Colliders}},
  \href{https://doi.org/10.1103/PhysRevD.37.3165}{\textit{Phys. Rev. D}
  {\bfseries 37} (1988) 3165}.

\bibitem{Hewett:1987bh}
J.~Hewett and T.~Rizzo, \textit{{Leptoquark Signals at $e^+ e^-$ Colliders}},
  \href{https://doi.org/10.1103/PhysRevD.36.3367}{\textit{Phys. Rev. D}
  {\bfseries 36} (1987) 3367}.

\bibitem{Cuypers:1995ax}
F.~Cuypers, \textit{{Leptoquark production in $e^-\gamma$ scattering}},
  \href{https://doi.org/10.1016/0550-3213(96)00270-2}{\textit{Nucl. Phys. B}
  {\bfseries 474} (1996) 57--71},
  [\href{https://arxiv.org/abs/hep-ph/9508397}{{\ttfamily hep-ph/9508397}}].

\bibitem{Blumlein:1996qp}
J.~Blumlein, E.~Boos and A.~Kryukov, \textit{{Leptoquark pair production in
  hadronic interactions}},
  \href{https://doi.org/10.1007/s002880050538}{\textit{Z. Phys. C} {\bfseries
  76} (1997) 137--153}, [\href{https://arxiv.org/abs/hep-ph/9610408}{{\ttfamily
  hep-ph/9610408}}].

\bibitem{Belyaev}
A.~Belyaev, C.~Leroy, R.~Mehdiyev and A.~Pukhov, \textit{{Leptoquark single and
  pair production at LHC with CalcHEP/CompHEP in the complete model}},
  \href{https://doi.org/10.1088/1126-6708/2005/09/005}{\textit{JHEP} {\bfseries
  09} (2005) 005}, [\href{https://arxiv.org/abs/hep-ph/0502067}{{\ttfamily
  hep-ph/0502067}}].

\bibitem{Kramer:1997hh}
M.~Kramer, T.~Plehn, M.~Spira and P.~Zerwas, \textit{{Pair production of scalar
  leptoquarks at the Tevatron}},
  \href{https://doi.org/10.1103/PhysRevLett.79.341}{\textit{Phys. Rev. Lett.}
  {\bfseries 79} (1997) 341--344},
  [\href{https://arxiv.org/abs/hep-ph/9704322}{{\ttfamily hep-ph/9704322}}].

\bibitem{Plehn:1997az}
T.~Plehn, H.~Spiesberger, M.~Spira and P.~Zerwas, \textit{{Formation and decay
  of scalar leptoquarks/squarks in ep collisions}},
  \href{https://doi.org/10.1007/s002880050426}{\textit{Z. Phys. C} {\bfseries
  74} (1997) 611--614}, [\href{https://arxiv.org/abs/hep-ph/9703433}{{\ttfamily
  hep-ph/9703433}}].

\bibitem{Eboli:1997fb}
O.~J.~P. Eboli, R.~Zukanovich~Funchal and T.~L. Lungov, \textit{{Signal and
  backgrounds for leptoquarks at the CERN LHC}},
  \href{https://doi.org/10.1103/PhysRevD.57.1715}{\textit{Phys. Rev. D}
  {\bfseries 57} (1998) 1715--1729},
  [\href{https://arxiv.org/abs/hep-ph/9709319}{{\ttfamily hep-ph/9709319}}].

\bibitem{Kramer:2004df}
M.~Kramer, T.~Plehn, M.~Spira and P.~M. Zerwas, \textit{{Pair production of
  scalar leptoquarks at the CERN LHC}},
  \href{https://doi.org/10.1103/PhysRevD.71.057503}{\textit{Phys. Rev. D}
  {\bfseries 71} (2005) 057503},
  [\href{https://arxiv.org/abs/hep-ph/0411038}{{\ttfamily hep-ph/0411038}}].

\bibitem{Hammett:2015sea}
J.~B. Hammett and D.~A. Ross, \textit{{NLO Leptoquark Production and Decay: The
  Narrow-Width Approximation and Beyond}},
  \href{https://doi.org/10.1007/JHEP07(2015)148}{\textit{JHEP} {\bfseries 07}
  (2015) 148}, [\href{https://arxiv.org/abs/1501.06719}{{\ttfamily
  1501.06719}}].

\bibitem{Mandal:2015vfa}
T.~Mandal, S.~Mitra and S.~Seth, \textit{{Single Productions of Colored
  Particles at the LHC: An Example with Scalar Leptoquarks}},
  \href{https://doi.org/10.1007/JHEP07(2015)028}{\textit{JHEP} {\bfseries 07}
  (2015) 028}, [\href{https://arxiv.org/abs/1503.04689}{{\ttfamily
  1503.04689}}].

\bibitem{Asadi:2021gah}
P.~Asadi, R.~Capdevilla, C.~Cesarotti and S.~Homiller, \textit{{Searching for
  Leptoquarks at Future Muon Colliders}},
  \href{https://arxiv.org/abs/2104.05720}{{\ttfamily 2104.05720}}.

\bibitem{Bandyopadhyay:2021pld}
P.~Bandyopadhyay, A.~Karan and R.~Mandal, \textit{{Distinguishing signatures of
  scalar leptoquarks at hadron and muon colliders}},
  \href{https://arxiv.org/abs/2108.06506}{{\ttfamily 2108.06506}}.

\bibitem{Bandyopadhyay:2020jez}
P.~Bandyopadhyay, S.~Dutta and A.~Karan, \textit{{Zeros of amplitude in the
  associated production of photon and leptoquark at $e-p$ collider}},
  \href{https://doi.org/10.1140/epjc/s10052-021-09090-z}{\textit{Eur. Phys. J.
  C} {\bfseries 81} (2021) 315},
  [\href{https://arxiv.org/abs/2012.13644}{{\ttfamily 2012.13644}}].

\bibitem{Bandyopadhyay:2020klr}
P.~Bandyopadhyay, S.~Dutta and A.~Karan, \textit{{Investigating the Production
  of Leptoquarks by Means of Zeros of Amplitude at Photon Electron Collider}},
  \href{https://doi.org/10.1140/epjc/s10052-020-8083-7}{\textit{Eur. Phys. J.
  C} {\bfseries 80} (2020) 573},
  [\href{https://arxiv.org/abs/2003.11751}{{\ttfamily 2003.11751}}].

\bibitem{Bandyopadhyay:2020wfv}
P.~Bandyopadhyay, S.~Dutta, M.~Jakkapu and A.~Karan, \textit{{Distinguishing
  Leptoquarks at the LHC/FCC}},
  \href{https://doi.org/10.1016/j.nuclphysb.2021.115524}{\textit{Nucl. Phys. B}
  {\bfseries 971} (2021) 115524},
  [\href{https://arxiv.org/abs/2007.12997}{{\ttfamily 2007.12997}}].

\bibitem{Dutta:2021wid}
S.~Dutta, P.~Bandyopadhyay and A.~Karan, \textit{{Distinguishing Different BSM
  Signatures at Present and Future Colliders}},
  \href{https://arxiv.org/abs/2105.00893}{{\ttfamily 2105.00893}}.

\bibitem{Behrend:1986jz}
{\scshape CELLO} collaboration, H.~Behrend et~al., \textit{{Search for Light
  Leptoquark Bosons}},
  \href{https://doi.org/10.1016/0370-2693(86)91410-3}{\textit{Phys. Lett. B}
  {\bfseries 178} (1986) 452--456}. [Addendum: Phys.Lett.B 184, 417 (1987)].

\bibitem{Bartel:1987de}
{\scshape JADE} collaboration, W.~Bartel et~al., \textit{{Search for
  Leptoquarks and Other New Particles With Lepton - Hadron Signature in $e^+
  e^-$ Interactions}}, \href{https://doi.org/10.1007/BF01556160}{\textit{Z.
  Phys. C} {\bfseries 36} (1987) 15}.

\bibitem{Kim:1989qz}
{\scshape AMY} collaboration, G.~Kim et~al., \textit{{A search for leptoquark
  and colored lepton pair production in $e^+e^-$ annihilations at TRISTAN}},
  \href{https://doi.org/10.1016/0370-2693(90)90442-9}{\textit{Phys. Lett. B}
  {\bfseries 240} (1990) 243--249}.

\bibitem{Abreu:1998fw}
{\scshape DELPHI} collaboration, P.~Abreu et~al., \textit{{Search for
  Leptoquarks and FCNC in $e^{+} e^{-}$ annihilations at $\sqrt {s}$ = 183
  GeV}}, \href{https://doi.org/10.1016/S0370-2693(98)01525-1}{\textit{Phys.
  Lett. B} {\bfseries 446} (1999) 62--74},
  [\href{https://arxiv.org/abs/hep-ex/9903072}{{\ttfamily hep-ex/9903072}}].

\bibitem{Collaboration:2011qaa}
{\scshape H1} collaboration, F.~Aaron et~al., \textit{{Search for first
  generation leptoquarks in $ep$ collisions at HERA}},
  \href{https://doi.org/10.1016/j.physletb.2011.09.017}{\textit{Phys. Lett. B}
  {\bfseries 704} (2011) 388--396},
  [\href{https://arxiv.org/abs/1107.3716}{{\ttfamily 1107.3716}}].

\bibitem{Abramowicz:2019uti}
{\scshape ZEUS} collaboration, H.~Abramowicz et~al., \textit{{Limits on contact
  interactions and leptoquarks at HERA}},
  \href{https://doi.org/10.1103/PhysRevD.99.092006}{\textit{Phys. Rev. D}
  {\bfseries 99} (2019) 092006},
  [\href{https://arxiv.org/abs/1902.03048}{{\ttfamily 1902.03048}}].

\bibitem{Alitti:1991dn}
{\scshape UA2} collaboration, J.~Alitti et~al., \textit{{A Search for scalar
  leptoquarks at the CERN $\bar{p}p$ collider}},
  \href{https://doi.org/10.1016/0370-2693(92)92024-B}{\textit{Phys. Lett. B}
  {\bfseries 274} (1992) 507--512}.

\bibitem{Abazov:2011qj}
{\scshape D0} collaboration, V.~M. Abazov et~al., \textit{{Search for first
  generation leptoquark pair production in the electron + missing energy + jets
  final state}},
  \href{https://doi.org/10.1103/PhysRevD.84.071104}{\textit{Phys. Rev. D}
  {\bfseries 84} (2011) 071104},
  [\href{https://arxiv.org/abs/1107.1849}{{\ttfamily 1107.1849}}].

\bibitem{Aaltonen:2007rb}
{\scshape CDF} collaboration, T.~Aaltonen et~al., \textit{{Search for Third
  Generation Vector Leptoquarks in $p \bar{p}$ Collisions at $\sqrt{s}$ = 1.96
  TeV}}, \href{https://doi.org/10.1103/PhysRevD.77.091105}{\textit{Phys. Rev.
  D} {\bfseries 77} (2008) 091105},
  [\href{https://arxiv.org/abs/0706.2832}{{\ttfamily 0706.2832}}].

\bibitem{Aad:2020iuy}
{\scshape ATLAS} collaboration, G.~Aad et~al., \textit{{Search for pairs of
  scalar leptoquarks decaying into quarks and electrons or muons in $ \sqrt{s}
  $ = 13 TeV $pp$ collisions with the ATLAS detector}},
  \href{https://doi.org/10.1007/JHEP10(2020)112}{\textit{JHEP} {\bfseries 10}
  (2020) 112}, [\href{https://arxiv.org/abs/2006.05872}{{\ttfamily
  2006.05872}}].

\bibitem{CMS:2020gru}
{\scshape CMS} collaboration, A.~M. Sirunyan et~al., \textit{{Search for singly
  and pair-produced leptoquarks coupling to third-generation fermions in
  proton-proton collisions at $\sqrt{s} =$ 13 TeV}},
  \href{https://doi.org/10.1016/j.physletb.2021.136446}{\textit{Phys. Lett. B}
  {\bfseries 819} (2021) 136446},
  [\href{https://arxiv.org/abs/2012.04178}{{\ttfamily 2012.04178}}].

\bibitem{Aad:2021rrh}
{\scshape ATLAS} collaboration, G.~Aad et~al., \textit{{Search for pair
  production of third-generation scalar leptoquarks decaying into a top quark
  and a \ensuremath{\tau}-lepton in pp collisions at $ \sqrt{s} $ = 13 TeV with
  the ATLAS detector}},
  \href{https://doi.org/10.1007/JHEP06(2021)179}{\textit{JHEP} {\bfseries 06}
  (2021) 179}, [\href{https://arxiv.org/abs/2101.11582}{{\ttfamily
  2101.11582}}].

\bibitem{CMS:2018qqq}
{\scshape CMS} collaboration, A.~M. Sirunyan et~al., \textit{{Constraints on
  models of scalar and vector leptoquarks decaying to a quark and a neutrino at
  $\sqrt{s}=$ 13 TeV}},
  \href{https://doi.org/10.1103/PhysRevD.98.032005}{\textit{Phys. Rev. D}
  {\bfseries 98} (2018) 032005},
  [\href{https://arxiv.org/abs/1805.10228}{{\ttfamily 1805.10228}}].

\bibitem{CMS:2020wzx}
{\scshape CMS} collaboration, A.~M. Sirunyan et~al., \textit{{Search for singly
  and pair-produced leptoquarks coupling to third-generation fermions in
  proton-proton collisions at s=13~TeV}},
  \href{https://doi.org/10.1016/j.physletb.2021.136446}{\textit{Phys. Lett. B}
  {\bfseries 819} (2021) 136446},
  [\href{https://arxiv.org/abs/2012.04178}{{\ttfamily 2012.04178}}].

\bibitem{Mandal:2019gff}
R.~Mandal and A.~Pich, \textit{{Constraints on scalar leptoquarks from lepton
  and kaon physics}},
  \href{https://doi.org/10.1007/JHEP12(2019)089}{\textit{JHEP} {\bfseries 12}
  (2019) 089}, [\href{https://arxiv.org/abs/1908.11155}{{\ttfamily
  1908.11155}}].

\bibitem{Davidson:1993qk}
S.~Davidson, D.~C. Bailey and B.~A. Campbell, \textit{{Model independent
  constraints on leptoquarks from rare processes}},
  \href{https://doi.org/10.1007/BF01552629}{\textit{Z. Phys. C} {\bfseries 61}
  (1994) 613--644}, [\href{https://arxiv.org/abs/hep-ph/9309310}{{\ttfamily
  hep-ph/9309310}}].

\bibitem{Isidori:2001bm}
G.~Isidori, G.~Ridolfi and A.~Strumia, \textit{{On the metastability of the
  standard model vacuum}},
  \href{https://doi.org/10.1016/S0550-3213(01)00302-9}{\textit{Nucl. Phys. B}
  {\bfseries 609} (2001) 387--409},
  [\href{https://arxiv.org/abs/hep-ph/0104016}{{\ttfamily hep-ph/0104016}}].

\bibitem{singletex}
M.~Gonderinger, H.~Lim and M.~J. Ramsey-Musolf, \textit{{Complex Scalar Singlet
  Dark Matter: Vacuum Stability and Phenomenology}},
  \href{https://doi.org/10.1103/PhysRevD.86.043511}{\textit{Phys. Rev. D}
  {\bfseries 86} (2012) 043511},
  [\href{https://arxiv.org/abs/1202.1316}{{\ttfamily 1202.1316}}].

\bibitem{Gonderinger:2009jp}
M.~Gonderinger, Y.~Li, H.~Patel and M.~J. Ramsey-Musolf, \textit{{Vacuum
  Stability, Perturbativity, and Scalar Singlet Dark Matter}},
  \href{https://doi.org/10.1007/JHEP01(2010)053}{\textit{JHEP} {\bfseries 01}
  (2010) 053}, [\href{https://arxiv.org/abs/0910.3167}{{\ttfamily 0910.3167}}].

\bibitem{Costa:2014qga}
R.~Costa, A.~P. Morais, M.~O.~P. Sampaio and R.~Santos, \textit{{Two-loop
  stability of a complex singlet extended Standard Model}},
  \href{https://doi.org/10.1103/PhysRevD.92.025024}{\textit{Phys. Rev. D}
  {\bfseries 92} (2015) 025024},
  [\href{https://arxiv.org/abs/1411.4048}{{\ttfamily 1411.4048}}].

\bibitem{Haba:2015rha}
N.~Haba and Y.~Yamaguchi, \textit{{Vacuum stability in the $U(1)_\chi$ extended
  model with vanishing scalar potential at the Planck scale}},
  \href{https://doi.org/10.1093/ptep/ptv121}{\textit{PTEP} {\bfseries 2015}
  (2015) 093B05}, [\href{https://arxiv.org/abs/1504.05669}{{\ttfamily
  1504.05669}}].

\bibitem{Guo:2010hq}
W.-L. Guo and Y.-L. Wu, \textit{{The Real singlet scalar dark matter model}},
  \href{https://doi.org/10.1007/JHEP10(2010)083}{\textit{JHEP} {\bfseries 10}
  (2010) 083}, [\href{https://arxiv.org/abs/1006.2518}{{\ttfamily 1006.2518}}].

\bibitem{Barger:2008jx}
V.~Barger, P.~Langacker, M.~McCaskey, M.~Ramsey-Musolf and G.~Shaughnessy,
  \textit{{Complex Singlet Extension of the Standard Model}},
  \href{https://doi.org/10.1103/PhysRevD.79.015018}{\textit{Phys. Rev. D}
  {\bfseries 79} (2009) 015018},
  [\href{https://arxiv.org/abs/0811.0393}{{\ttfamily 0811.0393}}].

\bibitem{Khan:2014kba}
N.~Khan and S.~Rakshit, \textit{{Study of electroweak vacuum metastability with
  a singlet scalar dark matter}},
  \href{https://doi.org/10.1103/PhysRevD.90.113008}{\textit{Phys. Rev. D}
  {\bfseries 90} (2014) 113008},
  [\href{https://arxiv.org/abs/1407.6015}{{\ttfamily 1407.6015}}].

\bibitem{Baek:2012uj}
S.~Baek, P.~Ko, W.-I. Park and E.~Senaha, \textit{{Vacuum structure and
  stability of a singlet fermion dark matter model with a singlet scalar
  messenger}}, \href{https://doi.org/10.1007/JHEP11(2012)116}{\textit{JHEP}
  {\bfseries 11} (2012) 116},
  [\href{https://arxiv.org/abs/1209.4163}{{\ttfamily 1209.4163}}].

\bibitem{HiggsDM1}
P.~Bandyopadhyay, E.~J. Chun, R.~Mandal and F.~S. Queiroz,
  \textit{{Scrutinizing Right-Handed Neutrino Portal Dark Matter With Yukawa
  Effect}}, \href{https://doi.org/10.1016/j.physletb.2018.12.003}{\textit{Phys.
  Lett. B} {\bfseries 788} (2019) 530--534},
  [\href{https://arxiv.org/abs/1807.05122}{{\ttfamily 1807.05122}}].

\bibitem{BLscalar}
P.~Bandyopadhyay, E.~J. Chun and R.~Mandal, \textit{{Implications of
  right-handed neutrinos in $B-L$ extended standard model with scalar dark
  matter}}, \href{https://doi.org/10.1103/PhysRevD.97.015001}{\textit{Phys.
  Rev. D} {\bfseries 97} (2018) 015001},
  [\href{https://arxiv.org/abs/1707.00874}{{\ttfamily 1707.00874}}].

\bibitem{2HDMs}
N.~Chakrabarty and B.~Mukhopadhyaya, \textit{{High-scale validity of a two
  Higgs doublet scenario: metastability included}},
  \href{https://doi.org/10.1140/epjc/s10052-017-4705-0}{\textit{Eur. Phys. J.
  C} {\bfseries 77} (2017) 153},
  [\href{https://arxiv.org/abs/1603.05883}{{\ttfamily 1603.05883}}].

\bibitem{Chakrabarty:2015yia}
N.~Chakrabarty, D.~K. Ghosh, B.~Mukhopadhyaya and I.~Saha, \textit{{Dark
  matter, neutrino masses and high scale validity of an inert Higgs doublet
  model}}, \href{https://doi.org/10.1103/PhysRevD.92.015002}{\textit{Phys. Rev.
  D} {\bfseries 92} (2015) 015002},
  [\href{https://arxiv.org/abs/1501.03700}{{\ttfamily 1501.03700}}].

\bibitem{Swiezewska:2015paa}
B.~Swiezewska, \textit{{Inert scalars and vacuum metastability around the
  electroweak scale}},
  \href{https://doi.org/10.1007/JHEP07(2015)118}{\textit{JHEP} {\bfseries 07}
  (2015) 118}, [\href{https://arxiv.org/abs/1503.07078}{{\ttfamily
  1503.07078}}].

\bibitem{Gopalakrishna:2015wwa}
S.~Gopalakrishna, T.~S. Mukherjee and S.~Sadhukhan, \textit{{Extra neutral
  scalars with vectorlike fermions at the LHC}},
  \href{https://doi.org/10.1103/PhysRevD.93.055004}{\textit{Phys. Rev. D}
  {\bfseries 93} (2016) 055004},
  [\href{https://arxiv.org/abs/1504.01074}{{\ttfamily 1504.01074}}].

\bibitem{Honorez:2010re}
L.~Lopez~Honorez and C.~E. Yaguna, \textit{{The inert doublet model of dark
  matter revisited}},
  \href{https://doi.org/10.1007/JHEP09(2010)046}{\textit{JHEP} {\bfseries 09}
  (2010) 046}, [\href{https://arxiv.org/abs/1003.3125}{{\ttfamily 1003.3125}}].

\bibitem{HiggsDM}
P.~Bandyopadhyay, E.~J. Chun and R.~Mandal, \textit{{Scalar Dark Matter in
  Leptophilic Two-Higgs-Doublet Model}},
  \href{https://doi.org/10.1016/j.physletb.2018.01.071}{\textit{Phys. Lett. B}
  {\bfseries 779} (2018) 201--205},
  [\href{https://arxiv.org/abs/1709.08581}{{\ttfamily 1709.08581}}].

\bibitem{khan1}
N.~Khan and S.~Rakshit, \textit{{Constraints on inert dark matter from the
  metastability of the electroweak vacuum}},
  \href{https://doi.org/10.1103/PhysRevD.92.055006}{\textit{Phys. Rev. D}
  {\bfseries 92} (2015) 055006},
  [\href{https://arxiv.org/abs/1503.03085}{{\ttfamily 1503.03085}}].

\bibitem{2HDMpheno}
A.~Datta, N.~Ganguly, N.~Khan and S.~Rakshit, \textit{{Exploring collider
  signatures of the inert Higgs doublet model}},
  \href{https://doi.org/10.1103/PhysRevD.95.015017}{\textit{Phys. Rev. D}
  {\bfseries 95} (2017) 015017},
  [\href{https://arxiv.org/abs/1610.00648}{{\ttfamily 1610.00648}}].

\bibitem{Tripletex}
S.~Yaser~Ayazi and S.~M. Firouzabadi, \textit{{Footprint of Triplet Scalar Dark
  Matter in Direct, Indirect Search and Invisible Higgs Decay}},
  \href{https://doi.org/10.1080/23311940.2015.1047559}{\textit{Cogent Phys.}
  {\bfseries 2} (2015) 1047559},
  [\href{https://arxiv.org/abs/1501.06176}{{\ttfamily 1501.06176}}].

\bibitem{Khan:2016sxm}
N.~Khan, \textit{{Exploring the hyperchargeless Higgs triplet model up to the
  Planck scale}},
  \href{https://doi.org/10.1140/epjc/s10052-018-5766-4}{\textit{Eur. Phys. J.
  C} {\bfseries 78} (2018) 341},
  [\href{https://arxiv.org/abs/1610.03178}{{\ttfamily 1610.03178}}].

\bibitem{exwfermion}
C.~Coriano, L.~Delle~Rose and C.~Marzo, \textit{{Vacuum Stability in U(1)-Prime
  Extensions of the Standard Model with TeV Scale Right Handed Neutrinos}},
  \href{https://doi.org/10.1016/j.physletb.2014.09.001}{\textit{Phys. Lett. B}
  {\bfseries 738} (2014) 13--19},
  [\href{https://arxiv.org/abs/1407.8539}{{\ttfamily 1407.8539}}].

\bibitem{Coriano:2015sea}
C.~Coriano, L.~Delle~Rose and C.~Marzo, \textit{{Constraints on abelian
  extensions of the Standard Model from two-loop vacuum stability and
  $U(1)_{B-L}$}}, \href{https://doi.org/10.1007/JHEP02(2016)135}{\textit{JHEP}
  {\bfseries 02} (2016) 135},
  [\href{https://arxiv.org/abs/1510.02379}{{\ttfamily 1510.02379}}].

\bibitem{DelleRose:2015bms}
L.~Delle~Rose, C.~Marzo and A.~Urbano, \textit{{On the stability of the
  electroweak vacuum in the presence of low-scale seesaw models}},
  \href{https://doi.org/10.1007/JHEP12(2015)050}{\textit{JHEP} {\bfseries 12}
  (2015) 050}, [\href{https://arxiv.org/abs/1506.03360}{{\ttfamily
  1506.03360}}].

\bibitem{Jangid:2020dqh}
S.~Jangid, P.~Bandyopadhyay, P.~S. Bhupal~Dev and A.~Kumar, \textit{{Vacuum
  stability in inert higgs doublet model with right-handed neutrinos}},
  \href{https://doi.org/10.1007/JHEP08(2020)154}{\textit{JHEP} {\bfseries 08}
  (2020) 154}, [\href{https://arxiv.org/abs/2001.01764}{{\ttfamily
  2001.01764}}].

\bibitem{Garg:2017iva}
I.~Garg, S.~Goswami, K.~N. Vishnudath and N.~Khan, \textit{{Electroweak vacuum
  stability in presence of singlet scalar dark matter in TeV scale seesaw
  models}}, \href{https://doi.org/10.1103/PhysRevD.96.055020}{\textit{Phys.
  Rev. D} {\bfseries 96} (2017) 055020},
  [\href{https://arxiv.org/abs/1706.08851}{{\ttfamily 1706.08851}}].

\bibitem{Bandyopadhyay:2020djh}
P.~Bandyopadhyay, S.~Jangid and M.~Mitra, \textit{{Scrutinizing Vacuum
  Stability in IDM with Type-III Inverse seesaw}},
  \href{https://doi.org/10.1007/JHEP02(2021)075}{\textit{JHEP} {\bfseries 02}
  (2021) 075}, [\href{https://arxiv.org/abs/2008.11956}{{\ttfamily
  2008.11956}}].

\bibitem{Bandyopadhyay:2016oif}
P.~Bandyopadhyay and R.~Mandal, \textit{{Vacuum stability in an extended
  standard model with a leptoquark}},
  \href{https://doi.org/10.1103/PhysRevD.95.035007}{\textit{Phys. Rev. D}
  {\bfseries 95} (2017) 035007},
  [\href{https://arxiv.org/abs/1609.03561}{{\ttfamily 1609.03561}}].

\bibitem{AristizabalSierra:2007nf}
D.~Aristizabal~Sierra, M.~Hirsch and S.~G. Kovalenko, \textit{{Leptoquarks:
  Neutrino masses and accelerator phenomenology}},
  \href{https://doi.org/10.1103/PhysRevD.77.055011}{\textit{Phys. Rev. D}
  {\bfseries 77} (2008) 055011},
  [\href{https://arxiv.org/abs/0710.5699}{{\ttfamily 0710.5699}}].

\bibitem{Dorsner:2017wwn}
I.~Dor\v{s}ner, S.~Fajfer and N.~Ko\v{s}nik, \textit{{Leptoquark mechanism of
  neutrino masses within the grand unification framework}},
  \href{https://doi.org/10.1140/epjc/s10052-017-4987-2}{\textit{Eur. Phys. J.
  C} {\bfseries 77} (2017) 417},
  [\href{https://arxiv.org/abs/1701.08322}{{\ttfamily 1701.08322}}].

\bibitem{Babu:2019mfe}
K.~S. Babu, P.~S.~B. Dev, S.~Jana and A.~Thapa, \textit{{Non-Standard
  Interactions in Radiative Neutrino Mass Models}},
  \href{https://doi.org/10.1007/JHEP03(2020)006}{\textit{JHEP} {\bfseries 03}
  (2020) 006}, [\href{https://arxiv.org/abs/1907.09498}{{\ttfamily
  1907.09498}}].

\bibitem{Pas:2015hca}
H.~P\"as and E.~Schumacher, \textit{{Common origin of $R_K$ and neutrino
  masses}}, \href{https://doi.org/10.1103/PhysRevD.92.114025}{\textit{Phys.
  Rev. D} {\bfseries 92} (2015) 114025},
  [\href{https://arxiv.org/abs/1510.08757}{{\ttfamily 1510.08757}}].

\bibitem{Chua:1999si}
C.-K. Chua, X.-G. He and W.-Y.~P. Hwang, \textit{{Neutrino mass induced
  radiatively by supersymmetric leptoquarks}},
  \href{https://doi.org/10.1016/S0370-2693(00)00325-7}{\textit{Phys. Lett. B}
  {\bfseries 479} (2000) 224--229},
  [\href{https://arxiv.org/abs/hep-ph/9905340}{{\ttfamily hep-ph/9905340}}].

\bibitem{Mahanta:1999xd}
U.~Mahanta, \textit{{Neutrino masses and mixing angles from leptoquark
  interactions}},
  \href{https://doi.org/10.1103/PhysRevD.62.073009}{\textit{Phys. Rev. D}
  {\bfseries 62} (2000) 073009},
  [\href{https://arxiv.org/abs/hep-ph/9909518}{{\ttfamily hep-ph/9909518}}].

\bibitem{Babu:2010vp}
K.~S. Babu and J.~Julio, \textit{{Two-Loop Neutrino Mass Generation through
  Leptoquarks}},
  \href{https://doi.org/10.1016/j.nuclphysb.2010.07.022}{\textit{Nucl. Phys. B}
  {\bfseries 841} (2010) 130--156},
  [\href{https://arxiv.org/abs/1006.1092}{{\ttfamily 1006.1092}}].

\bibitem{Arnold:2012sd}
J.~M. Arnold, B.~Fornal and M.~B. Wise, \textit{{Simplified models with baryon
  number violation but no proton decay}},
  \href{https://doi.org/10.1103/PhysRevD.87.075004}{\textit{Phys. Rev. D}
  {\bfseries 87} (2013) 075004},
  [\href{https://arxiv.org/abs/1212.4556}{{\ttfamily 1212.4556}}].

\bibitem{Kovalenko:2002eh}
S.~Kovalenko and I.~Schmidt, \textit{{Proton stability in leptoquark models}},
  \href{https://doi.org/10.1016/S0370-2693(03)00544-6}{\textit{Phys. Lett. B}
  {\bfseries 562} (2003) 104--108},
  [\href{https://arxiv.org/abs/hep-ph/0210187}{{\ttfamily hep-ph/0210187}}].

\bibitem{Barr:1989fi}
S.~M. Barr and E.~M. Freire, \textit{{$\epsilon^\prime / \epsilon$ in
  Leptoquark and Diquark Models of {CP} Violation}},
  \href{https://doi.org/10.1103/PhysRevD.41.2129}{\textit{Phys. Rev. D}
  {\bfseries 41} (1990) 2129}.

\bibitem{Nath:2006ut}
P.~Nath and P.~Fileviez~Perez, \textit{{Proton stability in grand unified
  theories, in strings and in branes}},
  \href{https://doi.org/10.1016/j.physrep.2007.02.010}{\textit{Phys. Rept.}
  {\bfseries 441} (2007) 191--317},
  [\href{https://arxiv.org/abs/hep-ph/0601023}{{\ttfamily hep-ph/0601023}}].

\bibitem{Dorsner:2005fq}
I.~Dorsner and P.~Fileviez~Perez, \textit{{Unification without supersymmetry:
  Neutrino mass, proton decay and light leptoquarks}},
  \href{https://doi.org/10.1016/j.nuclphysb.2005.06.016}{\textit{Nucl. Phys. B}
  {\bfseries 723} (2005) 53--76},
  [\href{https://arxiv.org/abs/hep-ph/0504276}{{\ttfamily hep-ph/0504276}}].

\bibitem{Staub:2013tta}
F.~Staub, \textit{{SARAH 4 : A tool for (not only SUSY) model builders}},
  \href{https://doi.org/10.1016/j.cpc.2014.02.018}{\textit{Comput. Phys.
  Commun.} {\bfseries 185} (2014) 1773--1790},
  [\href{https://arxiv.org/abs/1309.7223}{{\ttfamily 1309.7223}}].

\bibitem{Staub:2015kfa}
F.~Staub, \textit{{Exploring new models in all detail with SARAH}},
  \href{https://doi.org/10.1155/2015/840780}{\textit{Adv. High Energy Phys.}
  {\bfseries 2015} (2015) 840780},
  [\href{https://arxiv.org/abs/1503.04200}{{\ttfamily 1503.04200}}].

\bibitem{tHooft:1972tcz}
G.~'t~Hooft and M.~J.~G. Veltman, \textit{{Regularization and Renormalization
  of Gauge Fields}},
  \href{https://doi.org/10.1016/0550-3213(72)90279-9}{\textit{Nucl. Phys. B}
  {\bfseries 44} (1972) 189--213}.

\bibitem{Machacek:1983fi}
M.~E. Machacek and M.~T. Vaughn, \textit{{Two Loop Renormalization Group
  Equations in a General Quantum Field Theory. 2. Yukawa Couplings}},
  \href{https://doi.org/10.1016/0550-3213(84)90533-9}{\textit{Nucl. Phys. B}
  {\bfseries 236} (1984) 221--232}.

\bibitem{Machacek:1983tz}
M.~E. Machacek and M.~T. Vaughn, \textit{{Two Loop Renormalization Group
  Equations in a General Quantum Field Theory. 1. Wave Function
  Renormalization}},
  \href{https://doi.org/10.1016/0550-3213(83)90610-7}{\textit{Nucl. Phys. B}
  {\bfseries 222} (1983) 83--103}.

\bibitem{Machacek:1984zw}
M.~E. Machacek and M.~T. Vaughn, \textit{{Two Loop Renormalization Group
  Equations in a General Quantum Field Theory. 3. Scalar Quartic Couplings}},
  \href{https://doi.org/10.1016/0550-3213(85)90040-9}{\textit{Nucl. Phys. B}
  {\bfseries 249} (1985) 70--92}.

\bibitem{Martin:1993zk}
S.~P. Martin and M.~T. Vaughn, \textit{{Two loop renormalization group
  equations for soft supersymmetry breaking couplings}},
  \href{https://doi.org/10.1103/PhysRevD.50.2282}{\textit{Phys. Rev. D}
  {\bfseries 50} (1994) 2282},
  [\href{https://arxiv.org/abs/hep-ph/9311340}{{\ttfamily hep-ph/9311340}}].
  [Erratum: Phys.Rev.D 78, 039903 (2008)].

\bibitem{Chetyrkin:2012rz}
K.~G. Chetyrkin and M.~F. Zoller, \textit{{Three-loop
  \textbackslash{}beta-functions for top-Yukawa and the Higgs self-interaction
  in the Standard Model}},
  \href{https://doi.org/10.1007/JHEP06(2012)033}{\textit{JHEP} {\bfseries 06}
  (2012) 033}, [\href{https://arxiv.org/abs/1205.2892}{{\ttfamily 1205.2892}}].

\bibitem{JuarezW:2017tmo}
S.~R. Ju\'arez~W., P.~Kielanowski, G.~Mora and A.~Bohm,
  \textit{{Renormalization group: new relations between the parameters of the
  Standard Model}},
  \href{https://doi.org/10.1016/j.physletb.2017.06.059}{\textit{Phys. Lett. B}
  {\bfseries 772} (2017) 294--299},
  [\href{https://arxiv.org/abs/1703.01523}{{\ttfamily 1703.01523}}].

\bibitem{Degrassi:2012ry}
G.~Degrassi, S.~Di~Vita, J.~Elias-Miro, J.~R. Espinosa, G.~F. Giudice,
  G.~Isidori et~al., \textit{{Higgs mass and vacuum stability in the Standard
  Model at NNLO}}, \href{https://doi.org/10.1007/JHEP08(2012)098}{\textit{JHEP}
  {\bfseries 08} (2012) 098},
  [\href{https://arxiv.org/abs/1205.6497}{{\ttfamily 1205.6497}}].

\bibitem{Buttazzo:2013uya}
D.~Buttazzo, G.~Degrassi, P.~P. Giardino, G.~F. Giudice, F.~Sala, A.~Salvio
  et~al., \textit{{Investigating the near-criticality of the Higgs boson}},
  \href{https://doi.org/10.1007/JHEP12(2013)089}{\textit{JHEP} {\bfseries 12}
  (2013) 089}, [\href{https://arxiv.org/abs/1307.3536}{{\ttfamily 1307.3536}}].

\bibitem{Coleman:1973jx}
S.~Coleman and E.~Weinberg, \textit{Radiative corrections as the origin of
  spontaneous symmetry breaking},
  \href{https://doi.org/10.1103/PhysRevD.7.1888}{\textit{Phys. Rev. D}
  {\bfseries 7} (Mar, 1973) 1888--1910}.

\bibitem{EliasMiro:2011aa}
J.~Elias-Miro, J.~R. Espinosa, G.~F. Giudice, G.~Isidori, A.~Riotto and
  A.~Strumia, \textit{{Higgs mass implications on the stability of the
  electroweak vacuum}},
  \href{https://doi.org/10.1016/j.physletb.2012.02.013}{\textit{Phys. Lett. B}
  {\bfseries 709} (2012) 222--228},
  [\href{https://arxiv.org/abs/1112.3022}{{\ttfamily 1112.3022}}].

\bibitem{Casas:1994us}
J.~A. Casas, J.~R. Espinosa, M.~Quiros and A.~Riotto, \textit{{The Lightest
  Higgs boson mass in the minimal supersymmetric standard model}},
  \href{https://doi.org/10.1016/0550-3213(94)00508-C}{\textit{Nucl. Phys. B}
  {\bfseries 436} (1995) 3--29},
  [\href{https://arxiv.org/abs/hep-ph/9407389}{{\ttfamily hep-ph/9407389}}].
  [Erratum: Nucl.Phys.B 439, 466--468 (1995)].

\bibitem{Masina:2012tz}
I.~Masina, \textit{{Higgs boson and top quark masses as tests of electroweak
  vacuum stability}},
  \href{https://doi.org/10.1103/PhysRevD.87.053001}{\textit{Phys. Rev. D}
  {\bfseries 87} (2013) 053001},
  [\href{https://arxiv.org/abs/1209.0393}{{\ttfamily 1209.0393}}].

\bibitem{10.1093/ptep/ptaa104}
{Particle Data Group}, P.~A. Zyla et~al., \textit{{Review of Particle
  Physics}}, \href{https://doi.org/10.1093/ptep/ptaa104}{\textit{Prog. Theor.
  Exp. Phys.} {\bfseries 2020} 083C01 (2020)}.

\bibitem{Carpentier:2010ue}
M.~Carpentier and S.~Davidson, \textit{{Constraints on two-lepton, two quark
  operators}},
  \href{https://doi.org/10.1140/epjc/s10052-010-1482-4}{\textit{Eur. Phys. J.
  C} {\bfseries 70} (2010) 1071--1090},
  [\href{https://arxiv.org/abs/1008.0280}{{\ttfamily 1008.0280}}].

\bibitem{Davidson:2018rqt}
S.~Davidson and A.~Saporta, \textit{{Constraints on $2\ell 2q$ operators from
  $\mu - e$ flavour-changing meson decays}},
  \href{https://doi.org/10.1103/PhysRevD.99.015032}{\textit{Phys. Rev. D}
  {\bfseries 99} (2019) 015032},
  [\href{https://arxiv.org/abs/1807.10288}{{\ttfamily 1807.10288}}].

\bibitem{Crivellin:2021egp}
A.~Crivellin, D.~M\"uller and L.~Schnell, \textit{{Combined constraints on
  first generation leptoquarks}},
  \href{https://doi.org/10.1103/PhysRevD.103.115023}{\textit{Phys. Rev. D}
  {\bfseries 103} (2021) 115023},
  [\href{https://arxiv.org/abs/2104.06417}{{\ttfamily 2104.06417}}].

\bibitem{Cheung:2001ip}
K.-m. Cheung, \textit{{Muon anomalous magnetic moment and leptoquark
  solutions}}, \href{https://doi.org/10.1103/PhysRevD.64.033001}{\textit{Phys.
  Rev. D} {\bfseries 64} (2001) 033001},
  [\href{https://arxiv.org/abs/hep-ph/0102238}{{\ttfamily hep-ph/0102238}}].

\bibitem{Shanker:1982nd}
O.~U. Shanker, \textit{{$\pi \ell$ 2, $K \ell$ 3 and $K^0 - \bar{K}$0
  Constraints on Leptoquarks and Supersymmetric Particles}},
  \href{https://doi.org/10.1016/0550-3213(82)90196-1}{\textit{Nucl. Phys. B}
  {\bfseries 204} (1982) 375--386}.

\bibitem{Dorsner:2019itg}
I.~Dor\v{s}ner, S.~Fajfer and O.~Sumensari, \textit{{Muon $g-2$ and scalar
  leptoquark mixing}},
  \href{https://doi.org/10.1007/JHEP06(2020)089}{\textit{JHEP} {\bfseries 06}
  (2020) 089}, [\href{https://arxiv.org/abs/1910.03877}{{\ttfamily
  1910.03877}}].

\bibitem{Schmaltz:2018nls}
M.~Schmaltz and Y.-M. Zhong, \textit{{The leptoquark Hunter\textquoteright{}s
  guide: large coupling}},
  \href{https://doi.org/10.1007/JHEP01(2019)132}{\textit{JHEP} {\bfseries 01}
  (2019) 132}, [\href{https://arxiv.org/abs/1810.10017}{{\ttfamily
  1810.10017}}].

\bibitem{Buonocore:2020erb}
L.~Buonocore, U.~Haisch, P.~Nason, F.~Tramontano and G.~Zanderighi,
  \textit{{Lepton-Quark Collisions at the Large Hadron Collider}},
  \href{https://doi.org/10.1103/PhysRevLett.125.231804}{\textit{Phys. Rev.
  Lett.} {\bfseries 125} (2020) 231804},
  [\href{https://arxiv.org/abs/2005.06475}{{\ttfamily 2005.06475}}].

\bibitem{CMS:2018sxp}
{\scshape CMS} collaboration, \textit{{Search for pair production of first
  generation scalar leptoquarks at $\sqrt{s}=13~\mathrm{TeV}$}},
  {\textit{CMS-PAS-EXO-17-009} (2018) }.

\bibitem{CMS:2018ipm}
{\scshape CMS} collaboration, A.~M. Sirunyan et~al., \textit{{Search for
  high-mass resonances in dilepton final states in proton-proton collisions at
  $\sqrt{s}=$ 13 TeV}},
  \href{https://doi.org/10.1007/JHEP06(2018)120}{\textit{JHEP} {\bfseries 06}
  (2018) 120}, [\href{https://arxiv.org/abs/1803.06292}{{\ttfamily
  1803.06292}}].

\bibitem{CMS:2015xzc}
{\scshape CMS} collaboration, V.~Khachatryan et~al., \textit{{Search for single
  production of scalar leptoquarks in proton-proton collisions at $\sqrt{s} =
  8$ $TeV$}}, \href{https://doi.org/10.1103/PhysRevD.93.032005}{\textit{Phys.
  Rev. D} {\bfseries 93} (2016) 032005},
  [\href{https://arxiv.org/abs/1509.03750}{{\ttfamily 1509.03750}}]. [Erratum:
  Phys.Rev.D 95, 039906 (2017)].

\bibitem{CMS:2017zts}
{\scshape CMS} collaboration, A.~M. Sirunyan et~al., \textit{{Search for new
  physics in final states with an energetic jet or a hadronically decaying $W$
  or $Z$ boson and transverse momentum imbalance at $\sqrt{s}=13\text{ }\text{
  }\mathrm{TeV}$}},
  \href{https://doi.org/10.1103/PhysRevD.97.092005}{\textit{Phys. Rev. D}
  {\bfseries 97} (2018) 092005},
  [\href{https://arxiv.org/abs/1712.02345}{{\ttfamily 1712.02345}}].

\bibitem{CMS:2018sgp}
{\scshape CMS} collaboration, \textit{{Search for pair production of second
  generation leptoquarks at sqrt(s)=13 TeV}}, {\textit{CMS-PAS-EXO-17-003}
  (2018) }.

\bibitem{CMS:2018eud}
{\scshape CMS} collaboration, \textit{{Search for heavy neutrinos and
  third-generation leptoquarks in final states with two hadronically decaying
  $\tau$ leptons and two jets in proton-proton collisions at $\sqrt{s} =
  13~\mathrm{TeV}$}}, {\textit{CMS-PAS-EXO-17-016} (2018) }.

\bibitem{CMS:2018hjx}
{\scshape CMS} collaboration, \textit{{Search for singly produced
  third-generation leptoquarks decaying to a $\tau$ lepton and a b quark in
  proton-proton collisions at $\sqrt{s}=13~\mathrm{TeV}$}},
  {\textit{CMS-PAS-EXO-17-029} (2018) }.

\bibitem{ATLAS:2017eiz}
{\scshape ATLAS} collaboration, M.~Aaboud et~al., \textit{{Search for
  additional heavy neutral Higgs and gauge bosons in the ditau final state
  produced in 36 fb$^{-1}$ of pp collisions at $ \sqrt{s}=13 $ TeV with the
  ATLAS detector}},
  \href{https://doi.org/10.1007/JHEP01(2018)055}{\textit{JHEP} {\bfseries 01}
  (2018) 055}, [\href{https://arxiv.org/abs/1709.07242}{{\ttfamily
  1709.07242}}].

\bibitem{Qweak:2018tjf}
{\scshape Qweak} collaboration, D.~Androi\'c et~al., \textit{{Precision
  measurement of the weak charge of the proton}},
  \href{https://doi.org/10.1038/s41586-018-0096-0}{\textit{Nature} {\bfseries
  557} (2018) 207--211}, [\href{https://arxiv.org/abs/1905.08283}{{\ttfamily
  1905.08283}}].

\end{thebibliography}\endgroup

\end{document}